\documentclass[twocolumn, tighten, times]{aastex62}
\usepackage{natbib}
\usepackage{epsfig}
\usepackage{graphics}

\newcommand{\ha}{$\rm{H}{\alpha}$}
\newcommand{\hb}{$\rm{H}{\beta}$}
\newcommand{\Fhb}{F$_{\rm{H}{\beta}}$}
\newcommand{\Fha}{F$_{\rm{H}{\alpha}}$}

\newcommand{\hd}{$\rm{H}{\delta}$}

\newcommand{\hdF}{$\rm{H}{\delta}_F$}
\newcommand{\hdFerr}{$\sigma_{\rm{EW(H}{\delta}_F)}$}

\newcommand{\hgF}{$\rm{H}{\gamma}_F$}
\newcommand{\uberhdF}{$\overline{\rm{H}_{\delta \gamma \beta}}$}
\newcommand{\uberhdFerr}{$\sigma_{{\rm EW(}\overline{\rm{H}_{\delta \gamma \beta}})}$}
\newcommand{\dfour}{$\rm{D_N}4000$}
\newcommand{\logmstar}{${\rm log(}M_*/M_{\odot})$}

\newcommand{\Msolar}{${\rm M}_{\odot}$}
\newcommand{\oiid}{$\rm{[OII]}(\lambda \lambda 3726, 3729)$}
\newcommand{\oiiid}{$\rm{[OIII]}(\lambda \lambda 4959, 5007)$}
\newcommand{\oid}{$\rm{[OI]}(\lambda \lambda 6300, 6364)$}
\newcommand{\siid}{$\rm{[SII]}{(\lambda \lambda 6716, 6731)}$}
\newcommand{\niid}{$\rm{[NII]}{(\lambda 6548, 6583)}$}

\newcommand{\oiii}{$\rm{[OIII]}{(\lambda 5007)}$}
\newcommand{\sii}{$\rm{[SII]}{(\lambda \lambda 6716, 6731)}$}
\newcommand{\nii}{$\rm{[NII]}{(\lambda 6583)}$}
\newcommand{\niishort}{$\rm{[NII]}$}

\newcommand{\kms}{$\,\rm{km\,s^{-1}}$}
\newcommand{\msolar}{$M_{\odot}$}
\newcommand{\rtwo}{$R_{200}$}

\newcommand{\logmtwo}{log($M_{200}/M_{\odot}$)}

\submitjournal{ApJ}
\shorttitle{The SAMI-GS: Transition galaxies in clusters}
\shortauthors{M.S. Owers {\it et al.}}
\begin{document}
\title{The SAMI Galaxy Survey: Quenching of star formation in clusters I. Transition galaxies.}
\author{Matt S. Owers}
\affiliation{Department of Physics and Astronomy, Macquarie University, NSW, 2109, Australia; E-mail: matt.owers@mq.edu.au}
\affiliation{Astronomy, Astrophysics and Astrophotonics Research Centre, Macquarie University, Sydney, NSW 2109, Australia}
\author{Michael J. Hudson}
\affiliation{Department of Physics \& Astronomy, University of Waterloo, Waterloo, ON, N2L 3G1, Canada}
\affiliation{Perimeter Institute for Theoretical Physics, 31 Caroline St. N., Waterloo, ON, N2L 2Y5, Canada}
\author{Kyle A. Oman}
\affiliation{Kapteyn Astronomical Institute, University of Groningen, Postbus 800, NL-9700 AV Groningen, the Netherlands}
\author{Joss Bland-Hawthorn}
\affiliation{Sydney Institute for Astronomy (SIfA), School of Physics, The University of Sydney, NSW, 2006, Australia}
\affiliation{ARC Centre of Excellence for All Sky Astrophysics in 3 Dimensions (ASTRO 3D)}
\author{S. Brough}
\affiliation{School of Physics, University of New South Wales, NSW 2052, Australia}
\affiliation{ARC Centre of Excellence for All Sky Astrophysics in 3 Dimensions (ASTRO 3D)}
\author{Julia J. Bryant}
\affiliation{Sydney Institute for Astronomy (SIfA), School of Physics, The University of Sydney, NSW, 2006, Australia}
\affiliation{Australian Astronomical Optics, AAO-USydney, School of Physics, University of Sydney, NSW 2006, Australia}
\affiliation{ARC Centre of Excellence for All Sky Astrophysics in 3 Dimensions (ASTRO 3D)}
\author{Luca Cortese}
\affiliation{International Centre for Radio Astronomy Research, University of Western Australia, 35 Stirling Highway, Crawley WA 6009, Australia}
\affiliation{ARC Centre of Excellence for All Sky Astrophysics in 3 Dimensions (ASTRO 3D)}
\author{Warrick J. Couch}
\affiliation{Centre for Astrophysics and Supercomputing, Swinburne University of Technology, VIC 3122, Melbourne, Australia}
\author{Scott M. Croom}
\affiliation{Sydney Institute for Astronomy (SIfA), School of Physics, The University of Sydney, NSW, 2006, Australia}
\affiliation{ARC Centre of Excellence for All Sky Astrophysics in 3 Dimensions (ASTRO 3D)}
\author{Jesse van de Sande}
\affiliation{Sydney Institute for Astronomy (SIfA), School of Physics, The University of Sydney, NSW, 2006, Australia}
\affiliation{ARC Centre of Excellence for All Sky Astrophysics in 3 Dimensions (ASTRO 3D)}
\author{Christoph Federrath}
\affiliation{Research School for Astronomy \& Astrophysics, Australian National University, Canberra, ACT 2611, Australia}
\author{Brent Groves}
\affiliation{Research School for Astronomy \& Astrophysics, Australian National University, Canberra, ACT 2611, Australia}
\affiliation{ARC Centre of Excellence for All Sky Astrophysics in 3 Dimensions (ASTRO 3D)}
\author{A.~M. Hopkins}
\affiliation{Australian Astronomical Optics, Macquarie University, 105 Delhi Rd, North Ryde, NSW 2113, Australia}
\author{J.~S. Lawrence}
\affiliation{Australian Astronomical Optics, Macquarie University, 105 Delhi Rd, North Ryde, NSW 2113, Australia}
\author{Nuria P.~F. Lorente}
\affiliation{Australian Astronomical Optics, Macquarie University, 105 Delhi Rd, North Ryde, NSW 2113, Australia}
\author{Richard M. McDermid}
\affiliation{Department of Physics and Astronomy, Macquarie University, NSW, 2109, Australia; E-mail: matt.owers@mq.edu.au}
\affiliation{Astronomy, Astrophysics and Astrophotonics Research Centre, Macquarie University, Sydney, NSW 2109, Australia}
\author{Anne M. Medling}
\altaffiliation{Hubble Fellow}
\affiliation{Ritter Astrophysical Research Center, University of Toledo, Toledo, OH 43606, USA}
\affiliation{Research School for Astronomy \& Astrophysics, Australian National University, Canberra, ACT 2611, Australia}
\author{Samuel N. Richards}
\affiliation{SOFIA Science Center, USRA, NASA Ames Research Center, Building N232, M/S 232-12, P.O. Box 1, Moffett Field, CA 94035-0001, USA}
\author{Nicholas Scott}
\affiliation{Sydney Institute for Astronomy (SIfA), School of Physics, The University of Sydney, NSW, 2006, Australia}
\affiliation{ARC Centre of Excellence for All Sky Astrophysics in 3 Dimensions (ASTRO 3D)}
\author{Dan S. Taranu}
\affiliation{Department of Astrophysical Sciences, Princeton University, 4 Ivy Lane, Princeton, NJ 08544, USA}
\affiliation{International Centre for Radio Astronomy Research, University of Western Australia, 35 Stirling Highway, Crawley WA 6009, Australia}
\author{Charlotte Welker}
\affiliation{International Centre for Radio Astronomy Research, University of Western Australia, 35 Stirling Highway, Crawley WA 6009, Australia}
\author{Sukyoung K. Yi}
\affiliation{Department of Astronomy and Yonsei University Observatory, Yonsei University, Seoul 03722, Republic of Korea}

\begin{abstract}
We use integral field spectroscopy from the SAMI Galaxy Survey to identify galaxies that show evidence for recent quenching of star formation. The galaxies exhibit strong Balmer absorption in the absence of  ongoing star formation in more than 10\% of their spectra within the SAMI field of view. These \hd-strong galaxies (HDSGs) are rare, making up only $\sim 2$\% (25/1220) of galaxies with stellar mass \logmstar$>10$. The HDSGs make up a significant fraction of non-passive cluster galaxies (15\%; 17/115) and a smaller fraction (2.0\%; 8/387) of the non-passive population in low-density environments. The majority (9/17) of cluster HDSGs show evidence for star formation at their centers, with the HDS regions found in the outer parts of the galaxy. Conversely, the \hd-strong signal is more evenly spread across the galaxy for the majority (6/8) of HDSGs in low-density environments, and is often associated with emission lines that are not due to star formation. We investigate the location of the HDSGs in the clusters, finding that they are exclusively within 0.6\rtwo\, of the cluster centre, and have a significantly higher velocity dispersion relative to the cluster population. Comparing their distribution in projected-phase-space to those derived from cosmological simulations indicates that the cluster HDSGs are consistent with an infalling population that have entered the central 0.5$r_{200, 3D}$ cluster region within the last $\sim 1$\,Gyr. In the 8/9 cluster HDSGs with central star formation, the extent of star formation is consistent with that expected of outside-in quenching by ram-pressure stripping. Our results indicate that the cluster HDSGs are currently being quenched by ram-pressure stripping on their first passage through the cluster. 
\end{abstract}

\keywords{surveys: SAMI --- galaxies: clusters: }

\section{Introduction}\label{intro}

One of the key problems in modern astrophysics is understanding how galaxies evolve, with the process likely governed by both internal and external influences that manifest as well-defined correlations between galaxy properties, stellar mass, and external environment. The sense of the correlations are clear: the fraction of galaxies that are bulge-dominated and devoid of star formation increases with stellar mass and local density, while the fraction of disk-dominated star-forming galaxies increases towards lower stellar mass and lower local density \citep{dressler1980, lewis2002, kauffmann2003b}. The relative importance of internal and external influences that act to stop the star formation in galaxies has been the subject of much study, with significant advances made possible due to large surveys such as the 2-degree Field Galaxy Redshift Survey \citep[2dFGRS;][]{colless2001} and Sloan Digital Sky Survey \citep[SDSS;][]{york2000}. The large sample sizes provided by these surveys have helped to separate the effects of mass and environment, and indicate that the environment plays an important role in quenching star formation in galaxies \citep{balogh2004,blanton2009,peng2010}. However, the dominant physical mechanisms responsible for the environment-driven quenching is still the subject of intense debate.

The impact of environmental quenching should reveal itself most prominently in the relatively hostile environments that exist in clusters of galaxies. There are a number of physical mechanisms that may act in clusters to both trigger and truncate star formation in infalling galaxies \citep[see][for a review]{boselli2006}. The processes can be divided into two categories: (i) interactions between the gas bound to the galaxy and the hot ($10^7-10^8$ K), rarefied ($10^{-3}$ particles cm$^{-3}$) intra-cluster medium (ICM), and (ii) gravitational interactions between either the galaxy and the cluster's gravitational potential, or interactions with other cluster galaxies. 

Interactions with the ICM, such as ram-pressure and viscous stripping \citep{gunn1972, nulsen1982} can easily remove the hot gas halo reservoir, thereby leading to a gradual decline in star formation \citep[strangulation;][]{larson1980, bekki2002, bekki2009}. Strong ram-pressure stripping can also remove the cold disk gas that fuels star formation, leading to quenching of star formation on short timescales \citep{roediger2006, bekki2014, boselli2014b, lee2017}. These hydrodynamical interactions are able to affect galaxy star formation with little impact on the structure of the old stellar population.

The effect of gravitational interactions, through either tides due to the cluster potential, other galaxies, or the combined effect \citep[harrassment;][]{moore1996}, can disrupt both the distribution of old stars and the gas in a cluster galaxy. This disruption may lead to transformations in the morphological, kinematical, star forming, and AGN properties of cluster galaxies \citep{byrd1990,bekki1999}. Galaxy-galaxy mergers are less frequent in the cores of clusters due to the high relative velocities of the galaxies \citep{ghigna1998}. However,  both simulations \citep{mcgee2009} and observations \citep{haines2018} indicate that $40-50$\% of galaxies observed in massive clusters are accreted through smaller, group-scale halos \citep[although the exact fraction depends on both halo and galaxy mass;][]{delucia2012}. Within these less-massive halos, the relative velocities between galaxies are lower, and pre-processing due to mergers and slower tidal interactions may be important \citep{cortese2006, bianconi2018}. Clearly there are many mechanisms by which the cluster environment can act to quench the star formation in a galaxy. The outstanding challenge is to disentangle the impacts each of these mechanisms have, individually, on the star formation of recently accreted galaxies, and to understand the timescales required for them to transition from star forming into quiescence.

Along these lines, it has been shown that the star formation rate (SFR) of star-forming galaxies within the central \rtwo\, of clusters is systematically lower than that of star-forming galaxies in the field \citep[e.g.,][]{gavazzi2002, gavazzi2006, koopmann2004b, haines2013}. Furthermore, the mean SFR of star-forming galaxies is seen to decline steadily from the outskirts to the centres of clusters \citep{vonderlinden2010, paccagnella2016, barsanti2018}. The slow decline in SFR with radius, coupled with kinematical evidence revealing that star-forming galaxies are consistent with being drawn from an infalling population \citep{colless1996,haines2015}, indicate that the cluster environment acts to quench the star formation of infalling galaxies on timescales longer than a few billion years. Similar conclusions were reached by \citet{taranu2014}, where it was found that in order to match the reddening of disk colours towards the cluster centre observed by \citet{hudson2010}, quenching must occur on relatively long $\sim 3$\,Gyr timescales after infall. These relatively long timescales favor mild processes such as strangulation as being responsible for quenching. 

However, other studies have found that the properties of cluster star-forming galaxies do not differ markedly from their field counterparts \citep{balogh2004, wetzel2012, muzzin2012}. This finding has led to the proposal of the ``delayed-then-rapid'' quenching scenario by \citet{wetzel2013}, where star-forming galaxies are unaffected by the environment for several Gyrs after becoming a satellite of a massive halo, before rapidly quenching on timescales shorter than $\sim 1$\,Gyr. The rapid phase of quenching is required to explain the strong bimodality observed in the SFR of cluster galaxies; there is a dearth of ``green valley'' galaxies with intermediate SFRs that are expected to exist if quenching acts on long timescales. A similar conclusion was reached by \citet{oman2016}, where it was found that all galaxies become quenched on first infall, shortly after first pericentric passage. 

Studies involving large, statistically significant samples of cluster galaxies allow constraints to be placed on overall quenching timescales. While these constraints help to understand which quenching mechanisms may be important, they do not allow for a detailed investigation of the processes at play. A complementary approach in this regard is to identify galaxies that show evidence for environmental perturbation, or {\it transition galaxies} that show evidence for very recent quenching, and target them with more detailed investigations. This approach has been successfully applied to galaxies in the nearby Virgo cluster where \citet{chung2007, chung_A2009} have characterized the HI morphology of a sample of spiral galaxies. They find that galaxies within 0.5\,Mpc of M87 have much smaller HI disks when compared with the stellar disks, while many galaxies at larger cluster-centric radii show one-sided tails that point away from the cluster core, concluding that these galaxies are being influenced by ram-pressure stripping on first infall. 

Using \ha\, imaging, \citet{koopmann2004} found that the distribution of star formation is truncated with respect to the stellar disk in the majority of the Virgo spirals that they studied. Very few star-forming galaxies in Virgo show an overall disk-wide reduction in SFR, indicating that ram-pressure stripping is more important than strangulation for Virgo spirals. \citet{crowl2008} used integral-field spectroscopy to follow up on a sample of ten truncated spirals selected from the \citet{koopmann2004} sample. They found that in {\it all} cases, the stellar populations in the regions just outside the radius  of truncation were young ($<500$\,Myr), indicating that the cessation of star formation following the stripping of gas occurs on short timescales. While these observations point to the importance of ram-pressure stripping in quenching star formation in Virgo, it must be emphasised that the centres of the truncated spirals in Virgo generally show normal SFRs. Crucially, transition galaxies analogous to those observed in Virgo but that exist in higher redshift clusters would be characterized as normal star-forming galaxies in single-fibre surveys. Therefore, key questions remain as to whether the Virgo-specific results are representative of the general cluster population.   

Post-starburst galaxies are amongst the best candidates for galaxies that are in the process of transitioning from star forming to quiescent systems. They were first identified in the spectroscopic surveys of intermediate redshift clusters  as galaxies that exhibit strong Balmer absorption and an absence of emission lines excited by ongoing star formation \citep{dressler1983, couch1987}. Spectrophotometric modelling indicates that very strong Balmer absorption, i.e., EW(H$\delta) < -5$\AA, can only occur by the rapid truncation of a starburst within the last $\sim 1$\,Gyr \citep{couch1987}. The weaker Balmer absorption seen in H$\delta$-strong galaxies ($-5\,$\AA$ < {\rm EW(H}\delta{\rm )} < -3\,$\AA)  is likely associated with recent truncation of normal star formation \citep[also referred to as post-star-forming galaxies;][]{couch1987,poggianti1999}. Their transitioning state has made H$\delta$-strong (HDS) galaxies attractive targets for attempting to identify the mechanism/s associated with the rapid quenching of star formation. 

Comparisons between the environments and properties of HDSGs indicate that field HDSGs are likely the result of galaxy-galaxy mergers \citep{zabludoff1996, blake2004, yang2008, pracy2009}, while ICM-related stripping mechanisms are thought to be responsible for the quenching of cluster HDSGs \citep{poggianti1999, tran2003, muzzin2014, paccagnella2017}. Most previous studies rely on single-fibre or single-slit spectroscopy to identify the HDS spectral signature. Therefore, in order for HDSGs to be identified, either the entire galaxy must be completely quenched of star formation, or the aperture through which the spectrum is measured must be coincident with a post-starforming region \citep[e.g., as seen in ][]{pracy2014}. Thus, galaxies that are currently being transformed in an outside-in manner, such as those seen in Virgo by \citet{crowl2008}, will not be identified in such surveys due to aperture effects. Further, the unresolved nature of the spectra mean that contributions from HDS and star-forming regions cannot be disentangled.

Since many environment-related mechanisms modulate star formation in a spatially non-uniform way, the spatially resolved information provided by Integral Field Spectroscopy (IFS) makes it a powerful tool for understanding environment-related quenching. To date, the predominantly monolithic IFS instruments have meant that the focus of these observations has been on following-up galaxies that are preselected because they show evidence for recent quenching or for being perturbed by the environment  \citep{pracy2005, merluzzi2013, fumagalli2014, fossati2016, merluzzi2016, poggianti2017, bellhouse2017, fritz2017, gullieuszik2017, fossati2018}. The advent of multi-IFS instruments such as the Sydney-AAO Multi-Object IFS \citep[SAMI;][]{bland2011,croom2012,bryant2014} has opened up a new era for galaxy surveys where resolved spectroscopy can be collected for large, unbiased samples of galaxies. 

Our aim in this paper is to use data from the SAMI Galaxy Survey \citep[hereafter SAMI-GS;][]{bryant2015} to identify galaxies that exhibit evidence for recent quenching in their spatially resolved spectroscopy, and to understand how the environment may be acting to quench the star formation in these galaxies. We build upon previous studies that used the SAMI-GS to investigate quenching and environment \citep[e.g.,][]{schaefer2017, medling2018, schaefer2018}  by both expanding the sample size and focussing on the cluster regions. Furthermore, we use the resolved spectroscopy to characterize both the ongoing star-forming distribution and to identify HDS regions associated with recent quenching. Critically, the SAMI-GS probes a broad range in environmental densities. The main portion of the survey targeted the equatorial GAMA regions \citep[Galaxy and Mass Assembly;][]{driver2011} that contain low- to intermediate-density environments, and added eight massive clusters \citep[][]{owers2017}, allowing us to extend the work of \citet{crowl2008} to a larger, more representative sample of clusters. Towards that aim, we have used resolved spectroscopic classification maps from a sample of 1220 SAMI galaxies with \logmstar$ > 10$ and spanning all environments to identify 26 galaxies where more than $10\%$ of their classifiable spaxels\footnote{here and throughout this paper, the term spaxel refers to the spatial element of the IFS data cube.} exhibit strong Balmer absorption, indicating recent quenching in those regions. We investigate the properties of these galaxies, focusing mainly on the HDSGs found in the cluster regions.

The outline of this paper is as follows: In Section~\ref{data}, we describe the SAMI-GS and ancillary data used in this paper, as well as the sample selection. In Section~\ref{line_strengths}, we describe our emission and absorption line measurements, Section~\ref{spec_class} describes the spectroscopic classification maps, and Section~\ref{gal_class} describes our classification of galaxies as passive, star forming or HDS. In Section~\ref{results}, we investigate the demographics of the HDSGs, paying particular attention to the environments of the cluster HDSGs, which we find are significantly different from those found in the GAMA regions, as well as being spatially and kinematically distinct from the passive and star-forming cluster galaxies. In Section~\ref{discussion}, we interpret our results, showing that the cluster HDSGs are consistent with a recently accreted population of star-forming galaxies that are being quenched from the outside-in due to the effects of ram-pressure stripping. Finally, in Section~\ref{summary}, we summarise our results and present our conclusions. Throughout this paper, we assume a standard $\Lambda$CDM cosmology, with $\Omega_m=0.3$, $\Omega_{\Lambda}=0.7$ and a Hubble-Lema{\^i}tre constant $H_0=70 {\rm km\, s}^{-1}\, {\rm Mpc}^{-1}$.

\section{Data and Sample Selection}\label{data}

In this Section, we describe the SAMI-GS data, the ancillary data used, and the selection of the SAMI-GS galaxies used in this paper.

\subsection{The SAMI Galaxy Survey}

The SAMI-GS was conducted with the Sydney-AAO Multi-obect Integral field spectrograph \citep[SAMI;][]{croom2012}, which was mounted at the prime focus of the 3.9m Anglo-Australian Telescope and provided a 1 degree diameter field of view. SAMI uses 13 fused fibre bundles \citep[Hexabundles;][]{bland2011, bryant2014} with a high (75\%) fill factor. Each bundle contains 61 fibres of $1.6\arcsec$ diameter resulting in each IFU having a diameter of $15\arcsec$. The IFUs, as well as 26 sky fibres, are plugged into pre-drilled plates using magnetic connectors. SAMI fibres are fed to the double-beam AAOmega spectrograph \citep{saunders2004,sharp2006}. AAOmega allows a range of different resolutions and wavelength ranges. For the SAMI Galaxy survey we used the 570V grating at $3700-5700\,$\AA\, giving a central resolution of $R=$1812 in the blue-arm ($\sigma=70\,$\kms; FWHM$=2.65$\,\AA), and the 1000R grating from $6250-7350\,$\AA\, giving a central resolution of $R=$4263 in the red-arm \citep[$\sigma=30\,$\kms, FWHM$=1.61$\,\AA;][]{VDS2016}. 

The SAMI-GS involved the observation of 3071 galaxies between 2013-2018 in the stellar mass range \logmstar\,$=8-12$\, and with redshift $0.004 < z \le 0.115$. The SAMI-GS galaxies are primarily drawn from the equatorial G09, G12, and G15 GAMA I regions \citep[2153 observed galaxies;][]{driver2011,liske2015}, and also include galaxies selected from regions containing eight massive clusters with virial masses in the range \logmtwo\,$=14.25-15.19$ \citep[918 observed galaxies;][]{owers2017}. The input catalogues for the GAMA and cluster regions targeted during the SAMI-GS are described in detail in \citet{bryant2015} and \citet{owers2017}. Briefly, the primary SAMI-GS targets in the GAMA regions are selected from a series of redshift bins with an increasing stellar mass limit in higher redshift bins. In the cluster regions, primary targets are selected using similar redshift-dependent stellar mass cuts, although a lower limit is set at \logmstar\,$=9.5$. Furthermore, primary targets in the cluster regions are constrained to have projected clustercentric distance $R<$\rtwo, and peculiar velocity $|v_{\rm pec}| < 3.5 \sigma_{200}$ with respect to the cluster redshift, where $\sigma_{200}$ is the cluster velocity dispersion measured within \rtwo. For both the GAMA and cluster regions, a number of secondary targets with relaxed selection criteria are also included in the observations. The secondary objects are excluded from the analysis presented in this paper.

The observing procedure is detailed in \citet{green2018}. Briefly, each observed field involves a series of 7-dither pointings designed to provide both complete coverage over the 15\arcsec\, diameter FOV for each hexabundle, and to reduce the impact on image quality of the 1.6\arcsec\, diameter fibre size, which undersamples the seeing point-spread function. Each dither pointing has a duration of 1800s, for a total 12600s exposure, and the 7-dither series is bookended by flat field and arc frames.  When possible, twilight flats are observed for the purpose of fibre tracing, throughput, and flat-fielding. In cases where twilight flats could not be observed, dome flats are used in their place. Each plate observes 12 galaxies and one calibration star that is used for telluric and flux calibration. The data were reduced using the SAMI {\sf PYTHON} package \citep{allen2014}, which incorporates the {\sf 2DFDR} package \citep{2dfdr2015}.  The reduced and calibrated data cubes are sampled on a regular spatial grid with $0.5\arcsec \times 0.5\arcsec$ spaxels, and the spectra have pixel scales of 1.03\,\AA\, and 0.56\,\AA\, for the blue- and red-arm spectra, respectively. The full end-to-end description of reducing the data from raw frames to fully calibrated data cubes is described elsewhere \citep{sharp2015, allen2015, green2018, scott2018}.

\subsection{Ancillary Data}

We make use of several existing data products during our analysis. For the GAMA portion of the survey, the stellar masses, $M_*$, are determined using the approximation of \citet{taylor2011} as outlined in \citet{bryant2015}, and use the aperture-matched $g-$ and $i-$band colours determined by \citet{hill2011}. Structural parameters (effective radius, $r_e$, S{\'e}rsic index, $n_{ser}$, ellipticity, and position angle) for the GAMA regions are drawn from the S{\'e}rsic profile fitting of SDSS $r$-band images as described in \citet{kelvin2012}. In the cluster regions, the same stellar mass proxy described in \citet{bryant2015} is used to determine $M_*$, along with aperture-matched $g-$ and $i-$band magnitudes as described in \citet{owers2017}. We also make use of the cluster masses ($M_{200}$),  velocity dispersions ($\sigma_{200}$), cluster redshifts ($z_{clus}$), galaxy peculiar velocities ($v_{\rm pec}$), and overdensity radii ($R_{200}$) published in \citet{owers2017}.

\subsubsection{S{\'e}rsic fits for cluster galaxies}\label{cluster_sersic}

Structural parameters for the cluster galaxies were determined from S{\'e}rsic profile fitting using the {\sf PROFIT}\footnote{https://github.com/ICRAR/ProFit} code \citep{robotham2017}. The fitting was performed on $r$-band images taken from the SDSS \citep[DR9;][]{ahn2012} and VST/ATLAS \citep{shanks2015} surveys. The VST/ATLAS data were reprocessed as described in \citet{owers2017} using the {\sf ASTRO-WISE} pipeline as described in \citet{mcfarland2013} and \citet{dejong2015}. At the position of each SAMI target in the cluster input catalogues, a $400\arcsec \times 400\arcsec$ cutout image was generated in each of the available bands. Point-like objects were selected based on their position in the size-surface-brightness plane, and non-saturated point sources with magnitude $ 16 < r < 20$ were fitted with Moffat profiles using {\sf PROFIT}. The median of the best-fit parameter values is used to generate a PSF specific to each $r$-band cutout, and this PSF is used for convolution during the S{\'e}rsic profile fitting. 

Prior to fitting, we perform local sky subtraction on each cutout after aggressively masking detected sources. We use the {\sf PROFOUND}\footnote{https://github.com/asgr/ProFound} software package \citep{robotham2018} to generate a detection image from an inverse-variance-weighted stack of the $griz$-band images (or $gri$-band in the case of VST/ATLAS, where the $z$-band was not available). We then run {\sf PROFOUND} on the detection image to detect and characterise the shapes of sources in the field. The shape parameters derived with {\sf PROFOUND} (i.e., position, position-angle, and axial ratio) were used to generate a mask around each detected object as follows. We use the {\sf PROFIT} software to produce a S{\'e}rsic model for each object using the {\sf PROFOUND}-derived magnitude and shape parameters, and assuming S{\'e}rsic index $n_{ser}=4$, which is typical of the early-type galaxies found in the clusters. We use the model to mask all pixels within a constant surface brightness $\mu = 30$mag/arcsec$^2$ for each object. This aggressive masking ensures that the remaining pixels are not contaminated by the faint outer wings of galaxies. We then define a 10\arcsec$\times$10\arcsec\, grid and determine the local sky at each gridpoint using a box  with an adaptive size that is grown until the box contains $10000$ unmasked sky pixels. The sky and sky noise are determined from the distribution of values in the adaptive box. Masked object regions are then interpolated using inverse-distance-weighted means, and the gridded sky distribution is interpolated back to the full-resolution grid using bicubic interpolation. This final sky distribution is subtracted from the $r-$band image prior to fitting.

During S{\'e}rsic fitting, any galaxy for which R100 (the elliptical semi-major axis that contains 100\% of the flux as defined by {\sf PROFOUND}) overlaps with that of the primary galaxy of interest, and that also has an isophotal area greater than 5\% of the primary galaxy's isophotal area, is simultaneously fitted along with the primary galaxy. Stars and objects that do not meet these criteria are masked using the segmentation map derived with {\sf PROFOUND}. Initial input estimates for the S{\'e}rsic profile parameters are derived from the {\sf PROFOUND} outputs, and further optimized via the {\sf R optim} function using the ``L-BFGS-B'' algorithm. The final parameters are determined using the {\sf LaplaceDemon} package, where we run 10,000 Markov chain Monte Carlo iterations using the Componentwise Hit-And-Run method (CHARM). 

We checked our fit parameters for both internal and external consistency. Internal checks were performed on a subset of 143 SAMI galaxies in the cluster A85 that have both VST/ATLAS and SDSS imaging. We found that the distribution of the relative difference between the SDSS and VST/ATLAS position angle (PA) and ellipticity measurements were smaller than 1\%, with dispersion 3.3\% and 1\%, respectively, indicating very good agreement between the two imaging surveys for these two parameters. However, we found a systematic offset in $r_e$ and $n_{ser}$ of the order 3.5-4\%, with the SDSS values being larger on average than those derived using the VST/ATLAS data. We also found a larger scatter of 6\% and 10\% for the relative differences in the $r_e$ and $n_{ser}$ measurements. Dividing the 143 galaxies into those galaxies with $n_{ser}>2$ and $n_{ser}<2$, i.e., disk- and bulge-dominated galaxies, respectively, we found that the systematic offsets in the relative difference for both $r_e$ and $n_{ser}$ are due to differences in the bulge-dominated sample. These systematic offsets are likely due to the oversubtraction of the sky around these larger objects in the VST/ATLAS data, which leads to a steepening of the outer profile and, therefore, smaller $r_e$ and $n_{ser}$ values when compared with those derived from the SDSS imaging. These systematic offsets are small and do not affect our conclusions. External checks are performed for the SDSS fits by comparing our results for a sample of 557 galaxies matched to the single S{\'e}rsic fits from \citet{meert2015}. We found good agreement, with the relative differences between $r_e$, $n_{ser}$, axial ratio and PA differing by less than 2\%, and scatter smaller than 10\%.

\subsection{Sample selection}\label{sample_selection}

The sample of galaxies used in this paper is selected from the 2,526 SAMI-GS galaxies observed prior to September 2017 (internal team release version V0.10.1; 894 cluster and 1632 GAMA galaxies). 
In the cluster regions, we only consider the 714 primary target galaxies that are allocated as cluster members in \citet{owers2017}. In order to better match the stellar mass and redshift distributions of the cluster sample, the GAMA sample only includes the 791 primary targets that have $z<0.06$ and \logmstar\,$>9.5$. For galaxies with multiple observations, we use the data from the observation taken in the best seeing conditions. The completeness of the GAMA portion of the sample is lower than the cluster sample (61\% c.f. 87\%, respectively), although there is no apparent stellar mass bias in the completeness.

In addition to the selection described above, in our final sample we only include the 579 cluster and 649 GAMA galaxies with \logmstar\,$\geq 10$.
There are two reasons for this additional selection criterion: first, for the clusters in the survey with $z > 0.045$, galaxies with \logmstar$< 10$ were not observed as primary targets; below this cut-off only those galaxies in the blue cloud were included as secondary targets \citep{owers2017}. Removing objects with \logmstar$<10$\, therefore allows for a homogeneous selection both within the cluster sample, and across the GAMA and cluster samples. Second, we wish to perform resolved spectroscopic classification based on both absorption- and emission-line measurements (outlined below in Section~\ref{spec_class}). The absorption-line classification requires S/N(4100\AA)$>3$\,pix$^{-1}$ (see Sections~\ref{absorption_lines} and \ref{absorption_class}) so that we can reliably classify spaxels based on the strength of the Balmer line absorption. To perform the galaxy classifications outlined in Section~\ref{galaxy_class}, it is desirable  that more than 100 spaxels meet this S/N criteria. This 100 spaxel area corresponds to that contained within a circular region with diameter $\sim 5.6$\arcsec, which is substantially larger than the mean seeing FWHM=2.06\arcsec\, \citep{scott2018}. When comparing the GAMA and cluster samples we found that 57\% (77/135) of cluster galaxies with 9.5$\leq$\logmstar$<10$ had fewer than 100 spaxels that met the S/N criterion, whereas this was the case for only 14\% (20/142) of low-mass galaxies in the GAMA regions.  For \logmstar$\geq 10$, $\sim 86\%\, {\rm and}\, 90$\% of SAMI galaxies in the cluster and GAMA regions, respectively, have more than 100 S/N(4100\AA)$>3\,$pix$^{-1}$ spaxels. Therefore, selecting galaxies with \logmstar$\geq 10$ allows for relatively unbiased comparisons to be performed between the two samples. 

\begin{figure*}
\includegraphics[width=0.5\textwidth]{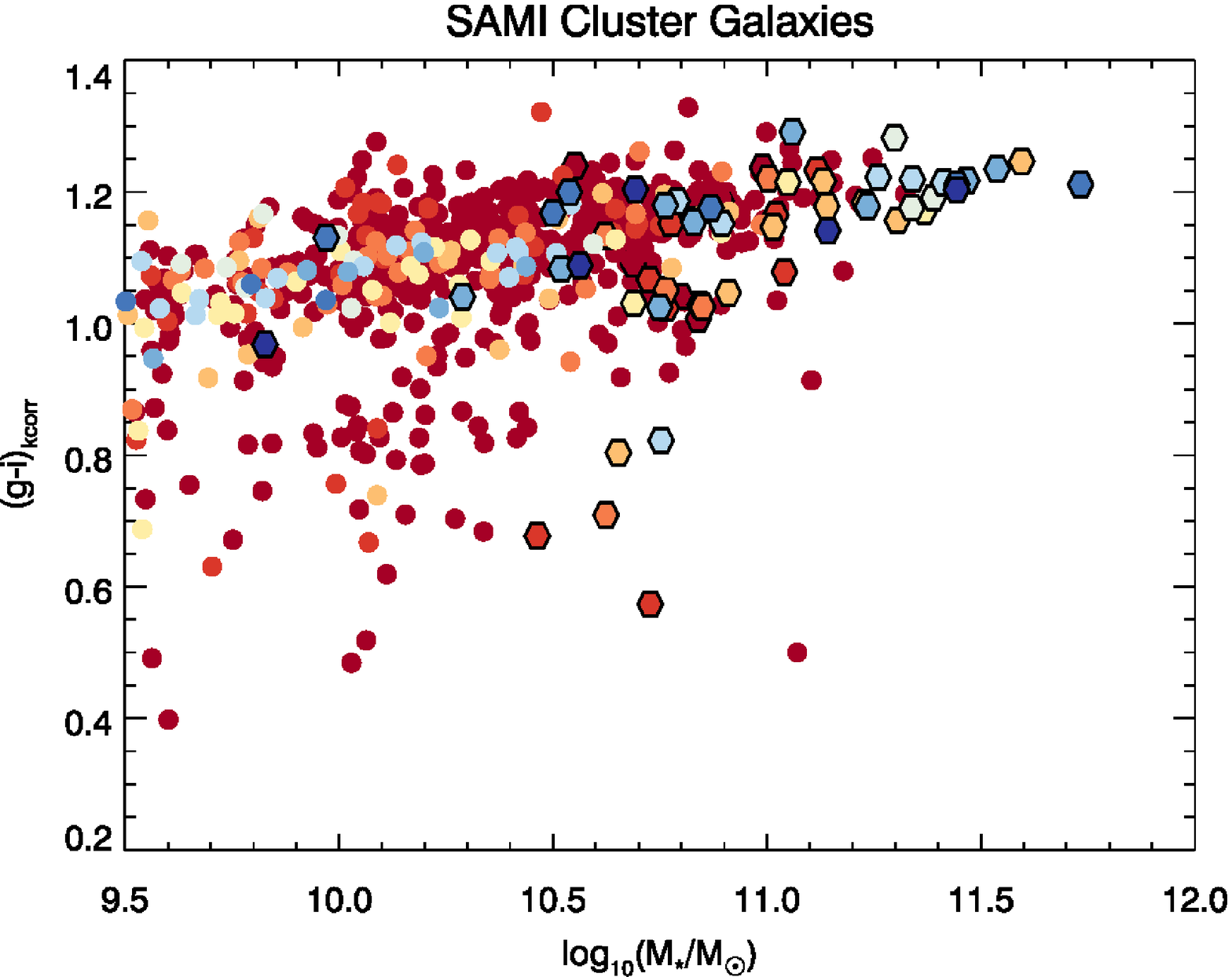}
\includegraphics[width=0.5\textwidth]{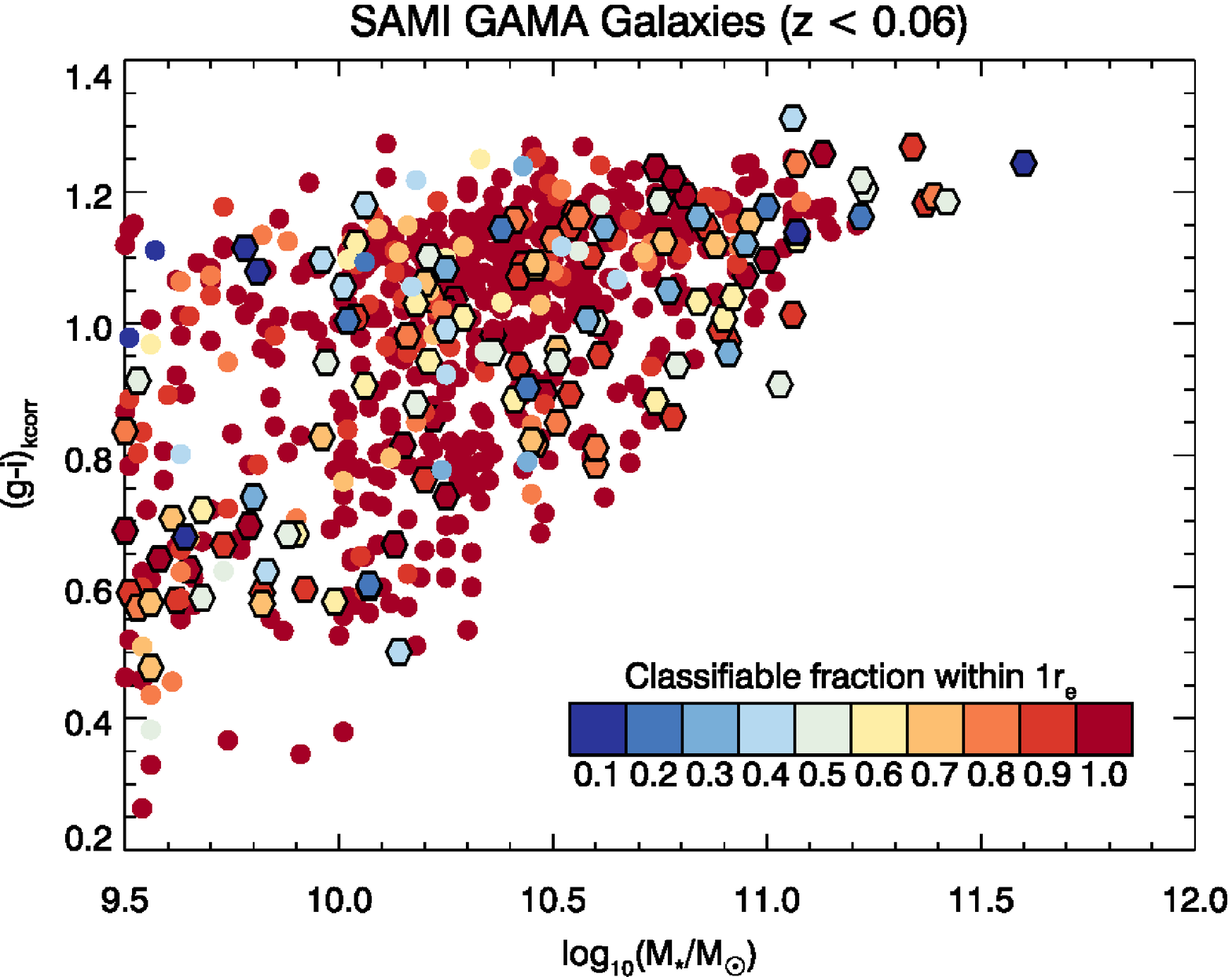}
\caption{Color-mass diagram for SAMI-GS galaxies in the cluster (left) and GAMA (right) region and that have $z<0.06$ and \logmstar$>9.5$. Each point is color-coded based on the fraction of the surface area within 1$r_e$ that is covered by classifiable spaxels (i.e., those with S/N(4100\AA)\,$>3$\,pix$^{-1}$). The color-bar shown in the right panel provides the key to convert between color and classifiable fraction. The galaxies that have an effective radius that is larger than the 7.5\arcsec\, SAMI hexabundle radius are plotted as hexagons.  \label{Frac_good}}
\end{figure*}

To investigate potential systematic biases in the spatial coverage of our spectral classification maps, we present Figure~\ref{Frac_good}, which shows the $(g-i)_{\rm kcorr}$ versus \logmstar\, color-mass plane for the galaxies in the cluster and GAMA regions (left and right panels, respectively). In Figure~\ref{Frac_good}, $(g-i)_{\rm kcorr}$ is the $k$-corrected color where the $k$-correction has been determined using the {\sf CALC\_KCOR} code\footnote{http://kcor.sai.msu.ru/getthecode/} from \citet{chilingarian2010}. Each point in Figure~\ref{Frac_good} is color-coded based on the fraction of the surface area contained within one effective radius for which there are spaxels with S/N(4100\AA)$>3\,$pix$^{-1}$, $f_{\rm class,r_e}$. A significant portion of the red-sequence cluster galaxies with 9.5$\leq$\logmstar$<10$ have a relatively low $f_{\rm class,r_e}$ when compared with blue cloud galaxies within the same \logmstar\, range. This systematic bias further justifies our decision to include only \logmstar$>10$ galaxies in our sample.  Within our sample of \logmstar$\geq 10$ galaxies, 86\% (87\%) of cluster (GAMA) galaxies have $f_{\rm class,r_e} \geq 0.7$. Of the galaxies that have  \logmstar$\geq 10$ and 
$f_{\rm class,r_e} < 0.7$, a significant fraction (48\% and 65\% in the cluster and GAMA regions, respectively) have an effective radius that is larger than the SAMI hexabundle size (i.e., they have $r_e > 7.5$\arcsec);  these galaxies are plotted as hexagons in Figure~\ref{Frac_good}. This effect is most prevalent at high masses (\logmstar$>11.2$), where almost all galaxies are affected. Aside from the systematic bias at large stellar masses, which equally affects both the cluster and GAMA samples, there do not appear to be any prominent biases in the $f_{\rm class,r_e}$ across the color-mass plane.

\section{Line strength measurements}\label{line_strengths}

The spectral classification scheme outlined in Section~\ref{spec_class} requires measurements of emission and absorption line strengths. In this Section we describe the procedure for defining the stellar continuum, and for measuring emission and absorption line fluxes and equivalent widths.

\subsection{Stellar continuum definition}\label{continuum_definition}

Accurate emission line flux measurements require that the stellar continuum be modelled and subtracted. This is particularly important for the Balmer lines, which can be significantly affected by underlying stellar absorption. The fidelity of the stellar continuum fit depends strongly on the S/N in the continuum of the spectrum. For accurate stellar continuum modelling, it is common to spatially bin spectra to reach a minimum S/N in the continuum \citep[e.g.][]{cappellari2003}. However, often the binning scheme used for continuum modelling is not suitable for emission lines, which can have good S/N in the unbinned data. For this reason, we employ a hybrid approach that uses a combination of binned and full spatial resolution data and is outlined below.

We use the penalised pixel fitting software \citep[pPXF;][]{cappellari2004,cappellari2017}, in combination with 73 Stellar Population Synthesis (SPS) templates drawn from the MILES \citep{vazdekis2010} and \citet{gonzalez2005} libraries, to fit the underlying stellar continuum for each spaxel. From the MILES SPS library, we select a subset of templates that contains four metallicities ([M/H]= -0.71, -0.40, 0.00, 0.22) and 13 logarithmically-spaced ages ranging from $0.0063-15.85$\,Gyrs. Following \citet{cidfernandes2013}, we also include the subset of \citet{gonzalez2005} SPS templates with metallicities [M/H] = -0.71, -0.40, 0.00 and ages $0.001-0.025$\,Gyr, which extends the MILES coverage to younger ages. The continuum for each spaxel is determined using the following multi-step process outlined below, and also in Figure~\ref{ppxf_eg}.  

\subsubsection{Refining template library using Voronoi binned data}\label{template_refine} 
\begin{figure*}
\includegraphics[width=0.99\textwidth, trim= 0 96 0 98, clip]{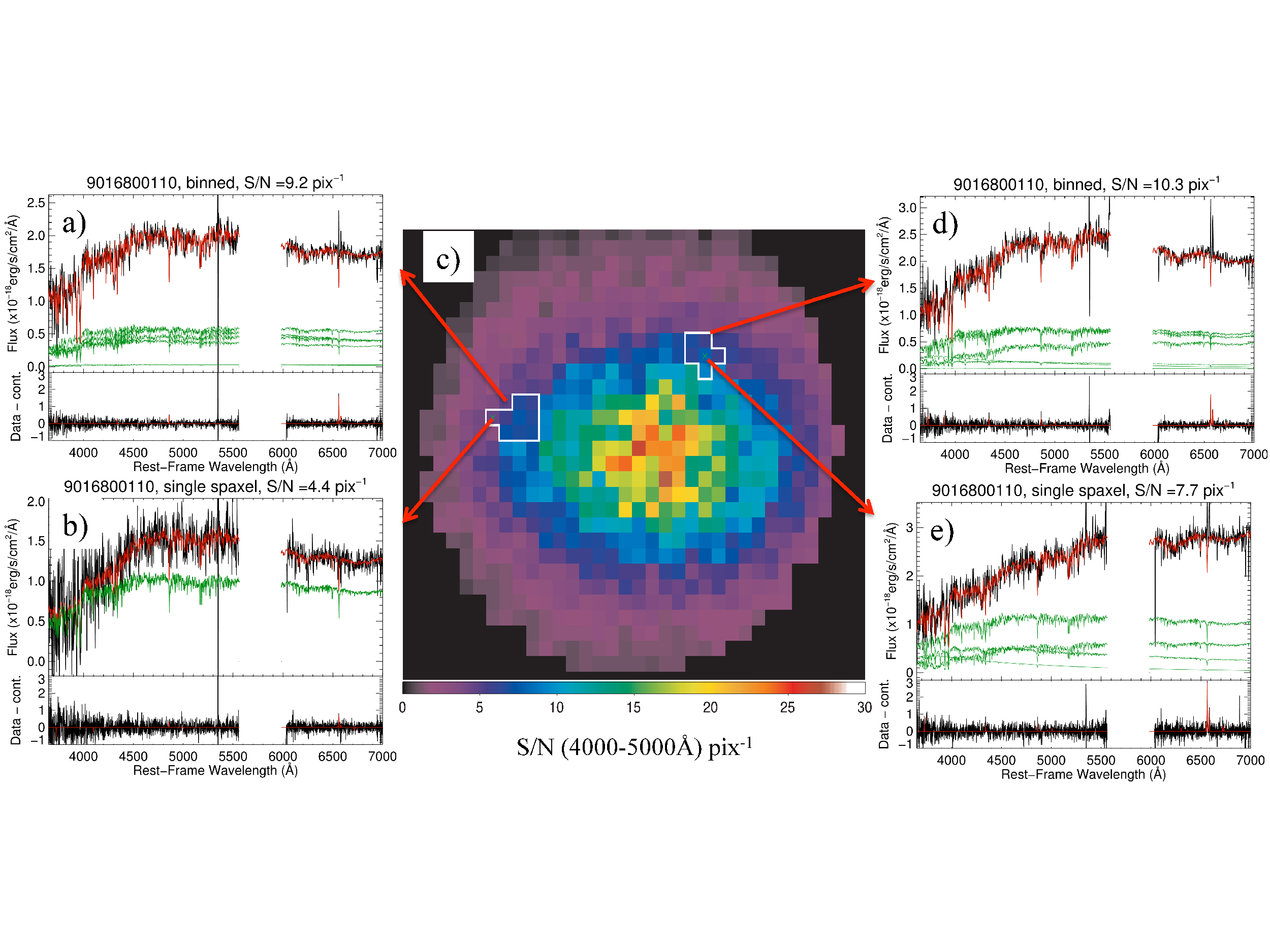}
\caption{This figure demonstrates the stellar continuum and emission line fitting process described in Sections~\ref{template_refine}, \ref{per_spaxel_fits} and \ref{emission_line_measurement}. Panel c) shows the spatial distribution of the median S/N in the wavelength range $4000-5000$\AA\, for the galaxy 9016800110. The white regions in panel c) show the spaxels that were combined using Voronoi binning to produce the S/N$\sim 10$\,pix$^{-1}$ spectra (black lines) shown in the top plots of panels a) and d). The red lines in panels a) and d) show the optimal template determined by {\sf pPXF} via weighted linear combination of the SPS templates shown in green, and modulated by a 12th order multiplicative polynomial (not shown). The bottom plots in panels a) and d) show the stellar-continuum subtracted, pure emission-line spectra as black lines, while the red line shows the best-fit emission line model determined using emission templates included during the {\sf pPXF} fits. The top plots in panels b) and e) show spectra from a single spaxel within the Voronoi bins. Again, the red lines in panels b) and e) show the optimal template determined with {\sf pPXF}, where the SPS template library includes only those templates with non-zero weighting shown in panels a) and d), respectively. The spectrum in panel b) has S/N$<5\,$pix$^{-1}$, so a single template determined by combining the templates shown in panel a) was included in the fit along with the multiplicative polynomial. The spectrum in panel e) has S/N$>5\,$pix$^{-1}$, so the weights applied to the restricted template set were allowed to vary during the {\sf pPXF} fits. The bottom plots in panels b) and e) show the pure emission-line spectra in black, while the red lines show the emission line models determined from the Gaussian fits described in Section~\ref{emission_line_measurement}.   \label{ppxf_eg}}
\end{figure*}

We follow a similar procedure to that outlined in \citet{VDS2016} where we use the higher signal-to-noise spatially binned data to select a subset of the 73 SPS templates to use in fitting the lower signal-to-noise single-spaxel data. This pre-selection of SPS templates helps to avoid overfitting of the noisier single-spaxel data. We use data that has been binned spatially to reach a S/N$\,\sim 10$ using Voronoi-binning \citep{cappellari2003}, where covariance between spaxels due to dithering has been accounted for when determining the variance of the combined spectrum \citep{sharp2015, allen2015}. Two examples of spectra resulting from the Voronoi-binning are shown in panels a) and d) of Figure~\ref{ppxf_eg}. Rather than re-fit the stellar kinematics, we use the existing two-moment (velocity and velocity dispersion) kinematic data that were described in \citet{VDS2016} to bring the spectra and templates to a common rest-frame and dispersion. The galaxy spectra are corrected to the rest-frame using the redshift $(1+z_{tot}) = (1+z_{gal})(1+v_{\rm pPXF}/c)$ where $z_{gal}$ is the galaxy redshift and $v_{\rm pPXF}$ is the velocity with respect to $z_{gal}$, and $c$ is the speed of light. We then convolve each SPS template using a Gaussian kernel with the wavelength-dependent width
\begin{equation}
\sigma_{tot}^2 = \left[ \left(\frac{\lambda \sigma_{\rm pPXF}}{c} \right)^2 + \left(\frac{\sigma_{inst}}{1+z} \right)^2 \right] - \sigma_{\rm MILES}^2
\end{equation}
where $\sigma_{\rm pPXF}$ is the velocity dispersion (in km/s) of the spectrum determined by \citet{VDS2016}, $\lambda$ is the wavelength of the pixel, $\sigma_{inst}=1.13\,{\rm \AA} (0.68\,{\rm \AA})$ is the instrument resolution for the blue (red) arm of the spectrograph \citep{VDS2016} and $\sigma_{\rm MILES}=1.06\,{\rm \AA}$ is the resolution of the MILES templates \citep{falconbarroso2011}. Both the SPS templates, the data, and the variance are rebinned onto a grid with constant velocity pixel size that is best matched to the blue-arm data (i.e., $\Delta v \sim 55$\,\kms), thereby undersampling the red-arm data. We then use pPXF to determine the optimal combination of the MILES templates while fixing $v_{\rm pPXF} =0\,$\kms\, and $\sigma_{pPXF}=0\,$\kms. A twelfth order multiplicative polynomial is used to correct for any effects due to data reduction artefacts, and also the effects of dust extinction. 

The above process is repeated twice. On the first iteration, the regions surrounding strong emission lines are masked. Following this first iteration, the error array associated with the spectrum is normalized by the ratio of the median absolute deviation of the residuals to the median of the error array. In the second iteration the emission lines are not masked and, in addition to the SPS templates, we include emission line templates for all Balmer lines from H$\zeta$ ($\lambda 3889$) in the blue to H$\alpha$ ($\lambda 6563$) in the red, as well as the strong forbidden lines [OII] ($\lambda \lambda 3726,\, 3729$), [OIII] ($\lambda \lambda 4959,\, 5007$), [OI] ($\lambda \lambda 6300, 6364$), [NII] ($\lambda \lambda 6548,\, 6583$), [SII] ($\lambda \lambda 6717,\, 6731$). We fit for the kinematics of the emission line templates, assuming the same kinematics for the Balmer and forbidden lines, and include the velocity, velocity dispersion and the higher-order $h_3$ and $h_4$ components. Example emission line fits are overplotted on the stellar continuum subtracted, pure emission line spectra shown in the lower plots of panels a) and d) in Figure~\ref{ppxf_eg}. We also use the {\sf CLEAN} keyword in order to reject outliers and to ensure the presence of weak emission lines do not impact the fit to the stellar continuum. Only those SPS templates with non-zero weights (shown in green in panels a) and d) of Figure~\ref{ppxf_eg}) in this final iteration are used for the per-spaxel fitting outlined in Section~\ref{per_spaxel_fits}. In addition, the emission-line kinematics derived here serve as initial estimates for the kinematics of the per-spaxel emission line fitting.

\subsubsection{Continuum definition for individual spaxels}\label{per_spaxel_fits}
Having refined the SPS library and determined initial estimates for the emission line kinematics, we now fit the spectrum of each spaxel contained in each of the Voronoi bins. The SPS templates, spectrum, and variance are rebinned to a pixel scale with constant velocity width that is tuned to best match the red-arm (i.e., $\Delta v \sim 25$\,\kms), thereby oversampling the blue-arm spectrum.
During fitting, the forbidden and Balmer emission-line species are assumed to have the same kinematics (velocity, dispersion, $h_3$ and $h_4$). The simultaneous fitting of the underlying stellar continuum and the emission lines allows for a better solution for the underlying stellar continuum to be found than if the emission lines were simply masked. This improvement is because important continuum regions surrounding emission lines can be included in the fit; in particular the age-sensitive Balmer lines bluer than H$\beta$ can now influence the fitted continuum. 

For spaxels with S/N\,$>5$ in the blue arm, the stellar kinematics can be determined reliably \citep{fogarty2015}. Therefore, those spaxels with S/N\,$>5$ have their stellar kinematics fixed to the per-spaxel value determined in \citet{VDS2016}. For S/N\,$>5$ spaxels, we also allow pPXF to fit for the weights of the refined SPS template library, as well as including a twelfth order multiplicative polynomial that corrects for residual differences in the SPS templates and the data (see panel e in Figure~\ref{ppxf_eg}). When the S/N\,$<5$, the stellar kinematics and SPS template weights are less well-constrained. For spaxels with S/N$<5$, we fix the stellar kinematics to the velocity and dispersion determined using the Voronoi-binned spectrum by \citet{VDS2016}. Furthermore, for S/N\,$<5$ spaxels, rather than fitting for the weights for the individual SPS templates, we use a single optimal template that is constructed using the weights determined in fitting the Voronoi-binned spectrum. Thus, for S/N\,$<5$ spaxels, the only free parameters used in fitting the stellar continuum are a single weight for the optimal template, as well as the coefficients of the twelfth-order multiplicative polynomial. This constrained fit allows for a more robust definition of the underlying stellar continuum even in lower S/N regimes (see panel b in Figure~\ref{ppxf_eg}).
 
\subsection{Emission line flux measurements}\label{emission_line_measurement}

While the per-spaxel continuum fitting procedure described in Section~\ref{continuum_definition} does produce emission line fluxes, the disparity between the pixel scales for the blue- and red-arm data means that the measurements are performed on heavily oversampled data in the blue. This oversampling may introduce inaccuracies in the measured fluxes, and the associated formal uncertainties. Instead, we use a Python implementation of {\sf mpfit}\footnote{This code was converted from IDL to Python by Mark Rivers, Sergey Koposov and Michele Cappellari.} \citep{markwardt2009, cappellari2017} to fit the Gaussians to the emission lines after subtracting the best-fitting stellar continuum. The best-fit model for the stellar continuum determined in Section~\ref{per_spaxel_fits} is redshifted to $z_{tot}$, interpolated onto the pixel scale of the blue and red arm data (1.03\AA\, and 0.56\AA, respectively), and subtracted from the data, leaving a pure emission-line spectrum as shown in the lower plots of panels b and e in Figure~\ref{ppxf_eg}. The fitting of this emission-line spectrum is outlined below. 
 
The line shapes often exhibit non-Gaussian profiles, meaning that fluxes determined from fitting a single Gaussian component may substantially underestimate the total line flux \citep[e.g.,][]{ho2014,ho2016}. To detect the presence of non-Gaussianity, we perform an initial fit to the \nii, \ha\, and \sii\, emission lines, which fall in the high resolution portion of the SAMI spectra. First, we fit a single Gaussian profile with velocity and velocity dispersion fixed for the different line species. A second fit is then performed with the addition of the higher-order Gauss-Hermite $h_3$ and $h_4$ terms, which parametrize asymmetric and symmetric departures from a Gaussian shape, respectively \citep{vandermarel1993}. We use the change in the Bayesian Information Criterion, $\Delta$BIC=${\rm BIC}_{\rm Gauss} - {\rm BIC}_{\rm Gauss-Hermite}$, to determine if the extra two parameters describing departures from a Gaussian shape are justified. Here, BIC=$\chi^2 + d(ln(N) - ln(2\pi))$, where $N$ is the number of data points and $d$ the number of free parameters in the fit. In the case that $\Delta$BIC $< 10$, a single Gaussian is deemed sufficient to describe the emission line shape. Where $\Delta$BIC $\geq 10$, we perform a third iteration of fitting where we use two Gaussian components. For the first Gaussian component, the velocity and dispersion determined in the first step are used as input guesses. We use the derivatives of the best-fitting Gaussian-plus-Gauss-Hermite model to determine initial estimates for the second Gaussian component using Equations~2a-4 in \citet{lindner2015}. 

Having determined whether a one- or two-Gaussian profile best describes the emission line shape, we then include the emission lines in the blue arm of the spectrum. We fit the \hd\, to \ha\, Balmer lines and the \oiid, \oiiid, \oid, \niid\, and \siid\, doublets. The velocity and velocity dispersion of the Balmer and forbidden lines are fixed to the same value, with the different instrument resolution of the blue- and red-arm spectra appropriately accounted for. The velocity and velocity dispersion determined during the initial fits to the \nii, \ha\, and \sii\, emission lines are used as initial inputs during the fitting to the full range of emission lines. The amplitudes of the \oiiid, \oid\, and \niid\, doublets are fixed to their expected values of 0.347, 0.333 and 0.339, respectively. Fluxes are determined for each line using the fitted amplitude and line dispersion. Uncertainties on the fluxes are determined by propagating the formal uncertainties on the amplitude and dispersion, and include covariance terms that can contribute significantly to the flux uncertainties obtained for the two-Gaussian cases. Throughout the remainder of this paper, we use the total emission-line flux determined from the one- or two-Gaussian profile that provides the best description of the emission-line shape.

\subsection{Absorption line equivalent widths}\label{absorption_lines}

Absorption line equivalent widths and uncertainties are determined using the direct summation method described in \citet{cardiel1998}. The bands used to define the line and continuum regions are shifted to the observed frame using the $z_{tot}$ determined in Section~\ref{continuum_definition}. Prior to measuring absorption line equivalent widths, the best-fitting emission line model is subtracted from the spectrum. This correction is only performed when the measured emission line kinematics are reliable, i.e., the velocity and dispersion have not hit a boundary in the parameter space, nor is the amplitude of the emission line negative. When the emission line model is subtracted, the uncertainty on the absorption line equivalent widths include a contribution due to the uncertainty in the emission line flux measurement. We measure the age-sensitive Balmer absorption lines \hdF\, and \hb\, using the definitions of \citet{worthey1997} and \citet{worthey1994}, respectively. The \hgF\, equivalent width is determined using the line bandpass and red continuum sideband definitions of \cite{worthey1997}, and the blue continuum sideband definition of \citet{fisher1998}. The shifting of the blue sideband helps to avoid contamination of the continuum measurement due to the G-band absorption at 4304\AA.

Many spaxels have S/N(4100\,\AA)\,$ \sim 3\,{\rm pix}^{-1}$, and this is particularly prevalent in the outer parts of galaxies where environmental quenching may
be more readily detected. The median uncertainty on EW(\hdF) measurements for spectra with S/N(4100\,\AA) = $3\,{\rm pix}^{-1}$ is 
\hdFerr\,$\sim 2$\,\AA, meaning that we cannot reliably distinguish passive and HDS spectra; the median uncertainty drops below 
$\sim 1$\AA\, only when the S/N(4100\,\AA)\,$ \sim 6\,{\rm pix}^{-1}$. Rather than removing all S/N(4100\,\AA)\,$ < 6\,{\rm pix}^{-1}$ spaxels, or binning
spatially to achieve a higher S/N in the continuum (which is generally not optimal for emission line measurements), we follow a similar procedure to
 \citet{blake2004} and use the correlation between the EW(\hdF), EW(\hgF) and EW(\hb) measurements to determine a higher S/N proxy for EW(\hdF). 

\begin{figure*}
\includegraphics[width=0.32\textwidth, trim= 0 0 0 1, clip]{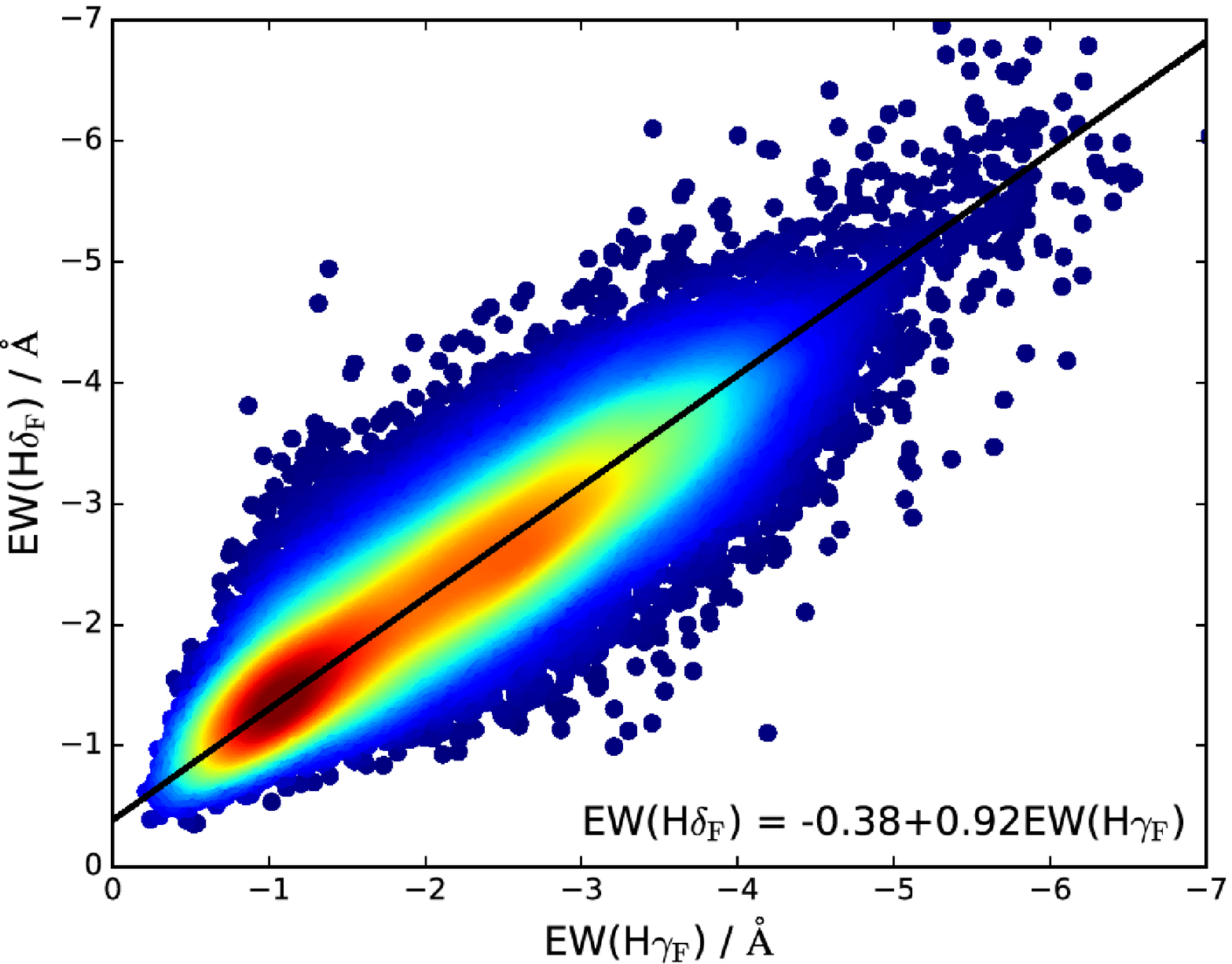}
\hspace{1mm}
\includegraphics[width=0.32\textwidth, trim= 0 0 0 1, clip]{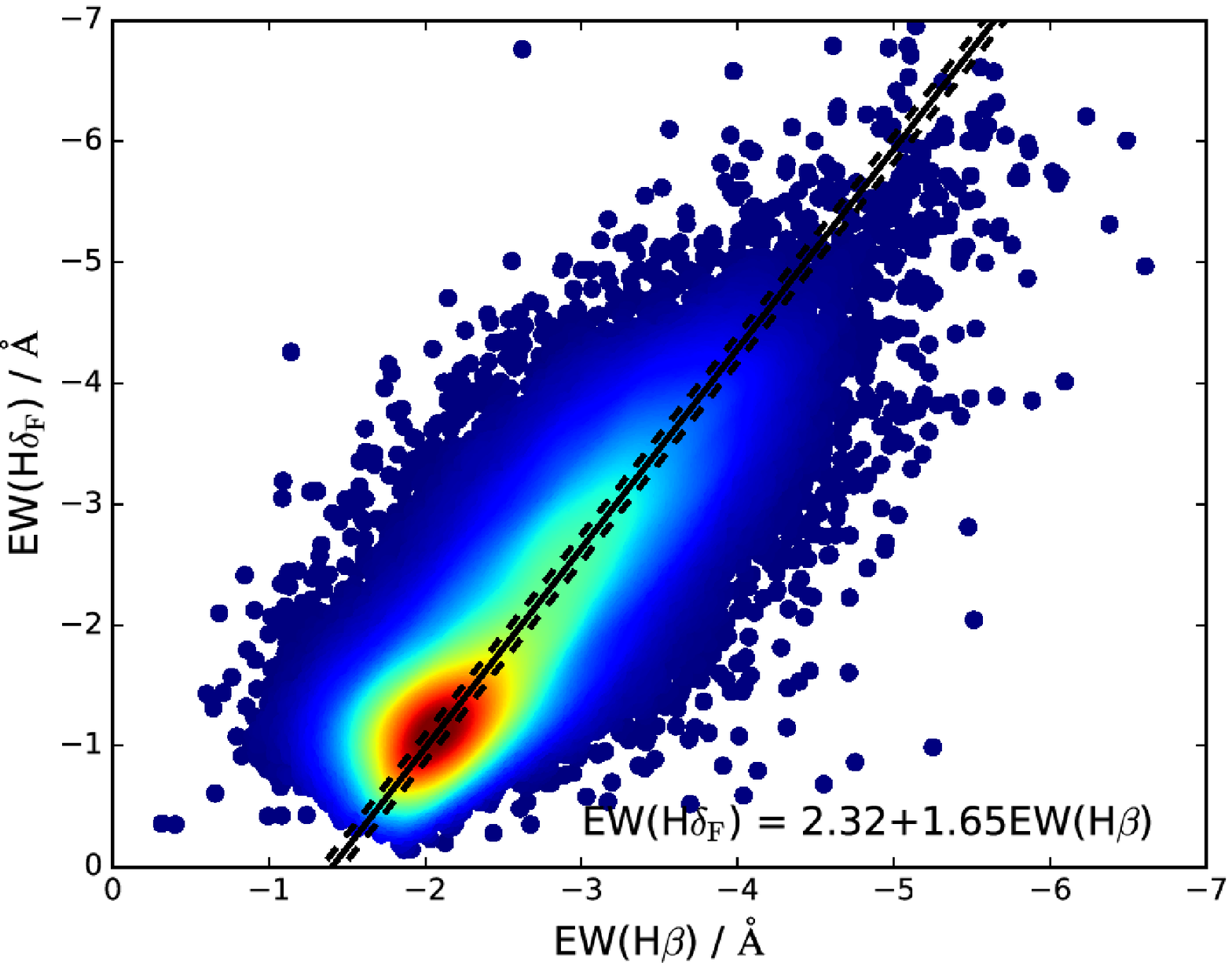}
\hspace{1mm}
\includegraphics[width=0.32\textwidth, trim= 0 0 0 1, clip]{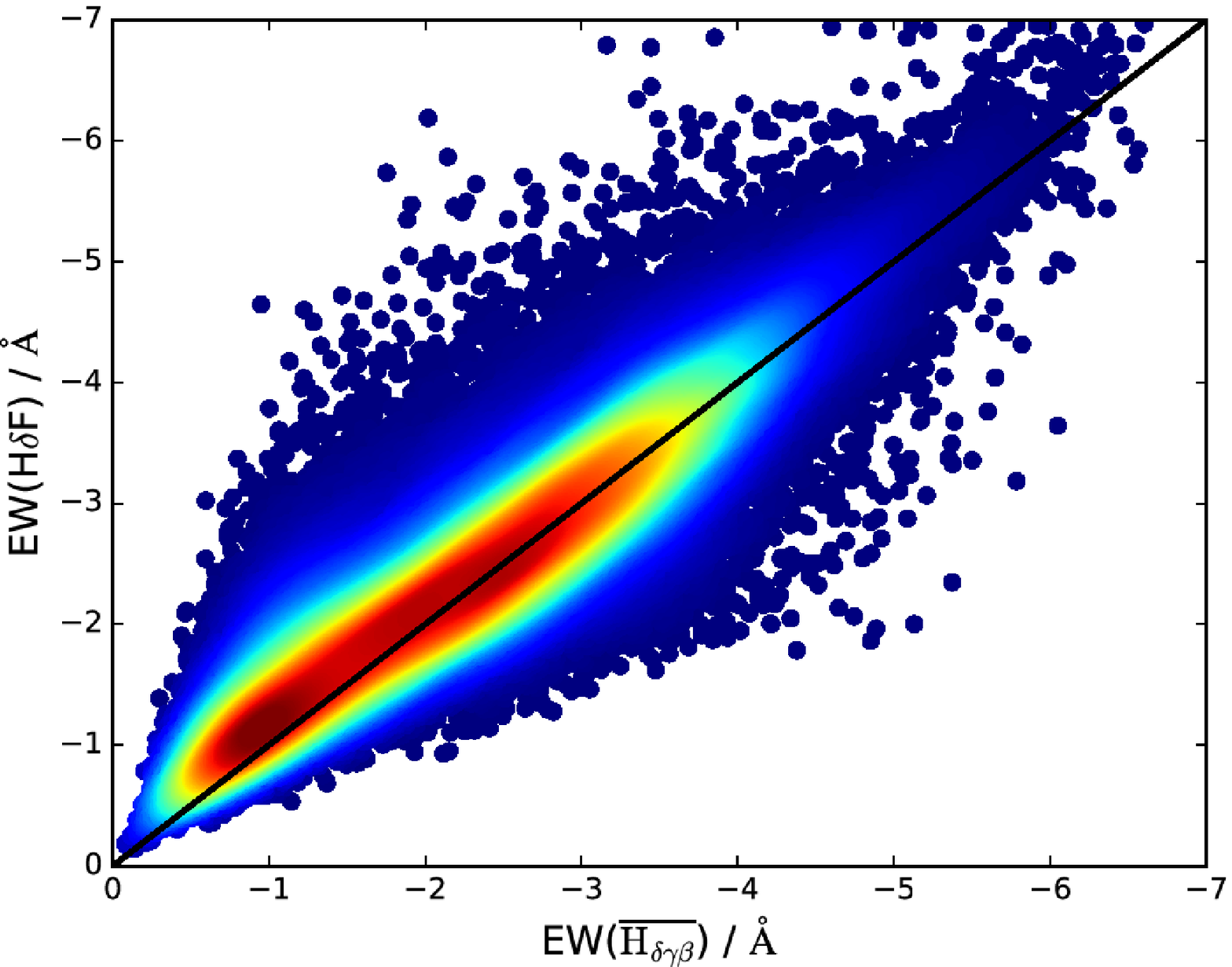}
\caption{The left and middle panels show the EW(\hdF) versus EW(\hgF), and EW(\hdF) versus EW(\hb) plots, revealing the strong correlations that exist between the Balmer line absorption strengths. In these two panels, the black lines show the best-fit linear relations determined with the {\sf HYPER-FIT} package and the best-fit parameters are shown on the lower right. We use these best-fit linear relations to determine a weighted average EW(\uberhdF), which is shown plotted against EW(\hdF) in the right panel. In all three panels, the colorscale shows smoothed density distribution of points, which ranges from 1\% of the peak density (dark blue) to 99\% of the peak density (dark red) on a linear scale. The right panel only includes spectra with S/N(4100\,\AA)\,$ > 6 {\rm pix}^{-1}$, EW(\ha)\,$<20$\,\AA, and where each EW is detected with S/N\,$> 2$. The black line shows the one-to-one relation. For absorption stronger than EW\,=\,-1\,\AA, there is a strong one-to-one correlation between EW(\uberhdF) and EW(\hdF), indicating that the weighted average used to determine EW(\uberhdF) does not introduce strong biases. \label{balmer_EW}}
\end{figure*}

Figure~\ref{balmer_EW} shows the strong correlations between EW(\hdF) and EW(\hgF) (left panel) and EW(\hdF) and EW(\hb) (middle panel). The correlations are fitted with a linear relation using the {\sf HYPER-FIT}\footnote{https://http://hyperfit.icrar.org} package \citep{robotham2015}, which accounts for uncertainties in both the x- and y-measurements. For the fitting, we only use measurements where S/N$(4100{\rm \AA}) > 10$, EW(\ha)\,$< 20\,$\AA\, and where S/N(EW)\,$ >2$ in absorption (where S/N(EW) = $|{\rm EW}|/\sigma_{\rm EW}$) for both EW measurements. The best-fit linear relations are shown in the lower right of both panels, and also plotted as black lines. We use the best-fit relations to produce a pseudo-EW(\hd) from the EW(\hgF) and EW(\hb) measurements. Uncertainties on the pseudo-EW(\hd) measurements are determined using standard error propagation, and include measurement uncertainty as well as contributions from the uncertainties on the  fitted parameters, and intrinsic scatter in the relations as determined by {\sf HYPER-FIT}. 

For each spectrum, the EW(\hdF) and pseudo-EW(\hdF) measurements are combined using a weighted average of the three measurements. The weighting includes an
inverse-variance term, as well as a term that de-weights the contribution from outlier measurements. The final weighted average of the three measurements,
hereafter EW(\uberhdF), is used for the classification described in Section~\ref{absorption_class}. In the right panel of Figure~\ref{balmer_EW} we show the
comparison of the EW(\uberhdF) and EW(\hdF) measurements for spectra with S/N$(4100{\rm \AA}) > 6 {\rm pix}^{-1}$, EW(\ha)\,$<20$\,\AA, and S/N(EW)\,$> 2$ in absorption. There is a strong one-to-one correlation between the two measurements, which indicates that the method for determining EW(\uberhdF) does not introduce strong biases into the estimates of the \hd\, strength. Moreover, for spectra with S/N(4100\AA)\,$\sim 3\,{\rm pix}^{-1}$, \uberhdFerr\,$\sim 1$\,\AA, indicating that we can now reliably distinguish passive and HDS spectra even at low S/N.

\section{Spectroscopic classification}\label{spec_class}

A key aim of this paper is to identify galaxies that are in the process of being quenched. This requires the identification of regions that contain young ($\lesssim 1.5$\,Gyr) stellar populations with no significant ongoing star formation. We identify these regions using a combination of emission and absorption line diagnostics as described below. The ten spectral classifications are summarized in Table~\ref{specclass_table}. We only include spaxels where the continuum signal-to-noise ratio is S/N(4100\,\AA)\,$ > 3\, {\rm pix}^{-1}$ to ensure that the continuum fits described in Section~\ref{continuum_definition} are reliable and that both emission- and absorption-line classification is possible. 

\begin{deluxetable*}{ccc}
\tabletypesize{\scriptsize}
\tablecaption{This Table summarizes the ten different spectroscopic classifications used in this paper. \label{specclass_table}}
\tablewidth{0pt}
\tablehead{\colhead{Spectral Class} & \colhead{expanded} & \colhead{Detailed description}}
\startdata
PAS & passive                  & Absorption line spectrum with no detected emission lines and EW(\uberhdF) $> -3$\,\AA,\\
    &                          &  indicating an old, passively evolving stellar population. \\
NSF & non-star forming         & Emission lines detected. Classified as outlined in Table~\ref{emline_table}.	Line ratios \\
    &                          & indicate excitation due to non-star forming radiation, e.g., shocks or AGN. \\
sNSF & strong non-star forming & As for NSF, but with EW(\ha)$\,>\, 6$\,\AA. \\
wNSF & weak non-star forming   & As for NSF, but with $3\, {\rm \AA} < \,$EW(\ha)$\,<\, 6$\,\AA. \\
rNSF & retired non-star forming& As for NSF, but with EW(\ha)$\,<\, 3$\,\AA. \\
SF & star forming              & Emission lines detected. Classified as outlined in Table~\ref{emline_table}.	Line ratios  \\
   &                           & indicate excitation due to ongoing star formation.\\ 
wSF & weak star forming        & As for SF, but with EW(\ha)$\,<\, 3$\,\AA. \\
INT & intermediate             & Emission lines detected. Line ratios are intermediate between the SF and NSF  \\
   &                           & diagnostic boundaries. Emission likely due to composite of star forming and \\
   &						      & non-star forming mechanisms.\\
rINT & retired intermediate             & As for INT, but with EW(\ha)$\,<\, 3$\,\AA.   \\
HDS & \hd -strong/post-star forming & Absorption line spectrum with no detected emission lines.\\
    &                          & Strong Balmer absorption with EW(\uberhdF) $\,< -3$\,\AA\, indicating truncation \\
    &						  & of star formation in last $\sim 1.5$\,Gyr.\\
NSF\_HDS & non-star forming    & As for HDS, but with detected emission lines that are classified as NSF. \\ 
&\hd-strong & 
\enddata
\end{deluxetable*}

\subsection{Emission line classification}\label{emission_class}

In order to be considered for emission line classification, a spaxel must have either \ha\, or \nii, plus one of \hb, \oiii, [SII]($\lambda 6716$) or [SII]$(\lambda 6730)$ lines detected with S/N\,$> 3$, where the S/N of the line measurement is estimated as the ratio of the measured flux and its formal uncertainty. For both of these two scenarios, we also require that the primary line (i.e., \ha\, or \nii) must have EW\,$> 1$\,\AA, which helps to reject spurious detections due to template mismatch.  The detection of at least two lines with S/N$>3$ guards against the bias towards false positive detections that are known to occur for single-line detections with S/N\,$< 5$ \citep{rola1994}. 

The standard procedure to classify emission line spectra is to use the line-ratio diagrams of \citet[][hereafter BPT]{baldwin1981} and \citet{veilleux1987}, which plot the flux ratios for \oiii/\hb\, versus \nii/\ha, \sii/\ha\, or [OI]($\lambda 6300$)/H$\alpha$. Generally, a S/N cut is made on each of the four lines involved in the line-ratio diagram so that their positions on BPT diagrams can be reliably measured \citep[e.g.,][]{kewley2006}. However, these more conservative cuts prohibit classification for a large number of emission-line spaxels where fewer than four lines are detected, and may bias against particular types of emission-line galaxies \citep{c.miller2003,cidfernandes2010}. Given these issues, and because our aim is to search for signatures of recent star formation in the absence of ongoing star formation, it is very important that we are able to characterise any emission detected in a spaxel as arising from a star-forming or non-star-forming ionizing source even when only a subset of the BPT lines are detected.  Spaxels that meet the emission-line classification may lie in five different categories depending on the combination of emission lines that are detected with S/N $> 3$:
\begin{itemize}
\item {\it Category A:} All four of the \ha, [NII] ($\lambda 6583$), H$\beta$ and [OIII]($\lambda 5007$) lines are detected;
\item{\it Category B:} \ha, [NII] ($\lambda 6583$), and [OIII]($\lambda 5007$) lines are detected, but H$\beta$ is not;
\item{\it Category C:} \ha, [NII] ($\lambda 6583$), and H$\beta$ lines are detected, but [OIII]($\lambda 5007$) is not;
\item{\it Category D:} \ha\, and [NII] ($\lambda 6583$) lines are detected, but  H$\beta$ and [OIII]($\lambda 5007$) are not;
\item{\it Category E:} \ha\, and one other line that is not [NII]($\lambda 6583$) are detected, or [NII]($\lambda 6583$) and one other line that is not \ha\, are detected.
\end{itemize}
The emission line classification scheme for each of these five categories is summarized in Table~\ref{emline_table}. A detailed explanation of the emission line classification scheme follows.

Spectra that fall into Categories A--C are classified in a probabilistic manner using the \nii/\ha\, versus \oiii/\hb\, BPT diagram \citep[similar to the methods  of ][]{carter2001, manzer2014, marziani2017}. We produce 5000 Monte-Carlo (MC) realizations of the [NII]/\ha\, and [OIII]/\hb\, ratios, assuming a Gaussian distribution centred at the measured line flux with dispersion equal to the flux uncertainty. We include a contribution due to uncertainty on the emission-line flux caused by the SPS template-based absorption correction for H$\beta$ and H$\alpha$, which is assumed to be $0.5$\,\AA\, in equivalent width, consistent with the typical uncertainty on the EW(\hb) measured in Section~\ref{absorption_lines}. We classify each MC realization based on its position in the BPT diagram using the regions defined by \citet{kewley2006}. MC realizations that have [NII]/H$\alpha$ and [OIII]/H$\beta$ ratios that place them: (i) below the empirically-based \citet{kauffmann2003a} demarcation are classified as star-forming (or SF), (ii) above the \citet{kauffmann2003a} and below the \citet{kewley2001} theoretical ``maximum starburst'' demarcation lines are classified as intermediate/composite (INT), and (iii) above the \citet{kewley2001} demarcation are classified as non-star-forming (NSF). We then use the fraction of MC realizations falling into the three separate BPT classifications to determine the probabilities P(SF), P(INT) and P(NSF). Spectra that fall into Category A are classified using the BPT class that has the highest probability.

\begin{deluxetable*}{ccccc}
\tabletypesize{\scriptsize}
\tablecaption{Summary of the emission line classification scheme described in Section~\ref{emission_class}. The probabilities listed for Category A, B, and C spectra are determined using 5000 Monte-Carlo realisations of the \nii/\ha\, and \oiii/\hb\, line ratios and determining the fraction that fall in the SF, INT and NSF regions of the BPT diagram defined by the demarcation lines of \citet{kauffman2003c} and \citet{kewley2001}. Category B, C, and D spectra may also be classified based on the two-line scheme of \citet{cidfernandes2010}. Subclasses based on EW(\ha) are also listed, where ``r'' stands for {\it retired}, ``w'' stands for {\it weak}, and ``s'' stands for {\it strong}. \label{emline_table}}
\tablewidth{0pt}
\tablehead{
\colhead{Category} & \colhead{Detected emission} & \colhead{} & \colhead{Classification method} & \colhead{}\\
\colhead{}& \colhead{lines (S/N $> 3$)} & \colhead{SF} & \colhead{INT} & \colhead{NSF} }
\startdata
A &\hb, \oiii,             & P(SF) $>$ P(INT), P(NSF)   & P(INT) $>$ P(SF), P(NSF)			 & P(NSF) $>$ P(SF), P(INT)\\
  &\ha, \niishort          &                            & 									 & \\\\
B & \oiii, \ha,            &P(NSF)$<0.9$ AND            & P(NSF)$<0.9$ AND 					 & P(NSF)$\geq 0.9$ OR  \\
  & \niishort              &log(\niishort/\ha)$ < -0.32$& $-0.32 <$\,log(\niishort/\ha)$ < 0.1$& (P(NSF)$<0.9$ AND  \\
  &                        &                            &										 &log(\niishort/\ha)$ > 0.1$)\\\\
C &\hb, \ha, \niishort     &P(SF)$\geq 0.9$ OR          &  P(SF)$< 0.9$ AND  					 & P(SF)$< 0.9$ AND \\
  &                        &(P(SF)$< 0.9$ AND            & $-0.32 <$\,log(\niishort/\ha)$ < 0.1$& log(\niishort/\ha)$ > 0.1$\\
  &                        & log(\niishort/\ha)$<-0.32$) & $-0.32 <$\,log(\niishort/\ha)$ < 0.1$  \\\\
D &\ha, \niishort          &log(\niishort/\ha)$<-0.32$  &$-0.32 <$\,log(\niishort/\ha)$ < 0.1$&log(\niishort/\ha)$ > 0.1$\\\\
E &(\ha, NOT \niishort) OR &IF \ha                      & -- 					         		& IF \niishort\\
  &(\niishort\, NOT \ha)   &                            & -- 						  			& \\\\
 & --            & wSF IF EW(\ha) $< 3$\,\AA  & rINT IF EW(\ha) $< 3$\,\AA 			& sNSF IF EW(\ha) $> 6$\,\AA \\
Subclasses  &                        &                            &							   			& wNSF IF $3 {\rm \AA} < $EW(\ha) $< 6$\,\AA \\
  &                        &                            &                            			& rNSF IF EW(\ha) $< 3$\,\AA
\enddata

\end{deluxetable*}

For the Category A spectra, the probabilistic classification is identical to classifying spectra based on the ratios derived from the measured emission line flux values, assuming the probability density distribution is symmetric about the measured line ratios. This method becomes more powerful when considering Category B and C spectra where judicious use of upper limits can enable a classification in the absence of a significant line detection for \hb\, or \oiii. For the Category B galaxies, we can place an upper limit on the \hb\, line flux based on the \ha\, line flux, and our knowledge of Case-B recombination which results in \Fhb$ < $\Fha$/2.86$. During the MC realizations, we enforce this upper limit. The upper limit on \Fhb\, enables a lower limit to be placed on the \oiii/\hb\, line ratio and allows us to robustly classify spectra as lying above the \citet{kewley2001} demarcation line, thereby ruling out INT and SF classifications. We can therefore classify any Category B spectrum as NSF, although we use a more conservative cut off of P(NSF)$ > 0.9$. Likewise, for Category C spectra we can determine upper limits on the \oiii\, line flux based on the flux uncertainties, which enables an upper limit on the \oiii/\hb\, line ratio to be determined.  Category C spectra are classified as lying below the \citet{kauffmann2003a} demarcation line when P(SF) $> 0.9$.

Category D spectra are classified based on the \nii/\ha\, ratio using the demarcation lines derived by \citet{cidfernandes2010}. The boundaries used for the SF, INT and NSF classifications are shown in Table~\ref{emline_table}.
The divisions at log(\nii/\ha)$=-0.32$ and log(\nii/\ha)$=-0.1$ correspond to the optimal dividing lines for the \citet{kauffmann2003a} and \citet{kewley2001} BPT demarcation lines, as determined in \citet{cidfernandes2010}. These divisions are chosen to be consistent with the classification scheme outlined for Category A galaxies. Those Category B and C spectra that could not be robustly classified as NSF or SF, respectively, were also classified using this method. Category E spectra, where the \ha\, line is detected and \nii\, is not, are classified as SF, while those where the \nii\, line is detected with no \ha\, detection are classified as NSF.

In the above classifications, we have thus far only made use of line-flux ratios. \citet{cidfernandes2011} advocated for the combined use of line ratios and the EW(\ha) when performing spectroscopic classification, particularly when only a subset of emission lines are detected. In particular, they classify spectra with EW(\ha)$< 3$\AA\, as being {\it retired} because the emission is likely powered by ionisation driven by post-AGB stars \citep{cidfernandes2011, singh2013,belfiore2016,belfiore2017}. We incorporate the EW(\ha) into our classifications in a similar manner to \citet{cidfernandes2011}. Spectra that have EW(\ha)\,$< 3$\,\AA, are categorized into the subcategories rINT, rNSF, and wSF where the ``r'' stands for {\it retired} (following the \cite{cidfernandes2011} nomenclature), and the ``w'' stands for {\it weak}, since the line ratios implies there may be star formation present but the low EW(\ha) indicates that it is relatively weak. We subcategorise those NSF galaxies with $3$\,\AA\,$<$\,EW(\ha)\,$<6 $\,\AA\, as wNSF and those with EW(\ha)\,$\geq 6$\,\AA\, as sNSF, where ``s'' stands for strong.

\subsection{Absorption line classification}\label{absorption_class}

Those spectra that do not have at least two emission lines detected as outlined in Section~\ref{emission_class} are classified as absorption-line spectra. Absorption-line spectra are further classified based on the strength of EW(\uberhdF). This classification is performed in a similar manner to that described in \citet{dressler1999} and \citet{poggianti1999}, where the strength of the age-sensitive \hd\, line was used as a proxy to identify passively evolving spectra, as well as those showing recently truncated star formation. We classify spectra with EW(\uberhdF)\,$ < -3$\,\AA\, and S/N(EW(\uberhdF))\,$ > 3$ as \hd-strong (HDS) spectra, and those absorption-line spectra not meeting this criteria as passive. We choose this limit in EW(\uberhdF) based on: (i) data limitations -- at our limiting S/N(4100\,\AA)\,=\,3\,pix$^{-1}$ the median error on EW(\uberhdF) is $\sim 1$\,\AA, meaning we can relatively reliably measure EW(\uberhdF) for our spectra, and (ii) spectra exhibiting \hd\, absorption stronger than this limit generally only occur due to a recent truncation of star formation (as opposed to a slow decline in star formation), as discussed in \citet{poggianti1999}. 

We stress here that we are not searching for {\it post-starburst} signatures, which would require a more stringent EW(\uberhdF)\,$< -5$\,\AA\, criterion as used in other studies \citep[e.g.,][]{couch1987,zabludoff1996,blake2004}. Only around 5\% of the HDS-classified spectra in our sample would meet this more stringent criterion. Rather, our criteria allow us to robustly identify spectra that are likely to have experienced a recent truncation of star formation within the last $\sim 1.5$\,Gyr, i.e., {\it post-starforming regions}, regardless of whether that truncation was preceded by a starburst. 

In addition to the absorption-line spectra that are classified as HDS, we add another HDS classification for those spectra that were classified as NSF in Section~\ref{emission_class}, but also have EW(\uberhdF) $<-3$\AA. These spectra are labelled NSF\_HDS; they meet the criteria of having evidence for young stellar populations with no ongoing star formation \citep[similar to the post-starburst galaxies in other studies][]{yan2006,alatalo2016}. Here, we add the additional criterion that the flux of \hd\, in emission must not exceed half the flux in absorption (as determined from the emission-line free spectrum). This criterion is somewhat arbitrary, but ensures that spectra where the strong emission completely masks the Balmer absorption are not classified as HDS. This complete masking can occur in regions of strong AGN emission, and in these cases the absorption-line measurements are strongly dependent on the correct modelling of the underlying stellar continuum and  may lead to spurious EW(\uberhdF) measurements.

\section{Galaxy classification scheme}\label{gal_class}

In this Section, we use the resolved spectroscopic classifications from Section~\ref{spec_class} to divide our sample into {\it passive}, {\it star-forming}, and \hd-strong galaxies (hereafter PASGs, SFGs, and HDSGs, respectively).

\subsection{What are passive and star-forming spaxels?}\label{passive_def}
Many of the spectroscopic classifications defined in Section~\ref{spec_class} are readily associated with passive stellar populations, ongoing star formation, or with recently truncated star formation.
For absorption-line spectra, the distinction is, by construction, trivial: the HDS spectra represent recently truncated, post-starforming regions, and the remainder, which show no strong Balmer absorption, are associated with older, passively evolving stellar populations. Similarly, spectra with strong emission lines with flux ratios placing them in the SF class are clearly associated with regions with ongoing star formation.

However, for other classes of emission line galaxies, e.g., INT, rINT or wSF classifications, it is not always obvious whether a spectrum should be classed as passive or star forming. To help with further classification, we investigate the distribution of the various spectral types in the EW(\uberhdF)-\dfour\, plane, where \dfour\, is the 4000\AA-break strength, which is determined using the definition of \cite{balogh1999}. The position on the EW(\uberhdF)-\dfour\, plane is a relatively reliable proxy for the luminosity-weighted age of the underlying stellar population. Young stellar populations inhabit regions with strong EW(\uberhdF) absorption and weaker breaks at \dfour, while older, passively evolving stellar populations inhabit regions with stronger \dfour\, and weaker EW(\uberhdF) \citep{balogh1999, kauffman2003c}. 

\begin{figure*}
\includegraphics[width=0.48\textwidth]{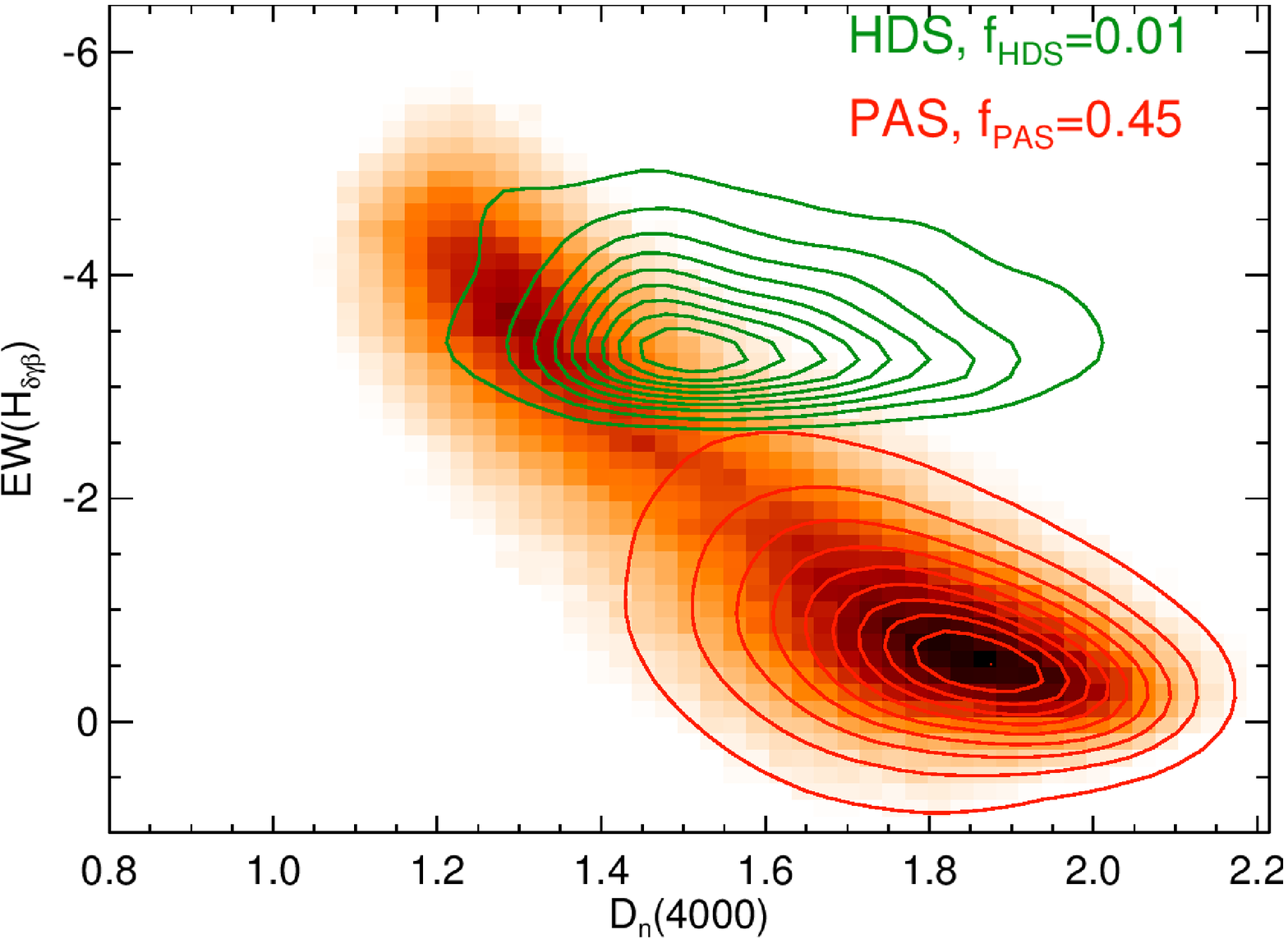}
\includegraphics[width=0.48\textwidth]{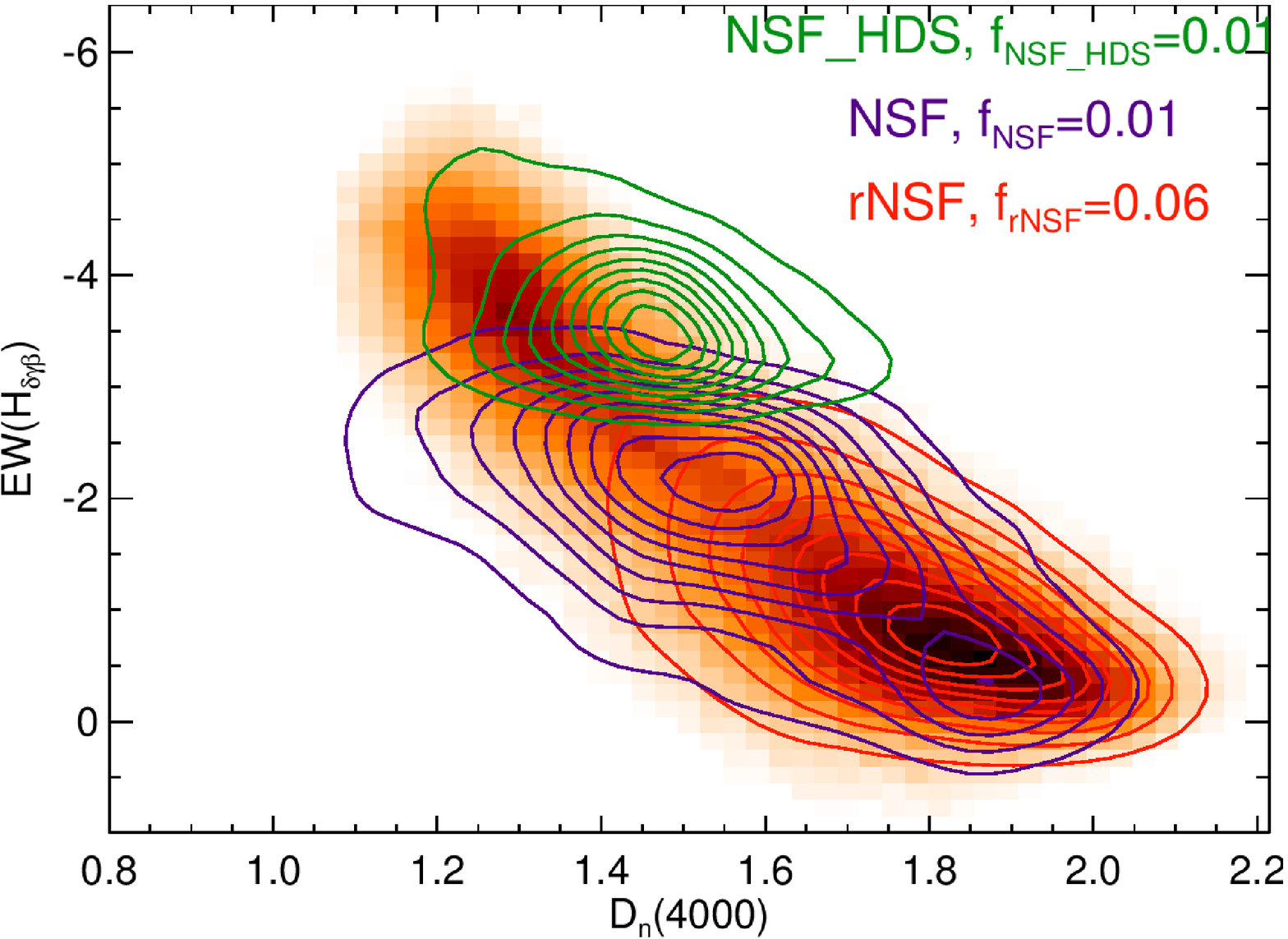}\\
\includegraphics[width=0.48\textwidth]{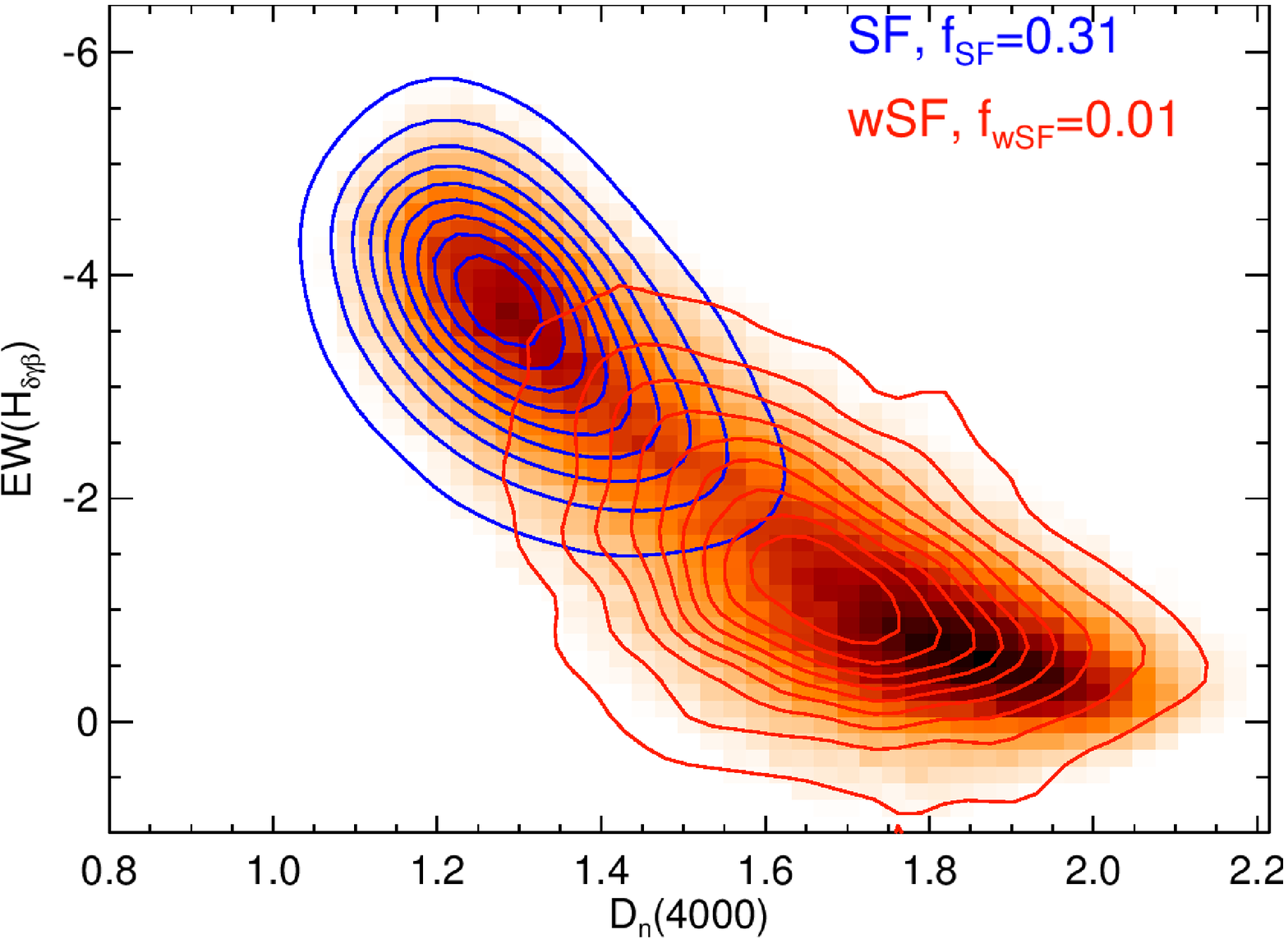}
\includegraphics[width=0.48\textwidth]{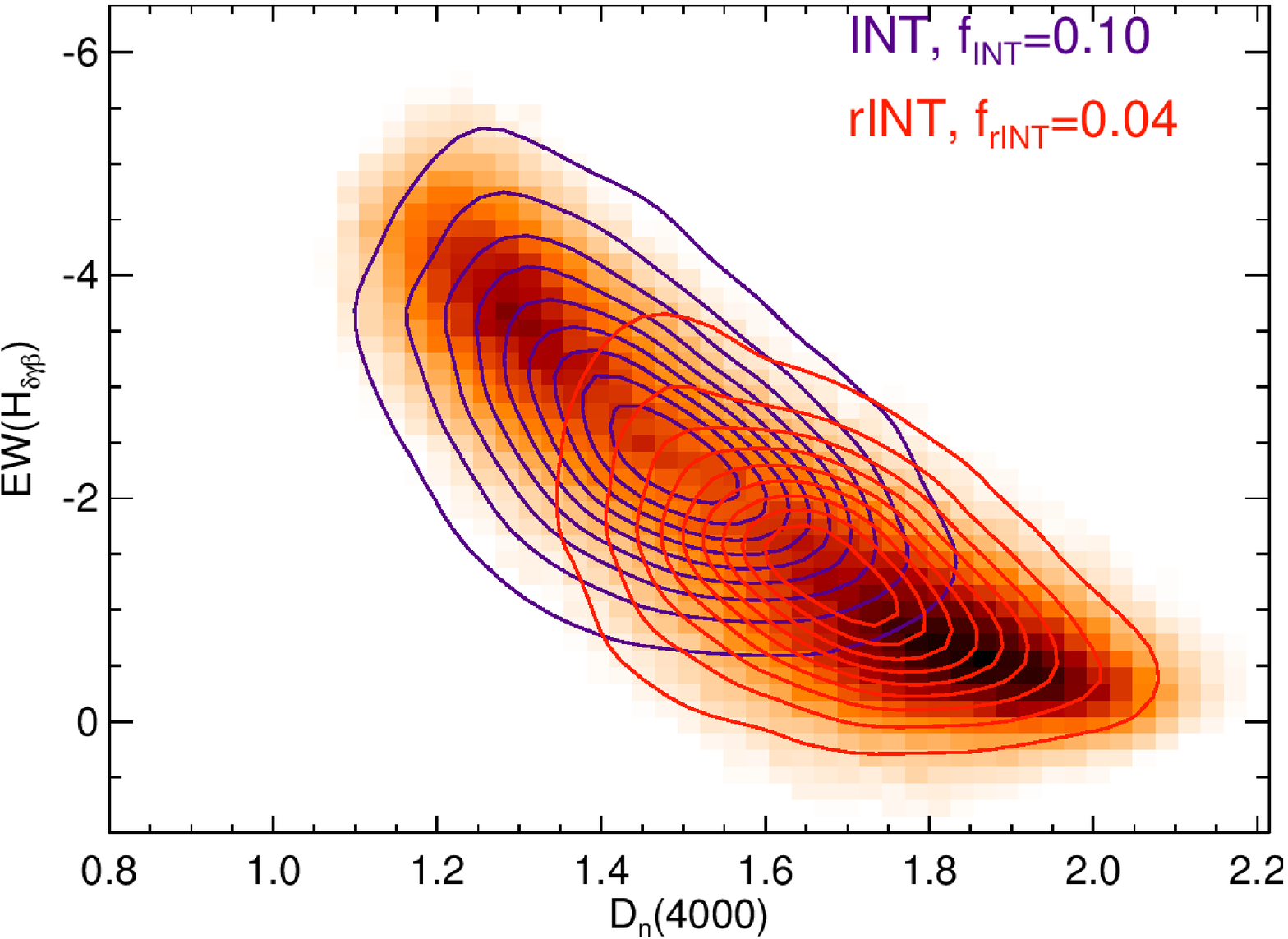}
\caption{The distribution of EW(\uberhdF) vs. \dfour\, for the spectral classifications defined in Sections~\ref{emission_class} and \ref{absorption_class}. The green and red contours in the top left panel show the distribution for spaxels classified as HDS and PAS, respectively. The top right panel shows the emission spaxels where the emission was classified as non-star-forming: green shows NSF\_HDS, purple NSF (sNSF plus wNSF) and red rNSF. The bottom left panel shows the distribution for emission line spaxels classified as either SF in (blue contours) or wSF (red contours). The bottom right panel shows the distributions for the rINT (red) and INT (purple) spectra. The heatmap in each plot shows the number density of the \dfour\, vs. EW(\uberhdF) distribution for all spaxels with S/N(4100\,\AA)\,$> 3$\,pix$^{-1}$ and for cluster and GAMA galaxies with \logmstar\,$\ge\,10$. The fractional contributon of each of the different spectral classifications is shown in the top right of each plot. The majority of spectra are classified as SF or PAS, which make up 31\%  and 45\% of classified spectra, respectively. The HDS and NSF\_HDS spectra are rare, contributing to only 2\% of the classified spectra. \label{dfour_plots}}
\end{figure*}

Figure~\ref{dfour_plots} shows the number-density distribution of the EW(\uberhdF) as a function of \dfour\, for all classifiable spectra in the cluster and GAMA samples that have S/N(4100\,\AA)$>3\,$pix$^{-1}$. The four panels in Figure~\ref{dfour_plots} show the EW(\uberhdF)-\dfour\, distribution for each of the spectroscopic sub-classifications in the absorption-line (top left panel), the NSF (top right panel), SF (lower left panel), and INT (lower right) classes. In these panels, the EW(\uberhdF)-\dfour\, distributions for the sub-classes are shown as nine equally-spaced contours that range from 10\% to 90\% of the peak in the smoothed number density for the spectral type of interest.  Figure~\ref{dfour_plots} reveals that there are two clear peaks in the EW(\uberhdF)-\dfour\, plane; one centred at EW(\uberhdF)$\simeq -0.5$\AA\, and \dfour$\simeq 1.85$ and the other at EW(\uberhdF)$\simeq -4.0$\AA\, and \dfour$\simeq 1.30$. The former peak is dominated by spaxels classified as PAS (red contours in the top left panel), which make up 45\% of all classified spaxels, while the latter peak primarily contains spaxels classified as SF (blue contours in the bottom left panel), which make up 31\% of classified spaxels.

The EW(\uberhdF)-\dfour\, distributions of the rNSF, wSF, and rINT classified emission line spaxels (all of which have weak EW(\ha) $<3$\AA) are shown as red contours in the top right, bottom left, and bottom right panels of Figure~\ref{dfour_plots}, respectively. The distributions of these weak \ha\, emitters are generally consistent with that of the PAS absorption-line galaxies, indicating that the stellar populations in these spectra are dominated by old, passive populations. It is interesting to note that even wSF classified spectra are more consistent with passively-evolving stellar populations, although there is a small fraction of wSF spectra that occupy regions consistent with recent star formation. We therefore conclude that the rNSF, wSF and rINT spectral types are to be considered alongside the PAS type as being dominated by passively evolving, old stellar populations. Due to their low numbers, wNSF and sNSF classified spaxels are combined and their EW(\uberhdF)-\dfour\, distribution is shown as purple contours in the top right panel of Figure~\ref{dfour_plots}. There is no strong indication that the NSF spaxels are dominated by young stellar populations. Given the non-star-forming origin of the emission in these spaxels, we also count them as passive spaxels. 

The purple contours in the lower right panel in Figure~\ref{dfour_plots} show the distribution of emission-line spaxels that are classified as INT.  For INT spectra, the peak in the distribution  is located between the SF and PAS peaks, and extends to encompass the peak associated with SF-classified spaxels, indicating that a large fraction of INT spectra harbour young stellar populations due to recent or ongoing star formation. INT-type spectra are often interpreted as being due to the combination of emission that has been ionised by both star formation and non-star-forming processes \citep[e.g., AGN;][]{kewley2006}. This interpretation is supported by the large fraction of INT spaxels that show evidence for young stellar population in Figure~\ref{dfour_plots}. We therefore include those INT spectra as star forming alongside the SF classified spaxels.

\subsection{How many spaxels define a galaxy class?}\label{galaxy_class}

\begin{figure}
\includegraphics[width=0.48\textwidth]{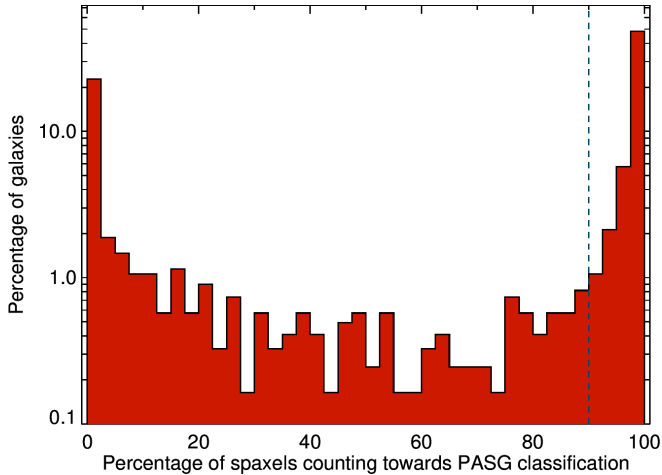}
\caption{The percentage of galaxies in the sample (on a log scale) as a function of the percentage of classifiable spaxels associated with passively evolving stellar populations (i.e., spectroscopic classes PAS, rNSF, wNSF, sNSF, rINT, wSF from Section~\ref{gal_class}). This plot shows that for 54\% of galaxies in the sample, $\geq 95$\% of spaxels with S/N(4100\,\AA)\,$> 3$\,pix$^{-1}$ have spectral types associated with passively evolving stellar populations. The dashed line shows our dividing line at 90\%, above which a galaxy is classified as PASG. The remaining galaxies with fewer than 90\% passive spaxels will be classified as either HDSG or SFGs.  \label{frac_passive}}
\end{figure}

We take a pragmatic approach to determining the fraction of spaxels associated with passively evolving stellar populations that are required for a galaxy to be classified as PASG. In Figure~\ref{frac_passive} we present a histogram that shows the relative frequency of the fraction of passive spaxels for SAMI galaxies in our sample. In determining the fraction of passive spaxels, only those spaxels with S/N(4100\,\AA)$>3$\,pix$^{-1}$ are used. 
 Figure~\ref{frac_passive} demonstrates that a large fraction of our sample are dominated by passive spaxels; 54\% of the galaxies have $\geq 95$\% of their spaxels belonging to spectroscopic classes that are associated with passively evolving stellar populations (i.e, those with PAS, NSF, rINT, and wSF). This fraction only increases to 57\% when considering galaxies with $\geq 90$\% of spaxels associated with passively evolving spectroscopic classes. This convergence at 90\% therefore sets a natural lower limit on the fraction of passive spaxels required for a galaxy to fall into the PASG class. Conversely, the limit for the PASG class also sets the lower limit of 10\% of spaxels classified as INT or SF for the SFG class. Likewise, a HDSG must have at least 10\% of its spaxels classified as HDS or NSF\_HDS. To summarize, our galaxy classes are defined as:
\begin{itemize}
\item PASG: passive galaxies that have more than 90\% of S/N(4100\,\AA)$>3$\,pix$^{-1}$ spaxels classified as PAS, rNSF, rINT, wNSF, sNSF or wSF.
\item SFG: star-forming galaxies have 10\% or more S/N(4100\,\AA)$>3$\,pix$^{-1}$ spaxels classified as either INT, SF.
\item HDSG: \hd-strong galaxies have 10\% or more S/N(4100\,\AA)$>3$\,pix$^{-1}$ spaxels classified as either HDS or NSF\_HDS.
\end{itemize}
In addition, for the SFG and HDSG classes we introduce a continuity criterion in order for a spaxel to contribute to the 10\% limit. For the SFG class, three of the six spaxels surrounding an INT or SF spaxel must also be classified as INT or SF. Likewise, to count towards HDSG classification, three of the six pixels surrounding an HDS or NSF\_HDS spaxel must be classified as HDS or NSF\_HDS. This guards against the contribution of isolated spaxels that can occur by chance in lower S/N spectra. We note that a galaxy may simultaneously meet the criteria for the SFG and HDSG classes, and in these cases the galaxy is included in the HDSG sample.

For the subset of 1220 SAMI targets used in this paper, the majority (88\%) of the sample contain 100 or more spaxels with S/N(4100\,\AA)\,$> 3$\,pix$^{-1}$. Therefore, a minimum of 10 spaxels are required to show evidence for recent/ongoing star formation in order for a galaxy to be classified as a SFG or HDSG. The median PSF of the SAMI survey has FWHM$\sim 2.06$\arcsec\, \citep{scott2018}, which corresponds to a 1-$\sigma$ surface area of 10 spaxels. Our criteria therefore ensures that for a galaxy to be classified as a SFG or HDSG the total area covered by the SF and HDS spaxels must be more extended than the PSF. 

\begin{figure*}
\includegraphics[width=0.45\textwidth]{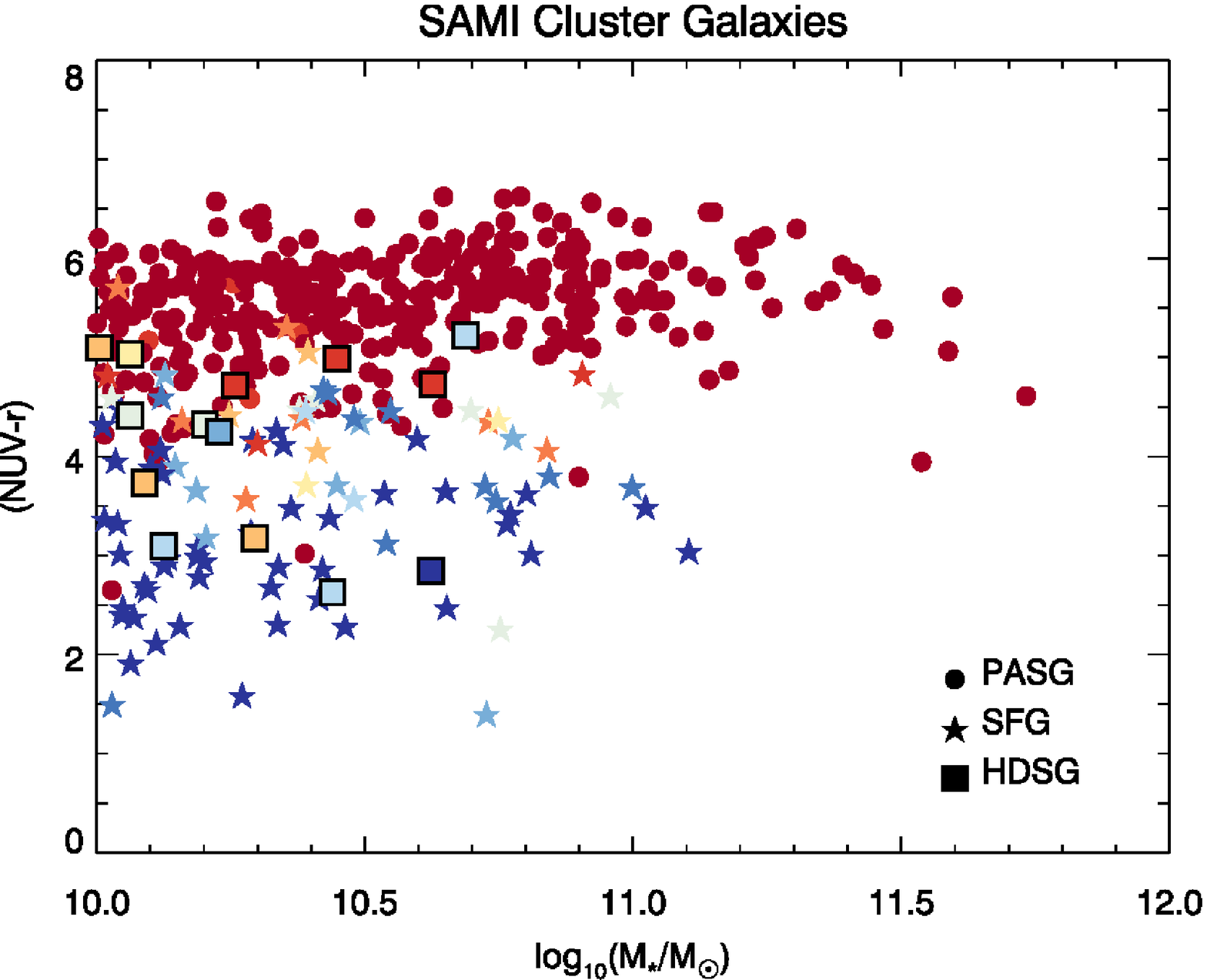}
\includegraphics[width=0.45\textwidth]{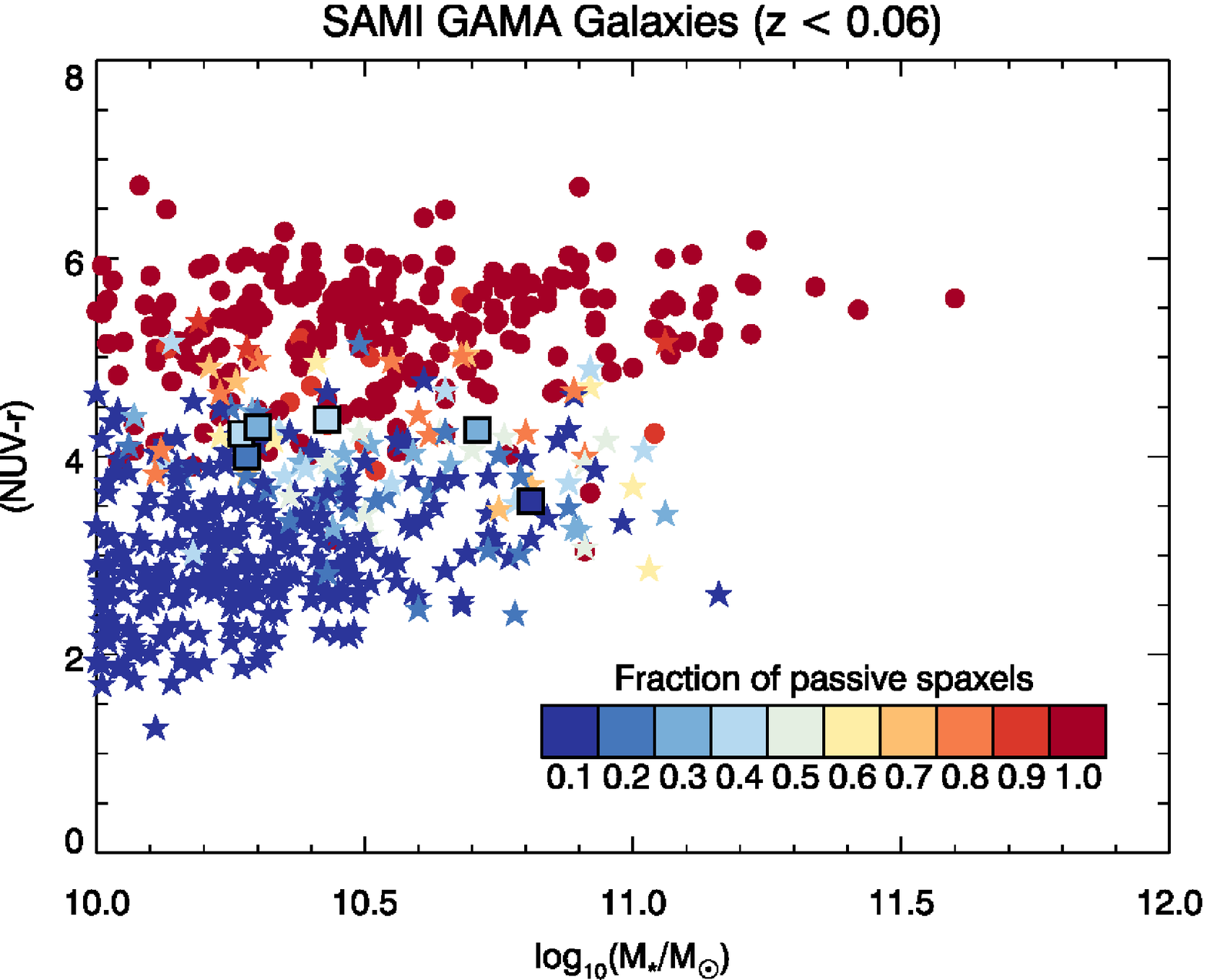}
\caption{Color-mass diagrams for SAMI galaxies in the clusters (left-panel) and GAMA regions (right-panel). The color-coding of each point indicates the fraction of classifiable spaxels that have passively-evolving stellar populations per the definition in Section~\ref{passive_def}. The three different galaxy classes are shown as different shapes, as indicated by the key in the bottom right of the left panel. Note that only the subset of cluster and GAMA galaxies that have NUV detections are plotted, and the NUV-$r$ colors are not $k$-corrected.  These two figures indicate that the galaxy classification scheme outlined in Section~\ref{galaxy_class} does a very good job of separating passive, red-sequence galaxies and blue-cloud galaxies that are actively forming stars. Also of note is that, while the HDSGs have bluer colours than the passive red-sequence galaxies, they are generally found in the green valley. \label{colour_mass_type}}
\end{figure*}

To check the veracity of our galaxy classification, we present Figure~\ref{colour_mass_type}, which shows the observed, Galactic extinction-corrected NUV$-r$ color versus stellar mass diagram  for the cluster regions (left panel) and the GAMA regions (right panel). The NUV-$r$ colors for the GAMA galaxies are obtained from  the {\sf LambdarPhotometry} catalogue released as part of the GAMA DR3 \citep{baldry2018}\footnote{http://www.gama-survey.org/dr3/}, which provides magnitudes measured via the aperture-matched and deblended photometry described in \citet{wright2016}. The cluster NUV magnitudes come from the catalogues produced by \citet{seibert2012}\footnote{https://archive.stsci.edu/prepds/gcat/}. 
Each point is colour-coded based on the fraction of S/N(4100\,\AA)\,$> 3$\,pix$^{-1}$ spaxels that are classified as PAS, with solid circle, stars, and hexagons showing galaxies classified as PASG, SFG, and HDSG, respectively. Both the cluster and GAMA regions show a well-defined red-sequence that is predominantly populated by PASGs, as well as a blue cloud that is dominated by SFGs. The HDSGs generally lie blueward of the red-sequence.

\section{Results}\label{results}
With the galaxy classifications at hand, we now focus on investigating the demographics of the HDSGs. Our primary aim here is to determine if there are any correlations with measures of environment that may indicate that external influences are responsible for the shut-down of star formation in these systems.  

\subsection{Comparison of HDSGs in the GAMA and cluster regions}\label{GAMA_vs_cluster}

As a first-order proxy for environment, we compare the fraction of HDSGs found in the GAMA and cluster regions. The GAMA regions are primarily comprised of galaxies that are either isolated or in groups with \logmtwo\,$\,<\,14$. The HDSGs are rare overall in both the GAMA and cluster regions of the SAMI-GS, making up only $1.2^{+0.6}_{-0.5}$\% (8/647) and $3.0^{+0.9}_{-0.6}$\% (17/575) of each sample, respectively. However, these fractions must be considered in light of the make-up of the galaxies in the two samples. The cluster sample is dominated by PASGs, which make up $80^{+2}_{-2}$\% (460/575) of the sample, while the GAMA sample is dominated by SFGs, which make up $59^{+2}_{-2}$\% (379/647) of the sample. If we consider the ``quenching efficiency'', similar to that defined by \citet{poggianti2009}, which measures the fractional contribution of HDSGs  to the population of galaxies that show evidence for recent star formation, i.e., $Q_{eff}= N_{\rm HDSG}/(N_{\rm SFG} + N_{\rm HDSG})$\footnote{We note that this definition for quenching efficiency differs from that used in other studies, where the excess of completely quenched galaxies is measured relative to the field \citep[e.g.,][]{peng2010, darvish2016, vanderberg2018}}, for the cluster regions we find $Q_{eff}=15^{+4}_{-3}\%$, which is significantly higher than the $Q_{eff}= 2.0^{+1.0}_{-0.4}\%$ found in the GAMA regions. All quoted uncertainties are 68 percent confidence intervals determined using the method described in \citet{cameron2011}. This result strongly indicates that the cluster environment is much more efficient at quenching star formation when compared with the lower density environments found in the GAMA regions.

Aside from the difference in the quenching efficiency, there are three striking differences between the HDSGs found in the cluster regions when compared with those found in the GAMA regions that point to a distinct origin for the two populations. First, the GAMA HDSGs are not associated with massive groups; only one HDSG
resides in a group with 6 members and \logmtwo\,$ \sim\, 13$. Of the remaining seven GAMA HDSGs, five have one or two neighboring galaxies within 100\,kpc and 100\,\kms, and two are completely isolated \citep[i.e., there are no associated neighbors in the GAMA group catalogues of ][]{robotham2011}. That the GAMA HDSGs are not associated with massive groups in the GAMA regions strongly suggests that they are not being quenched by processes associated with cluster- or group-scale environmental influences. 

\begin{figure*}
\includegraphics[width=.43\textwidth]{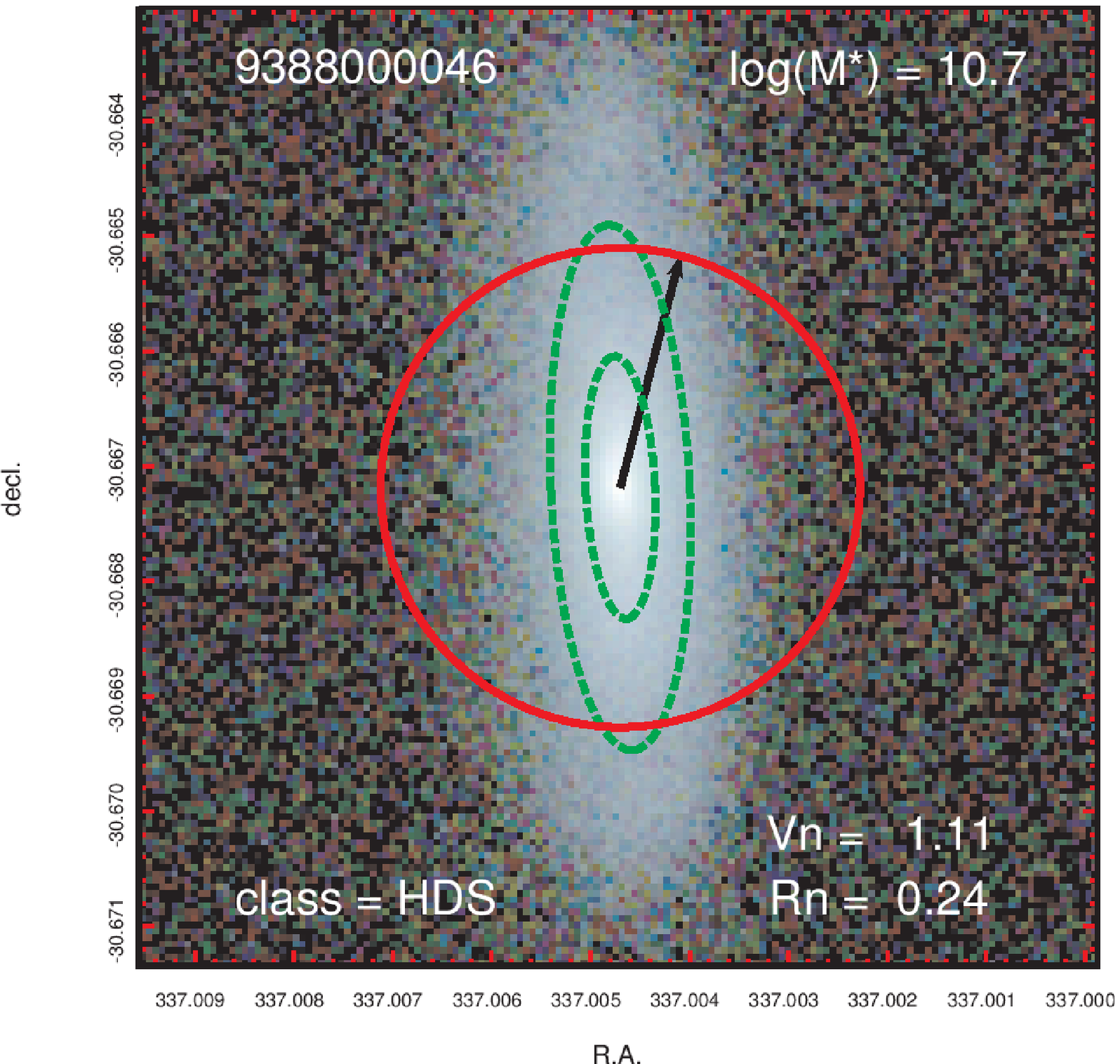}
\includegraphics[width=.55\textwidth]{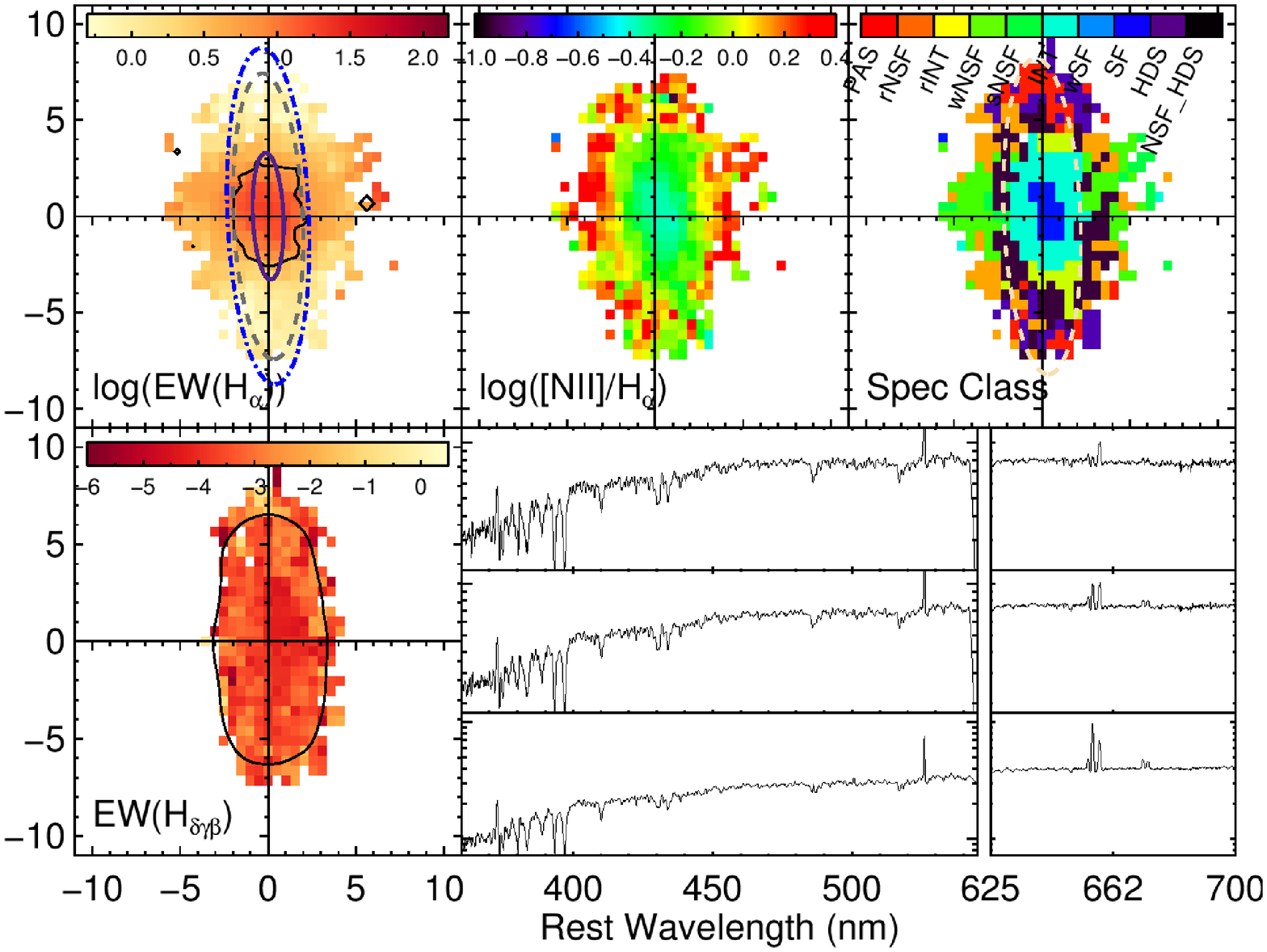}
\\
\includegraphics[width=.43\textwidth]{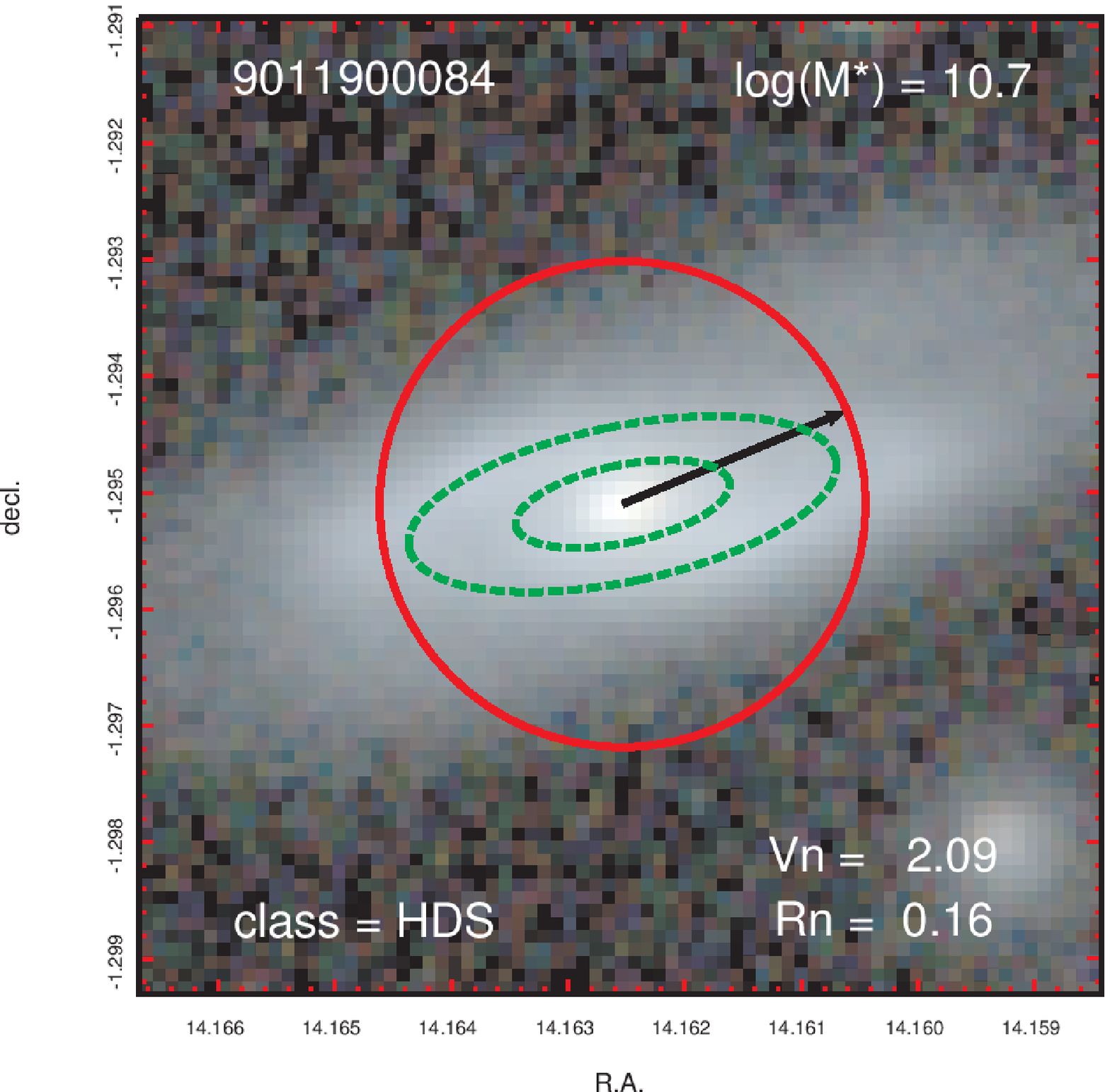}
\includegraphics[width=.55\textwidth]{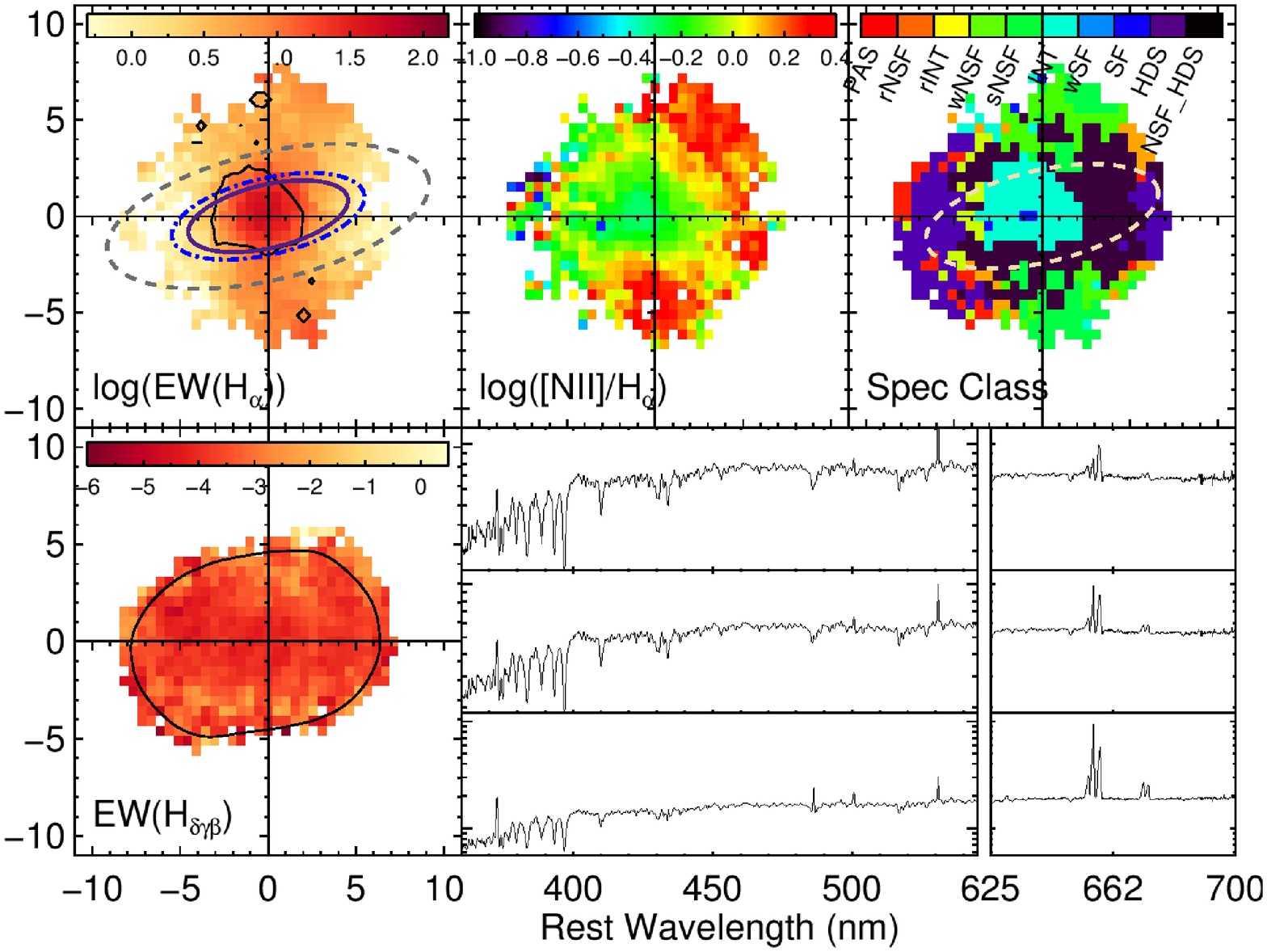}
\caption{The left-most panel shows the VST/ATLAS or SDSS $gri$ color image for the two most massive cluster HDSGs. The red circle shows the size of a SAMI hexabundle, the dashed green ellipses have major axis radii of 0.5$r_e$ and $r_e$, respectively, the black arrow points to the cluster center, and the SAMI-ID number for each galaxies is listed at the top left. The top row of the right-most small panels show the EW(\ha), \nii/\ha\, ratio and the spectroscopic classification maps, respectively. The bottom row of the right-most small panels shows the EW(\uberhdF) distribution, as well as three coadded spectra taken from the aperture defined by the smaller ellipse shown in in the left-panel (bottom spectrum), the annulus defined by the region between 0.5-1$r_e$ (middle spectrum), and all spaxels defined as HDS or NSF\_HDS shown in spectroscopic classification map (top spectrum). Overlaid on the EW(\ha) map are three ellipses with major axis equal to the stripping radius, $R_{strip}$, as defined by Equation~\ref{rstrip} for the three estimates of $P_{ram}$ described in Section~\ref{out_in_quenching}: $P_{ram}[R, v_{inf}(R)]$ (solid purple ellipse), $P_{ram}[R, v_{pec}]$ (dot-dashed blue ellipse) and $P_{ram}[0.5$\rtwo$]$. The position angle and ellipticity are set to the same values as those used to define the ellipses shown in the left-most panel. The black contours on the EW(\ha) map shows the extent of the star-forming emission where EW(\ha)$> 3$\AA. The 1$r_e$ ellipse is also shown as a dashed yellow line on the spectroscopic classification map. The remaining 15 cluster  HDSGs are shown in the Appendix.
\label{cluster_HDS_galaxies}}
\end{figure*}

\begin{figure*}
\includegraphics[width=.43\textwidth]{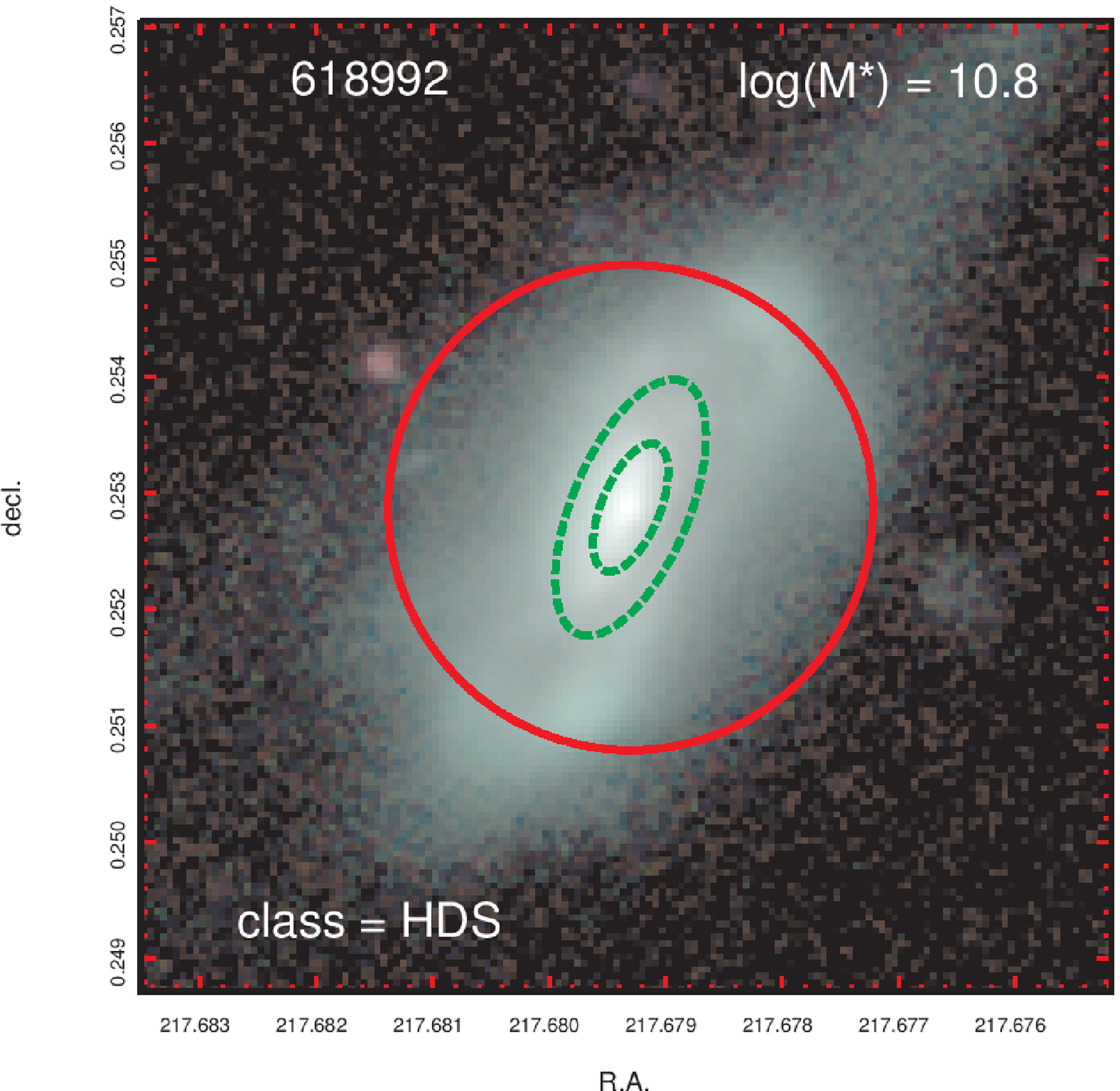}
\includegraphics[width=.55\textwidth]{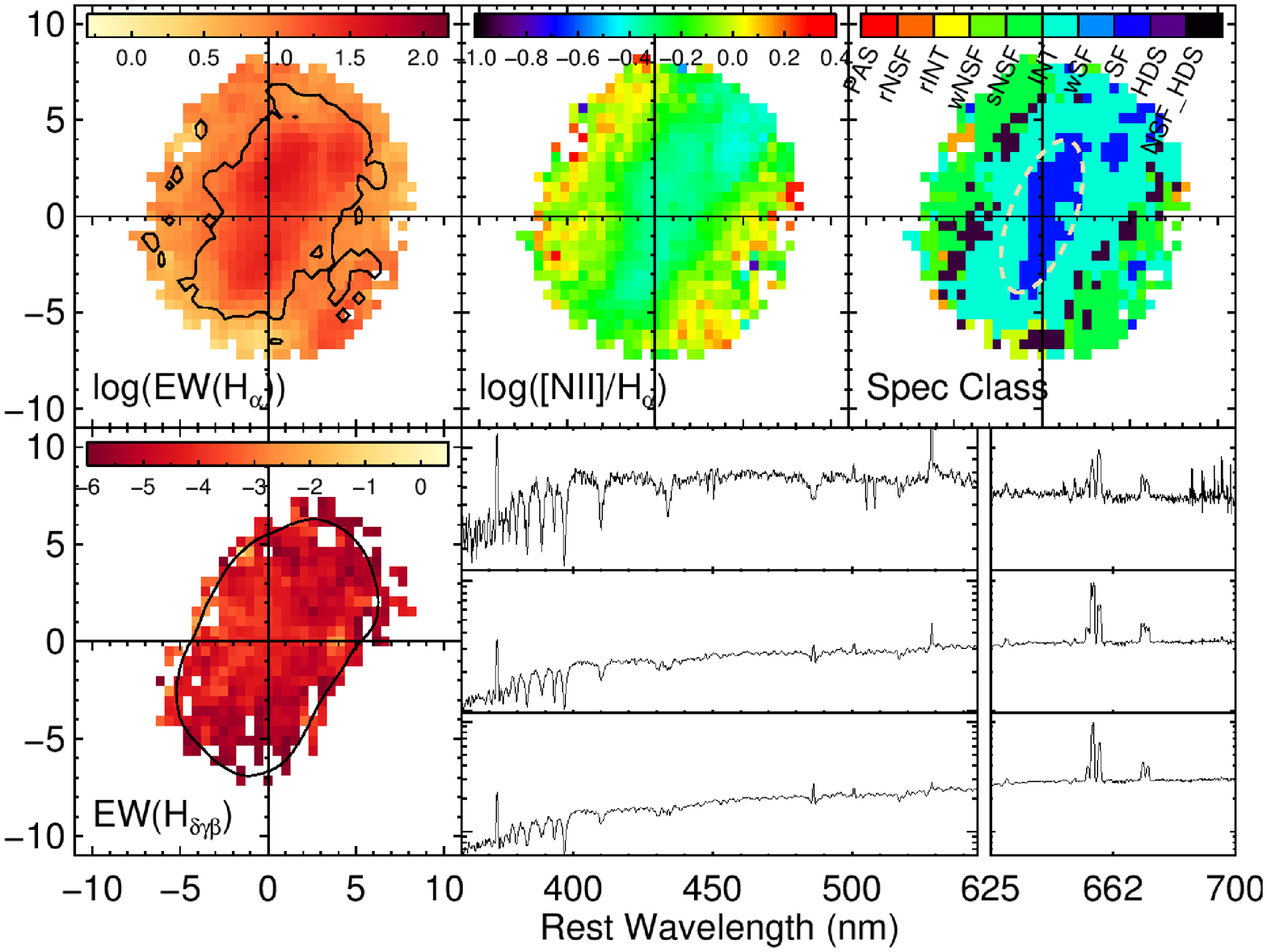}
\\
\includegraphics[width=.43\textwidth]{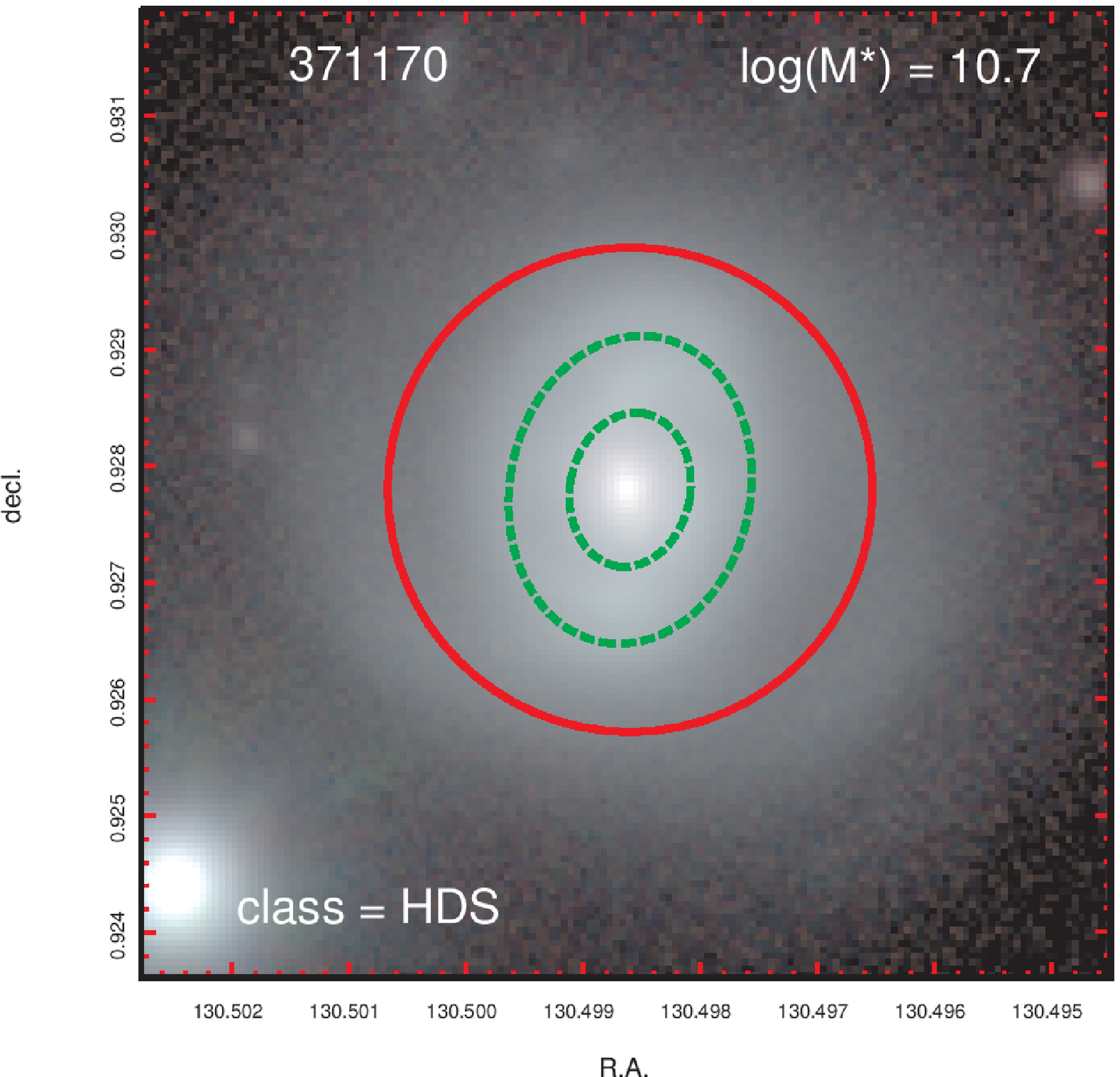}
\includegraphics[width=.55\textwidth]{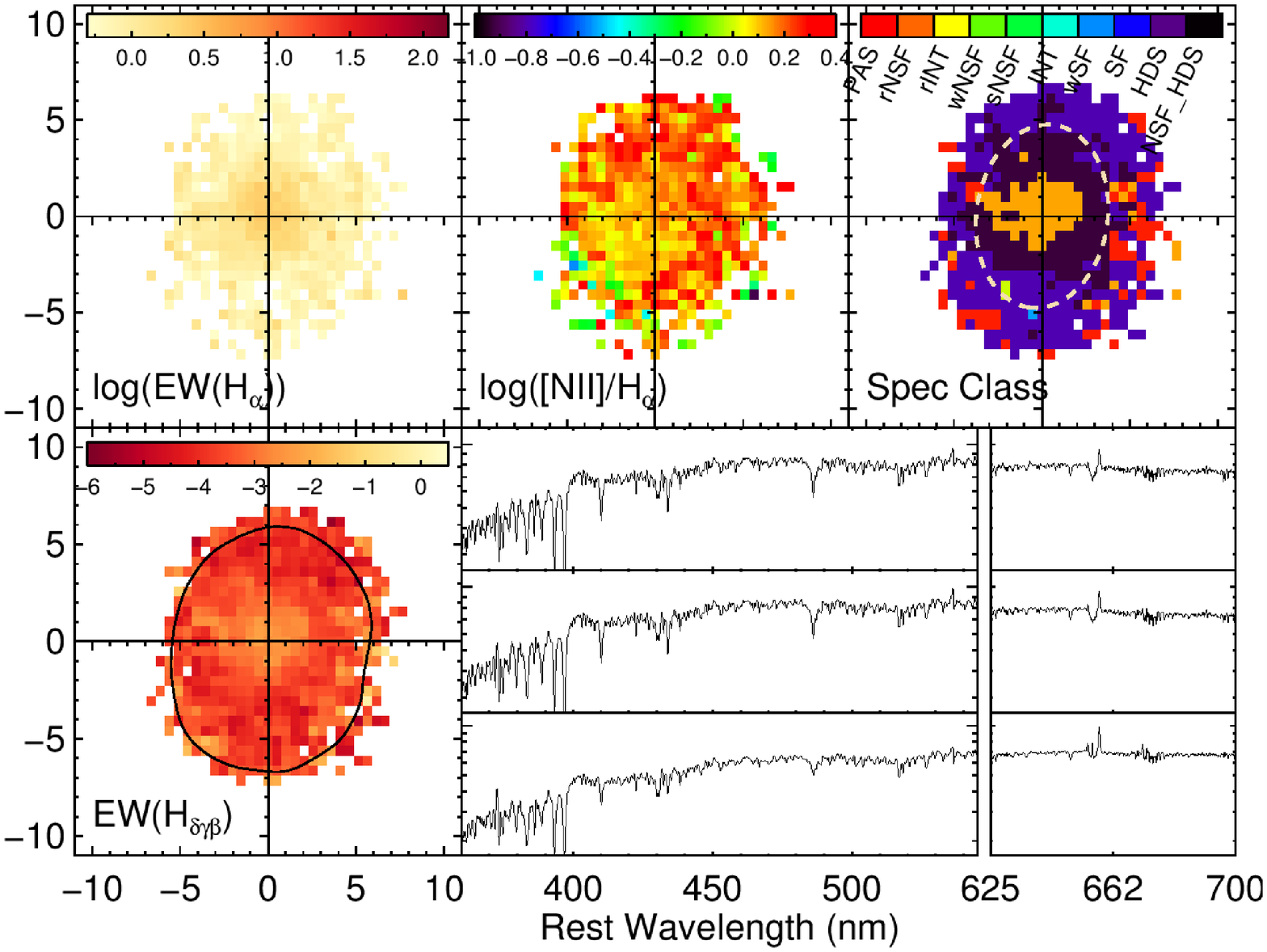}
\caption{The same as Figure~\ref{cluster_HDS_galaxies}, but for the two most massive HDSGs selected from the GAMA portion of the survey. The remaining six GAMA HDSGs are shown in the Appendix.
\label{GAMA_HDS_galaxies}}
\end{figure*}

Second, both the spatial distribution of the HDS regions and the nature of the emission lines differ when comparing the cluster and GAMA HDSGs. The differences are highlighted in Figures~\ref{cluster_HDS_galaxies} (clusters) and \ref{GAMA_HDS_galaxies} (GAMA), where the left-most panel shows the $gri$-band composite RGB images, the top row of the right-most panel show maps of the EW(\ha), log(\nii/\ha), and  spectroscopic classification (top row), and the bottom row of the right-most panel shows the map of the EW(\uberhdF), as well as three example spectra. The example spectra are formed from the co-addition of the spaxels within 0.5$r_e$ (bottom spectrum), 0.5-1$r_e$ (middle spectrum), and from those spaxels classified as HDS or NSF\_HDS (top spectrum). 

The spectroscopic classification maps in Figures~\ref{cluster_HDS_galaxies} reveal that more than half (9/17) of the HDSGs in the clusters harbour evidence for \ha\ emission due to ongoing star formation within the central 0.5$r_e$ of the galaxy. On the other hand, the spectroscopic classification maps in Figure~\ref{GAMA_HDS_galaxies} show that only one of the eight GAMA HDSGs has evidence for ongoing star formation in its centre. The emission associated with the other seven GAMA HDSGs often classified as being due to AGN or shock ionization that is not associated with ongoing star formation, similar to those described in \citet{alatalo2016}. Considering the distribution of the HDS regions, Figure~\ref{cluster_HDS_galaxies} shows that in 14/17 cluster HDSGs the HDS regions are found in the outer parts of the galaxy beyond 0.5$r_e$. Inspection of Figure~\ref{GAMA_HDS_galaxies} reveals that the HDS regions are far more evenly distributed throughout the GAMA HDSGs, where often the central 1$r_e$ is dominated by NSF\_HDS-classified spaxels. The fact that the cluster HDSGs often exhibit central star formation, with HDS regions found in the outer parts of the galaxies, indicates that their star formation is being quenched in an outside-in manner. Contrastingly, the more evenly spread HDS regions found in the GAMA HDSGs, coupled with the evidence for shock-like and AGN emission associated with the HDS regions, indicates that the quenching of star formation may be a galaxy-wide event.

\begin{figure*}
\includegraphics[width=0.48\textwidth]{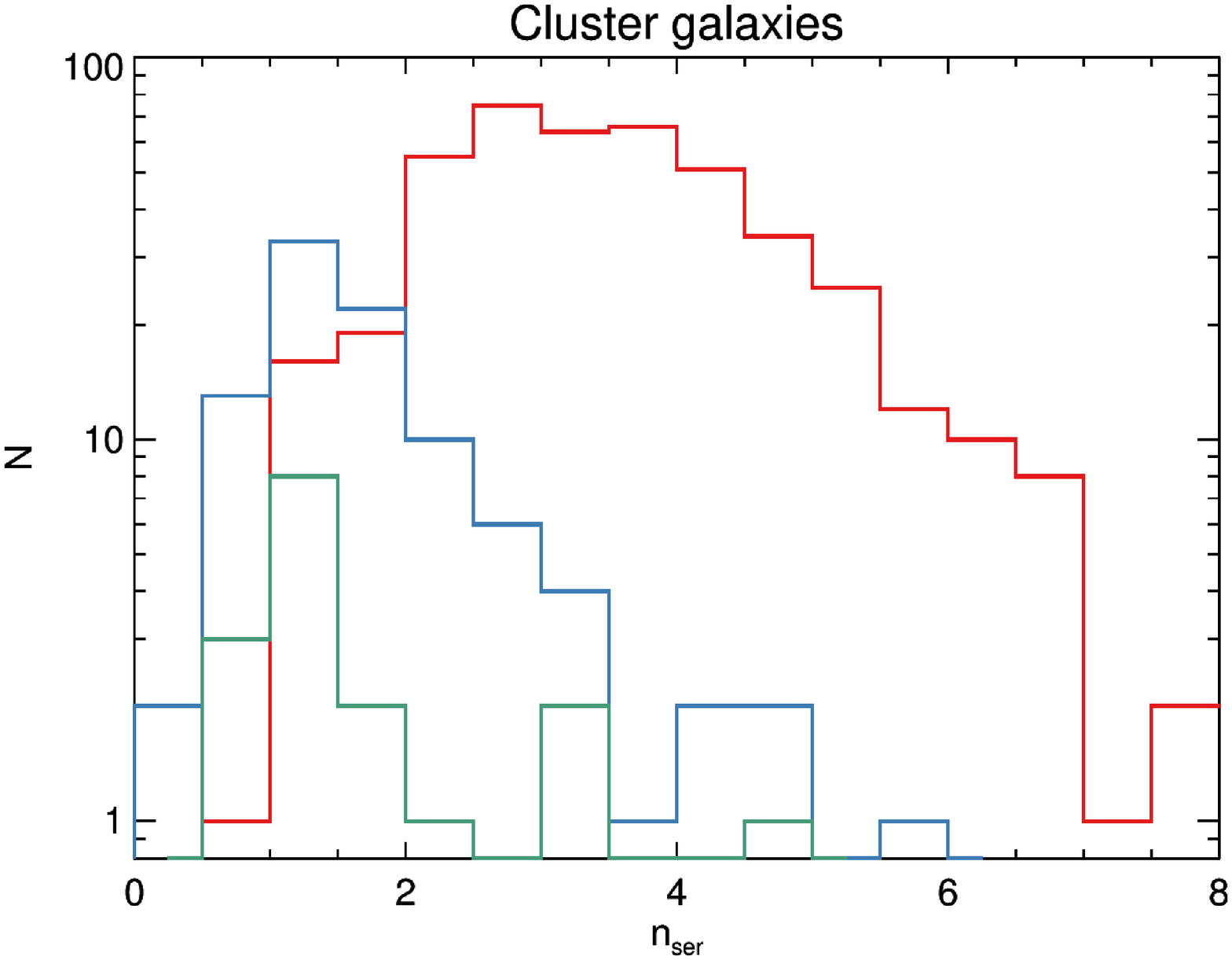}
\includegraphics[width=0.48\textwidth]{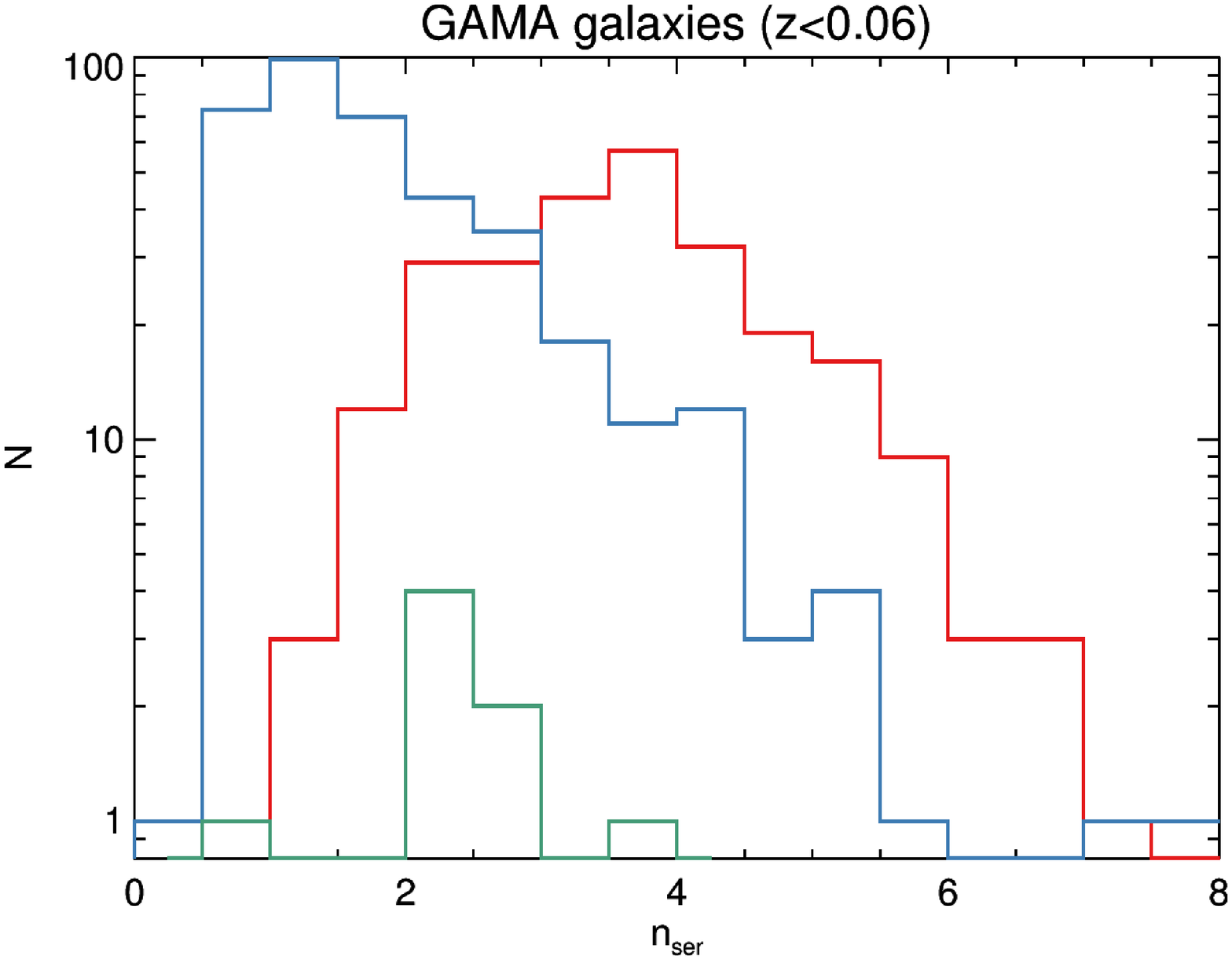}
\caption{The distribution of S{\'e}rsic index $n_{ser}$ for SAMI galaxies in the clusters (left-panel) and GAMA regions (right-panel) separated by galaxy class where the red, blue, and green histograms show the $n_{ser}$ distribution of the PASG, SFG, and HDSG samples, respectively. The majority of the cluster HDSGs have $n_{ser} <2$ and their distribution is statistically consistent with the $n_{ser}$ distribution of the cluster SGFs. The majority of the GAMA HDSGs have $n_{ser}>2$ and the KS-test indicates that the GAMA HDSG and SFG $n_{ser}$ distributions are not drawn from the same parent population. \label{nser_dist}}
\end{figure*}

Third, the structure of the GAMA HDSGs is different from that of the cluster HDSGs. Figure~\ref{nser_dist} shows the distribution of the S{\'e}rsic index, $n_{ser}$, for the cluster and GAMA galaxies divided into the three galaxy classes. For both the GAMA and cluster samples, the distribution of $n_{ser}$ for SFGs and PASGs peaks at $n_{ser}\simeq 1-1.5$ and $n_{ser}\simeq 3-4$, respectively. The distributions of $n_{ser}$ are consistent with the expectation that the SFGs are disk-dominated, while the PASGs are bulge-dominated.  Of the cluster HDSGs, $\sim 76$\% have $n_{ser}<2$ indicating that the majority of cluster HDSGs are disk-dominated. The $n_{ser}$ distribution for the cluster HDSGs is consistent with the cluster SFGs; a Kolmogorov-Smirnov (KS) test does not reject the null hypothesis that the two distributions are drawn from the same parent population, returning a probability $P=0.15$. On the other hand, the majority (7/8) of the GAMA HDSGs have $n_{ser} > 2$. On comparing the GAMA HDSG and SFG $n_{ser}$ distributions, the KS test returns a probability $P=0.006$, rejecting the null hypothesis that they are drawn from the same parent population. Directly comparing the $n_{ser}$ distributions of the cluster and GAMA HDSGs, the KS test returns $P=0.012$, indicating that the two distributions are unlikely to be drawn from the same parent distribution. The differences in the $n_{ser}$ distributions suggest that the GAMA HDSGs may harbor larger bulge-to-total fractions than their cluster counterparts.

The differences in the environments, spectral properties, and structure of the cluster and GAMA HDSGs indicate that the GAMA HDSGs are not being quenched in the same manner as the cluster HDSGs. While the GAMA HDSGs are an interesting subset of the HDSGs selected here, further detailed analysis of their properties is beyond the scope of this paper. We will instead analyse a larger sample of HDSGs drawn from the full SAMI-GS sample in a future paper. For the remainder of this paper, we will focus on investigating the environments and properties of the cluster HDSGs.

\subsection{Demographics of the cluster HDSGs}\label{cluster_HDS_demographics}

Having identified several significant differences between the cluster and GAMA HDSGs, we now focus on the cluster regions. Our aim here is to identify correlations with cluster-specific environment metrics in order to understand which, if any, environment-related quenching processes may be at play. In Sections~\ref{radial_distribution}, \ref{vpec_distribution}, and \ref{PPS} we investigate the variations in the radial, velocity, and projected phase space (PPS) distributions for the PASGs, SFGs and HDSGs. Because the HDSG sample is relatively small, we produce an ensemble cluster by stacking the normalized coordinates $R/R_{200}$ and $v_{pec}/\sigma_{200}$ across the eight SAMI-GS clusters.

\subsubsection{Star-forming properties}\label{SF_properties}

\begin{figure}
\includegraphics[width=0.48\textwidth]{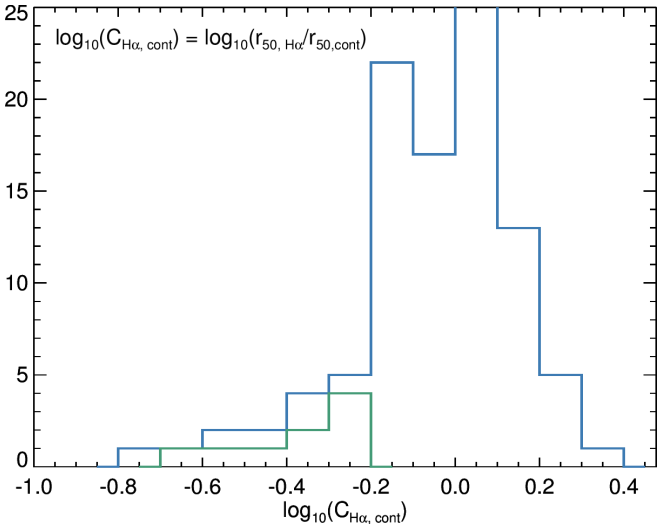}
\caption{The concentration of H$_\alpha$ flux relative to continuum flux, $C_{{\rm H}\alpha,cont}$ for the cluster HDSGs with central star formation (green histogram). Also shown is the distribution for the cluster SFGs. The star formation in the HDSGs is much more concentrated when compared with the SFGs, indicating that the cluster HDSGs are being quenched from  the outside-in. \label{ha_conc}}
\end{figure}

We noted in Section~\ref{GAMA_vs_cluster} that many of the cluster HDSGs show evidence of ongoing star formation at their centers, implying that the star formation in the cluster HDSGs is being quenched in an outside-in fashion. In Figure~\ref{ha_conc}, we quantify this outside-in quenching by showing the distribution of the concentration of \ha\, flux relative to the continuum, $C_{H\alpha,cont}$, for the cluster HDSGs (green histogram) with central star formation, along with the cluster SFGs (blue histogram). The $C_{H\alpha,cont}$ values are determined in a similar fashion to that described in \citet{schaefer2017}. Briefly, we measure the cumulative flux profile in elliptical apertures centered on each galaxy using the ellipticity and position angle determined by the S{\'e}rsic fits to the $r$-band data described in Section~\ref{cluster_sersic}. For both the \ha\, and continuum flux (where the continuum flux level is determined in emission-free bands surrounding the H$_\alpha$ line), the radius containing 50\% of the flux, $r_{50, H\alpha}$ and $r_{50, cont}$, respectively, is measured and the concentration is determined as $C_{\rm H\alpha,cont}=r_{50, {\rm H}\alpha}/r_{50, cont}$. Note that in determining cumulative flux used to measure $r_{50, {\rm H}\alpha}$, only spaxels that are classified as INT, SF, or wSF are included. Thus, H$_\alpha$ flux that is due to non-starforming ionisation processes is not included in the $C_{{\rm H}\alpha}$ measurement. Figure~\ref{ha_conc} shows that $C_{{\rm H}\alpha,cont} < 1$ for all HDSGs with central star formation, and that their $C_{{\rm H}\alpha,cont}$ values are much lower when compared with the majority of the SFGs. A KS test returns $P \ll 0.001$, thereby rejecting the hypothesis that the HDSG and SFG $C_{{\rm H}\alpha,cont}$ distributions are drawn from the same parent distribution. We note that while the majority (68\%) of the cluster SFGs show evidence of ongoing star formation at their centers, a substantial fraction do not. We therefore repeated the comparison between the HDSG and SFG $C_{{\rm H}\alpha,cont}$ distributions including only those SFGs with central star formation, finding that our main result remains unchanged.

\begin{figure}
\includegraphics[width=0.48\textwidth]{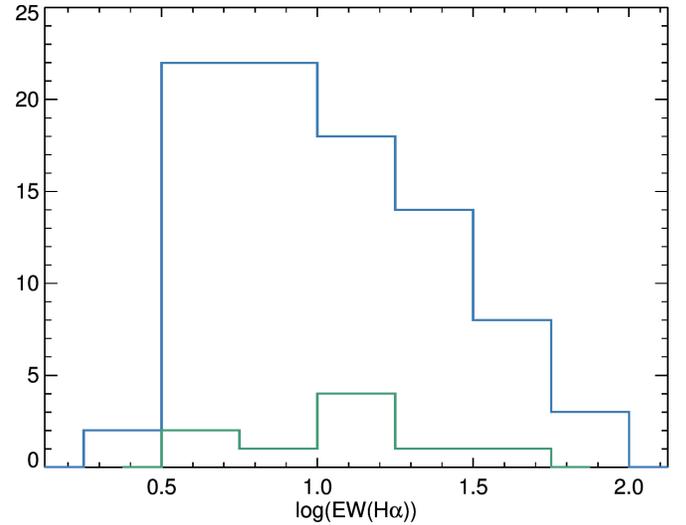}
\caption{The distribution of the median emission-line EW(H$\alpha$) measured within 0.5$r_e$ for the cluster HDSGs with central star formation (green histogram) and the cluster SFGs (blue histogram). The distributions do not differ significantly, indicating that the mode of star formation in the centres of the HDSGs and SFGs is not different. \label{ha_EW}}
\end{figure}

Many of the environmental processes introduced in Section~\ref{intro} predict enhanced star formation at the centers of affected galaxies, which may in turn lead to the more concentrated H$\alpha$ flux revealed for the HDSGs in Figure~\ref{ha_conc}. We test for evidence of central starbursts in Figure~\ref{ha_EW} where we show the distribution of the median EW(\ha) of the spaxels within 0.5$r_e$ for each of the HDSGs with central star formation, as well as the cluster SFGs. Again, only spaxels that are classified as SF, wSF or INT are used in determining the median EW(\ha). We find no significant difference when comparing the EW(H$\alpha$) distribution for the HDSGs and SFGs; a KS test does not reject the hypothesis that the two distributions are drawn from the same parent population, returning a probability $P=0.68$. This similarity in the EW(\ha) distributions indicates that, while the spatial distribution of star formation in a large portion of the HDSGs is more concentrated than that seen in the cluster SFGs, the mode of star formation does not appear to be dramatically different.

\subsubsection{Projected cluster-centric distance}\label{radial_distribution}

\begin{figure*}
\includegraphics[width=0.48\textwidth]{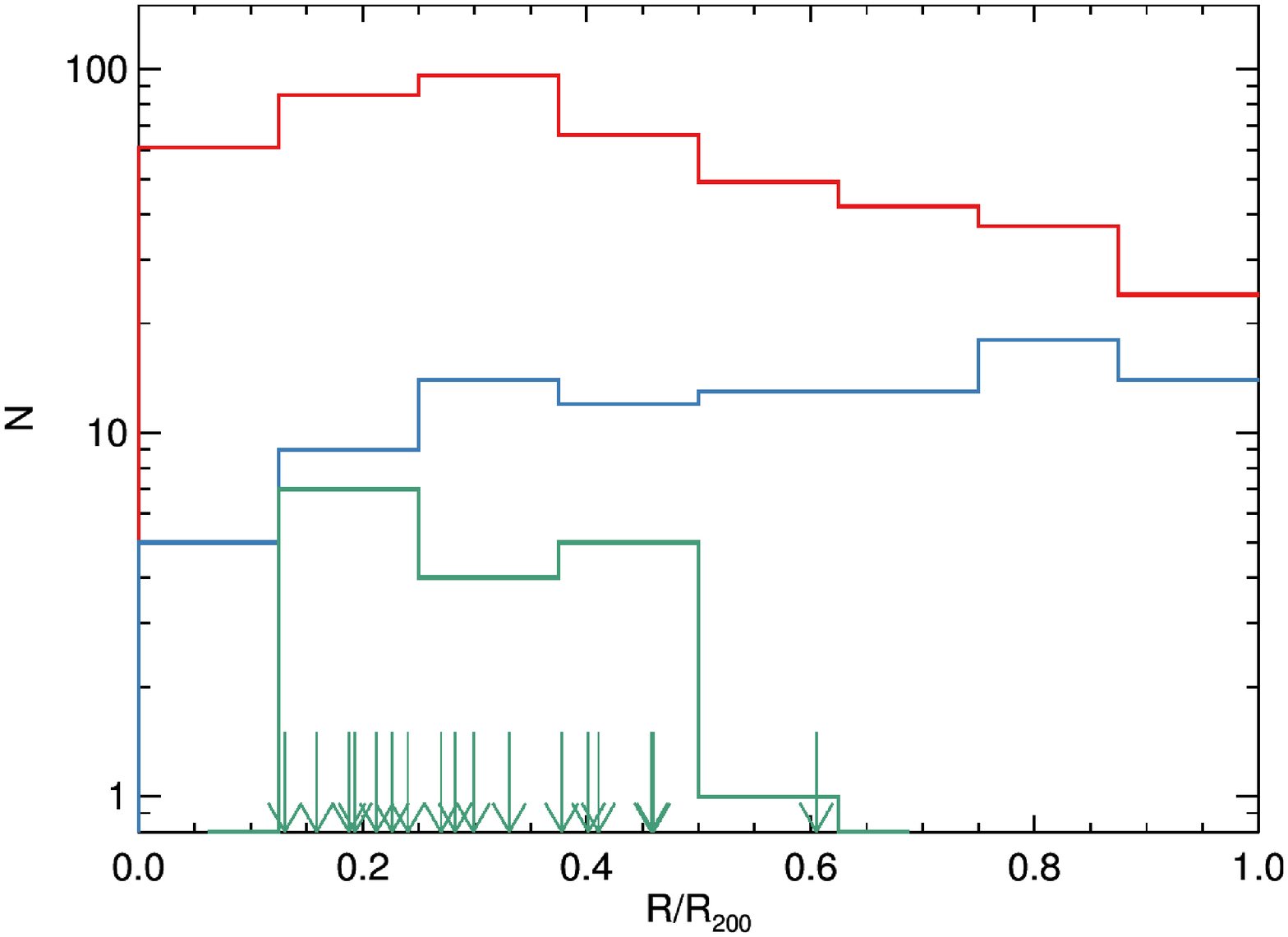}
\includegraphics[width=0.48\textwidth]{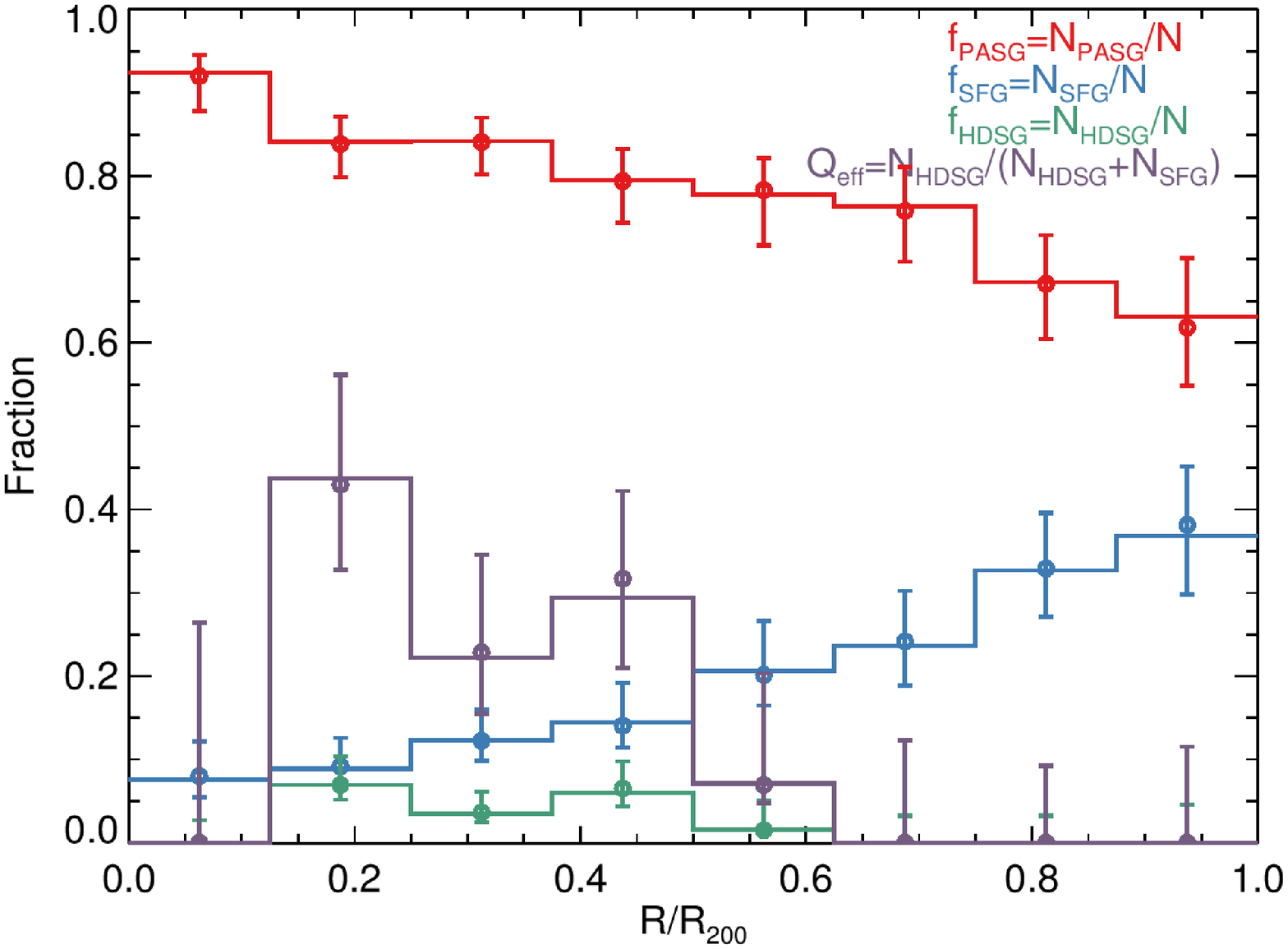}
\caption{The left panel shows the binned distribution of cluster galaxies with different spectral types as a function of $R/$\rtwo. The positions of the HDSGs are indicated by green arrows. The histograms in the right panel shows the corresponding binned fractions for each of the PASG, SFG, and HDSG samples. For both panels the PASGs, SFGs, and HDSGs are shown as red, blue and green lines, respectively. In the right panel, the circles show the completeness-corrected fractions, which do not differ significantly from the non-corrected fractions (shown as histograms). The purple line in the right-panel shows the quenching efficiency, $Q_{eff}$, defined as the fraction of HDSGs relative to the total (SFGs+HDSGs) active galaxies. The HDSGs all have $R/$\rtwo$<0.6$, and are more concentrated toward the cluster centers than the cluster SFG distribution. \label{frac_rad}}
\end{figure*}

It is well established that the fraction of passive cluster galaxies increases with decreasing cluster-centric distance, while the fraction of star-forming galaxies increases with increasing clustercentric distance \citep{lewis2002, vonderlinden2010, haines2015, barsanti2018}. More recently, \citet{paccagnella2017} have found that the fraction of \hd-strong galaxies increases by a factor of $\sim 1.7$ going from the outskirts to the centre of low-redshift clusters, although their selection is based on single-fibre spectroscopy. The left and right panels of Figure~\ref{frac_rad} show, respectively, the distribution and fractions of the PASGs, SFGs, and HDSGs (red, blue, and green lines, respectively) as a function of normalized cluster-centric distance. The normalized projected cluster-centric distances, $R/$\rtwo, are measured from the cluster centres listed in Table~1 of \citet{owers2017}. The corresponding 68 percent confidence intervals shown in the right panel of Figure~\ref{frac_rad} were calculated per the method described by \citet{cameron2011}. The fractions shown as histograms in Figure~\ref{frac_rad} are not corrected for the radial- and stellar-mass-dependent incompleteness of the sample. 
The completeness-corrected fractions are shown as open circles and are calculated by determining a weighting for each galaxy in the sample that accounts for the radius- and stellar-mass-dependent completeness. The corrected fractions do not differ significantly from the uncorrected values. 

We find that {\it the vast majority} (16/17) of the HDSGs are within the radial range $0.125 < R/$\rtwo$ < 0.5$, where $f_{\rm HDSG}\sim 5$\%; only galaxy 9008500492 has $R/$\rtwo$=0.6$. Inspection of the spectrum derived by stacking the HDS regions shown in Figure~\ref{cluster_HDS_galaxies} indicates that 9008500492 has only relatively weak Balmer absorption and is the least convincing HDSG in our sample; only 23/199 spaxels are classified as HDS or NSF\_HDS. In Section~\ref{GAMA_vs_cluster} we found that $Q_{eff}=2.0$\% for the GAMA portion of the survey, which is assumed to be representative of the field population that is accreted onto clusters. Given this, we may expect 1-2 star-forming galaxies that are falling into the clusters to be undergoing similar quenching to that observed in the GAMA HDSGs. This may explain the relatively large projected clustercentric distance of 9008500492.

The small number of HDSGs found outside 0.5\rtwo\, indicates that the true 3D location of the HDSGs is also within $0.5 r_{200,3D}$, and not due to projection effects. The left panel of Figure~\ref{frac_rad} shows that the number of PASGs and SFGs increase with radius to $R$/\rtwo$=0.3$, but show differing behaviour thereafter with the number of PASGs declining with radius, and the number of SFGs remaining relatively flat with increasing radius. Both the two-sample KS and Anderson-Darling (AD) tests return a probability $P \ll 0.001$, strongly rejecting the hypothesis that the differences in the cumulative distribution functions of the radii of the PASGs and SFGs can occur by random chance if the two samples were drawn from the same parent distribution. Similarly, both the AD and KS tests return $P \ll 0.001$ for the comparison between the HDSG and SFG radial distributions. The comparison between the HDSG and PASG returns $P=0.06$ and $P=0.08$ for the KS and AD tests, respectively, indicating that we cannot reject the hypothesis that the two radial distributions are drawn from the same parent population. We note that repeating the comparisons between the PASG and HDSG $R/R_{200}$ distributions after removing galaxy 9008500492 from the HDSG sample returns $P=0.019$ and $P=0.043$ for the KS and AD tests, respectively. These tests confirm what can be deduced by inspection of the left panel of Figure~\ref{frac_rad}: the HDSG sample is significantly more concentrated towards the cluster centers than the SFG sample and does not appear to follow the same radial distribution as the PASG sample, although the latter result is not as statistically robust as the former.

Figure~\ref{frac_rad} also shows that the quenching efficiency, $Q_{eff}$, (purple line; defined as in Section~\ref{GAMA_vs_cluster}) is largest in the $R/$\rtwo $\simeq 0.2$ bin, where it is 44\%, and decreases to $\sim 20-30$\% in the next two larger radial bins. 
However, we note that the 68 percent confidence intervals overlap between the bins with $R/$\rtwo$<0.5$, so the increase is not statistically significant. Both the PASG and SFG fractions follow the expected radial trends, with the PASG fraction declining from $f_{\rm PASG}=92$\% at $R/$\rtwo $< 0.125$ to $f_{\rm PASG}=63$\% in the $R/$\rtwo=0.875-1 bin, and the SFG fraction increasing from $f_{\rm SFG}=8$\% at $R/$\rtwo $< 0.125$ to $f_{\rm SFG}= 37$\% in the $R/$\rtwo\,=0.875-1 bin.

\subsubsection{Velocity Distribution}\label{vpec_distribution}

\begin{figure}
\includegraphics[width=0.45\textwidth]{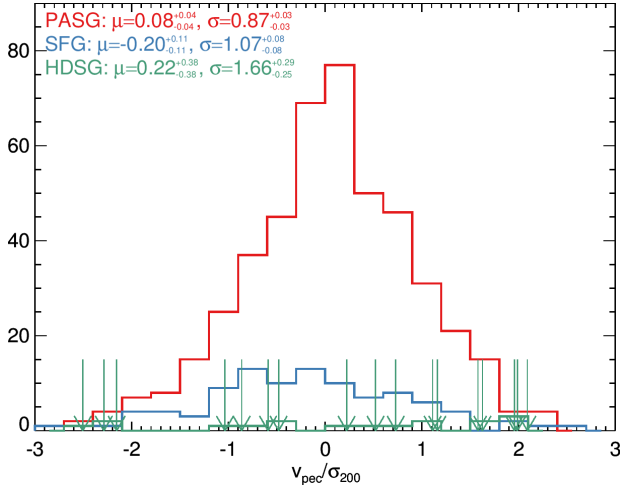}
\caption{The $v_{\rm pec}/\sigma_{200}$ distribution of the PASGs (red), SFGs (blue) and HDSGs (green). The locations of the HDSGs are highlighted by green arrows. The biweight estimates for the mean, $\mu$, and standard deviation, $\sigma$, are shown for each sample at the top left of the panel. \label{vpec_hist}}
\end{figure}

Figure~\ref{vpec_hist} shows the relative LOS velocity distribution, $v_{\rm pec}/\sigma_{200}$, for the PASG (red), SFG (blue) and HDSG (green) samples. In the upper left of Figure~\ref{vpec_hist}, for each sample we list the mean, $\mu$, and standard deviation, $\sigma$, determined using biweight estimators \citep{beers1990}, and their associated uncertainties, which are determined using jack-knife resampling. The distributions do not appear to depart significantly from a Gaussian shape; both the KS and AD tests fail to reject the hypothesis that any of the PASG, SFG or HDSG  distributions are drawn from a Gaussian parent distributions. 

Velocity segregation is readily observed between the passive and star-forming cluster populations, regardless of the proxy used to distinguish the two populations \citep[e.g., morphology, color or the presence/absence of emission lines;][]{colless1996, biviano1997, barsanti2016}. This velocity segregation is attributed to the different dynamical states of the passive and star-forming galaxies. The passive galaxies form the bulk of the virialized cluster population, whereas the star-forming spirals are the dominant population of infalling galaxies and follow more radial orbits \citep{biviano2004}. We also find notable differences when comparing the PASG with both the SFG and HDSG $v_{\rm pec}/\sigma_{200}$ distributions. The $\sigma$ values measured for the SFG and HDSG samples indicate that they have larger dispersions than the PASGs. This is particularly true for the HDSGs compared with the PASGs, where the measured uncertainties indicate that the difference in $\sigma$ is significant with $99.9$\% confidence, while the SFG $\sigma$ is larger than that of the PASG with $>95$\% confidence. The measured $\mu$ for the HDSG and SFG samples are offset from the PASG value, although the uncertainties indicate that the differences are not statistically significant. 

We note that the dispersion of the PASGs is $\sigma = 0.87^{+0.03}_{-0.03}\,\sigma_{200}$, which implies that their dispersion is significantly lower than that expected of a virialised population. This lower dispersion is likely due to the different samples used to define the $\sigma_{200}$ values determined by \citet{owers2017}, which used all spectroscopically confirmed cluster members within \rtwo, and therefore includes a contribution from a larger fraction of less massive, star-forming galaxies. Their inclusion leads to an overall higher estimate of $\sigma_{200}$ and, therefore, lower normalized velocities; indeed the dispersion measure using all SAMI-GS cluster galaxies is $\sigma = 0.94^{+0.03}_{-0.03}$.

Comparing the samples against each other using the two-sample KS test, the hypothesis that the PASG and SFG $v_{\rm pec}/\sigma_{200}$ distributions are drawn from the same parent distribution is rejected ($P < 0.01$), but the KS test fails to reject this hypothesis for the HDSG versus SFG and HDSG versus PASG comparisons. The KS test is most sensitive to differences that occur around the centres of the distributions whereas the AD test has better sensitivity to departures in the tails of two distributions. Using the two-sample AD test, we find that we can reject the hypothesis that the PASG, HDSG, and SFG $v_{\rm pec}/\sigma_{200}$ distributions are drawn from the same parent distribution ($P=0.009$, $P=0.04$, and $P=0.005$ for the HDSG versus PASG, HDSG versus SFG, and SFG versus PASG, respectively). Taking the results of the AD tests in concert with their different $\sigma$ values, it is likely that the SFG, HDSG, and PASG samples are kinematically distinct from each other.

\subsubsection{Projected Phase Space distribution}\label{PPS}

In Sections~\ref{radial_distribution} and \ref{vpec_distribution}, we showed that there are differences in both the radial and velocity distributions of the PASG, HDSG, and SFG samples. While these differences in radial and velocity distributions are informative when considered separately, the combination of line-of-sight velocity and radius can be a more powerful discriminant of populations with different accretion histories. 
Simulations of clusters reveal that infalling and recently accreted galaxies occupy projected-phase-space (PPS) regions that are relatively distinct from the virialized population even when knowledge of the full 6D phase-space information is confused by projection effects \citep{gill2005, oman2013, haines2015, rhee2017}. Therefore, the PPS is recognised as a powerful diagnostic for understanding quenching in clusters \citep{biviano1997,solanes2001, vollmer2001, mahajan2011, noble2013, muzzin2014, jaffe2015, oman2016, barsanti2018}.

\begin{figure*}

\includegraphics[width=0.95\textwidth]{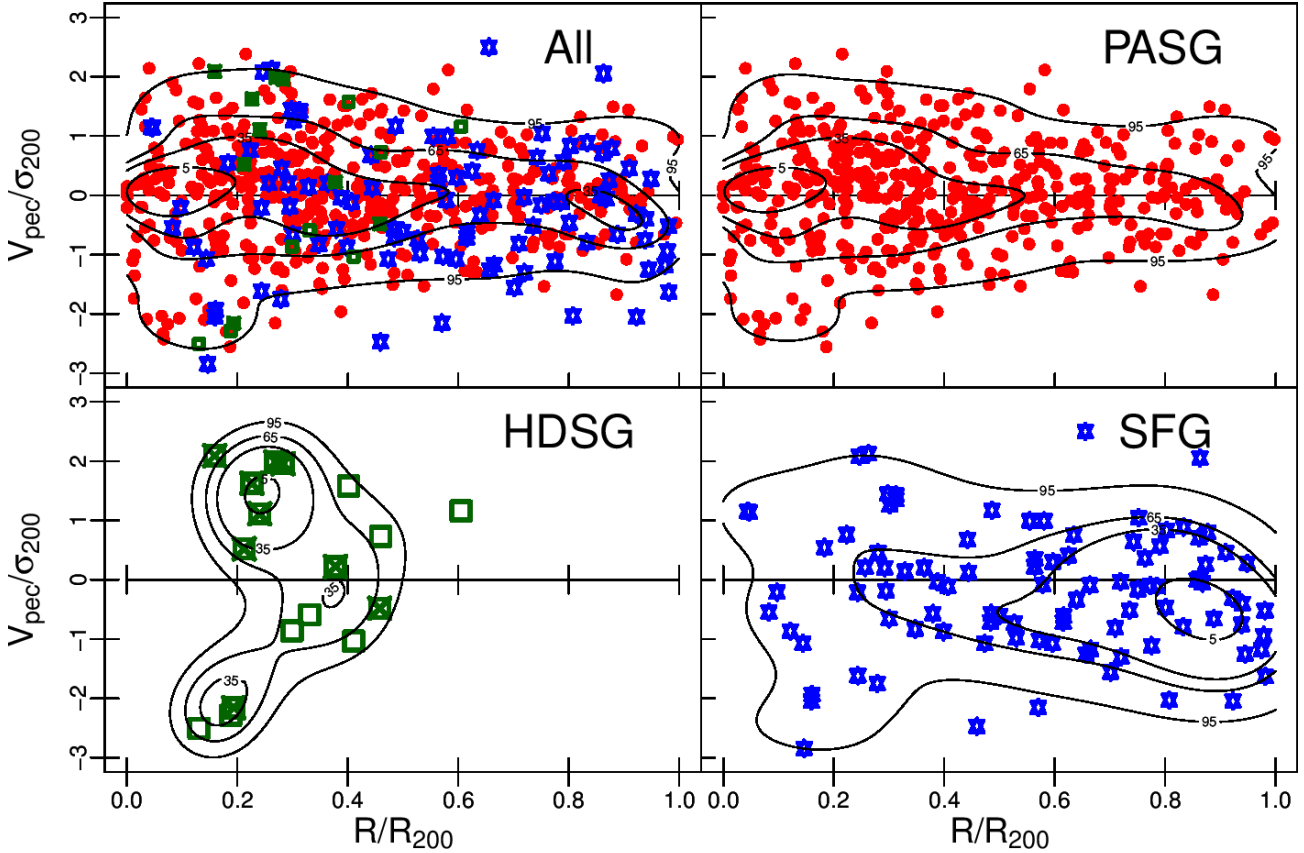}

\caption{The projected-phase space distribution of SAMI-GS cluster galaxies. The red circles, blue stars and green squares show the positions of PASGs, SFGs, and HDSGs, respectively. HDSGs with central star formation are highlighted with green crosses. The black contours show the smoothed density distribution corresponding to the sample named in the upper right of each plot. The HDSGs occupy a coherent, arc-shaped region in the PPS that is very different to the distribution of SFGs.\label{PPS_plot}}
\end{figure*}

In the top left panel of Figure~\ref{PPS_plot}, we show the stacked PPS for SAMI galaxies in the eight clusters. 
Each of the PASGs, SFGs, and HDSGs are plotted as filled red circles, blue stars, and green squares, respectively. The top right, bottom left, and bottom right panels in Figure~\ref{PPS_plot} show, separately, the PPS distributions for the PASGs, HDSGs, and SFGs, respectively. Those HDSGs that show evidence for star formation at their centers are highlighted by green crosses. In each of the four panels, contours are generated from the smoothed kernel density estimate (KDE) for the distribution of points, which is determined using the {\sf ks} software\footnote{https://cran.r-project.org/web/packages/ks/index.html} \citep{duong2007}.

Figure~\ref{PPS_plot} shows that the distribution of HDSGs at small radius is strikingly different from both the PASGs and SFGs. The distribution of HDSGs forms an arc-like shape in PPS, with the smallest projected radius (R/\rtwo $\lesssim 0.3$) having large velocities $|v_{pec}|/\sigma_{200} > 1.0$ whereas those with R/\rtwo\, $\gtrsim 0.3$ have lower velocities $|v_{pec}|/\sigma_{200} <1.0$. The HDSGs are absent from the low velocity and small radius ($R/$\rtwo$<0.2$) part of the PPS, which is dominated by PASGs. Interestingly, the HDSGs that do not have evidence for star formation at their centers primarily occupy the small velocity, large radius region of the arc-like shape in PPS. The KDE contours reveal that the PASGs and SFGs exhibit a mirror-flipped distribution about the $R/R_{200}$ axis: the density of SFGs increases with radius and the velocity spread stays relatively constant, whereas the density and velocity spread of PASGs decrease with radius.

We test the significance of the difference observed in the PPS distributions for the PASG, SFG, and HDSG samples using the multivariate two-sample KDE test developed by \citet{duong2012} for the purpose of comparing cell morphologies, and recently applied to PPS distributions by \citet{lopes2017} and \citet{decarvalho2017}. The KDE test is nonparametric and uses the integrated square error as a measure of the discrepancy between two KDEs to test the hypothesis that two distributons are drawn from the same underlying density distribution \citep[see][for details]{duong2012}. The test returns $P \ll 0.01$ for both of the PASG versus SFG and SFG versus HDSG comparisons, rejecting the hypothesis that these distributions are drawn from the same underlying density distribution. Similarly, a two-sample 2D KS-test \citep{peacock1983} returns $P \ll 0.001$ for these two comparisons. The comparison of the PASG and HDSG distributions yields less significant results, with the KDE test returning $P=0.02$ and the 2D KS-test returning $P=0.12$. Removing galaxy 9008500492 from the sample and rerunning the KDE and KS-tests return $P=0.01$ and $P=0.08$, respectively. We also use the 2D KS-test to compare the three PPS distributions using the absolute value of the velocity, $|v_{pec}|/\sigma_{200}$. For each of the three comparisons, the 2D KS-test returns $P \leq 0.001$.  Therefore, we can confidently conclude that the SFG distribution is significantly distinct from both the PASG and HDSG distributions, while the HDSG and PASG distributions appear distinct, but with lower significance.

\begin{deluxetable*}{cccccccc}
\tabletypesize{\scriptsize}
\tablecaption{Fraction of spectral types in various PPS regions\label{PPS_table}}
\tablewidth{0pt}
\tablehead{
\colhead{$R/R_{200}$} & \colhead{$|V_{pec}|/\sigma_{200}$} & \colhead{N$_{\rm SAMI}$} & \colhead{Completeness} & \colhead{$f_{\rm PASG}$} & \colhead{$f_{\rm SFG}$} & \colhead{$f_{\rm HDSG}$} & \colhead{$Q_{eff}$}}
\startdata
0.0$-$0.5&0.0- 1.5&309 & 89$_{- 2}^{+ 1} $& 87$_{- 2}^{+ 2} $& 10$_{- 1}^{+ 2} $&  3$_{- 1}^{+ 1} $& 20$_{- 5}^{+ 8}$ \\
0.0$-$0.5&1.5- 3.0& 55 & 81$_{- 6}^{+ 4} $& 71$_{- 7}^{+ 5} $& 15$_{- 4}^{+ 6} $& 15$_{- 4}^{+ 6} $& 50$_{-12}^{+12}$ \\
0.5$-$1.0&0.0- 1.5&199 & 83$_{- 3}^{+ 2} $& 74$_{- 3}^{+ 3} $& 26$_{- 3}^{+ 3} $&  1$_{- 0}^{+ 1} $&  2$_{- 1}^{+ 4}$ \\
0.5$-$1.0&1.5- 3.0& 12 & 75$_{-13}^{+ 8} $& 42$_{-12}^{+14} $& 58$_{-14}^{+12} $&  0$_{- 0}^{+13} $&  0$_{- 0}^{+21}$ 
\enddata


\end{deluxetable*}

In Table~\ref{PPS_table}, we show the fractions for the different galaxy types as a function of position in PPS. The PPS distributions are relatively symmetric about the velocity axis, which means that the absolute velocity, $|v_{pec}|/\sigma_{200}$, can be used to compare the PPS region, thereby enhancing the number of objects in each region without losing information. We divide the PPS into quadrants with boundaries listed in Table~\ref{PPS_table}. Eight out of 17 HDSGs are found in the $|v_{pec}|/\sigma_{200} > 1.5$, R/\rtwo\, $< 0.5$ quadrant, and in this region the fractional contribution of the HDSGs to the total active (SFG+HDSG) galaxy samples is 15\% and 50\%, respectively, which is significantly higher than elsewhere in the PPS. The differences in the PPS distributions for the HDSGs, PASGs, and SFGs indicates that the three populations are at different stages in their accretion histories.

\subsubsection{Dependence on cluster mass}

\begin{figure}
\includegraphics[width=0.48\textwidth]{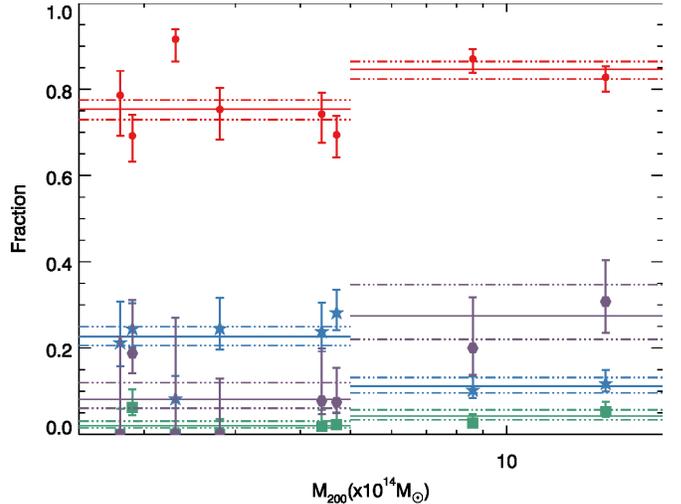}
\caption{The fraction of PASGs, SFGs, and HDSGs as a function of cluster mass shown for the individual clusters as red circles, blue stars, and green squares, respectively. The quenching efficiencies ($Q_{eff} =  N_{\rm HDSG}/(N_{\rm HDSG}+N_{SFG})$) for each cluster are shown as purple hexagons. The solid lines show the fractions of PASGs, SFGs, HDSGs, and $Q_{eff}$ in two mass bins for clusters with $14 < $\logmstar$ < 14.7$ and \logmstar $\geq 14.7$ with the same color scheme as for the individual points. Dash-dotted lines show the corresponding 68 percent confidence intervals. This suggests that clusters with \logmtwo$ > 14.7$ have a higher $Q_{eff}$ when compared with lower mass clusters, although cluster-to-cluster variations may affect these results. \label{mass_fracs}}
\end{figure}

We note that the majority (11/17) of the cluster HDSGs are found in the two most massive clusters in the sample, A119 (N$_{\rm HDSG}=3$) and A85 (N$_{\rm HDSG}=8$). This majority may be expected given that these two clusters account for a large fraction of the total number of galaxies in the sample (119 and 150 galaxies for A119 and A85, respectively). However, simulations predict that more massive clusters may be more efficient at quenching star formation \citep[e.g., due to stronger ram-pressure][]{bekki2014}, and \citet{paccagnella2017} found that the H$\delta$-strong-to-active galaxy fraction increases with increasing X-ray luminosity of their WINGS clusters. Figure~\ref{mass_fracs} shows the completeness-corrected fraction of the PASGs, SFGs, and HDSGs as a function of cluster mass (red circles, blue stars, and green squares, respectively). The cluster mass is determined from the caustics method \citep[outlined in][]{owers2017}, plotted for each of the eight SAMI clusters, along with the quenching efficiency (purple hexagons) determined as in Section~\ref{GAMA_vs_cluster}. The solid lines with the same color-coding show the fractions measured in two bins that divide the cluster halos at $M_{200}=5\times 10^{14}$\msolar. The binned results suggest that the fraction of PASGs is higher in the M$_{200}>5\times 10^{14}$\msolar\, bin, along with a commensurate decrease in the fraction of SFGs. Interestingly, the value of $Q_{eff}$ does increase going from the low mass to high mass bin, where $Q_{eff}=29^{+7}_{-6}$\%, compared with $Q_{eff}=8^{+4}_{-2}$\% in the low mass bin. This increase in $Q_{eff}$ with cluster mass suggests that the more massive clusters are more efficient at quenching galaxies, although a larger sample of clusters is required to rule out halo-to-halo variations and confirm this result. 

\section{Discussion}\label{discussion}
The principal drivers of this study were to first identify galaxies that show evidence for ongoing quenching of star formation and then to characterize their environments in order to help understand which environmental quenching mechanisms are most germane. As a first step towards the latter goal, in Section~\ref{GAMA_vs_cluster} we compared the HDSGs (\hd-strong galaxies) found in the lower density GAMA regions with those found in the cluster regions. We found that the frequency of galaxies with HDS signatures was significantly higher amongst the non-passive cluster population ($15^{+4}_{-3}$\%) relative to that found in the GAMA regions ($2^{+1.0}_{-0.4}$\%). This indicates that the cluster environment is more efficiently quenching the SFGs (star-forming galaxies).

Furthermore, we found three notable differences between the GAMA and cluster HDSGs that indicate that the quenching mechanisms acting in the two samples are different. First, {\it none} of the GAMA HDSGs are associated with groups more massive than $M = 10^{13}$\msolar. Therefore, external processes related to cluster- or group-scale environments are not responsible for the quenching of the GAMA HDSGs. Second, in 7/8 GAMA HDSGs the distribution of HDS spaxels cover a large portion of the galaxy, including the central regions, and there is often evidence for emission lines associated with shocks or AGN activity (Figure~\ref{GAMA_HDS_galaxies}). In stark contrast, more than half (9/17) of cluster HDSGs show evidence for star formation in their centers and the HDS spaxels are found in the outer regions of the galaxy (Figure~\ref{cluster_HDS_galaxies}). Third, the structure of the GAMA and cluster HDSGs differs significantly. The HDSGs in the clusters are more disk-like, consistent with the cluster SFGs, whereas those found in the GAMA regions tend to have S{\'e}rsic indices intermediate between the SFG and PASG (passive galaxy) populations (Figure~\ref{nser_dist}). These three differences indicate that the mechanism that has quenched the star formation in the GAMA HDSGs has acted on a galaxy-wide scale, may have altered the galaxy structure, and may be internally driven. Given our aims here are to investigate {\it environment-related} quenching, and that the GAMA HDSGs are likely quenched by internal mechanisms, a full investigation of the GAMA HDSGs is beyond the scope of this paper.

On the other hand, in the cluster regions the quenching tends to occur more locally, starting in the outer parts of the galaxy, consistent with external environment-driven quenching. Our interpretation is that the cluster HDSGs are drawn from a population of newly accreted SFGs that have very recently entered the central 0.5\rtwo\, region close to the cluster core. During the first passage through the cluster, ram-pressure stripping removes gas from the outer parts of the galaxy, leading to the outside-in quenching of star formation in the galaxies. For the remainder of this Section, we outline the evidence supporting this interpretation. 

\subsection{A recently accreted population}\label{recent_accrete}

In Section~\ref{cluster_HDS_demographics}, we compared the velocity, radial and PPS distribution of the PASGs, SFGs, and HDSGs. The analysis revealed that HDSGs reside in a tight range in projected clustercentric distance ($0.15 < R/R_{200} < 0.6$), have a significantly larger spread in velocity relative to the overall galaxy population ($\sigma_{\rm HDS} =1.66\sigma_{200}$), and occupy an arc-shaped region in PPS with the low velocity HDSGs found at larger projected radii and the high velocity HDSGs at smaller projected radii. Since many of the cluster HDSGs harbor evidence for ongoing star formation, it is intriguing that their environments differ so markedly from that of the cluster SFGs, which are more evenly distributed in radius and have a similar velocity dispersion to the general cluster population (although a factor of 1.2 larger than the PASGs). 

These differences can be explained in a self-consistent way by considering the HDSGs as a subset of the infalling SFG population that have crossed 0.5\rtwo\, within the last $\sim 1$\,Gyr, while the PASGs form a virialised population that has existed within the cluster for many Gyrs. During infall, the velocity of a galaxy on a radial orbit increases with decreasing radius, peaking at pericentre, and decreasing thereafter prior to reaching zero velocity at apocentre where it spends a significant fraction of its orbit \citep{gill2005}. Thus, galaxies that have recently passed 0.5$r_{200, 3D}$ on their first passage are more likely to be found in the higher velocity, small radius region of the PPS diagram. Due to projection, a subset of the infalling galaxies will have the majority of their radial velocity vector aligned perpendicular to our line of sight and will therefore appear in the low velocity, small radius region of PPS. 

To test the validity of this interpretation we make use of the orbit libraries of \citet{oman2013} and \citet{oman2016}, which are used to derive probability distribution functions (PDF) for infall times as a function of position in PPS. The orbit libraries are derived from satellites associated with cluster-scale halos in the Multidark Run 1 N-body cosmological simulation \citep{prada2012}. For a detailed description of the extraction of the orbit libraries, and the conversion of the full 6D phase-space quantities onto 2D PPS, we refer to \citet{oman2013} and \citet{oman2016}. We rescale the simulation PPS coordinates to match our observed values as follows: for the velocity $\sqrt{3}|V_{pec}|/\sigma_{\rm 3D} = |V_{pec}|/\sigma_{1D} \simeq |V_{pec}|/\sigma_{200}$ and for the radius $1.3{\rm R}/r_{\rm vir} \simeq {\rm R}/$\rtwo. We note that the simulated and observed $V_{pec}$ values are determined in a consistent manner, while $r_{\rm vir}$ is determined as outlined in \citet{bryan1998}. We stack orbit libraries from host halos with $M_{\rm host, vir} > 10^{13}$\msolar, and include satellites with halo masses $M_{\rm sat, vir} > 10^{11.9}$\msolar. The limit in $M_{\rm sat, vir}$ helps to guard against incompleteness due to artificial disruption, and corresponds to a stellar mass \logmstar\,$\sim\, 10.3$, which is marginally higher than the limit imposed here on the SAMI galaxies \citep[][]{oman2013}. 

Motivated by our aim of understanding if the distribution of HDSGs in PPS is consistent with an infalling population that has recently crossed 0.5$r_{200, 3D}$, infall times, $t_{cross}$, are measured from the time a satellite first crosses 0.5$r_{\rm 200, 3D}$ in bins of size 0.2\,Gyr, which matches the redshift zero time-stamp resolution of the simulations. We generate PDFs on a grid that spans the radial and velocity range of the SAMI galaxies used in this paper, i.e., $0 \leq R/$\rtwo $\leq 1$ and $|V_{\rm pec}|/\sigma_{200} \leq 3$, respectively, with bins of size $\Delta$\rtwo\,$=0.1$ and $\Delta |V_{\rm pec}|/\sigma_{200} = 0.2$. Within each PPS bin, the fraction of interlopers is also determined. An interloper is defined as a halo that has $V_{pec} < 2.0 \sigma_{\rm 3D}$ and $R < 2.5r_{\rm vir}$, but has never entered the region with $r_{\rm 3D} < 0.5 r_{\rm 200, 3D}$ \citep[similar to that described in][]{oman2016}. The interloper population therefore contains satellites that are first-infallers as well as halos that will never enter within 0.5$r_{200,3D}$.

\begin{figure*}

\includegraphics[width=0.32\textwidth]{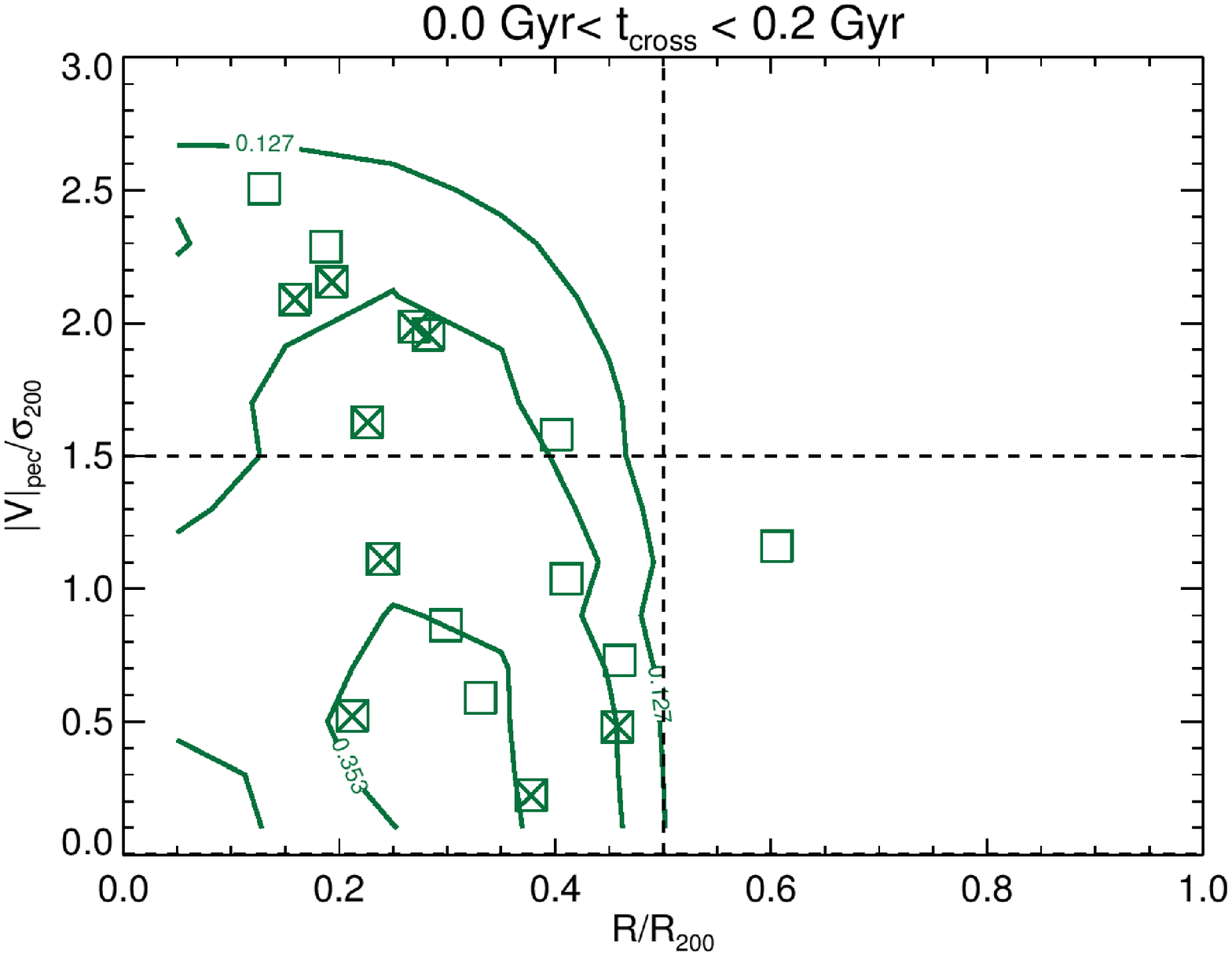}
\includegraphics[width=0.32\textwidth]{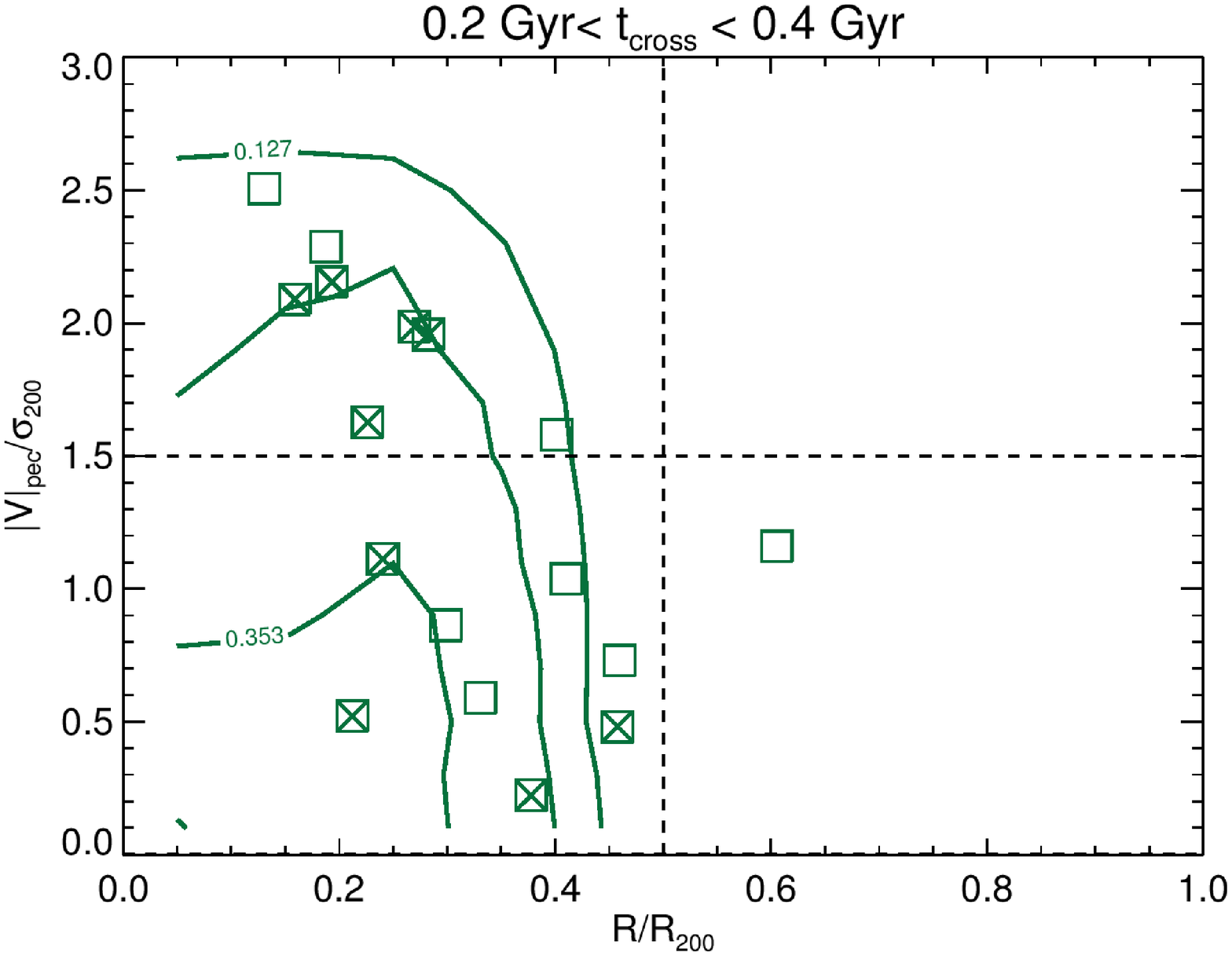}
\includegraphics[width=0.32\textwidth]{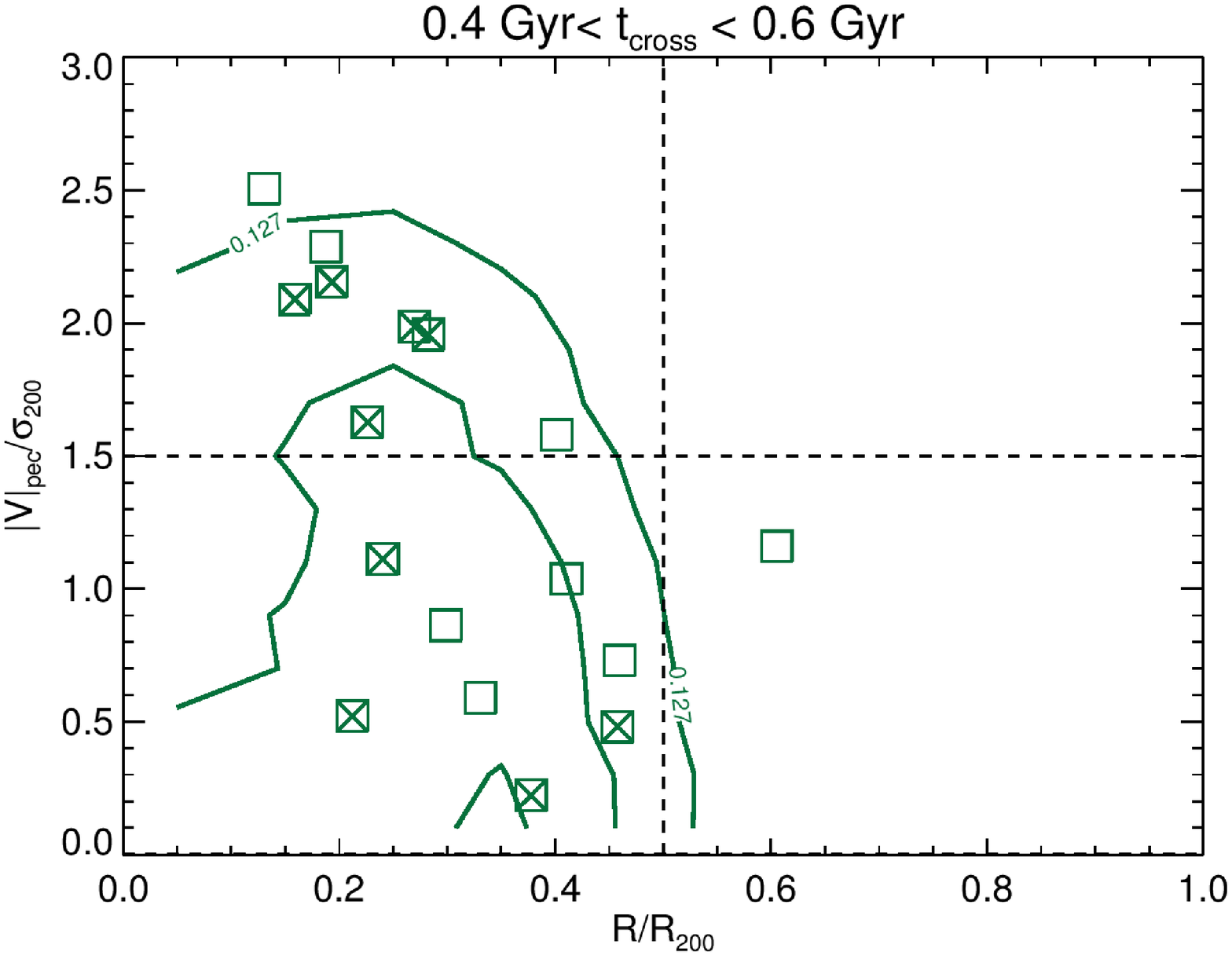}\\
\includegraphics[width=0.32\textwidth]{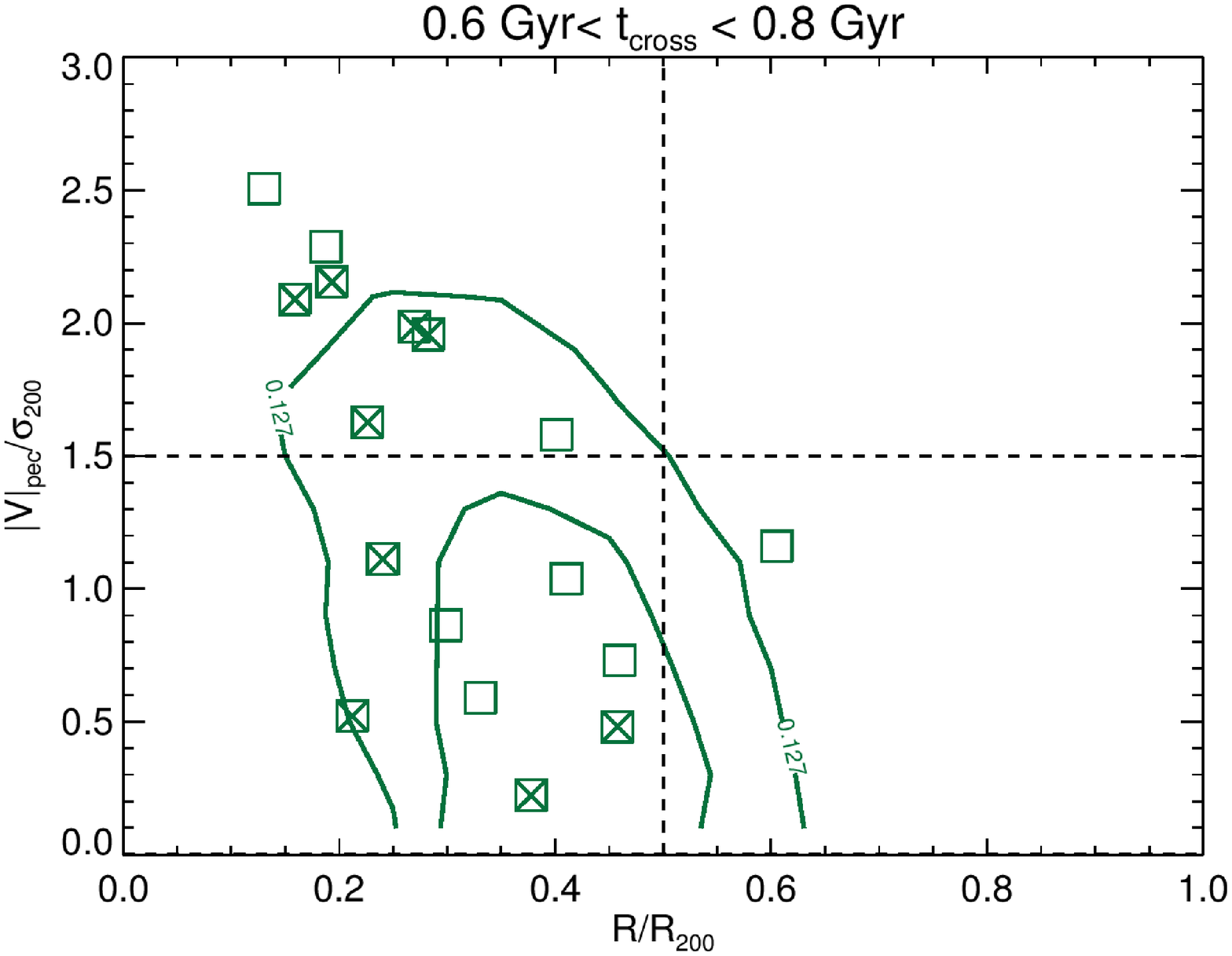}
\includegraphics[width=0.32\textwidth]{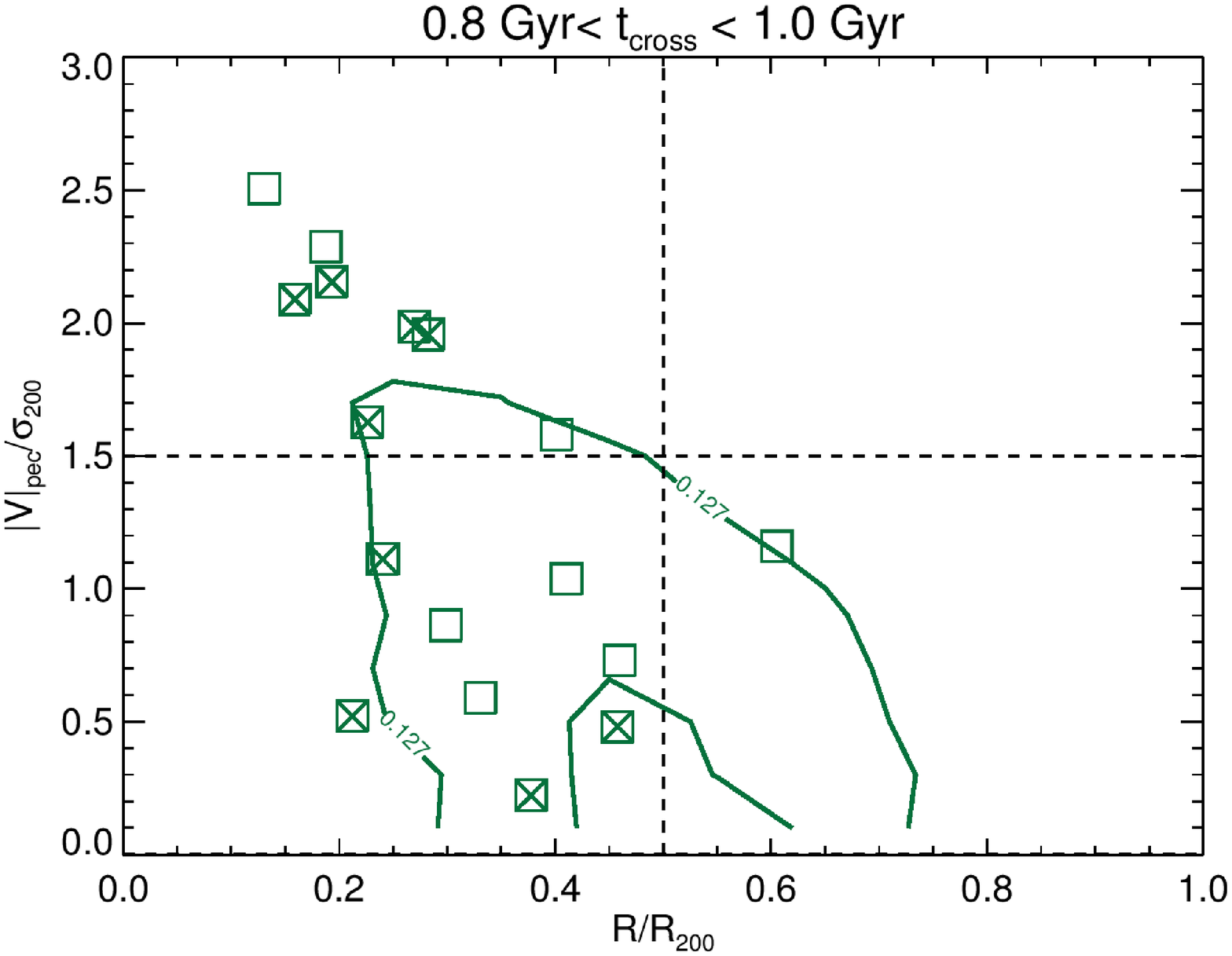}
\includegraphics[width=0.32\textwidth]{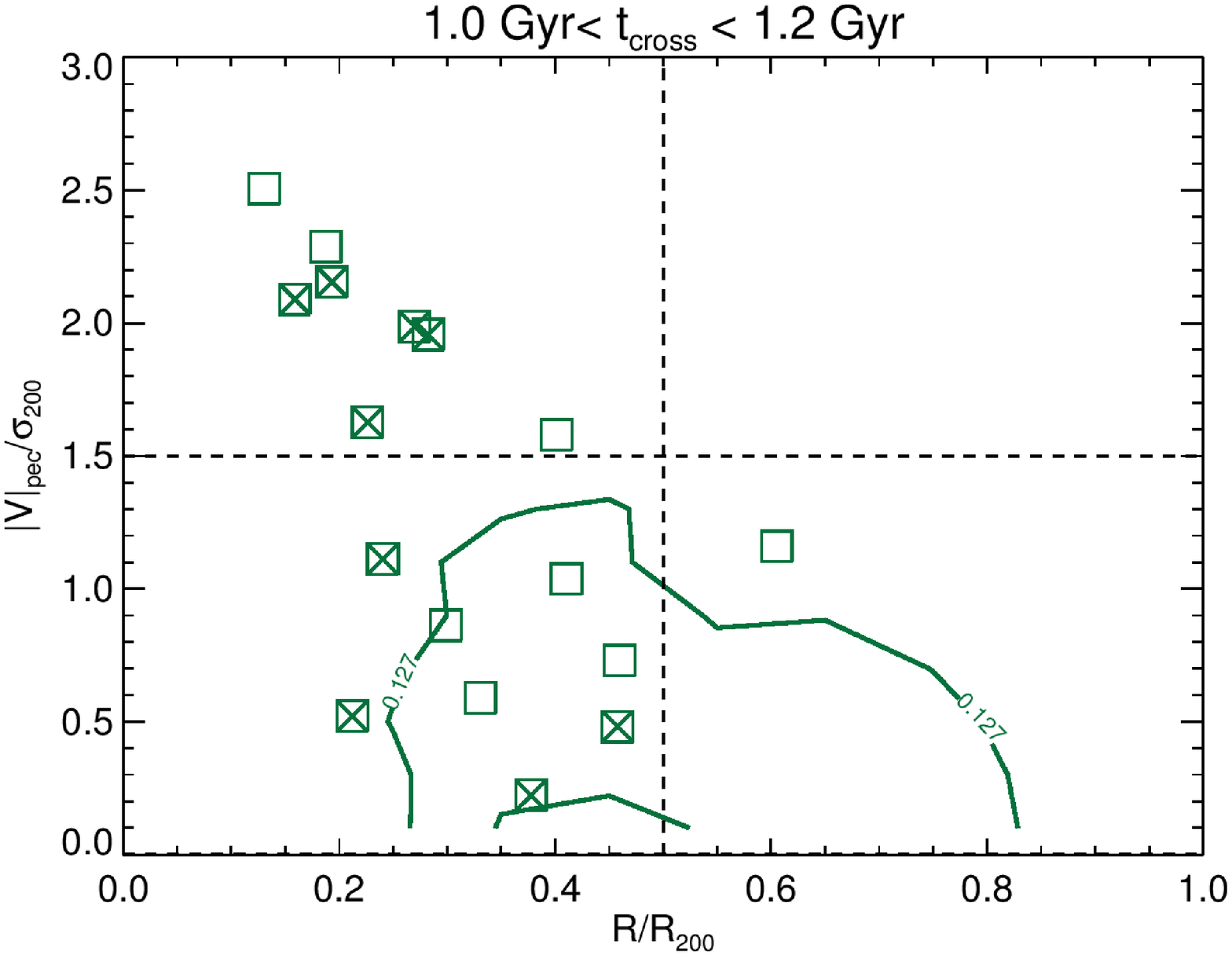}
\caption{The predicted PPS distribution of satellites as a function of time since crossing 0.5$r_{200, 3D}$. The bins used for $t_{cross}$ are shown at the top of each panel. The 
green contours are determined by combining the observed PDF for the ensemble of 8 clusters in PPS with the PDF determined from orbit libraries drawn from simulated halos, as described in the text. Contours are determined for 10 logarithmically spaced levels between one and 100 percent of the maximum of the smoothed density PPS distribution for the SAMI galaxies. The squares show the position of the 17 HDSGs, and those marked with crosses have central star formation. The distribution of the predicted PPS position for objects that have $t_{cross} < 1\,$Gyr is in good agrement with the distribution of HDSGs in PPS. \label{SIM_PPS}}
\end{figure*}

In Figure~\ref{SIM_PPS} the green contours show the predicted number density distribution in PPS for a population of satellites with $0 \leq t_{\rm cross} \leq 1.2$\,Gyr in six time steps. The contours are generated from a grid where the simulated PDFs are multiplied by the observed density distribution of SAMI galaxies in PPS giving the predicted number of galaxies in each PPS pixel for the time step listed on each panel. Given the relatively small sample size, to determine the observed galaxy density we adaptively smooth the observed PPS following the method outlined in \citet{owers2017}; this smoothing helps to minimise the impact of shot-noise for the bins with few galaxies. 

Comparing the contours with the distribution of HDSGs (green squares), the closest match occurs at two timesteps: $0\leq t_{cross} \leq 0.2$\,Gyr and $0.4 \leq t_{cross} \leq 0.6$\,Gyr (top left and top right panel in Figure~\ref{SIM_PPS}, respectively). The timestep in the $0.2 \leq t_{cross} \leq 0.4$\,Gyr range encompass pericentric passage; this period predicts a larger number of galaxies with small radius and large velocity. However, this timestep predicts too many galaxies in the $R/R_{200} < 0.2, |V_{\rm pec}|/\sigma_{200}< 1.5$ corner of PPS to be compatible with the observed distribution, which is completely devoid of HDSGs. At later stages when $0.6 \leq t_{cross} \leq 0.8$\,Gyr, the contours mainly coincide with the low velocity HDSGs with $0.3 \leq R/$\rtwo $\leq 0.5$. At later times ($t_{cross} > 0.8$\,Gyr), the contours move towards larger radius, lower velocity, and become less coherent. The approximate symmetry of the velocity evolution of infalling galaxies about pericentre causes the degeneracy in the PPS distribution for the $0\leq t_{cross} \leq 0.2$\,Gyr and $0.4 \leq t_{cross} \leq 0.6$\,Gyr timesteps. This degeneracy means that using PPS alone cannot distinguish whether the HDSGs in our sample belong to a population that is observed just prior to, or just following a pericenter passage. However, we can conclude that the distribution of HDSGs is consistent with that of a population that has been accreted within 0.5$r_{200, 3D}$ within the last 1\,Gyr.

\begin{figure*}
\includegraphics[width=0.32\textwidth]{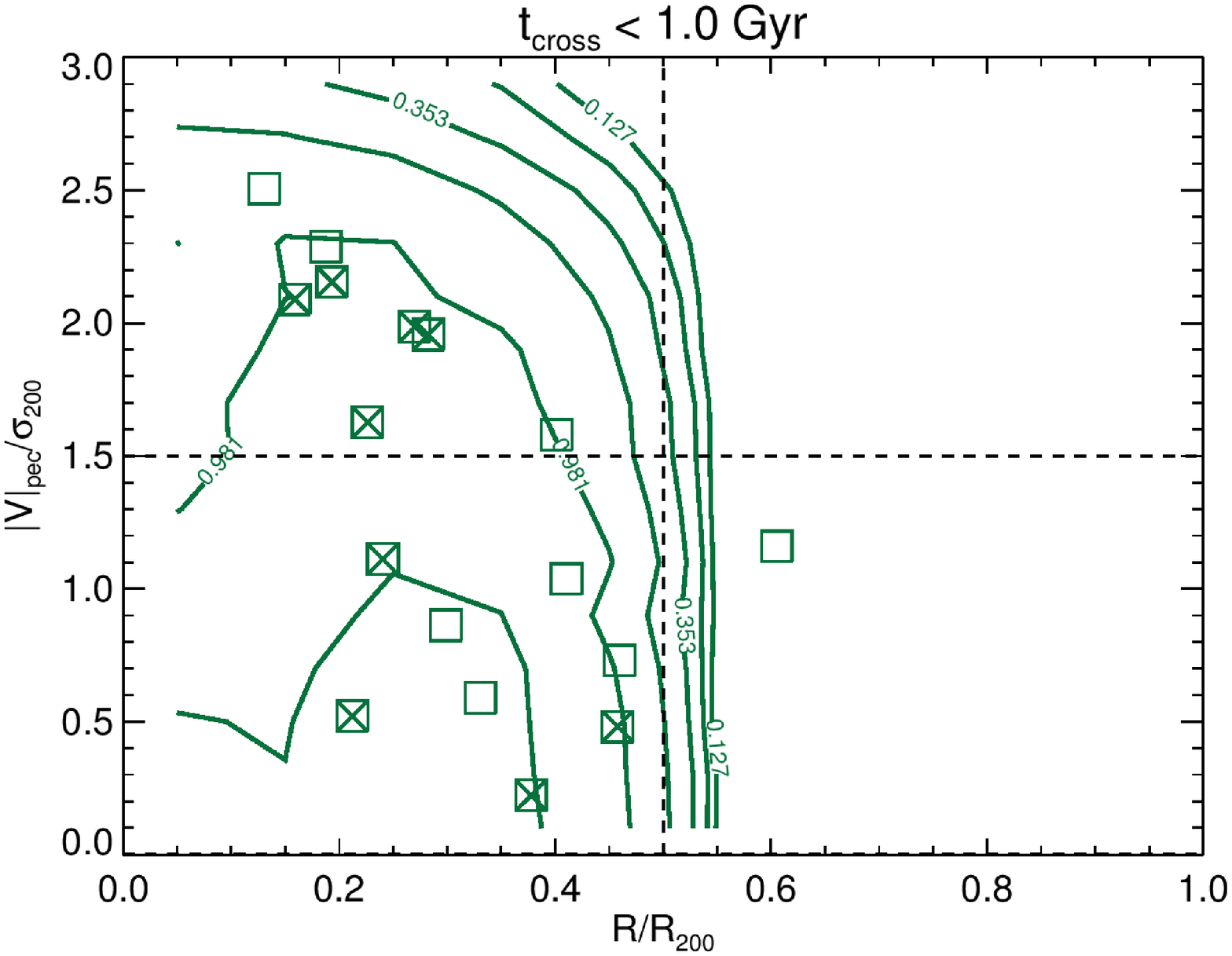}
\includegraphics[width=0.32\textwidth]{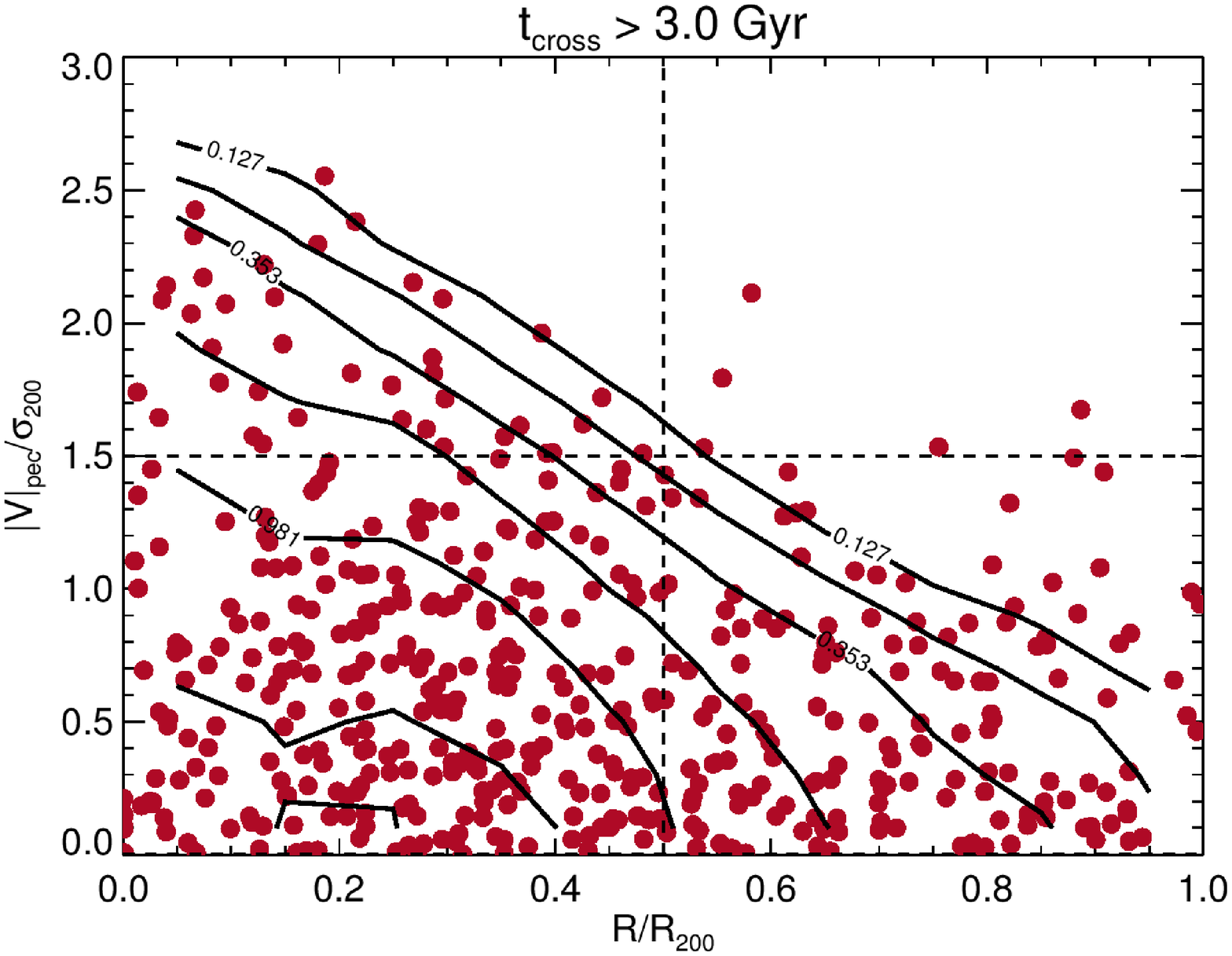}
\includegraphics[width=0.32\textwidth]{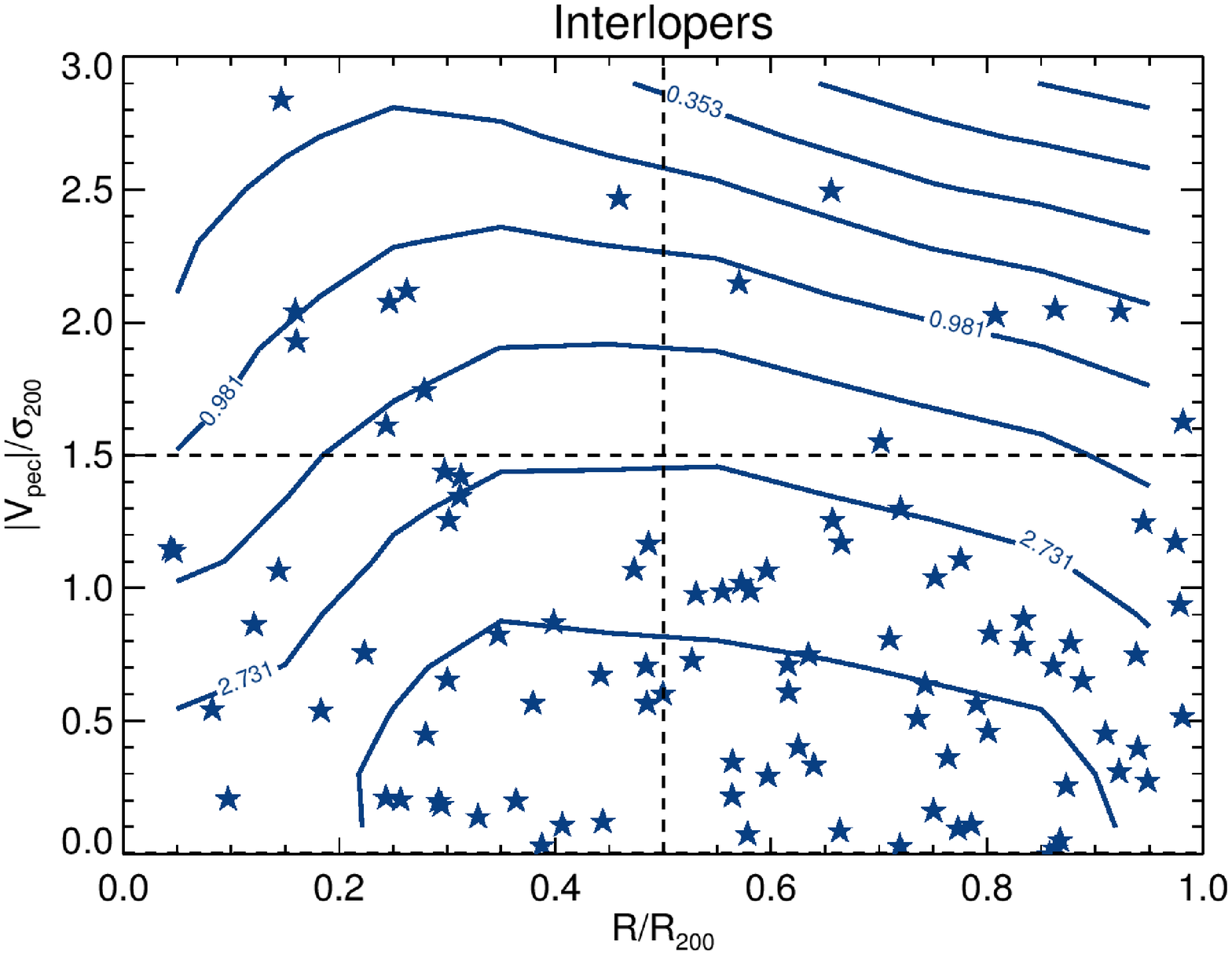}
\caption{Comparison of the distribution of HDSGs (green squares, left panel), PASGs (red circles, middle panel), and SFGs (blue stars, right panel) with the predicted distribution of satellites with $t_{cross}<1$\,Gyr (green contours, left panel), $t_{cross}>3$\,Gyrs (black contours, middle panel), and objects that have yet to cross 0.5$r_{200, 3D}$ (blue contours, right panel). The agreement between the observed and predicted distributions provides support for the interpretation that the PASGs form a virialised population, the SFGs form an infalling population that have yet to pass pericenter, while the HDSGs are infalling objects that are observed at close to pericenter.\label{PPS_PMF}}
\end{figure*}

In Figure~\ref{PPS_PMF}, the contours overlaid on the left, middle and right panels are generated as described above, but now show the predicted distribution of satellites that have $t_{cross} < 1$\,Gyr, $t_{cross} > 3$\,Gyr, and interlopers that have yet to cross 0.5$r_{200, 3D}$, respectively. Also shown are the PPS distributions of the HDSGs, PASGs, and SFGs (green squares in left panel, red filled circles in middle panel, and blue filled stars in right panel, respectively). The three timesteps are chosen to show the PPS distribution expected of a recently accreted, virialised, and infalling population. Qualitatively, the match between the predicted and observed distributions for the three subsets is very good. This match supports our interpretation that the SFGs form an infalling population that have not yet passed within 0.5$r_{200, 3D}$, while the HDSGs are consistent with being drawn from the population of infalling galaxies that are very close to pericenter, having recently passed 0.5$r_{200, 3D}$. The PASGs form the bulk of the virialised population that has resided in the cluster for several crossing times, supporting previous results \citep[e.g.,][]{biviano2004}.

\subsection{The outside-in quenching of infalling galaxies}\label{out_in_quenching}

In Section~\ref{GAMA_vs_cluster}, we found that around half of the cluster HDSGs show evidence for central star formation, with the HDS signature found in the outer parts of those galaxies. This strongly suggests that these cluster HDSGs are being quenched in an outside-in manner. In attempting to understand the dominant process that is responsible for this outside-in quenching, there are three important pieces of evidence to consider. First, the preceding analysis suggests that the HDSGs are recent additions to the cluster environment, and are very close to pericentric passage where both infall velocity and intracluster medium density peak. Second, the analysis in Section~\ref{GAMA_vs_cluster} indicates that the structure of the galaxies is consistent with them being disk-like systems, while inspection of the images in Figure~\ref{cluster_HDS_galaxies} reveals very little visual evidence for disturbances in the stellar distribution of the galaxies. Finally, the existence of strong Balmer absorption in the outer regions suggests that the quenching must have occurred on a relatively short timescale ($<1$\,Gyr).  Taken in concert, the evidence supports a quenching process related to hydrodynamical interactions between the galaxy's gas and the intracluster medium (ICM) that is capable of removing the gas required for star formation without disturbing the stellar component of the disk. Our interpretation is that stripping of the cold atomic HI and molecular gas disk due to ram-pressure stripping \citep{gunn1972} is the most likely mechanism responsible for the quenching. However, it is worth considering why other processes commonly attributed to cluster-related quenching of star formation are less likely.

We can immediately rule out strangulation as being responsible for the production of the HDSGs. Strangulation quenches star formation via the removal of the hot gas halo reservoir that replenishes the cold gas required for star formation \citep{larson1980}. This process is slow-acting, requiring several Gyrs to quench a galaxy, and simulations show that the quenching is a gradual disk-wide process resulting in passive-spiral-like galaxies \citep{bekki2002}. These predictions are inconsistent with the existence of the young, localised HDS regions observed here that indicate a more rapid removal of gas. While the hot gas reservoirs may have been removed from the HDSGs during the early phases of infall, it is not the dominant quenching mechanism currently at play. 

Similarly, we can rule out quenching due to major galaxy-galaxy interactions and mergers. First, the large relative velocities between the galaxies in the cores of clusters mean that major mergers are rare in cluster cores, and are more likely to occur in the cluster outskirts \citep[$R>$\rtwo;][]{ghigna1998,moran2007}. Second, inspection of the galaxy images shown in Figure~\ref{cluster_HDS_galaxies} reveals that only one galaxy (9016800216) shows obvious tidal-like features in the stellar distribution, whereas the majority of the cluster HDSGs appear to have relatively unperturbed stellar morphologies with no evidence for mergers. Multiple, high-speed tidal interactions between galaxies \citep[harrassment;][]{moore1996} and tidal forces due to the cluster potential \citep{byrd1990}, can enhance central star formation and disrupt galaxy disks. However, we do not see any evidence for abnormal star formation rates (as determined by their EW(H$\alpha$) in Section~\ref{SF_properties}). Furthermore, it is unlikely that tidal interactions will preferentially remove gas without also affecting the stellar disk, while \citet{boselli2006} find that tidal interactions with the cluster potential are unlikely to remove large amounts of HI gas. While these processes are unlikely to be responsible for the appearance of the HDSGs features, their cumulative effects may well have played a role in aiding the process of gas stripping that led to the current state.

On the other hand, stripping of the cold gas via ICM interactions naturally explain the observations because they can remove gas in an outside-in manner without significantly affecting the stellar structure of the galaxy \citep{boselli2006}. Simulations predict that stripping can be a multi-stage process with a continuous phase that slowly removes gas via viscous stripping, and a more rapid phase that generally occurs near a pericentric passage where ram-pressure stripping (RPS) can displace the HI gas from the stellar disk on timescales of less than a few hundred Myrs \citep{nulsen1982, quilis2000, schulz2001, roediger2006, roediger2007, roediger2009, bekki2014, jung2018}. Recent observational evidence has indicated that, as well as stripping the atomic HI gas component, strong ram-pressure stripping may also remove the molecular components of the galaxy within the optical disk, leading to quenching of star formation \citep{cortese2012, boselli2014b, lee2017, noble2019}. This rapid local stripping would lead to localized HDS signatures in the outer parts of the disk, as observed in our HDSG sample.

We can estimate the expected effects of RPS using the analytical prescription of \citet{gunn1972}, and following a procedure similar to that performed by \citep[][see also \citealp{boselli2006, hernandez2014, jaffe2015, jaffe2018}]{cortese2007}, which estimates the stripping radius, $R_{\rm strip}$, by determining the radius at which ram pressure overcomes the gravitational restoring force per unit area of the galaxy. The ram pressure is estimated as
\begin{equation}\label{Pram}
P_{\rm ram} = \rho_{\rm ICM} v_{\rm gal}^2 
\end{equation}
where $\rho_{ICM}$ is the ICM density, and $v_{\rm gal}$ is the relative velocity difference between the ICM and the galaxy. 
To determine $\rho_{ICM}$, we make use of the results obtained by  \citet{ghirardini2018}, who fitted the scaled electron density, $n_e(r/r_{500})/E(z)^2 $, profiles of 12 clusters using the analytical form defined in Eq. 3 of \citet{vikhlinin2006}. We use the best-fitting parameter estimates given in Table~3 of \citet{ghirardini2018} to determine the ICM density at radius $r$ as $\rho_{\rm ICM}(r)=\mu_{\rm H} m_p n_e(r)/\nu_{\rm H}$, where $m_p$ is the proton mass, $\nu_{\rm H}=1.17366$ is the number of electrons per Hydrogen atom, $\mu_{\rm H}=1.34732$ is the mean particle weight per Hydrogen atom all assuming an ICM with metallicity $Z=0.3Z_{\odot}$ \citep{grevesse1998}. Since we normalise our projected radii using \rtwo, we rescale as $R_{500}\simeq0.65$\rtwo\, \citep{reiprich2013}.

It is not possible to determine $v_{\rm gal}$ from observations; the LOS velocity that we measure, $v_{\rm pec}$, provides a lower limit on this value. Likewise, the projected radius, $R$, also provides a lower limit on the 3D position $r_{\rm 3D}$, meaning that both the estimate for $\rho_{\rm ICM}$ and $v_{\rm gal}$ have some uncertainty associated with them due to projection effects. However, as outlined in Section~\ref{recent_accrete}, the distribution of the HDSGs in PPS is consistent with an infalling population that recently crossed 0.5\rtwo, and are likely to be on close to radial orbits that follow the infall velocity profile of the cluster. We can therefore approximate $v_{\rm gal}$ using the infall velocity $v_{inf}(r)=\sqrt{2GM(<r)/r}$, where $M(<r)$ is the mass within radius $r$, which is determined from the caustics mass profile derived in \citet{owers2017}. We produce three estimates of $P_{\rm ram}$ for each of the HDSGs: the first estimates $P_{\rm ram}(R, v_{\rm pec}) = \rho_{\rm ICM}(R) v_{\rm pec}^2$ at the observed position in PPS, the second $P_{\rm ram}(R, v_{inf}(R))=\rho_{\rm ICM}(R) v_{inf}(R)^2$ assumes the projected $R=r$ and estimates the infall velocity, $v_{inf}(R)$ at that position, and the third $P_{\rm ram}(0.5$\rtwo$)= \rho_{\rm ICM}(0.5$\rtwo$) v_{inf}(0.5$\rtwo$)^2$ estimates ram-pressure experienced upon entering 0.5\rtwo\, at the infall velocity. These estimates span a range of $P_{\rm ram}$ values that a galaxy is expected to experience having entered within 0.5\rtwo; the minimum occurs at $P_{\rm ram}(0.5$\rtwo$)$, and an upper limit at the position in PPS of $P_{\rm ram}(R, v_{inf}(R))$, assuming that the galaxy has not yet past pericenter. 

The restoring force per unit area for each HDS galaxy is:
\begin{equation}\label{restore}
\Pi=2 \pi G \Sigma_{stars} \Sigma_{gas}
\end{equation}
where $\Sigma_{stars}$ and $\Sigma_{gas}$ are the surface density profiles for the stars and gas, respectively, which are assumed to follow an exponentially declining profile   estimated following \citet{domainko2006}:
\begin{equation}
\Sigma(r)=\Sigma_{0} e^{-r_g/R_d}
\end{equation}
where $R_d$ is the disk scale-length, $\Sigma_0$ is the central surface density, and $r_g$ is the distance from the centre of the galaxy. For the stars, an exponentially declining profile is justified given that the majority of the HDSGs have $r$-band S{\'e}rsic indices $n_{ser}\simeq 1$, and are therefore disk-dominated systems (Figure~\ref{nser_dist}). The stellar central density is $\Sigma_0^*=M_{*}/(2\pi R_d^2)$, where we have assumed that the stellar mass proxy, $M_{*}$ is dominated by the disk component of the galaxy. The stellar disk scale-length, $R_d^*=0.59r_e$ where $r_e$ is the effective radius measured in Section~\ref{cluster_sersic}. Estimates for the gas scale length, $R_{d, gas}$, and central density, $\Sigma_{0, gas}=M_{gas}/(2\pi R_{d, gas}^2)$, are less well-constrained for the HDSGs. For the scale length, we follow \citet{boselli2006} and assume that $R_{d, gas}\simeq 1.8R_d^*$. For the gas mass, we use the $M_{gas}/M^*$ scaling relations provided by \citet{catinella2018} to estimate the total $M_{gas}$. We assume that our HDSGs have originated from SFGs on the blue-cloud, which has $NUV - r \sim 3$ in Figure~\ref{colour_mass_type}. According to Table~2 in \citet{catinella2018}, galaxies with this colour have $M_{gas} \simeq 0.4M^*$, and we therefore use this for our estimates of $M_{gas}$ for each HDS galaxy.

\begin{figure}
\includegraphics[width=0.48\textwidth]{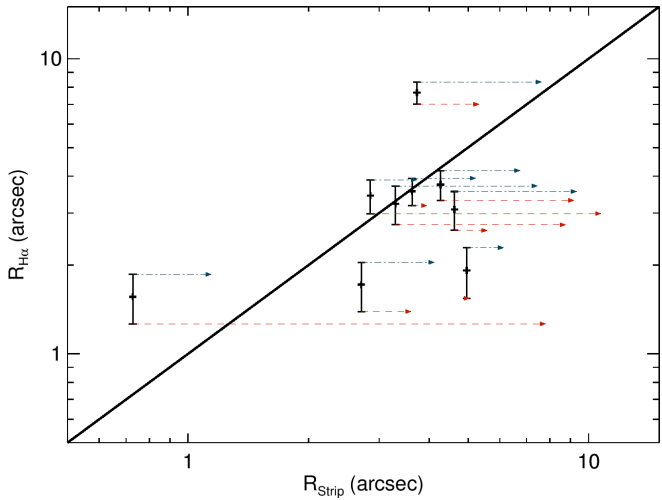}
\caption{The extent of the \ha\, emission driven by star formation, ${\rm R}_{{\rm H}\alpha}$, versus predictions for the stripping radius $R_{strip}$ determined from Equation~\ref{rstrip}. The black crosses show the ${\rm R}_{{\rm H}\alpha}$ versus $R_{strip}$ for the $P_{\rm ram}(v_{inf}(R), R)$ ram pressure estimate at the projected position of the galaxy at the infall velocity. The blue dot-dashed and red dashed arrows show $R_{strip}$ values estimated for the ram pressure estimated at $P_{\rm ram}(0.5$\rtwo$)$ and $P_{\rm ram}(R, v_{\rm pec})$, which are the ram-pressure estimates using the infall velocity at 0.5\rtwo, and the observed position in PPS, respectively. The black solid line shows the one-to-one relation. In 8/9 cases, the $R_{strip}$ estimated for $P_{\rm ram}(v_{inf}(R), R)$ is within a factor $\lesssim 1.5$ of ${\rm R}_{{\rm H}\alpha}$. \label{ha_rad_vs_rstrip}}
\end{figure}

We can now combine Equations~\ref{Pram} and \ref{restore} to measure the stripping radius
\begin{equation}\label{rstrip}
R_{strip}= \frac{r_e}{2.64}\times \ln \left ( \frac{0.4 G M^{2}_*}{2.47 P_{\rm ram} r_e^4} \right ) 
\end{equation}
for each of our three different $P_{ram}$ values and for each HDSG. We note that the \citet{gunn1972} analytical approximation for $R_{strip}$ assumes that the galaxy is traversing the ICM face-on. However, \citet{roediger2007} showed that there is good agreement between analytical and simulated $R_{strip}$ estimates, except for galaxies moving close to edge-on though the ICM. The results are shown as ellipses overlaid on the EW(H$_\alpha$) maps in Figure~\ref{cluster_HDS_galaxies}, where $R_{strip}$ is represented as the major axis of the ellipse with PA and ellipticity determined from the S{\'e}rsic fits in Section~\ref{cluster_sersic}. The gray dashed, blue dot-dashed, and purple solid ellipses show the results for $P_{\rm ram}(0.5$\rtwo$)$, $P_{\rm ram}(v_{\rm pec}, R)$, and $P_{\rm ram}(v_{inf}(R), R)$, respectively. Also overlaid on the EW(H$_\alpha$) maps in Figure~\ref{cluster_HDS_galaxies} are black contour levels drawn at EW(H$_\alpha)=3$\AA\, including only spaxels that are classified as SF, INT or wSF, and therefore show the region at which the star formation is truncated. In Figure~\ref{ha_rad_vs_rstrip} we show the semi-major axis of the EW(H$_\alpha)=3$\AA\, region, $R_{{\rm H}\alpha}$, which is determined by fitting an ellipse to the black contours shown in Figure~\ref{cluster_HDS_galaxies}, versus the three measures of $R_{strip}$. The $R_{strip}$ estimates for the $P_{\rm ram}(0.5$\rtwo$)$ and $P_{\rm ram}(R, v_{\rm pec})$ (blue dot-dashed and red dashed lines in Figure~\ref{ha_rad_vs_rstrip}) values are generally much larger than $R_{{\rm H}\alpha}$. On the other hand, for 8/9 HDSGs with central star formation the $R_{strip}$ values estimated using $P_{\rm ram}(v_{inf}(R), R)$ (shown as black crosses in Figure~\ref{ha_rad_vs_rstrip}) are within a factor $\lesssim 1.5$ of the $R_{{\rm H}\alpha}$. This agreement supports the hypothesis that the ram-pressure stripping encountered by these HDSGs on first-infall is capable of removing the gas disk leading to the observed truncation of star formation. 

The above analysis provides support for ram-pressure stripping in the HDSGs with central star formation. However, inspection of the $R_{strip}$ ellipses for the 8 HDSGs with no evidence for ongoing star formation indicates that the predicted ram-pressure at their current positions is insufficient to completely strip their gas. In this regard, it is important to note that the stripping radius estimates given by Equation~\ref{rstrip} do not account for past stripping, e.g., if those galaxies have previously passed pericenter where they experienced peak ram pressure. Figure~\ref{SIM_PPS} shows that 6/8 of the non-star-forming HDSGs are located at larger projected distances and smaller LOS velocities when compared with the remaining HDSGs. We speculate that the majority of these galaxies have passed pericenter and are currently outbound toward apocenter. This interpretation is supported by the simulations presented in Figure~\ref{SIM_PPS}, which shows that the expected position of the post-pericenter, outbound satellites in the $0.6-1.0$\,Gyr panels is consistent with the positions of the non-star-forming HDSGs. In this scenario, the strong ram-pressure experienced during pericenter would have completely stripped the gas from the galaxies, leading to their completely quenched state. 

\subsection{Comparison with previous work}

The results and analysis presented in this paper lead us to conclude that ram-pressure stripping plays an important role in quenching infalling star-forming disk galaxies on their first passage through the cluster core. This conclusion is supported by observations in the nearby Virgo cluster, where \citet{koopmann2004} found that around half of the spiral galaxies in the Virgo cluster exhibited truncated H$\alpha$ disks with little evidence for any disturbance in the stellar disks, and that the majority of the truncated spirals were found close to the core of the cluster. Moreover, \citet{chung2009} and \citet{chung2007} found that the HI distributions of galaxies in the core of Virgo \citep[R/\rtwo\,$\lesssim 0.3$, where \rtwo $=1550$\,kpc;][]{ferrarese2012} are often asymmetric and always truncated with respect to the stellar disk, while many galaxies in the $0.4 \leq $R/\rtwo$\leq 0.7$ range exhibit one-sided HI tails that point away from the center of Virgo. These observations provide evidence that ram-pressure begins to strip the HI disk of galaxies as they fall into Virgo, and removes much of the HI disk during pericenter, leading to truncated star formation. 

Our results are broadly consistent with those of \citet{crowl2008}, where the stellar population ages in the outer disks of around half of their sample of truncated Virgo spirals were consistent with the timescales expected due to ram-pressure stripping experienced near pericenter. However, \citet{crowl2008} also find evidence that a subset of their galaxies must have been stripped outside of the cluster core; their interpretation being that tidal interactions and enhanced RPS due to bulk motions and ICM substructure may be responsible for stripping at larger clustercentric distances. In contrast with the findings of \citet{crowl2008}, we find only one HDSG outside its cluster core ($R> 0.5$\rtwo). However, it is worth noting that only a handful of the \citet{crowl2008} galaxies would pass our EW(H$\delta$)$<-3$\AA\, selection criteria, so care needs to be taken in directly comparing our results with theirs. It is possible that our selection criteria bias the selection to include only those objects undergoing the most rapid stripping, and that there are galaxies with truncated star formation that are not included in our sample. Indeed, Figure~\ref{ha_conc} reveals that there are 13 SFGs that have $C_{\rm H\alpha, cont}$ in the same range as the HDSGs with central star formation; half of these SFGs have $R>0.5$\rtwo. We will investigate the properties of these SFGs with centrally concentrated star formation in a forthcoming paper.

\citet{cortese2009} investigated a sample of local transition galaxies that are located in the green valley in the NUV-H diagram. They found that the majority of the transition galaxies are HI-deficient and located in the Virgo cluster. The Virgo transition galaxies are disc-like, have radial and velocity distribution that are consistent with that of an infalling population, and around half show evidence for star formation at their centers. These pieces of evidence led \citet{cortese2009} to conclude that ram-pressure stripping was responsible for the quenching of star formation in the Virgo transition galaxies, consistent with the results presented here for our HDSGs. The importance of ram-pressure stripping in quenching star formation in Virgo has found much support from many other avenues \citep[][]{cortese2011, cortese2012, boselli2014,boselli2014b,boselli2016}. The results presented here provide support for these previous results, and extend them to environments beyond the nearby Virgo cluster.

The radial distribution of the HDSGs within our cluster ensemble is consistent with that found in \citet{paccagnella2017}, who investigated the environments of a large sample of H$\delta$-strong\footnote{\citet{paccagnella2017} use the nomenclature ``post-starburst'' and ``strong post-starburst'' to describe their samples. We note that their post-starburst sample is selected to have $3 < $\,EW(H$\delta$)\,$ < 8$\,\AA, while galaxies with EW(H$\delta$)\,$ > 8$\,\AA are strong post-starbursts} galaxies selected from the WIde-Field Nearby Galaxy-cluster Survey \citep[WINGS;][]{fasano2006}. In agreement with the results in Figure~\ref{frac_rad}, \citet{paccagnella2017} found that the ratio of H$\delta$-strong galaxies to the total number of active galaxies (emission line and H$\delta$-strong galaxies) increases markedly towards the cluster center. In contrast with our results, \citet{paccagnella2017} did not find significant differences when comparing the kinematics and PPS distribution of the passive and H$\delta$-strong sample. However, their sample of H$\delta$-strong galaxies with EW(H$\delta$)$>8$\AA\, did show both a significantly higher velocity dispersion than the passive galaxies, as well as a tendency to be clustered at $|v_{pec}|/\sigma_{200} \simeq 1.5$ and $R/$\rtwo$\simeq0.25$, in agreement with the results shown in Figures~\ref{vpec_hist} and \ref{PPS_plot}. 

In comparing the \citet{paccagnella2017} results with those presented here, a crucial caveat is that their selection relies on single-fiber spectroscopy. At the redshifts of the WINGS clusters \citep[$0.04 < z < 0.07$][]{moretti2017}, the spectra cover an area that subtends a physical scale of only $1.3 - 2.8$\,kpc in diameter. This means that in order to be included in their \hd-strong sample, the strong Balmer absorption signature must be present within the very central parts of the galaxy. Furthermore, \citet{paccagnella2017} exclude from their \hd-strong sample all galaxies that have [OII], \oiii, \hb\, or \ha\, detected in emission. Inspection of Figure~\ref{cluster_HDS_galaxies} reveals that only two of our cluster HDSGs have central spectra that meet these criteria. Further investigation of the spectral properties within the central 2\arcsec\, for the remaining SAMI cluster galaxies reveals that no other objects would meet the \citet{paccagnella2017} \hd-strong criteria. Thus, fewer than 0.5\% of our sample would be classified as \hd-strong in single-fiber surveys, consistent with previous results for local clusters where \hd-strong galaxies with \logmstar\,$>10$ are rare \citep{poggianti2004,gavazzi2010, boselli2014}. This fraction is, however, substantially lower than that reported by \citet{paccagnella2017}, who found 7.3\% of \logmstar$ > 9.8$ cluster galaxies are classified as \hd-strong. We also note that a large fraction of the \hd-strong galaxies in \citet{paccagnella2017} lie on the red-sequence (see their Figure~3), whereas the majority of our sample lie blueward of the red-sequence (Figure~\ref{colour_mass_type}). Thus, while there are some consistencies between the trends with environment seen in our HDSGs and those reported by \citet{paccagnella2017}, it is clear that the two samples are unlikely to be tracing the same population of transition galaxies.

The coherent arc-like distribution of HDSGs in PPS shown in Figure~\ref{PPS_plot} is strikingly similar to that found in a sample of $z\sim1$ clusters by \citet{muzzin2014}. Comparing the distribution of their HDSGs to four simulated clusters, \citet{muzzin2014} find that the galaxies were likely quenched within 500\,Myr after crossing 0.5\rtwo, consistent with our interpretation that the SAMI HDSGs are recent arrivals to the region within 0.5\rtwo. \citet{muzzin2014} favour the fast-acting ram-pressure stripping of the cold gas as the likely quenching mechanism, although they could not conclusively rule out slower processes such as starvation/strangulation due to the removal of the hot gas reservoir. If the HDSGs presented here are local analogues of those in \citet{muzzin2014}, then they may offer a valuable and more accessible insight into the quenching processes operating in higher redshift clusters. In this vein, we note that the HDSGs in \citet{muzzin2014} are selected to have weak [OII] emission indicating that they are fully quenched, whereas many of our HDSGs are only partially quenched. This partial quenching indicates that our HDSGs are at an earlier quenching phase than those detected in \citet{muzzin2014}. If so, then the relative consistency of the coherent phase-space positions between the two samples supports a quenching scenario driven by ram-pressure stripping during the core passage phase in both high- and low-redshift clusters. 

The rapid phase of the ``delayed-then-rapid'' quenching scenario \citep{wetzel2012} has been attributed to very efficient, complete quenching that occurs within $\sim 1$\,Gyr of pericentric passage, likely due to the effects of ram-pressure stripping \citep{oman2016}. The HDSGs observed here provide support for a substantial fraction of infalling star-forming galaxies being affected by ram-pressure stripping at close to pericenter. In particular, Table~\ref{PPS_table} shows that the fraction of HDSGs is highest in the $R/$\rtwo$< 0.5, |V_{\rm pec}/\sigma_{200}> 1.5$ portion of the PPS diagram, where they make up 15\% of the total population, and 50\% of the population that show evidence for recent or ongoing star formation. The simulations in Section~\ref{recent_accrete} indicate that this PPS region contains the highest fraction of recently accreted satellites, with 25\% having $t_{cross} \leq 1$\,Gyr, 17\% with $t_{cross}>1$\,Gyr, while the remaining 58\% of satellites have not yet crossed 0.5$r_{200, 3D}$. Therefore, the star-forming galaxies seen in this region of PPS that do not show evidence for quenching can be accounted for by projection effects, indicating that the majority of star-forming galaxies that pass within 0.5$r_{200, 3D}$ will begin the quenching process. 

While many of the HDSGs are completely quenched, thereby supporting a rapid quenching phase within 1\,Gyr of pericenter, a substantial fraction of the HDSGs are not yet {\it completely} quenched. These partially quenched galaxies may be either caught just prior to peak ram-pressure experienced at pericenter, or they are outgoing and were not fully quenched at pericenter. In principle, these two scenarios can be distinguished by deep, high resolution narrow-band \ha\, or UV imaging that can reveal one-sided, extra-planar tails of ionised gas and young stars, such as those seen in galaxies in nearby clusters \citep{smith2010,yagi2010,gavazzi2018,boselli2018}. The kinematics and radial distributions of these tailed galaxies indicate that they are an infalling population at close to pericenter, consistent with our HDSGs. The additional information provided by the orientation of the tails, which are found to generally point away from the cluster center, strongly indicates that the galaxies are observed prior to pericenter, and that stripping is occurring on the first passage. We note that there is some tentative evidence for one-sided or extra-planar \ha\, emission in several of our HDSGs (e.g., 9016800074, 9008500100, 9008500107, 9011900084 in Figure~\ref{cluster_HDS_galaxies}). However, it is difficult to draw strong conclusions based on those few galaxies because the emission is often close to the edge of the SAMI FoV, meaning that it is not clear if the \ha\, morphology is tail-like. Moreover, for a large fraction of our sample the SAMI FoV is too small to probe extra-planar emission and, therefore, to detect tails.

\section{Summary and Conclusions}\label{summary}

The spatially resolved spectroscopy provided by the SAMI-GS (SAMI Galaxy Survey) has allowed for a new method of selecting galaxies that show evidence for ongoing or very recent cessation of star formation, or quenching. Our method outlined in Section~\ref{gal_class} uses the fraction of spaxels that show strong Balmer absorption in the absence of emission lines associated with ongoing star formation. This selection allows for the detection of localized quenching that, for many of the galaxies in our sample, would not be observed in single-fibre spectroscopic surveys. Using this new technique, we selected 25 HDSGs (\hd-strong galaxies) from the SAMI-GS, with 17 found in the cluster regions and 8 found in the GAMA regions. We have focussed on investigating the environments of the cluster HDSGs with the primary aim of disentangling which environment-related mechanisms were most important in shutting down star formation. Our key results can be summarized as follows:
\begin{enumerate}
\item Galaxies with HDS signatures are rare overall, making up only $\sim 2$\% of the \logmstar\,$>\,10$ galaxy population. However, they are significantly more common amongst the population of galaxies that show evidence for recent or ongoing star formation in the cluster regions (15\%) when compared with the lower density GAMA regions (2\%). Notably, only 2/17 of the cluster HDSGs would be identified in single-fibre surveys; the majority of the HDS regions are found in the outer parts of the galaxy, away from the center.

\item The HDSGs found in the GAMA regions are different from those found in the cluster regions in several important ways. First, they are not associated with massive ($M_{200} > 10^{13}$\Msolar) groups in the GAMA regions. Second, the majority (7/8) of the GAMA HDSGs show little evidence for ongoing star formation, whereas many of the cluster HDSGs show central star formation (9/17). Third, the structure of the cluster HDSGs is generally disk-like ($n_{ser} < 2$), whereas the majority of the GAMA HDSGs have $n_{ser}>2$.

\item Focusing on the cluster regions, we find that there are significant differences between the radial, velocity, and PPS (projected-phase-space) distributions of the HDSGs when compared to the PASGs (passive galaxies) and SFGs (star-forming galaxies). The cluster HDSGs are exclusively found to have clustercentric distances $R/$\rtwo$\leq 0.6$, and have a larger velocity dispersion than the general cluster population ($\sigma_{\rm HDS}=1.66^{+0.29}_{-0.25} \sigma_{200})$. The distribution of HDSGs in the PPS reveals a coherent arc-like structure, where the HDSGs with smaller clustercentric distances have higher velocities, and those at larger clustercentric distances have smaller velocities.

\item By comparing with the simulated orbit libraries derived from clusters selected from cosmological N-body simulations by \citet{oman2013}, we find that the distribution of HDSGs in PPS is consistent with that expected of a population of infalling galaxies that have entered within 0.5$r_{200, 3D}$ within the last 1\,Gyr. We find that the SFG PPS distribution is consistent with an infalling population that are yet to pass 0.5$r_{200, 3D}$, while the PASG PPS distribution is consistent with a virialised population.

\item For the 8/9 of the cluster HDSGs with central star formation, the extent of the EW(\ha) emission can be explained by outside-in quenching due to ram-pressure stripping. 

\end{enumerate}

We conclude that the HDSGs in the cluster regions consist of a population of infalling star-forming galaxies that are close to pericenter, and are currently being quenched due to ram-pressure stripping. On the other hand, the quenching of the star formation in the GAMA HDSGs is unlikely to be associated with large-scale environment processes, and may be internally driven. We note that by definition our selection biases us towards selecting objects that have had their star formation quenched within the last $\sim 1.5$\,Gyr, and therefore towards selecting for processes that quench star formation on relatively short timescales. 

While our results show that ram-pressure stripping is likely a very important mechanism for quenching in clusters, more subtle effects on star formation due to, e.g., starvation, may also gradually lower star formation during the early phases of infall. There is also the key question of the future evolution of the HDSGs. Do they maintain some star formation at their centers as they head to apocenter, or are they completely stripped during pericenter? In future work, we will address these important questions by using the cluster portion of the SAMI-GS to perform a more comprehensive investigation of the current and recent star formation of cluster galaxies, and the relation to the cluster environment. We note that the next-generation HECTOR Galaxy Survey \citep{bryant2016} will extend the SAMI-GS by providing resolved spectroscopy for a much larger sample of cluster galaxies, which will include more high-mass halos ($M_{200} > 10^{14.5}$\Msolar), and also galaxies in the cluster outskirts (up to 2\rtwo). This extended survey will enable the study of important regions where preprocessing may be important, as well as probing the regions where ``backsplash'' galaxies are most commonly found \citep{balogh2000,	gill2005}.

\section*{Acknowledgements}

We thank the anonymous referee for their comments that have helped to improve this paper. MSO acknowledges the funding support from the Australian Research Council through a Future Fellowship (FT140100255). MH acknowledges support from an NSERC Discovery Grant, from the Australian Astronomical Observatory Distinguished Visitor Scheme and from the Australian Research Council Centre of Excellence for All Sky Astrophysics in 3 Dimensions (ASTRO 3D) also as a Distinguished Visitor.
KO received support from VICI grant 016.130.338 of the Netherlands Foundation for Scientific Research (NWO).
JBH is supported by an ARC Laureate Fellowship that funds Jesse van de Sande and an ARC Federation Fellowship that funded the SAMI prototype.
SB acknowledges the funding support from the Australian Research Council through a Future Fellowship (FT140101166). 
JJB acknowledges support of an Australian Research Council Future Fellowship (FT180100231). 
LC is the recipient of an Australian Research Council Future Fellowship (FT180100066) funded by the Australian Government. 
JvdS is funded under Bland-Hawthorn's ARC Laureate Fellowship (FL140100278). 
C.F. acknowledges funding provided by the Australian Research Council (Discovery Projects DP150104329 and DP170100603, and Future Fellowship FT180100495), and the Australia-Germany Joint Research Cooperation Scheme (UA-DAAD). 
BG  is  the  recipient  of  an  Australian  Research Council Future Fellowship (FT140101202). 
Support for AMM is provided by NASA through Hubble Fellowship grant \#HST-HF2-51377 awarded by the Space Telescope Science Institute, which is operated by the Association of Universities for Research in Astronomy, Inc., for NASA, under contract NAS5-26555.
NS acknowledges support of a University of Sydney Postdoctoral Research Fellowship. 
S.K.Y. acknowledges support from the Korean National Research Foundation (2017R1A2A1A05001116) and by the Yonsei University Future Leading Research Initiative (2015-22-0064). This study was performed under the umbrella of the joint collaboration between Yonsei University Observatory and the Korean Astronomy and Space Science Institute.  

The SAMI Galaxy Survey is based on observations made at the Anglo-Australian Telescope. The Sydney-AAO Multi-object Integral field spectrograph (SAMI) was developed jointly by the University of Sydney and the Australian Astronomical Observatory. The SAMI input catalogue is based on data taken from the Sloan Digital Sky Survey, the GAMA Survey and the VST/ATLAS Survey. The SAMI Galaxy Survey is supported by the Australian Research Council Centre of Excellence for All Sky Astrophysics in 3 Dimensions (ASTRO 3D), through project number CE170100013, the Australian Research Council Centre of Excellence for All-sky Astrophysics (CAASTRO), through project number CE110001020, and other participating institutions. The SAMI Galaxy Survey website is http://sami-survey.org/ .

GAMA is a joint European-Australasian project based around a spectroscopic campaign using the Anglo-Australian Telescope. The GAMA input catalogue is based on data taken from the Sloan Digital Sky Survey and the UKIRT Infrared Deep Sky Survey. Complementary imaging of the GAMA regions is being obtained by a number of independent survey programmes including GALEX MIS, VST KiDS, VISTA VIKING, WISE, Herschel-ATLAS, GMRT and ASKAP providing UV to radio coverage. GAMA is funded by the STFC (UK), the ARC (Australia), the AAO, and the participating institutions. The GAMA website is http://www.gama-survey.org/. Based on observations made with ESO Telescopes at the La Silla Paranal Observatory under programme ID 177.A-3016. Based on data products (VST/ATLAS) from observations made with ESO Telescopes at the La Silla Paranal Observatory under programme ID 177.A-3011(AJ). This paper includes data that have been provided by AAO Data Central  (datacentral.org.au).






\begin{thebibliography}{}
\expandafter\ifx\csname natexlab\endcsname\relax\def\natexlab#1{#1}\fi
\providecommand{\url}[1]{\href{#1}{#1}}
\providecommand{\dodoi}[1]{doi:~\href{http://doi.org/#1}{\nolinkurl{#1}}}
\providecommand{\doeprint}[1]{\href{http://ascl.net/#1}{\nolinkurl{http://ascl.net/#1}}}
\providecommand{\doarXiv}[1]{\href{https://arxiv.org/abs/#1}{\nolinkurl{https://arxiv.org/abs/#1}}}

\bibitem[{{AAO software team}(2015)}]{2dfdr2015}
{AAO software team}. 2015, {2dfdr: Data reduction software}, Astrophysics
  Source Code Library.
\newblock \doeprint{1505.015}

\bibitem[{{Ahn} {et~al.}(2012){Ahn}, {Alexandroff}, {Allende Prieto},
  {Anderson}, {Anderton}, {Andrews}, {Aubourg}, {Bailey}, {Balbinot}, {Barnes},
  \& et~al.}]{ahn2012}
{Ahn}, C.~P., {Alexandroff}, R., {Allende Prieto}, C., {et~al.} 2012, \apjs,
  203, 21, \dodoi{10.1088/0067-0049/203/2/21}

\bibitem[{{Alatalo} {et~al.}(2016){Alatalo}, {Cales}, {Rich}, {Appleton},
  {Kewley}, {Lacy}, {Lanz}, {Medling}, \& {Nyland}}]{alatalo2016}
{Alatalo}, K., {Cales}, S.~L., {Rich}, J.~A., {et~al.} 2016, \apjs, 224, 38,
  \dodoi{10.3847/0067-0049/224/2/38}

\bibitem[{{Allen} {et~al.}(2014){Allen}, {Green}, {Fogarty}, {Sharp},
  {Nielsen}, {Konstantopoulos}, {Taylor}, {Scott}, {Cortese}, {Richards},
  {Croom}, {Owers}, {Bauer}, {Sweet}, \& {Bryant}}]{allen2014}
{Allen}, J.~T., {Green}, A.~W., {Fogarty}, L.~M.~R., {et~al.} 2014, {SAMI:
  Sydney-AAO Multi-object Integral field spectrograph pipeline}, Astrophysics
  Source Code Library.
\newblock \doeprint{1407.006}

\bibitem[{{Allen} {et~al.}(2015){Allen}, {Croom}, {Konstantopoulos}, {Bryant},
  {Sharp}, {Cecil}, {Fogarty}, {Foster}, {Green}, {Ho}, {Owers}, {Schaefer},
  {Scott}, {Bauer}, {Baldry}, {Barnes}, {Bland-Hawthorn}, {Bloom}, {Brough},
  {Colless}, {Cortese}, {Couch}, {Drinkwater}, {Driver}, {Goodwin},
  {Gunawardhana}, {Hampton}, {Hopkins}, {Kewley}, {Lawrence}, {Leon-Saval},
  {Liske}, {L{\'o}pez-S{\'a}nchez}, {Lorente}, {McElroy}, {Medling}, {Mould},
  {Norberg}, {Parker}, {Power}, {Pracy}, {Richards}, {Robotham}, {Sweet},
  {Taylor}, {Thomas}, {Tonini}, \& {Walcher}}]{allen2015}
{Allen}, J.~T., {Croom}, S.~M., {Konstantopoulos}, I.~S., {et~al.} 2015,
  \mnras, 446, 1567, \dodoi{10.1093/mnras/stu2057}

\bibitem[{{Baldry} {et~al.}(2018){Baldry}, {Liske}, {Brown}, {Robotham},
  {Driver}, {Dunne}, {Alpaslan}, {Brough}, {Cluver}, {Eardley}, {Farrow},
  {Heymans}, {Hildebrandt}, {Hopkins}, {Kelvin}, {Loveday}, {Moffett},
  {Norberg}, {Owers}, {Taylor}, {Wright}, {Bamford}, {Bland-Hawthorn},
  {Bourne}, {Bremer}, {Colless}, {Conselice}, {Croom}, {Davies}, {Foster},
  {Grootes}, {Holwerda}, {Jones}, {Kafle}, {Kuijken}, {Lara-Lopez},
  {L{\'o}pez-S{\'a}nchez}, {Meyer}, {Phillipps}, {Sutherland}, {van Kampen}, \&
  {Wilkins}}]{baldry2018}
{Baldry}, I.~K., {Liske}, J., {Brown}, M.~J.~I., {et~al.} 2018, \mnras, 474,
  3875, \dodoi{10.1093/mnras/stx3042}

\bibitem[{{Baldwin} {et~al.}(1981){Baldwin}, {Phillips}, \&
  {Terlevich}}]{baldwin1981}
{Baldwin}, J.~A., {Phillips}, M.~M., \& {Terlevich}, R. 1981, \pasp, 93, 5

\bibitem[{{Balogh} {et~al.}(2004){Balogh}, {Eke}, {Miller}, {Lewis}, {Bower},
  {Couch}, {Nichol}, {Bland-Hawthorn}, {Baldry}, {Baugh}, {Bridges}, {Cannon},
  {Cole}, {Colless}, {Collins}, {Cross}, {Dalton}, {de Propris}, {Driver},
  {Efstathiou}, {Ellis}, {Frenk}, {Glazebrook}, {Gomez}, {Gray}, {Hawkins},
  {Jackson}, {Lahav}, {Lumsden}, {Maddox}, {Madgwick}, {Norberg}, {Peacock},
  {Percival}, {Peterson}, {Sutherland}, \& {Taylor}}]{balogh2004}
{Balogh}, M., {Eke}, V., {Miller}, C., {et~al.} 2004, \mnras, 348, 1355,
  \dodoi{10.1111/j.1365-2966.2004.07453.x}

\bibitem[{{Balogh} {et~al.}(1999){Balogh}, {Morris}, {Yee}, {Carlberg}, \&
  {Ellingson}}]{balogh1999}
{Balogh}, M.~L., {Morris}, S.~L., {Yee}, H.~K.~C., {Carlberg}, R.~G., \&
  {Ellingson}, E. 1999, \apj, 527, 54, \dodoi{10.1086/308056}

\bibitem[{{Balogh} {et~al.}(2000){Balogh}, {Navarro}, \& {Morris}}]{balogh2000}
{Balogh}, M.~L., {Navarro}, J.~F., \& {Morris}, S.~L. 2000, \apj, 540, 113,
  \dodoi{10.1086/309323}

\bibitem[{{Barsanti} {et~al.}(2016){Barsanti}, {Girardi}, {Biviano}, {Borgani},
  {Annunziatella}, \& {Nonino}}]{barsanti2016}
{Barsanti}, S., {Girardi}, M., {Biviano}, A., {et~al.} 2016, \aap, 595, A73,
  \dodoi{10.1051/0004-6361/201629012}

\bibitem[{{Barsanti} {et~al.}(2018){Barsanti}, {Owers}, {Brough}, {Davies},
  {Driver}, {Gunawardhana}, {Holwerda}, {Liske}, {Loveday}, {Pimbblet},
  {Robotham}, \& {Taylor}}]{barsanti2018}
{Barsanti}, S., {Owers}, M.~S., {Brough}, S., {et~al.} 2018, ArXiv e-prints.
\newblock \doarXiv{1803.05076}

\bibitem[{{Beers} {et~al.}(1990){Beers}, {Flynn}, \& {Gebhardt}}]{beers1990}
{Beers}, T.~C., {Flynn}, K., \& {Gebhardt}, K. 1990, \aj, 100, 32,
  \dodoi{10.1086/115487}

\bibitem[{{Bekki}(1999)}]{bekki1999}
{Bekki}, K. 1999, \apjl, 510, L15, \dodoi{10.1086/311796}

\bibitem[{{Bekki}(2009)}]{bekki2009}
---. 2009, \mnras, 399, 2221, \dodoi{10.1111/j.1365-2966.2009.15431.x}

\bibitem[{{Bekki}(2014)}]{bekki2014}
---. 2014, \mnras, 438, 444, \dodoi{10.1093/mnras/stt2216}

\bibitem[{{Bekki} {et~al.}(2002){Bekki}, {Couch}, \& {Shioya}}]{bekki2002}
{Bekki}, K., {Couch}, W.~J., \& {Shioya}, Y. 2002, \apj, 577, 651,
  \dodoi{10.1086/342221}

\bibitem[{{Belfiore} {et~al.}(2016){Belfiore}, {Maiolino}, {Maraston},
  {Emsellem}, {Bershady}, {Masters}, {Yan}, {Bizyaev}, {Boquien}, {Brownstein},
  {Bundy}, {Drory}, {Heckman}, {Law}, {Roman-Lopes}, {Pan}, {Stanghellini},
  {Thomas}, {Weijmans}, \& {Westfall}}]{belfiore2016}
{Belfiore}, F., {Maiolino}, R., {Maraston}, C., {et~al.} 2016, \mnras, 461,
  3111, \dodoi{10.1093/mnras/stw1234}

\bibitem[{{Belfiore} {et~al.}(2017){Belfiore}, {Maiolino}, {Maraston},
  {Emsellem}, {Bershady}, {Masters}, {Bizyaev}, {Boquien}, {Brownstein},
  {Bundy}, {Diamond-Stanic}, {Drory}, {Heckman}, {Law}, {Malanushenko},
  {Oravetz}, {Pan}, {Roman-Lopes}, {Thomas}, {Weijmans}, {Westfall}, \&
  {Yan}}]{belfiore2017}
---. 2017, \mnras, 466, 2570, \dodoi{10.1093/mnras/stw3211}

\bibitem[{{Bellhouse} {et~al.}(2017){Bellhouse}, {Jaff{\'e}}, {Hau}, {McGee},
  {Poggianti}, {Moretti}, {Gullieuszik}, {Bettoni}, {Fasano}, {D'Onofrio},
  {Fritz}, {Omizzolo}, {Sheen}, \& {Vulcani}}]{bellhouse2017}
{Bellhouse}, C., {Jaff{\'e}}, Y.~L., {Hau}, G.~K.~T., {et~al.} 2017, \apj, 844,
  49, \dodoi{10.3847/1538-4357/aa7875}

\bibitem[{{Bianconi} {et~al.}(2018){Bianconi}, {Smith}, {Haines}, {McGee},
  {Finoguenov}, \& {Egami}}]{bianconi2018}
{Bianconi}, M., {Smith}, G.~P., {Haines}, C.~P., {et~al.} 2018, \mnras, 473,
  L79, \dodoi{10.1093/mnrasl/slx167}

\bibitem[{{Biviano} \& {Katgert}(2004)}]{biviano2004}
{Biviano}, A., \& {Katgert}, P. 2004, \aap, 424, 779,
  \dodoi{10.1051/0004-6361:20041306}

\bibitem[{{Biviano} {et~al.}(1997){Biviano}, {Katgert}, {Mazure}, {Moles}, {den
  Hartog}, {Perea}, \& {Focardi}}]{biviano1997}
{Biviano}, A., {Katgert}, P., {Mazure}, A., {et~al.} 1997, \aap, 321, 84

\bibitem[{{Blake} {et~al.}(2004){Blake}, {Pracy}, {Couch}, {Bekki}, {Lewis},
  {Glazebrook}, {Baldry}, {Baugh}, {Bland-Hawthorn}, {Bridges}, {Cannon},
  {Cole}, {Colless}, {Collins}, {Dalton}, {De Propris}, {Driver}, {Efstathiou},
  {Ellis}, {Frenk}, {Jackson}, {Lahav}, {Lumsden}, {Maddox}, {Madgwick},
  {Norberg}, {Peacock}, {Peterson}, {Sutherland}, \& {Taylor}}]{blake2004}
{Blake}, C., {Pracy}, M.~B., {Couch}, W.~J., {et~al.} 2004, \mnras, 355, 713,
  \dodoi{10.1111/j.1365-2966.2004.08351.x}

\bibitem[{{Bland-Hawthorn} {et~al.}(2011){Bland-Hawthorn}, {Bryant},
  {Robertson}, {Gillingham}, {O'Byrne}, {Cecil}, {Haynes}, {Croom}, {Ellis},
  {Maack}, {Skovgaard}, \& {Noordegraaf}}]{bland2011}
{Bland-Hawthorn}, J., {Bryant}, J., {Robertson}, G., {et~al.} 2011, Optics
  Express, 19, 2649, \dodoi{10.1364/OE.19.002649}

\bibitem[{{Blanton} \& {Moustakas}(2009)}]{blanton2009}
{Blanton}, M.~R., \& {Moustakas}, J. 2009, \araa, 47, 159,
  \dodoi{10.1146/annurev-astro-082708-101734}

\bibitem[{{Boselli} {et~al.}(2014{\natexlab{a}}){Boselli}, {Cortese},
  {Boquien}, {Boissier}, {Catinella}, {Gavazzi}, {Lagos}, \&
  {Saintonge}}]{boselli2014b}
{Boselli}, A., {Cortese}, L., {Boquien}, M., {et~al.} 2014{\natexlab{a}}, \aap,
  564, A67, \dodoi{10.1051/0004-6361/201322313}

\bibitem[{{Boselli} \& {Gavazzi}(2006)}]{boselli2006}
{Boselli}, A., \& {Gavazzi}, G. 2006, \pasp, 118, 517, \dodoi{10.1086/500691}

\bibitem[{{Boselli} {et~al.}(2014{\natexlab{b}}){Boselli}, {Voyer}, {Boissier},
  {Cucciati}, {Consolandi}, {Cortese}, {Fumagalli}, {Gavazzi}, {Heinis},
  {Roehlly}, \& {Toloba}}]{boselli2014}
{Boselli}, A., {Voyer}, E., {Boissier}, S., {et~al.} 2014{\natexlab{b}}, \aap,
  570, A69, \dodoi{10.1051/0004-6361/201424419}

\bibitem[{{Boselli} {et~al.}(2016){Boselli}, {Roehlly}, {Fossati}, {Buat},
  {Boissier}, {Boquien}, {Burgarella}, {Ciesla}, {Gavazzi}, \&
  {Serra}}]{boselli2016}
{Boselli}, A., {Roehlly}, Y., {Fossati}, M., {et~al.} 2016, \aap, 596, A11,
  \dodoi{10.1051/0004-6361/201629221}

\bibitem[{{Boselli} {et~al.}(2018){Boselli}, {Fossati}, {Ferrarese},
  {Boissier}, {Consolandi}, {Longobardi}, {Amram}, {Balogh}, {Barmby},
  {Boquien}, {Boulanger}, {Braine}, {Buat}, {Burgarella}, {Combes}, {Contini},
  {Cortese}, {C{\^o}t{\'e}}, {C{\^o}t{\'e}}, {Cuillandre}, {Drissen}, {Epinat},
  {Fumagalli}, {Gallagher}, {Gavazzi}, {Gomez-Lopez}, {Gwyn}, {Harris},
  {Hensler}, {Koribalski}, {Marcelin}, {McConnachie}, {Miville-Deschenes},
  {Navarro}, {Patton}, {Peng}, {Plana}, {Prantzos}, {Robert}, {Roediger},
  {Roehlly}, {Russeil}, {Salome}, {Sanchez-Janssen}, {Serra}, {Spekkens},
  {Sun}, {Taylor}, {Tonnesen}, {Vollmer}, {Willis}, {Wozniak}, {Burdullis},
  {Devost}, {Mahoney}, {Manset}, {Petric}, {Prunet}, \&
  {Withington}}]{boselli2018}
{Boselli}, A., {Fossati}, M., {Ferrarese}, L., {et~al.} 2018, \aap, 614, A56,
  \dodoi{10.1051/0004-6361/201732407}

\bibitem[{{Bryan} \& {Norman}(1998)}]{bryan1998}
{Bryan}, G.~L., \& {Norman}, M.~L. 1998, \apj, 495, 80, \dodoi{10.1086/305262}

\bibitem[{{Bryant} {et~al.}(2014){Bryant}, {Bland-Hawthorn}, {Fogarty},
  {Lawrence}, \& {Croom}}]{bryant2014}
{Bryant}, J.~J., {Bland-Hawthorn}, J., {Fogarty}, L.~M.~R., {Lawrence}, J.~S.,
  \& {Croom}, S.~M. 2014, \mnras, 438, 869, \dodoi{10.1093/mnras/stt2254}

\bibitem[{{Bryant} {et~al.}(2015){Bryant}, {Owers}, {Robotham}, {Croom},
  {Driver}, {Drinkwater}, {Lorente}, {Cortese}, {Scott}, {Colless}, {Schaefer},
  {Taylor}, {Konstantopoulos}, {Allen}, {Baldry}, {Barnes}, {Bauer},
  {Bland-Hawthorn}, {Bloom}, {Brooks}, {Brough}, {Cecil}, {Couch}, {Croton},
  {Davies}, {Ellis}, {Fogarty}, {Foster}, {Glazebrook}, {Goodwin}, {Green},
  {Gunawardhana}, {Hampton}, {Ho}, {Hopkins}, {Kewley}, {Lawrence},
  {Leon-Saval}, {Leslie}, {McElroy}, {Lewis}, {Liske}, {L{\'o}pez-S{\'a}nchez},
  {Mahajan}, {Medling}, {Metcalfe}, {Meyer}, {Mould}, {Obreschkow}, {O'Toole},
  {Pracy}, {Richards}, {Shanks}, {Sharp}, {Sweet}, {Thomas}, {Tonini}, \&
  {Walcher}}]{bryant2015}
{Bryant}, J.~J., {Owers}, M.~S., {Robotham}, A.~S.~G., {et~al.} 2015, \mnras,
  447, 2857, \dodoi{10.1093/mnras/stu2635}

\bibitem[{{Bryant} {et~al.}(2016){Bryant}, {Bland-Hawthorn}, {Lawrence},
  {Croom}, {Brown}, {Venkatesan}, {Gillingham}, {Zhelem}, {Content},
  {Saunders}, {Staszak}, {van de Sande}, {Couch}, {Leon-Saval}, {Tims},
  {McDermid}, \& {Schaefer}}]{bryant2016}
{Bryant}, J.~J., {Bland-Hawthorn}, J., {Lawrence}, J., {et~al.} 2016, in
  \procspie, Vol. 9908, Ground-based and Airborne Instrumentation for Astronomy
  VI, 99081F

\bibitem[{{Byrd} \& {Valtonen}(1990)}]{byrd1990}
{Byrd}, G., \& {Valtonen}, M. 1990, \apj, 350, 89, \dodoi{10.1086/168362}

\bibitem[{{Cameron}(2011)}]{cameron2011}
{Cameron}, E. 2011, \pasa, 28, 128, \dodoi{10.1071/AS10046}

\bibitem[{{Cappellari}(2017)}]{cappellari2017}
{Cappellari}, M. 2017, \mnras, 466, 798, \dodoi{10.1093/mnras/stw3020}

\bibitem[{{Cappellari} \& {Copin}(2003)}]{cappellari2003}
{Cappellari}, M., \& {Copin}, Y. 2003, \mnras, 342, 345,
  \dodoi{10.1046/j.1365-8711.2003.06541.x}

\bibitem[{{Cappellari} \& {Emsellem}(2004)}]{cappellari2004}
{Cappellari}, M., \& {Emsellem}, E. 2004, \pasp, 116, 138,
  \dodoi{10.1086/381875}

\bibitem[{{Cardiel} {et~al.}(1998){Cardiel}, {Gorgas}, {Cenarro}, \&
  {Gonzalez}}]{cardiel1998}
{Cardiel}, N., {Gorgas}, J., {Cenarro}, J., \& {Gonzalez}, J.~J. 1998, \aaps,
  127, 597, \dodoi{10.1051/aas:1998123}

\bibitem[{{Carter} {et~al.}(2001){Carter}, {Fabricant}, {Geller}, {Kurtz}, \&
  {McLean}}]{carter2001}
{Carter}, B.~J., {Fabricant}, D.~G., {Geller}, M.~J., {Kurtz}, M.~J., \&
  {McLean}, B. 2001, \apj, 559, 606, \dodoi{10.1086/322349}

\bibitem[{{Catinella} {et~al.}(2018){Catinella}, {Saintonge}, {Janowiecki},
  {Cortese}, {Dav{\'e}}, {Lemonias}, {Cooper}, {Schiminovich}, {Hummels},
  {Fabello}, {Ger{\'e}b}, {Kilborn}, \& {Wang}}]{catinella2018}
{Catinella}, B., {Saintonge}, A., {Janowiecki}, S., {et~al.} 2018, \mnras, 476,
  875, \dodoi{10.1093/mnras/sty089}

\bibitem[{{Chilingarian} {et~al.}(2010){Chilingarian}, {Melchior}, \&
  {Zolotukhin}}]{chilingarian2010}
{Chilingarian}, I.~V., {Melchior}, A.-L., \& {Zolotukhin}, I.~Y. 2010, \mnras,
  405, 1409, \dodoi{10.1111/j.1365-2966.2010.16506.x}

\bibitem[{{Chung} {et~al.}(2009{\natexlab{a}}){Chung}, {van Gorkom}, {Kenney},
  {Crowl}, \& {Vollmer}}]{chung_A2009}
{Chung}, A., {van Gorkom}, J.~H., {Kenney}, J.~D.~P., {Crowl}, H., \&
  {Vollmer}, B. 2009{\natexlab{a}}, \aj, 138, 1741,
  \dodoi{10.1088/0004-6256/138/6/1741}

\bibitem[{{Chung} {et~al.}(2007){Chung}, {van Gorkom}, {Kenney}, \&
  {Vollmer}}]{chung2007}
{Chung}, A., {van Gorkom}, J.~H., {Kenney}, J.~D.~P., \& {Vollmer}, B. 2007,
  \apjl, 659, L115, \dodoi{10.1086/518034}

\bibitem[{{Chung} {et~al.}(2009{\natexlab{b}}){Chung}, {Gonzalez}, {Clowe},
  {Zaritsky}, {Markevitch}, \& {Jones}}]{chung2009}
{Chung}, S.~M., {Gonzalez}, A.~H., {Clowe}, D., {et~al.} 2009{\natexlab{b}},
  \apj, 691, 963, \dodoi{10.1088/0004-637X/691/2/963}

\bibitem[{{Cid Fernandes} {et~al.}(2011){Cid Fernandes}, {Stasi{\'n}ska},
  {Mateus}, \& {Vale Asari}}]{cidfernandes2011}
{Cid Fernandes}, R., {Stasi{\'n}ska}, G., {Mateus}, A., \& {Vale Asari}, N.
  2011, \mnras, 413, 1687, \dodoi{10.1111/j.1365-2966.2011.18244.x}

\bibitem[{{Cid Fernandes} {et~al.}(2010){Cid Fernandes}, {Stasi{\'n}ska},
  {Schlickmann}, {Mateus}, {Vale Asari}, {Schoenell}, \&
  {Sodr{\'e}}}]{cidfernandes2010}
{Cid Fernandes}, R., {Stasi{\'n}ska}, G., {Schlickmann}, M.~S., {et~al.} 2010,
  \mnras, 403, 1036, \dodoi{10.1111/j.1365-2966.2009.16185.x}

\bibitem[{{Cid Fernandes} {et~al.}(2013){Cid Fernandes}, {P{\'e}rez},
  {Garc{\'{\i}}a Benito}, {Gonz{\'a}lez Delgado}, {de Amorim}, {S{\'a}nchez},
  {Husemann}, {Falc{\'o}n Barroso}, {S{\'a}nchez-Bl{\'a}zquez}, {Walcher}, \&
  {Mast}}]{cidfernandes2013}
{Cid Fernandes}, R., {P{\'e}rez}, E., {Garc{\'{\i}}a Benito}, R., {et~al.}
  2013, \aap, 557, A86, \dodoi{10.1051/0004-6361/201220616}

\bibitem[{{Colless} \& {Dunn}(1996)}]{colless1996}
{Colless}, M., \& {Dunn}, A.~M. 1996, \apj, 458, 435, \dodoi{10.1086/176827}

\bibitem[{{Colless} {et~al.}(2001){Colless}, {Dalton}, {Maddox}, {Sutherland},
  {Norberg}, {Cole}, {Bland-Hawthorn}, {Bridges}, {Cannon}, {Collins}, {Couch},
  {Cross}, {Deeley}, {De Propris}, {Driver}, {Efstathiou}, {Ellis}, {Frenk},
  {Glazebrook}, {Jackson}, {Lahav}, {Lewis}, {Lumsden}, {Madgwick}, {Peacock},
  {Peterson}, {Price}, {Seaborne}, \& {Taylor}}]{colless2001}
{Colless}, M., {Dalton}, G., {Maddox}, S., {et~al.} 2001, \mnras, 328, 1039,
  \dodoi{10.1046/j.1365-8711.2001.04902.x}

\bibitem[{{Cortese} {et~al.}(2011){Cortese}, {Catinella}, {Boissier},
  {Boselli}, \& {Heinis}}]{cortese2011}
{Cortese}, L., {Catinella}, B., {Boissier}, S., {Boselli}, A., \& {Heinis}, S.
  2011, \mnras, 415, 1797, \dodoi{10.1111/j.1365-2966.2011.18822.x}

\bibitem[{{Cortese} {et~al.}(2006){Cortese}, {Gavazzi}, {Boselli}, {Franzetti},
  {Kennicutt}, {O'Neil}, \& {Sakai}}]{cortese2006}
{Cortese}, L., {Gavazzi}, G., {Boselli}, A., {et~al.} 2006, \aap, 453, 847,
  \dodoi{10.1051/0004-6361:20064873}

\bibitem[{{Cortese} \& {Hughes}(2009)}]{cortese2009}
{Cortese}, L., \& {Hughes}, T.~M. 2009, \mnras, 400, 1225,
  \dodoi{10.1111/j.1365-2966.2009.15548.x}

\bibitem[{{Cortese} {et~al.}(2007){Cortese}, {Marcillac}, {Richard},
  {Bravo-Alfaro}, {Kneib}, {Rieke}, {Covone}, {Egami}, {Rigby}, {Czoske}, \&
  {Davies}}]{cortese2007}
{Cortese}, L., {Marcillac}, D., {Richard}, J., {et~al.} 2007, \mnras, 376, 157,
  \dodoi{10.1111/j.1365-2966.2006.11369.x}

\bibitem[{{Cortese} {et~al.}(2012){Cortese}, {Ciesla}, {Boselli}, {Bianchi},
  {Gomez}, {Smith}, {Bendo}, {Eales}, {Pohlen}, {Baes}, {Corbelli}, {Davies},
  {Hughes}, {Hunt}, {Madden}, {Pierini}, {di Serego Alighieri}, {Zibetti},
  {Boquien}, {Clements}, {Cooray}, {Galametz}, {Magrini}, {Pappalardo},
  {Spinoglio}, \& {Vlahakis}}]{cortese2012}
{Cortese}, L., {Ciesla}, L., {Boselli}, A., {et~al.} 2012, \aap, 540, A52,
  \dodoi{10.1051/0004-6361/201118499}

\bibitem[{{Couch} \& {Sharples}(1987)}]{couch1987}
{Couch}, W.~J., \& {Sharples}, R.~M. 1987, \mnras, 229, 423

\bibitem[{{Croom} {et~al.}(2012){Croom}, {Lawrence}, {Bland-Hawthorn},
  {Bryant}, {Fogarty}, {Richards}, {Goodwin}, {Farrell}, {Miziarski}, {Heald},
  {Jones}, {Lee}, {Colless}, {Brough}, {Hopkins}, {Bauer}, {Birchall}, {Ellis},
  {Horton}, {Leon-Saval}, {Lewis}, {L{\'o}pez-S{\'a}nchez}, {Min}, {Trinh}, \&
  {Trowland}}]{croom2012}
{Croom}, S.~M., {Lawrence}, J.~S., {Bland-Hawthorn}, J., {et~al.} 2012, \mnras,
  421, 872, \dodoi{10.1111/j.1365-2966.2011.20365.x}

\bibitem[{{Crowl} \& {Kenney}(2008)}]{crowl2008}
{Crowl}, H.~H., \& {Kenney}, J.~D.~P. 2008, \aj, 136, 1623,
  \dodoi{10.1088/0004-6256/136/4/1623}

\bibitem[{{Darvish} {et~al.}(2016){Darvish}, {Mobasher}, {Sobral}, {Rettura},
  {Scoville}, {Faisst}, \& {Capak}}]{darvish2016}
{Darvish}, B., {Mobasher}, B., {Sobral}, D., {et~al.} 2016, \apj, 825, 113,
  \dodoi{10.3847/0004-637X/825/2/113}

\bibitem[{{de Carvalho} {et~al.}(2017){de Carvalho}, {Ribeiro}, {Stalder},
  {Rosa}, {Costa}, \& {Moura}}]{decarvalho2017}
{de Carvalho}, R.~R., {Ribeiro}, A.~L.~B., {Stalder}, D.~H., {et~al.} 2017,
  \aj, 154, 96, \dodoi{10.3847/1538-3881/aa7f2b}

\bibitem[{{de Jong} {et~al.}(2015){de Jong}, {Verdoes Kleijn}, {Boxhoorn},
  {Buddelmeijer}, {Capaccioli}, {Getman}, {Grado}, {Helmich}, {Huang},
  {Irisarri}, {Kuijken}, {La Barbera}, {McFarland}, {Napolitano}, {Radovich},
  {Sikkema}, {Valentijn}, {Begeman}, {Brescia}, {Cavuoti}, {Choi}, {Cordes},
  {Covone}, {Dall'Ora}, {Hildebrandt}, {Longo}, {Nakajima}, {Paolillo},
  {Puddu}, {Rifatto}, {Tortora}, {van Uitert}, {Buddendiek},
  {Harnois-D{\'e}raps}, {Erben}, {Eriksen}, {Heymans}, {Hoekstra}, {Joachimi},
  {Kitching}, {Klaes}, {Koopmans}, {K{\"o}hlinger}, {Roy}, {Sif{\'o}n},
  {Schneider}, {Sutherland}, {Viola}, \& {Vriend}}]{dejong2015}
{de Jong}, J.~T.~A., {Verdoes Kleijn}, G.~A., {Boxhoorn}, D.~R., {et~al.} 2015,
  \aap, 582, A62, \dodoi{10.1051/0004-6361/201526601}

\bibitem[{{De Lucia} {et~al.}(2012){De Lucia}, {Weinmann}, {Poggianti},
  {Arag{\'o}n-Salamanca}, \& {Zaritsky}}]{delucia2012}
{De Lucia}, G., {Weinmann}, S., {Poggianti}, B.~M., {Arag{\'o}n-Salamanca}, A.,
  \& {Zaritsky}, D. 2012, \mnras, 423, 1277,
  \dodoi{10.1111/j.1365-2966.2012.20983.x}

\bibitem[{{Domainko} {et~al.}(2006){Domainko}, {Mair}, {Kapferer}, {van
  Kampen}, {Kronberger}, {Schindler}, {Kimeswenger}, {Ruffert}, \&
  {Mangete}}]{domainko2006}
{Domainko}, W., {Mair}, M., {Kapferer}, W., {et~al.} 2006, \aap, 452, 795,
  \dodoi{10.1051/0004-6361:20053921}

\bibitem[{{Dressler}(1980)}]{dressler1980}
{Dressler}, A. 1980, \apj, 236, 351, \dodoi{10.1086/157753}

\bibitem[{{Dressler} \& {Gunn}(1983)}]{dressler1983}
{Dressler}, A., \& {Gunn}, J.~E. 1983, \apj, 270, 7, \dodoi{10.1086/161093}

\bibitem[{{Dressler} {et~al.}(1999){Dressler}, {Smail}, {Poggianti}, {Butcher},
  {Couch}, {Ellis}, \& {Oemler}}]{dressler1999}
{Dressler}, A., {Smail}, I., {Poggianti}, B.~M., {et~al.} 1999, \apjs, 122, 51,
  \dodoi{10.1086/313213}

\bibitem[{{Driver} {et~al.}(2011){Driver}, {Hill}, {Kelvin}, {Robotham},
  {Liske}, {Norberg}, {Baldry}, {Bamford}, {Hopkins}, {Loveday}, {Peacock},
  {Andrae}, {Bland-Hawthorn}, {Brough}, {Brown}, {Cameron}, {Ching}, {Colless},
  {Conselice}, {Croom}, {Cross}, {de Propris}, {Dye}, {Drinkwater}, {Ellis},
  {Graham}, {Grootes}, {Gunawardhana}, {Jones}, {van Kampen}, {Maraston},
  {Nichol}, {Parkinson}, {Phillipps}, {Pimbblet}, {Popescu}, {Prescott},
  {Roseboom}, {Sadler}, {Sansom}, {Sharp}, {Smith}, {Taylor}, {Thomas},
  {Tuffs}, {Wijesinghe}, {Dunne}, {Frenk}, {Jarvis}, {Madore}, {Meyer},
  {Seibert}, {Staveley-Smith}, {Sutherland}, \& {Warren}}]{driver2011}
{Driver}, S.~P., {Hill}, D.~T., {Kelvin}, L.~S., {et~al.} 2011, \mnras, 413,
  971, \dodoi{10.1111/j.1365-2966.2010.18188.x}

\bibitem[{{Duong}(2007)}]{duong2007}
{Duong}, T. 2007, Journal of Statistical Software, 21, 1,
  \dodoi{10.18637/jss.v021.i07}

\bibitem[{{Duong} {et~al.}(2012){Duong}, {Goud}, \& {Schauer}}]{duong2012}
{Duong}, T., {Goud}, B., \& {Schauer}, K. 2012, Proceedings of the National
  Academy of Science, 109, 8382, \dodoi{10.1073/pnas.1117796109}

\bibitem[{{Falc{\'o}n-Barroso} {et~al.}(2011){Falc{\'o}n-Barroso},
  {S{\'a}nchez-Bl{\'a}zquez}, {Vazdekis}, {Ricciardelli}, {Cardiel}, {Cenarro},
  {Gorgas}, \& {Peletier}}]{falconbarroso2011}
{Falc{\'o}n-Barroso}, J., {S{\'a}nchez-Bl{\'a}zquez}, P., {Vazdekis}, A.,
  {et~al.} 2011, \aap, 532, A95, \dodoi{10.1051/0004-6361/201116842}

\bibitem[{{Fasano} {et~al.}(2006){Fasano}, {Marmo}, {Varela}, {D'Onofrio},
  {Poggianti}, {Moles}, {Pignatelli}, {Bettoni}, {Kj{\ae}rgaard}, {Rizzi},
  {Couch}, \& {Dressler}}]{fasano2006}
{Fasano}, G., {Marmo}, C., {Varela}, J., {et~al.} 2006, \aap, 445, 805,
  \dodoi{10.1051/0004-6361:20053816}

\bibitem[{{Ferrarese} {et~al.}(2012){Ferrarese}, {C{\^o}t{\'e}}, {Cuillandre},
  {Gwyn}, {Peng}, {MacArthur}, {Duc}, {Boselli}, {Mei}, {Erben}, {McConnachie},
  {Durrell}, {Mihos}, {Jord{\'a}n}, {Lan{\c c}on}, {Puzia}, {Emsellem},
  {Balogh}, {Blakeslee}, {van Waerbeke}, {Gavazzi}, {Vollmer}, {Kavelaars},
  {Woods}, {Ball}, {Boissier}, {Courteau}, {Ferriere}, {Gavazzi},
  {Hildebrandt}, {Hudelot}, {Huertas-Company}, {Liu}, {McLaughlin}, {Mellier},
  {Milkeraitis}, {Schade}, {Balkowski}, {Bournaud}, {Carlberg}, {Chapman},
  {Hoekstra}, {Peng}, {Sawicki}, {Simard}, {Taylor}, {Tully}, {van Driel},
  {Wilson}, {Burdullis}, {Mahoney}, \& {Manset}}]{ferrarese2012}
{Ferrarese}, L., {C{\^o}t{\'e}}, P., {Cuillandre}, J.-C., {et~al.} 2012, \apjs,
  200, 4, \dodoi{10.1088/0067-0049/200/1/4}

\bibitem[{{Fisher} {et~al.}(1998){Fisher}, {Fabricant}, {Franx}, \& {van
  Dokkum}}]{fisher1998}
{Fisher}, D., {Fabricant}, D., {Franx}, M., \& {van Dokkum}, P. 1998, \apj,
  498, 195, \dodoi{10.1086/305553}

\bibitem[{{Fogarty} {et~al.}(2015){Fogarty}, {Scott}, {Owers}, {Croom},
  {Bekki}, {Houghton}, {van de Sande}, {D'Eugenio}, {Cecil}, {Colless},
  {Bland-Hawthorn}, {Brough}, {Cortese}, {Davies}, {Jones}, {Pracy}, {Allen},
  {Bryant}, {Goodwin}, {Green}, {Konstantopoulos}, {Lawrence}, {Lorente},
  {Richards}, \& {Sharp}}]{fogarty2015}
{Fogarty}, L.~M.~R., {Scott}, N., {Owers}, M.~S., {et~al.} 2015, \mnras, 454,
  2050, \dodoi{10.1093/mnras/stv2060}

\bibitem[{{Fossati} {et~al.}(2016){Fossati}, {Fumagalli}, {Boselli}, {Gavazzi},
  {Sun}, \& {Wilman}}]{fossati2016}
{Fossati}, M., {Fumagalli}, M., {Boselli}, A., {et~al.} 2016, \mnras, 455,
  2028, \dodoi{10.1093/mnras/stv2400}

\bibitem[{{Fossati} {et~al.}(2018){Fossati}, {Mendel}, {Boselli}, {Cuillandre},
  {Vollmer}, {Boissier}, {Consolandi}, {Ferrarese}, {Gwyn}, {Amram}, {Boquien},
  {Buat}, {Burgarella}, {Cortese}, {C{\^o}t{\'e}}, {C{\^o}t{\'e}}, {Durrell},
  {Fumagalli}, {Gavazzi}, {Gomez-Lopez}, {Hensler}, {Koribalski}, {Longobardi},
  {Peng}, {Roediger}, {Sun}, \& {Toloba}}]{fossati2018}
{Fossati}, M., {Mendel}, J.~T., {Boselli}, A., {et~al.} 2018, \aap, 614, A57,
  \dodoi{10.1051/0004-6361/201732373}

\bibitem[{{Fritz} {et~al.}(2017){Fritz}, {Moretti}, {Gullieuszik}, {Poggianti},
  {Bruzual}, {Vulcani}, {Nicastro}, {Jaff{\'e}}, {Cervantes Sodi}, {Bettoni},
  {Biviano}, {Fasano}, {Charlot}, {Bellhouse}, \& {Hau}}]{fritz2017}
{Fritz}, J., {Moretti}, A., {Gullieuszik}, M., {et~al.} 2017, \apj, 848, 132,
  \dodoi{10.3847/1538-4357/aa8f51}

\bibitem[{{Fumagalli} {et~al.}(2014){Fumagalli}, {Fossati}, {Hau}, {Gavazzi},
  {Bower}, {Sun}, \& {Boselli}}]{fumagalli2014}
{Fumagalli}, M., {Fossati}, M., {Hau}, G.~K.~T., {et~al.} 2014, \mnras, 445,
  4335, \dodoi{10.1093/mnras/stu2092}

\bibitem[{{Gavazzi} {et~al.}(2006){Gavazzi}, {Boselli}, {Cortese}, {Arosio},
  {Gallazzi}, {Pedotti}, \& {Carrasco}}]{gavazzi2006}
{Gavazzi}, G., {Boselli}, A., {Cortese}, L., {et~al.} 2006, \aap, 446, 839,
  \dodoi{10.1051/0004-6361:20053843}

\bibitem[{{Gavazzi} {et~al.}(2002){Gavazzi}, {Boselli}, {Pedotti}, {Gallazzi},
  \& {Carrasco}}]{gavazzi2002}
{Gavazzi}, G., {Boselli}, A., {Pedotti}, P., {Gallazzi}, A., \& {Carrasco}, L.
  2002, \aap, 396, 449, \dodoi{10.1051/0004-6361:20021403}

\bibitem[{{Gavazzi} {et~al.}(2018){Gavazzi}, {Consolandi}, {Gutierrez},
  {Boselli}, \& {Yoshida}}]{gavazzi2018}
{Gavazzi}, G., {Consolandi}, G., {Gutierrez}, M.~L., {Boselli}, A., \&
  {Yoshida}, M. 2018, \aap, 618, A130, \dodoi{10.1051/0004-6361/201833427}

\bibitem[{{Gavazzi} {et~al.}(2010){Gavazzi}, {Fumagalli}, {Cucciati}, \&
  {Boselli}}]{gavazzi2010}
{Gavazzi}, G., {Fumagalli}, M., {Cucciati}, O., \& {Boselli}, A. 2010, \aap,
  517, A73, \dodoi{10.1051/0004-6361/201014153}

\bibitem[{{Ghigna} {et~al.}(1998){Ghigna}, {Moore}, {Governato}, {Lake},
  {Quinn}, \& {Stadel}}]{ghigna1998}
{Ghigna}, S., {Moore}, B., {Governato}, F., {et~al.} 1998, \mnras, 300, 146,
  \dodoi{10.1046/j.1365-8711.1998.01918.x}

\bibitem[{{Ghirardini} {et~al.}(2018){Ghirardini}, {Eckert}, {Ettori},
  {Pointecouteau}, {Molendi}, {Gaspari}, {Rossetti}, {De Grandi}, {Roncarelli},
  {Bourdin}, {Mazzotta}, {Rasia}, \& {Vazza}}]{ghirardini2018}
{Ghirardini}, V., {Eckert}, D., {Ettori}, S., {et~al.} 2018, ArXiv e-prints.
\newblock \doarXiv{1805.00042}

\bibitem[{{Gill} {et~al.}(2005){Gill}, {Knebe}, \& {Gibson}}]{gill2005}
{Gill}, S.~P.~D., {Knebe}, A., \& {Gibson}, B.~K. 2005, \mnras, 356, 1327,
  \dodoi{10.1111/j.1365-2966.2004.08562.x}

\bibitem[{{Gonz{\'a}lez Delgado} {et~al.}(2005){Gonz{\'a}lez Delgado},
  {Cervi{\~n}o}, {Martins}, {Leitherer}, \& {Hauschildt}}]{gonzalez2005}
{Gonz{\'a}lez Delgado}, R.~M., {Cervi{\~n}o}, M., {Martins}, L.~P.,
  {Leitherer}, C., \& {Hauschildt}, P.~H. 2005, \mnras, 357, 945,
  \dodoi{10.1111/j.1365-2966.2005.08692.x}

\bibitem[{{Green} {et~al.}(2018){Green}, {Croom}, {Scott}, {Cortese},
  {Medling}, {D'Eugenio}, {Bryant}, {Bland-Hawthorn}, {Allen}, {Sharp}, {Ho},
  {Groves}, {Drinkwater}, {Mannering}, {Harischandra}, {van de Sande},
  {Thomas}, {O'Toole}, {McDermid}, {Vuong}, {Sealey}, {Bauer}, {Brough},
  {Catinella}, {Cecil}, {Colless}, {Couch}, {Driver}, {Federrath}, {Foster},
  {Goodwin}, {Hampton}, {Hopkins}, {Jones}, {Konstantopoulos}, {Lawrence},
  {Leon-Saval}, {Liske}, {L{\'o}pez-S{\'a}nchez}, {Lorente}, {Mould},
  {Obreschkow}, {Owers}, {Richards}, {Robotham}, {Schaefer}, {Sweet}, {Taranu},
  {Tescari}, {Tonini}, \& {Zafar}}]{green2018}
{Green}, A.~W., {Croom}, S.~M., {Scott}, N., {et~al.} 2018, \mnras, 475, 716,
  \dodoi{10.1093/mnras/stx3135}

\bibitem[{{Grevesse} \& {Sauval}(1998)}]{grevesse1998}
{Grevesse}, N., \& {Sauval}, A.~J. 1998, \ssr, 85, 161,
  \dodoi{10.1023/A:1005161325181}

\bibitem[{{Gullieuszik} {et~al.}(2017){Gullieuszik}, {Poggianti}, {Moretti},
  {Fritz}, {Jaff{\'e}}, {Hau}, {Bischko}, {Bellhouse}, {Bettoni}, {Fasano},
  {Vulcani}, {D'Onofrio}, \& {Biviano}}]{gullieuszik2017}
{Gullieuszik}, M., {Poggianti}, B.~M., {Moretti}, A., {et~al.} 2017, \apj, 846,
  27, \dodoi{10.3847/1538-4357/aa8322}

\bibitem[{{Gunn} \& {Gott}(1972)}]{gunn1972}
{Gunn}, J.~E., \& {Gott}, J.~R.~I. 1972, \apj, 176, 1

\bibitem[{{Haines} {et~al.}(2013){Haines}, {Pereira}, {Smith}, {Egami},
  {Sanderson}, {Babul}, {Finoguenov}, {Merluzzi}, {Busarello}, {Rawle}, \&
  {Okabe}}]{haines2013}
{Haines}, C.~P., {Pereira}, M.~J., {Smith}, G.~P., {et~al.} 2013, \apj, 775,
  126, \dodoi{10.1088/0004-637X/775/2/126}

\bibitem[{{Haines} {et~al.}(2015){Haines}, {Pereira}, {Smith}, {Egami},
  {Babul}, {Finoguenov}, {Ziparo}, {McGee}, {Rawle}, {Okabe}, \&
  {Moran}}]{haines2015}
---. 2015, \apj, 806, 101, \dodoi{10.1088/0004-637X/806/1/101}

\bibitem[{{Haines} {et~al.}(2018){Haines}, {Finoguenov}, {Smith}, {Babul},
  {Egami}, {Mazzotta}, {Okabe}, {Pereira}, {Bianconi}, {McGee}, {Ziparo},
  {Campusano}, \& {Loyola}}]{haines2018}
{Haines}, C.~P., {Finoguenov}, A., {Smith}, G.~P., {et~al.} 2018, \mnras, 477,
  4931, \dodoi{10.1093/mnras/sty651}

\bibitem[{{Hern{\'a}ndez-Fern{\'a}ndez}
  {et~al.}(2014){Hern{\'a}ndez-Fern{\'a}ndez}, {Haines}, {Diaferio},
  {Iglesias-P{\'a}ramo}, {Mendes de Oliveira}, \& {Vilchez}}]{hernandez2014}
{Hern{\'a}ndez-Fern{\'a}ndez}, J.~D., {Haines}, C.~P., {Diaferio}, A., {et~al.}
  2014, \mnras, 438, 2186, \dodoi{10.1093/mnras/stt2354}

\bibitem[{{Hill} {et~al.}(2011){Hill}, {Kelvin}, {Driver}, {Robotham},
  {Cameron}, {Cross}, {Andrae}, {Baldry}, {Bamford}, {Bland-Hawthorn},
  {Brough}, {Conselice}, {Dye}, {Hopkins}, {Liske}, {Loveday}, {Norberg},
  {Peacock}, {Croom}, {Frenk}, {Graham}, {Jones}, {Kuijken}, {Madore},
  {Nichol}, {Parkinson}, {Phillipps}, {Pimbblet}, {Popescu}, {Prescott},
  {Seibert}, {Sharp}, {Sutherland}, {Thomas}, {Tuffs}, \& {van
  Kampen}}]{hill2011}
{Hill}, D.~T., {Kelvin}, L.~S., {Driver}, S.~P., {et~al.} 2011, \mnras, 412,
  765, \dodoi{10.1111/j.1365-2966.2010.17950.x}

\bibitem[{{Ho} {et~al.}(2014){Ho}, {Kewley}, {Dopita}, {Medling}, {Allen},
  {Bland-Hawthorn}, {Bloom}, {Bryant}, {Croom}, {Fogarty}, {Goodwin}, {Green},
  {Konstantopoulos}, {Lawrence}, {L{\'o}pez-S{\'a}nchez}, {Owers}, {Richards},
  \& {Sharp}}]{ho2014}
{Ho}, I.-T., {Kewley}, L.~J., {Dopita}, M.~A., {et~al.} 2014, \mnras, 444,
  3894, \dodoi{10.1093/mnras/stu1653}

\bibitem[{{Ho} {et~al.}(2016){Ho}, {Medling}, {Groves}, {Rich}, {Rupke},
  {Hampton}, {Kewley}, {Bland-Hawthorn}, {Croom}, {Richards}, {Schaefer},
  {Sharp}, \& {Sweet}}]{ho2016}
{Ho}, I.-T., {Medling}, A.~M., {Groves}, B., {et~al.} 2016, \apss, 361, 280,
  \dodoi{10.1007/s10509-016-2865-2}

\bibitem[{{Hudson} {et~al.}(2010){Hudson}, {Stevenson}, {Smith}, {Wegner},
  {Lucey}, \& {Simard}}]{hudson2010}
{Hudson}, M.~J., {Stevenson}, J.~B., {Smith}, R.~J., {et~al.} 2010, \mnras,
  409, 405, \dodoi{10.1111/j.1365-2966.2010.17318.x}

\bibitem[{{Jaff{\'e}} {et~al.}(2015){Jaff{\'e}}, {Smith}, {Candlish},
  {Poggianti}, {Sheen}, \& {Verheijen}}]{jaffe2015}
{Jaff{\'e}}, Y.~L., {Smith}, R., {Candlish}, G.~N., {et~al.} 2015, \mnras, 448,
  1715, \dodoi{10.1093/mnras/stv100}

\bibitem[{{Jaff{\'e}} {et~al.}(2018){Jaff{\'e}}, {Poggianti}, {Moretti},
  {Gullieuszik}, {Smith}, {Vulcani}, {Fasano}, {Fritz}, {Tonnesen}, {Bettoni},
  {Hau}, {Biviano}, {Bellhouse}, \& {McGee}}]{jaffe2018}
{Jaff{\'e}}, Y.~L., {Poggianti}, B.~M., {Moretti}, A., {et~al.} 2018, \mnras,
  476, 4753, \dodoi{10.1093/mnras/sty500}

\bibitem[{{Jung} {et~al.}(2018){Jung}, {Choi}, {Wong}, {Kimm}, {Chung}, \&
  {Yi}}]{jung2018}
{Jung}, S.~L., {Choi}, H., {Wong}, O.~I., {et~al.} 2018, \apj, 865, 156,
  \dodoi{10.3847/1538-4357/aadda2}

\bibitem[{{Kauffmann} {et~al.}(2003{\natexlab{a}}){Kauffmann}, {Heckman},
  {White}, {Charlot}, {Tremonti}, {Peng}, {Seibert}, {Brinkmann}, {Nichol},
  {SubbaRao}, \& {York}}]{kauffmann2003b}
{Kauffmann}, G., {Heckman}, T.~M., {White}, S.~D.~M., {et~al.}
  2003{\natexlab{a}}, \mnras, 341, 54, \dodoi{10.1046/j.1365-8711.2003.06292.x}

\bibitem[{{Kauffmann} {et~al.}(2003{\natexlab{b}}){Kauffmann}, {Heckman},
  {Tremonti}, {Brinchmann}, {Charlot}, {White}, {Ridgway}, {Brinkmann},
  {Fukugita}, {Hall}, {Ivezi{\'c}}, {Richards}, \&
  {Schneider}}]{kauffmann2003a}
{Kauffmann}, G., {Heckman}, T.~M., {Tremonti}, C., {et~al.} 2003{\natexlab{b}},
  \mnras, 346, 1055, \dodoi{10.1111/j.1365-2966.2003.07154.x}

\bibitem[{{Kauffmann} {et~al.}(2003{\natexlab{c}}){Kauffmann}, {Heckman},
  {White}, {Charlot}, {Tremonti}, {Brinchmann}, {Bruzual}, {Peng}, {Seibert},
  {Bernardi}, {Blanton}, {Brinkmann}, {Castander}, {Cs{\'a}bai}, {Fukugita},
  {Ivezic}, {Munn}, {Nichol}, {Padmanabhan}, {Thakar}, {Weinberg}, \&
  {York}}]{kauffman2003c}
{Kauffmann}, G., {Heckman}, T.~M., {White}, S.~D.~M., {et~al.}
  2003{\natexlab{c}}, \mnras, 341, 33, \dodoi{10.1046/j.1365-8711.2003.06291.x}

\bibitem[{{Kelvin} {et~al.}(2012){Kelvin}, {Driver}, {Robotham}, {Hill},
  {Alpaslan}, {Baldry}, {Bamford}, {Bland-Hawthorn}, {Brough}, {Graham},
  {H{\"a}ussler}, {Hopkins}, {Liske}, {Loveday}, {Norberg}, {Phillipps},
  {Popescu}, {Prescott}, {Taylor}, \& {Tuffs}}]{kelvin2012}
{Kelvin}, L.~S., {Driver}, S.~P., {Robotham}, A.~S.~G., {et~al.} 2012, \mnras,
  421, 1007, \dodoi{10.1111/j.1365-2966.2012.20355.x}

\bibitem[{{Kewley} {et~al.}(2001){Kewley}, {Dopita}, {Sutherland}, {Heisler},
  \& {Trevena}}]{kewley2001}
{Kewley}, L.~J., {Dopita}, M.~A., {Sutherland}, R.~S., {Heisler}, C.~A., \&
  {Trevena}, J. 2001, \apj, 556, 121, \dodoi{10.1086/321545}

\bibitem[{{Kewley} {et~al.}(2006){Kewley}, {Groves}, {Kauffmann}, \&
  {Heckman}}]{kewley2006}
{Kewley}, L.~J., {Groves}, B., {Kauffmann}, G., \& {Heckman}, T. 2006, \mnras,
  372, 961, \dodoi{10.1111/j.1365-2966.2006.10859.x}

\bibitem[{{Koopmann} \& {Kenney}(2004{\natexlab{a}})}]{koopmann2004b}
{Koopmann}, R.~A., \& {Kenney}, J.~D.~P. 2004{\natexlab{a}}, \apj, 613, 851,
  \dodoi{10.1086/423190}

\bibitem[{{Koopmann} \& {Kenney}(2004{\natexlab{b}})}]{koopmann2004}
---. 2004{\natexlab{b}}, \apj, 613, 866, \dodoi{10.1086/423191}

\bibitem[{{Larson} {et~al.}(1980){Larson}, {Tinsley}, \&
  {Caldwell}}]{larson1980}
{Larson}, R.~B., {Tinsley}, B.~M., \& {Caldwell}, C.~N. 1980, \apj, 237, 692,
  \dodoi{10.1086/157917}

\bibitem[{{Lee} {et~al.}(2017){Lee}, {Chung}, {Tonnesen}, {Kenney}, {Wong},
  {Vollmer}, {Petitpas}, {Crowl}, \& {van Gorkom}}]{lee2017}
{Lee}, B., {Chung}, A., {Tonnesen}, S., {et~al.} 2017, \mnras, 466, 1382,
  \dodoi{10.1093/mnras/stw3162}

\bibitem[{{Lewis} {et~al.}(2002){Lewis}, {Balogh}, {De Propris}, {Couch},
  {Bower}, {Offer}, {Bland-Hawthorn}, {Baldry}, {Baugh}, {Bridges}, {Cannon},
  {Cole}, {Colless}, {Collins}, {Cross}, {Dalton}, {Driver}, {Efstathiou},
  {Ellis}, {Frenk}, {Glazebrook}, {Hawkins}, {Jackson}, {Lahav}, {Lumsden},
  {Maddox}, {Madgwick}, {Norberg}, {Peacock}, {Percival}, {Peterson},
  {Sutherland}, \& {Taylor}}]{lewis2002}
{Lewis}, I., {Balogh}, M., {De Propris}, R., {et~al.} 2002, \mnras, 334, 673,
  \dodoi{10.1046/j.1365-8711.2002.05558.x}

\bibitem[{{Lindner} {et~al.}(2015){Lindner}, {Vera-Ciro}, {Murray},
  {Stanimirovi{\'c}}, {Babler}, {Heiles}, {Hennebelle}, {Goss}, \&
  {Dickey}}]{lindner2015}
{Lindner}, R.~R., {Vera-Ciro}, C., {Murray}, C.~E., {et~al.} 2015, \aj, 149,
  138, \dodoi{10.1088/0004-6256/149/4/138}

\bibitem[{{Liske} {et~al.}(2015){Liske}, {Baldry}, {Driver}, {Tuffs},
  {Alpaslan}, {Andrae}, {Brough}, {Cluver}, {Grootes}, {Gunawardhana},
  {Kelvin}, {Loveday}, {Robotham}, {Taylor}, {Bamford}, {Bland-Hawthorn},
  {Brown}, {Drinkwater}, {Hopkins}, {Meyer}, {Norberg}, {Peacock}, {Agius},
  {Andrews}, {Bauer}, {Ching}, {Colless}, {Conselice}, {Croom}, {Davies}, {De
  Propris}, {Dunne}, {Eardley}, {Ellis}, {Foster}, {Frenk}, {H{\"a}u{\ss}ler},
  {Holwerda}, {Howlett}, {Ibarra}, {Jarvis}, {Jones}, {Kafle}, {Lacey},
  {Lange}, {Lara-L{\'o}pez}, {L{\'o}pez-S{\'a}nchez}, {Maddox}, {Madore},
  {McNaught-Roberts}, {Moffett}, {Nichol}, {Owers}, {Palamara}, {Penny},
  {Phillipps}, {Pimbblet}, {Popescu}, {Prescott}, {Proctor}, {Sadler},
  {Sansom}, {Seibert}, {Sharp}, {Sutherland}, {V{\'a}zquez-Mata}, {van Kampen},
  {Wilkins}, {Williams}, \& {Wright}}]{liske2015}
{Liske}, J., {Baldry}, I.~K., {Driver}, S.~P., {et~al.} 2015, \mnras, 452,
  2087, \dodoi{10.1093/mnras/stv1436}

\bibitem[{{Lopes} {et~al.}(2017){Lopes}, {Ribeiro}, \& {Rembold}}]{lopes2017}
{Lopes}, P.~A.~A., {Ribeiro}, A.~L.~B., \& {Rembold}, S.~B. 2017, \mnras, 472,
  409, \dodoi{10.1093/mnras/stx2046}

\bibitem[{{Mahajan} {et~al.}(2011){Mahajan}, {Mamon}, \&
  {Raychaudhury}}]{mahajan2011}
{Mahajan}, S., {Mamon}, G.~A., \& {Raychaudhury}, S. 2011, \mnras, 416, 2882,
  \dodoi{10.1111/j.1365-2966.2011.19236.x}

\bibitem[{{Manzer} \& {De Robertis}(2014)}]{manzer2014}
{Manzer}, L.~H., \& {De Robertis}, M.~M. 2014, \apj, 788, 140,
  \dodoi{10.1088/0004-637X/788/2/140}

\bibitem[{{Markwardt}(2009)}]{markwardt2009}
{Markwardt}, C.~B. 2009, in Astronomical Society of the Pacific Conference
  Series, Vol. 411, Astronomical Data Analysis Software and Systems XVIII, ed.
  D.~A. {Bohlender}, D.~{Durand}, \& P.~{Dowler}, 251

\bibitem[{{Marziani} {et~al.}(2017){Marziani}, {D'Onofrio}, {Bettoni},
  {Poggianti}, {Moretti}, {Fasano}, {Fritz}, {Cava}, {Varela}, \&
  {Omizzolo}}]{marziani2017}
{Marziani}, P., {D'Onofrio}, M., {Bettoni}, D., {et~al.} 2017, \aap, 599, A83,
  \dodoi{10.1051/0004-6361/201628941}

\bibitem[{{McFarland} {et~al.}(2013){McFarland}, {Verdoes-Kleijn}, {Sikkema},
  {Helmich}, {Boxhoorn}, \& {Valentijn}}]{mcfarland2013}
{McFarland}, J.~P., {Verdoes-Kleijn}, G., {Sikkema}, G., {et~al.} 2013,
  Experimental Astronomy, 35, 45, \dodoi{10.1007/s10686-011-9266-x}

\bibitem[{{McGee} {et~al.}(2009){McGee}, {Balogh}, {Bower}, {Font}, \&
  {McCarthy}}]{mcgee2009}
{McGee}, S.~L., {Balogh}, M.~L., {Bower}, R.~G., {Font}, A.~S., \& {McCarthy},
  I.~G. 2009, \mnras, 400, 937, \dodoi{10.1111/j.1365-2966.2009.15507.x}

\bibitem[{{Medling} {et~al.}(2018){Medling}, {Cortese}, {Croom}, {Green},
  {Groves}, {Hampton}, {Ho}, {Davies}, {Kewley}, {Moffett}, {Schaefer},
  {Taylor}, {Zafar}, {Bekki}, {Bland-Hawthorn}, {Bloom}, {Brough}, {Bryant},
  {Catinella}, {Cecil}, {Colless}, {Couch}, {Drinkwater}, {Driver},
  {Federrath}, {Foster}, {Goldstein}, {Goodwin}, {Hopkins}, {Lawrence},
  {Leslie}, {Lewis}, {Lorente}, {Owers}, {McDermid}, {Richards}, {Sharp},
  {Scott}, {Sweet}, {Taranu}, {Tescari}, {Tonini}, {van de Sande}, {Walcher},
  \& {Wright}}]{medling2018}
{Medling}, A.~M., {Cortese}, L., {Croom}, S.~M., {et~al.} 2018, \mnras, 475,
  5194, \dodoi{10.1093/mnras/sty127}

\bibitem[{{Meert} {et~al.}(2015){Meert}, {Vikram}, \& {Bernardi}}]{meert2015}
{Meert}, A., {Vikram}, V., \& {Bernardi}, M. 2015, \mnras, 446, 3943,
  \dodoi{10.1093/mnras/stu2333}

\bibitem[{{Merluzzi} {et~al.}(2016){Merluzzi}, {Busarello}, {Dopita}, {Haines},
  {Steinhauser}, {Bourdin}, \& {Mazzotta}}]{merluzzi2016}
{Merluzzi}, P., {Busarello}, G., {Dopita}, M.~A., {et~al.} 2016, \mnras, 460,
  3345, \dodoi{10.1093/mnras/stw1198}

\bibitem[{{Merluzzi} {et~al.}(2013){Merluzzi}, {Busarello}, {Dopita}, {Haines},
  {Steinhauser}, {Mercurio}, {Rifatto}, {Smith}, \& {Schindler}}]{merluzzi2013}
---. 2013, \mnras, 429, 1747, \dodoi{10.1093/mnras/sts466}

\bibitem[{{Miller} {et~al.}(2003){Miller}, {Nichol}, {G{\'o}mez}, {Hopkins}, \&
  {Bernardi}}]{c.miller2003}
{Miller}, C.~J., {Nichol}, R.~C., {G{\'o}mez}, P.~L., {Hopkins}, A.~M., \&
  {Bernardi}, M. 2003, \apj, 597, 142, \dodoi{10.1086/378383}

\bibitem[{{Moore} {et~al.}(1996){Moore}, {Katz}, {Lake}, {Dressler}, \&
  {Oemler}}]{moore1996}
{Moore}, B., {Katz}, N., {Lake}, G., {Dressler}, A., \& {Oemler}, A. 1996,
  \nat, 379, 613, \dodoi{10.1038/379613a0}

\bibitem[{{Moran} {et~al.}(2007){Moran}, {Ellis}, {Treu}, {Smith}, {Rich}, \&
  {Smail}}]{moran2007}
{Moran}, S.~M., {Ellis}, R.~S., {Treu}, T., {et~al.} 2007, \apj, 671, 1503,
  \dodoi{10.1086/522303}

\bibitem[{{Moretti} {et~al.}(2017){Moretti}, {Gullieuszik}, {Poggianti},
  {Paccagnella}, {Couch}, {Vulcani}, {Bettoni}, {Fritz}, {Cava}, {Fasano},
  {D'Onofrio}, \& {Omizzolo}}]{moretti2017}
{Moretti}, A., {Gullieuszik}, M., {Poggianti}, B., {et~al.} 2017, ArXiv
  e-prints.
\newblock \doarXiv{1701.02590}

\bibitem[{{Muzzin} {et~al.}(2012){Muzzin}, {Wilson}, {Yee}, {Gilbank},
  {Hoekstra}, {Demarco}, {Balogh}, {van Dokkum}, {Franx}, {Ellingson}, {Hicks},
  {Nantais}, {Noble}, {Lacy}, {Lidman}, {Rettura}, {Surace}, \&
  {Webb}}]{muzzin2012}
{Muzzin}, A., {Wilson}, G., {Yee}, H.~K.~C., {et~al.} 2012, \apj, 746, 188,
  \dodoi{10.1088/0004-637X/746/2/188}

\bibitem[{{Muzzin} {et~al.}(2014){Muzzin}, {van der Burg}, {McGee}, {Balogh},
  {Franx}, {Hoekstra}, {Hudson}, {Noble}, {Taranu}, {Webb}, {Wilson}, \&
  {Yee}}]{muzzin2014}
{Muzzin}, A., {van der Burg}, R.~F.~J., {McGee}, S.~L., {et~al.} 2014, \apj,
  796, 65, \dodoi{10.1088/0004-637X/796/1/65}

\bibitem[{{Noble} {et~al.}(2013){Noble}, {Webb}, {Muzzin}, {Wilson}, {Yee}, \&
  {van der Burg}}]{noble2013}
{Noble}, A.~G., {Webb}, T.~M.~A., {Muzzin}, A., {et~al.} 2013, \apj, 768, 118,
  \dodoi{10.1088/0004-637X/768/2/118}

\bibitem[{{Noble} {et~al.}(2019){Noble}, {Muzzin}, {McDonald}, {Rudnick},
  {Matharu}, {Cooper}, {Demarco}, {Lidman}, {Nantais}, {van Kampen}, {Webb},
  {Wilson}, \& {Yee}}]{noble2019}
{Noble}, A.~G., {Muzzin}, A., {McDonald}, M., {et~al.} 2019, \apj, 870, 56,
  \dodoi{10.3847/1538-4357/aaf1c6}

\bibitem[{{Nulsen}(1982)}]{nulsen1982}
{Nulsen}, P.~E.~J. 1982, \mnras, 198, 1007

\bibitem[{{Oman} \& {Hudson}(2016)}]{oman2016}
{Oman}, K.~A., \& {Hudson}, M.~J. 2016, \mnras, 463, 3083,
  \dodoi{10.1093/mnras/stw2195}

\bibitem[{{Oman} {et~al.}(2013){Oman}, {Hudson}, \& {Behroozi}}]{oman2013}
{Oman}, K.~A., {Hudson}, M.~J., \& {Behroozi}, P.~S. 2013, \mnras, 431, 2307,
  \dodoi{10.1093/mnras/stt328}

\bibitem[{{Owers} {et~al.}(2017){Owers}, {Allen}, {Baldry}, {Bryant}, {Cecil},
  {Cortese}, {Croom}, {Driver}, {Fogarty}, {Green}, {Helmich}, {de Jong},
  {Kuijken}, {Mahajan}, {McFarland}, {Pracy}, {Robotham}, {Sikkema}, {Sweet},
  {Taylor}, {Verdoes Kleijn}, {Bauer}, {Bland-Hawthorn}, {Brough}, {Colless},
  {Couch}, {Davies}, {Drinkwater}, {Goodwin}, {Hopkins}, {Konstantopoulos},
  {Foster}, {Lawrence}, {Lorente}, {Medling}, {Metcalfe}, {Richards}, {van de
  Sande}, {Scott}, {Shanks}, {Sharp}, {Thomas}, \& {Tonini}}]{owers2017}
{Owers}, M.~S., {Allen}, J.~T., {Baldry}, I., {et~al.} 2017, \mnras, 468, 1824,
  \dodoi{10.1093/mnras/stx562}

\bibitem[{{Paccagnella} {et~al.}(2016){Paccagnella}, {Vulcani}, {Poggianti},
  {Moretti}, {Fritz}, {Gullieuszik}, {Couch}, {Bettoni}, {Cava}, {D'Onofrio},
  \& {Fasano}}]{paccagnella2016}
{Paccagnella}, A., {Vulcani}, B., {Poggianti}, B.~M., {et~al.} 2016, \apjl,
  816, L25, \dodoi{10.3847/2041-8205/816/2/L25}

\bibitem[{{Paccagnella} {et~al.}(2017){Paccagnella}, {Vulcani}, {Poggianti},
  {Fritz}, {Fasano}, {Moretti}, {Jaff{\'e}}, {Biviano}, {Gullieuszik},
  {Bettoni}, {Cava}, {Couch}, \& {D'Onofrio}}]{paccagnella2017}
---. 2017, \apj, 838, 148, \dodoi{10.3847/1538-4357/aa64d7}

\bibitem[{{Peacock}(1983)}]{peacock1983}
{Peacock}, J.~A. 1983, \mnras, 202, 615, \dodoi{10.1093/mnras/202.3.615}

\bibitem[{{Peng} {et~al.}(2010){Peng}, {Lilly}, {Kova{\v c}}, {Bolzonella},
  {Pozzetti}, {Renzini}, {Zamorani}, {Ilbert}, {Knobel}, {Iovino}, {Maier},
  {Cucciati}, {Tasca}, {Carollo}, {Silverman}, {Kampczyk}, {de Ravel},
  {Sanders}, {Scoville}, {Contini}, {Mainieri}, {Scodeggio}, {Kneib}, {Le
  F{\`e}vre}, {Bardelli}, {Bongiorno}, {Caputi}, {Coppa}, {de la Torre},
  {Franzetti}, {Garilli}, {Lamareille}, {Le Borgne}, {Le Brun}, {Mignoli},
  {Perez Montero}, {Pello}, {Ricciardelli}, {Tanaka}, {Tresse}, {Vergani},
  {Welikala}, {Zucca}, {Oesch}, {Abbas}, {Barnes}, {Bordoloi}, {Bottini},
  {Cappi}, {Cassata}, {Cimatti}, {Fumana}, {Hasinger}, {Koekemoer},
  {Leauthaud}, {Maccagni}, {Marinoni}, {McCracken}, {Memeo}, {Meneux}, {Nair},
  {Porciani}, {Presotto}, \& {Scaramella}}]{peng2010}
{Peng}, Y.-j., {Lilly}, S.~J., {Kova{\v c}}, K., {et~al.} 2010, \apj, 721, 193,
  \dodoi{10.1088/0004-637X/721/1/193}

\bibitem[{{Poggianti} {et~al.}(2004){Poggianti}, {Bridges}, {Komiyama}, {Yagi},
  {Carter}, {Mobasher}, {Okamura}, \& {Kashikawa}}]{poggianti2004}
{Poggianti}, B.~M., {Bridges}, T.~J., {Komiyama}, Y., {et~al.} 2004, \apj, 601,
  197, \dodoi{10.1086/380195}

\bibitem[{{Poggianti} {et~al.}(1999){Poggianti}, {Smail}, {Dressler}, {Couch},
  {Barger}, {Butcher}, {Ellis}, \& {Oemler}}]{poggianti1999}
{Poggianti}, B.~M., {Smail}, I., {Dressler}, A., {et~al.} 1999, \apj, 518, 576,
  \dodoi{10.1086/307322}

\bibitem[{{Poggianti} {et~al.}(2009){Poggianti}, {Arag{\'o}n-Salamanca},
  {Zaritsky}, {De Lucia}, {Milvang-Jensen}, {Desai}, {Jablonka}, {Halliday},
  {Rudnick}, {Varela}, {Bamford}, {Best}, {Clowe}, {Noll}, {Saglia},
  {Pell{\'o}}, {Simard}, {von der Linden}, \& {White}}]{poggianti2009}
{Poggianti}, B.~M., {Arag{\'o}n-Salamanca}, A., {Zaritsky}, D., {et~al.} 2009,
  \apj, 693, 112, \dodoi{10.1088/0004-637X/693/1/112}

\bibitem[{{Poggianti} {et~al.}(2017){Poggianti}, {Moretti}, {Gullieuszik},
  {Fritz}, {Jaff{\'e}}, {Bettoni}, {Fasano}, {Bellhouse}, {Hau}, {Vulcani},
  {Biviano}, {Omizzolo}, {Paccagnella}, {D'Onofrio}, {Cava}, {Sheen}, {Couch},
  \& {Owers}}]{poggianti2017}
{Poggianti}, B.~M., {Moretti}, A., {Gullieuszik}, M., {et~al.} 2017, \apj, 844,
  48, \dodoi{10.3847/1538-4357/aa78ed}

\bibitem[{{Pracy} {et~al.}(2005){Pracy}, {Couch}, {Blake}, {Bekki}, {Harrison},
  {Colless}, {Kuntschner}, \& {de Propris}}]{pracy2005}
{Pracy}, M.~B., {Couch}, W.~J., {Blake}, C., {et~al.} 2005, \mnras, 359, 1421,
  \dodoi{10.1111/j.1365-2966.2005.08983.x}

\bibitem[{{Pracy} {et~al.}(2009){Pracy}, {Kuntschner}, {Couch}, {Blake},
  {Bekki}, \& {Briggs}}]{pracy2009}
{Pracy}, M.~B., {Kuntschner}, H., {Couch}, W.~J., {et~al.} 2009, \mnras, 396,
  1349, \dodoi{10.1111/j.1365-2966.2009.14836.x}

\bibitem[{{Pracy} {et~al.}(2014){Pracy}, {Owers}, {Zwaan}, {Couch},
  {Kuntschner}, {Croom}, \& {Sadler}}]{pracy2014}
{Pracy}, M.~B., {Owers}, M.~S., {Zwaan}, M., {et~al.} 2014, \mnras, 443, 388,
  \dodoi{10.1093/mnras/stu1103}

\bibitem[{{Prada} {et~al.}(2012){Prada}, {Klypin}, {Cuesta}, {Betancort-Rijo},
  \& {Primack}}]{prada2012}
{Prada}, F., {Klypin}, A.~A., {Cuesta}, A.~J., {Betancort-Rijo}, J.~E., \&
  {Primack}, J. 2012, \mnras, 423, 3018,
  \dodoi{10.1111/j.1365-2966.2012.21007.x}

\bibitem[{{Quilis} {et~al.}(2000){Quilis}, {Moore}, \& {Bower}}]{quilis2000}
{Quilis}, V., {Moore}, B., \& {Bower}, R. 2000, Science, 288, 1617,
  \dodoi{10.1126/science.288.5471.1617}

\bibitem[{{Reiprich} {et~al.}(2013){Reiprich}, {Basu}, {Ettori}, {Israel},
  {Lovisari}, {Molendi}, {Pointecouteau}, \& {Roncarelli}}]{reiprich2013}
{Reiprich}, T.~H., {Basu}, K., {Ettori}, S., {et~al.} 2013, \ssr, 177, 195,
  \dodoi{10.1007/s11214-013-9983-8}

\bibitem[{{Rhee} {et~al.}(2017){Rhee}, {Smith}, {Choi}, {Yi}, {Jaff{\'e}},
  {Candlish}, \& {S{\'a}nchez-J{\'a}nssen}}]{rhee2017}
{Rhee}, J., {Smith}, R., {Choi}, H., {et~al.} 2017, \apj, 843, 128,
  \dodoi{10.3847/1538-4357/aa6d6c}

\bibitem[{{Robotham} {et~al.}(2018){Robotham}, {Davies}, {Driver}, {Koushan},
  {Taranu}, {Casura}, \& {Liske}}]{robotham2018}
{Robotham}, A.~S.~G., {Davies}, L.~J.~M., {Driver}, S.~P., {et~al.} 2018,
  \mnras, 476, 3137, \dodoi{10.1093/mnras/sty440}

\bibitem[{{Robotham} \& {Obreschkow}(2015)}]{robotham2015}
{Robotham}, A.~S.~G., \& {Obreschkow}, D. 2015, \pasa, 32, e033,
  \dodoi{10.1017/pasa.2015.33}

\bibitem[{{Robotham} {et~al.}(2017){Robotham}, {Taranu}, {Tobar}, {Moffett}, \&
  {Driver}}]{robotham2017}
{Robotham}, A.~S.~G., {Taranu}, D.~S., {Tobar}, R., {Moffett}, A., \& {Driver},
  S.~P. 2017, \mnras, 466, 1513, \dodoi{10.1093/mnras/stw3039}

\bibitem[{{Robotham} {et~al.}(2011){Robotham}, {Norberg}, {Driver}, {Baldry},
  {Bamford}, {Hopkins}, {Liske}, {Loveday}, {Merson}, {Peacock}, {Brough},
  {Cameron}, {Conselice}, {Croom}, {Frenk}, {Gunawardhana}, {Hill}, {Jones},
  {Kelvin}, {Kuijken}, {Nichol}, {Parkinson}, {Pimbblet}, {Phillipps},
  {Popescu}, {Prescott}, {Sharp}, {Sutherland}, {Taylor}, {Thomas}, {Tuffs},
  {van Kampen}, \& {Wijesinghe}}]{robotham2011}
{Robotham}, A.~S.~G., {Norberg}, P., {Driver}, S.~P., {et~al.} 2011, \mnras,
  416, 2640, \dodoi{10.1111/j.1365-2966.2011.19217.x}

\bibitem[{{Roediger}(2009)}]{roediger2009}
{Roediger}, E. 2009, Astronomische Nachrichten, 330, 888,
  \dodoi{10.1002/asna.200911256}

\bibitem[{{Roediger} \& {Br{\"u}ggen}(2006)}]{roediger2006}
{Roediger}, E., \& {Br{\"u}ggen}, M. 2006, \mnras, 369, 567,
  \dodoi{10.1111/j.1365-2966.2006.10335.x}

\bibitem[{{Roediger} \& {Br{\"u}ggen}(2007)}]{roediger2007}
---. 2007, \mnras, 380, 1399, \dodoi{10.1111/j.1365-2966.2007.12241.x}

\bibitem[{{Rola} \& {Pelat}(1994)}]{rola1994}
{Rola}, C., \& {Pelat}, D. 1994, \aap, 287, 676

\bibitem[{{Saunders} {et~al.}(2004){Saunders}, {Bridges}, {Gillingham},
  {Haynes}, {Smith}, {Whittard}, {Churilov}, {Lankshear}, {Croom}, {Jones}, \&
  {Boshuizen}}]{saunders2004}
{Saunders}, W., {Bridges}, T., {Gillingham}, P., {et~al.} 2004, in Presented at
  the Society of Photo-Optical Instrumentation Engineers (SPIE) Conference,
  Vol. 5492, Ground-based Instrumentation for Astronomy. Edited by Alan F. M.
  Moorwood and Iye Masanori. Proceedings of the SPIE, Volume 5492, pp. 389-400
  (2004)., ed. A.~F.~M. {Moorwood} \& M.~{Iye}, 389--400

\bibitem[{{Schaefer} {et~al.}(2017){Schaefer}, {Croom}, {Allen}, {Brough},
  {Medling}, {Ho}, {Scott}, {Richards}, {Pracy}, {Gunawardhana}, {Norberg},
  {Alpaslan}, {Bauer}, {Bekki}, {Bland-Hawthorn}, {Bloom}, {Bryant}, {Couch},
  {Driver}, {Fogarty}, {Foster}, {Goldstein}, {Green}, {Hopkins},
  {Konstantopoulos}, {Lawrence}, {L{\'o}pez-S{\'a}nchez}, {Lorente}, {Owers},
  {Sharp}, {Sweet}, {Taylor}, {van de Sande}, {Walcher}, \&
  {Wong}}]{schaefer2017}
{Schaefer}, A.~L., {Croom}, S.~M., {Allen}, J.~T., {et~al.} 2017, \mnras, 464,
  121, \dodoi{10.1093/mnras/stw2289}

\bibitem[{{Schaefer} {et~al.}(2018){Schaefer}, {Croom}, {Scott}, {Brough},
  {Allen}, {Bekki}, {Bland-Hawthorn}, {Bloom}, {Bryant}, {Cortese}, {Davies},
  {Federrath}, {Fogarty}, {Green}, {Groves}, {Hopkins}, {Konstantopoulos},
  {L{\'o}pez-S{\'a}nchez}, {Lawrence}, {McElroy}, {Medling}, {Owers}, {Pracy},
  {Richards}, {Robotham}, {van de Sande}, {Tonini}, \& {Yi}}]{schaefer2018}
{Schaefer}, A.~L., {Croom}, S.~M., {Scott}, N., {et~al.} 2018, \mnras,
  \dodoi{10.1093/mnras/sty3258}

\bibitem[{{Schulz} \& {Struck}(2001)}]{schulz2001}
{Schulz}, S., \& {Struck}, C. 2001, \mnras, 328, 185,
  \dodoi{10.1046/j.1365-8711.2001.04847.x}

\bibitem[{{Scott} {et~al.}(2018){Scott}, {van de Sande}, {Croom}, {Groves},
  {Owers}, {Poetrodjojo}, {D'Eugenio}, {Medling}, {Barat}, {Barone},
  {Bland-Hawthorn}, {Brough}, {Bryant}, {Cortese}, {Foster}, {Green}, {Oh},
  {Colless}, {Drinkwater}, {Driver}, {Goodwin}, {Gunawardhana}, {Federrath},
  {Harischandra}, {Jin}, {Lawrence}, {Lorente}, {Mannering}, {O'Toole},
  {Richards}, {Sanchez}, {Schaefer}, {Sealey}, {Sharp}, {Sweet}, {Taranu}, \&
  {Varidel}}]{scott2018}
{Scott}, N., {van de Sande}, J., {Croom}, S.~M., {et~al.} 2018, \mnras,
  \dodoi{10.1093/mnras/sty2355}

\bibitem[{{Seibert} {et~al.}(2012){Seibert}, {Wyder}, {Neill}, {Madore},
  {Bianchi}, {Smith}, {Shiao}, {Schiminovich}, {Rich}, {Conrow}, {Martin}, \&
  {GALEX Catalog Team}}]{seibert2012}
{Seibert}, M., {Wyder}, T., {Neill}, J., {et~al.} 2012, in American
  Astronomical Society Meeting Abstracts, Vol. 219, American Astronomical
  Society Meeting Abstracts \#219, 340.01

\bibitem[{{Shanks} {et~al.}(2015){Shanks}, {Metcalfe}, {Chehade}, {Findlay},
  {Irwin}, {Gonzalez-Solares}, {Lewis}, {Yoldas}, {Mann}, {Read}, {Sutorius},
  \& {Voutsinas}}]{shanks2015}
{Shanks}, T., {Metcalfe}, N., {Chehade}, B., {et~al.} 2015, \mnras, 451, 4238,
  \dodoi{10.1093/mnras/stv1130}

\bibitem[{{Sharp} {et~al.}(2006){Sharp}, {Saunders}, {Smith}, {Churilov},
  {Correll}, {Dawson}, {Farrel}, {Frost}, {Haynes}, {Heald}, {Lankshear},
  {Mayfield}, {Waller}, \& {Whittard}}]{sharp2006}
{Sharp}, R., {Saunders}, W., {Smith}, G., {et~al.} 2006, in Presented at the
  Society of Photo-Optical Instrumentation Engineers (SPIE) Conference, Vol.
  6269, Ground-based and Airborne Instrumentation for Astronomy. Edited by
  McLean, Ian S.; Iye, Masanori. Proceedings of the SPIE, Volume 6269, pp.
  62690G (2006).

\bibitem[{{Sharp} {et~al.}(2015){Sharp}, {Allen}, {Fogarty}, {Croom},
  {Cortese}, {Green}, {Nielsen}, {Richards}, {Scott}, {Taylor}, {Barnes},
  {Bauer}, {Birchall}, {Bland-Hawthorn}, {Bloom}, {Brough}, {Bryant}, {Cecil},
  {Colless}, {Couch}, {Drinkwater}, {Driver}, {Foster}, {Goodwin},
  {Gunawardhana}, {Ho}, {Hampton}, {Hopkins}, {Jones}, {Konstantopoulos},
  {Lawrence}, {Leslie}, {Lewis}, {Liske}, {L{\'o}pez-S{\'a}nchez}, {Lorente},
  {McElroy}, {Medling}, {Mahajan}, {Mould}, {Parker}, {Pracy}, {Obreschkow},
  {Owers}, {Schaefer}, {Sweet}, {Thomas}, {Tonini}, \& {Walcher}}]{sharp2015}
{Sharp}, R., {Allen}, J.~T., {Fogarty}, L.~M.~R., {et~al.} 2015, \mnras, 446,
  1551, \dodoi{10.1093/mnras/stu2055}

\bibitem[{{Singh} {et~al.}(2013){Singh}, {van de Ven}, {Jahnke}, {Lyubenova},
  {Falc{\'o}n-Barroso}, {Alves}, {Cid Fernandes}, {Galbany},
  {Garc{\'{\i}}a-Benito}, {Husemann}, {Kennicutt}, {Marino}, {M{\'a}rquez},
  {Masegosa}, {Mast}, {Pasquali}, {S{\'a}nchez}, {Walcher}, {Wild}, {Wisotzki},
  \& {Ziegler}}]{singh2013}
{Singh}, R., {van de Ven}, G., {Jahnke}, K., {et~al.} 2013, \aap, 558, A43,
  \dodoi{10.1051/0004-6361/201322062}

\bibitem[{{Smith} {et~al.}(2010){Smith}, {Lucey}, {Hammer}, {Hornschemeier},
  {Carter}, {Hudson}, {Marzke}, {Mouhcine}, {Eftekharzadeh}, {James},
  {Khosroshahi}, {Kourkchi}, \& {Karick}}]{smith2010}
{Smith}, R.~J., {Lucey}, J.~R., {Hammer}, D., {et~al.} 2010, \mnras, 408, 1417,
  \dodoi{10.1111/j.1365-2966.2010.17253.x}

\bibitem[{{Solanes} {et~al.}(2001){Solanes}, {Manrique},
  {Garc{\'{\i}}a-G{\'o}mez}, {Gonz{\'a}lez-Casado}, {Giovanelli}, \&
  {Haynes}}]{solanes2001}
{Solanes}, J.~M., {Manrique}, A., {Garc{\'{\i}}a-G{\'o}mez}, C., {et~al.} 2001,
  \apj, 548, 97, \dodoi{10.1086/318672}

\bibitem[{{Taranu} {et~al.}(2014){Taranu}, {Hudson}, {Balogh}, {Smith},
  {Power}, {Oman}, \& {Krane}}]{taranu2014}
{Taranu}, D.~S., {Hudson}, M.~J., {Balogh}, M.~L., {et~al.} 2014, \mnras, 440,
  1934, \dodoi{10.1093/mnras/stu389}

\bibitem[{{Taylor} {et~al.}(2011){Taylor}, {Hopkins}, {Baldry}, {Brown},
  {Driver}, {Kelvin}, {Hill}, {Robotham}, {Bland-Hawthorn}, {Jones}, {Sharp},
  {Thomas}, {Liske}, {Loveday}, {Norberg}, {Peacock}, {Bamford}, {Brough},
  {Colless}, {Cameron}, {Conselice}, {Croom}, {Frenk}, {Gunawardhana},
  {Kuijken}, {Nichol}, {Parkinson}, {Phillipps}, {Pimbblet}, {Popescu},
  {Prescott}, {Sutherland}, {Tuffs}, {van Kampen}, \&
  {Wijesinghe}}]{taylor2011}
{Taylor}, E.~N., {Hopkins}, A.~M., {Baldry}, I.~K., {et~al.} 2011, \mnras, 418,
  1587, \dodoi{10.1111/j.1365-2966.2011.19536.x}

\bibitem[{{Tran} {et~al.}(2003){Tran}, {Franx}, {Illingworth}, {Kelson}, \&
  {van Dokkum}}]{tran2003}
{Tran}, K.-V.~H., {Franx}, M., {Illingworth}, G., {Kelson}, D.~D., \& {van
  Dokkum}, P. 2003, \apj, 599, 865, \dodoi{10.1086/379804}

\bibitem[{{van de Sande} {et~al.}(2017){van de Sande}, {Bland-Hawthorn},
  {Fogarty}, {Cortese}, {d'Eugenio}, {Croom}, {Scott}, {Allen}, {Brough},
  {Bryant}, {Cecil}, {Colless}, {Couch}, {Davies}, {Elahi}, {Foster},
  {Goldstein}, {Goodwin}, {Groves}, {Ho}, {Jeong}, {Jones}, {Konstantopoulos},
  {Lawrence}, {Leslie}, {L{\'o}pez-S{\'a}nchez}, {McDermid}, {McElroy},
  {Medling}, {Oh}, {Owers}, {Richards}, {Schaefer}, {Sharp}, {Sweet}, {Taranu},
  {Tonini}, {Walcher}, \& {Yi}}]{VDS2016}
{van de Sande}, J., {Bland-Hawthorn}, J., {Fogarty}, L.~M.~R., {et~al.} 2017,
  \apj, 835, 104, \dodoi{10.3847/1538-4357/835/1/104}

\bibitem[{{van der Burg} {et~al.}(2018){van der Burg}, {McGee}, {Aussel},
  {Dahle}, {Arnaud}, {Pratt}, \& {Muzzin}}]{vanderberg2018}
{van der Burg}, R.~F.~J., {McGee}, S., {Aussel}, H., {et~al.} 2018, \aap, 618,
  A140, \dodoi{10.1051/0004-6361/201833572}

\bibitem[{{van der Marel} \& {Franx}(1993)}]{vandermarel1993}
{van der Marel}, R.~P., \& {Franx}, M. 1993, \apj, 407, 525,
  \dodoi{10.1086/172534}

\bibitem[{{Vazdekis} {et~al.}(2010){Vazdekis}, {S{\'a}nchez-Bl{\'a}zquez},
  {Falc{\'o}n-Barroso}, {Cenarro}, {Beasley}, {Cardiel}, {Gorgas}, \&
  {Peletier}}]{vazdekis2010}
{Vazdekis}, A., {S{\'a}nchez-Bl{\'a}zquez}, P., {Falc{\'o}n-Barroso}, J.,
  {et~al.} 2010, \mnras, 404, 1639, \dodoi{10.1111/j.1365-2966.2010.16407.x}

\bibitem[{{Veilleux} \& {Osterbrock}(1987)}]{veilleux1987}
{Veilleux}, S., \& {Osterbrock}, D.~E. 1987, \apjs, 63, 295,
  \dodoi{10.1086/191166}

\bibitem[{{Vikhlinin} {et~al.}(2006){Vikhlinin}, {Kravtsov}, {Forman}, {Jones},
  {Markevitch}, {Murray}, \& {Van Speybroeck}}]{vikhlinin2006}
{Vikhlinin}, A., {Kravtsov}, A., {Forman}, W., {et~al.} 2006, \apj, 640, 691,
  \dodoi{10.1086/500288}

\bibitem[{{Vollmer} {et~al.}(2001){Vollmer}, {Cayatte}, {Balkowski}, \&
  {Duschl}}]{vollmer2001}
{Vollmer}, B., {Cayatte}, V., {Balkowski}, C., \& {Duschl}, W.~J. 2001, \apj,
  561, 708, \dodoi{10.1086/323368}

\bibitem[{{von der Linden} {et~al.}(2010){von der Linden}, {Wild}, {Kauffmann},
  {White}, \& {Weinmann}}]{vonderlinden2010}
{von der Linden}, A., {Wild}, V., {Kauffmann}, G., {White}, S.~D.~M., \&
  {Weinmann}, S. 2010, \mnras, 404, 1231,
  \dodoi{10.1111/j.1365-2966.2010.16375.x}

\bibitem[{{Wetzel} {et~al.}(2012){Wetzel}, {Tinker}, \& {Conroy}}]{wetzel2012}
{Wetzel}, A.~R., {Tinker}, J.~L., \& {Conroy}, C. 2012, \mnras, 424, 232,
  \dodoi{10.1111/j.1365-2966.2012.21188.x}

\bibitem[{{Wetzel} {et~al.}(2013){Wetzel}, {Tinker}, {Conroy}, \& {van den
  Bosch}}]{wetzel2013}
{Wetzel}, A.~R., {Tinker}, J.~L., {Conroy}, C., \& {van den Bosch}, F.~C. 2013,
  \mnras, 432, 336, \dodoi{10.1093/mnras/stt469}

\bibitem[{{Worthey} {et~al.}(1994){Worthey}, {Faber}, {Gonzalez}, \&
  {Burstein}}]{worthey1994}
{Worthey}, G., {Faber}, S.~M., {Gonzalez}, J.~J., \& {Burstein}, D. 1994,
  \apjs, 94, 687, \dodoi{10.1086/192087}

\bibitem[{{Worthey} \& {Ottaviani}(1997)}]{worthey1997}
{Worthey}, G., \& {Ottaviani}, D.~L. 1997, \apjs, 111, 377,
  \dodoi{10.1086/313021}

\bibitem[{{Wright} {et~al.}(2016){Wright}, {Robotham}, {Bourne}, {Driver},
  {Dunne}, {Maddox}, {Alpaslan}, {Andrews}, {Bauer}, {Bland-Hawthorn},
  {Brough}, {Brown}, {Clarke}, {Cluver}, {Davies}, {Grootes}, {Holwerda},
  {Hopkins}, {Jarrett}, {Kafle}, {Lange}, {Liske}, {Loveday}, {Moffett},
  {Norberg}, {Popescu}, {Smith}, {Taylor}, {Tuffs}, {Wang}, \&
  {Wilkins}}]{wright2016}
{Wright}, A.~H., {Robotham}, A.~S.~G., {Bourne}, N., {et~al.} 2016, \mnras,
  460, 765, \dodoi{10.1093/mnras/stw832}

\bibitem[{{Yagi} {et~al.}(2010){Yagi}, {Yoshida}, {Komiyama}, {Kashikawa},
  {Furusawa}, {Okamura}, {Graham}, {Miller}, {Carter}, {Mobasher}, \&
  {Jogee}}]{yagi2010}
{Yagi}, M., {Yoshida}, M., {Komiyama}, Y., {et~al.} 2010, \aj, 140, 1814,
  \dodoi{10.1088/0004-6256/140/6/1814}

\bibitem[{{Yan} {et~al.}(2006){Yan}, {Newman}, {Faber}, {Konidaris}, {Koo}, \&
  {Davis}}]{yan2006}
{Yan}, R., {Newman}, J.~A., {Faber}, S.~M., {et~al.} 2006, \apj, 648, 281,
  \dodoi{10.1086/505629}

\bibitem[{{Yang} {et~al.}(2008){Yang}, {Zabludoff}, {Zaritsky}, \&
  {Mihos}}]{yang2008}
{Yang}, Y., {Zabludoff}, A.~I., {Zaritsky}, D., \& {Mihos}, J.~C. 2008, \apj,
  688, 945, \dodoi{10.1086/591656}

\bibitem[{{York} {et~al.}(2000){York}, {Adelman}, {Anderson}, {Anderson},
  {Annis}, {Bahcall}, {Bakken}, {Barkhouser}, {Bastian}, {Berman}, {Boroski},
  {Bracker}, {Briegel}, {Briggs}, {Brinkmann}, {Brunner}, {Burles}, {Carey},
  {Carr}, {Castander}, {Chen}, {Colestock}, {Connolly}, {Crocker}, {Csabai},
  {Czarapata}, {Davis}, {Doi}, {Dombeck}, {Eisenstein}, {Ellman}, {Elms},
  {Evans}, {Fan}, {Federwitz}, {Fiscelli}, {Friedman}, {Frieman}, {Fukugita},
  {Gillespie}, {Gunn}, {Gurbani}, {de Haas}, {Haldeman}, {Harris}, {Hayes},
  {Heckman}, {Hennessy}, {Hindsley}, {Holm}, {Holmgren}, {Huang}, {Hull},
  {Husby}, {Ichikawa}, {Ichikawa}, {Ivezi{\'c}}, {Kent}, {Kim}, {Kinney},
  {Klaene}, {Kleinman}, {Kleinman}, {Knapp}, {Korienek}, {Kron}, {Kunszt},
  {Lamb}, {Lee}, {Leger}, {Limmongkol}, {Lindenmeyer}, {Long}, {Loomis},
  {Loveday}, {Lucinio}, {Lupton}, {MacKinnon}, {Mannery}, {Mantsch}, {Margon},
  {McGehee}, {McKay}, {Meiksin}, {Merelli}, {Monet}, {Munn}, {Narayanan},
  {Nash}, {Neilsen}, {Neswold}, {Newberg}, {Nichol}, {Nicinski}, {Nonino},
  {Okada}, {Okamura}, {Ostriker}, {Owen}, {Pauls}, {Peoples}, {Peterson},
  {Petravick}, {Pier}, {Pope}, {Pordes}, {Prosapio}, {Rechenmacher}, {Quinn},
  {Richards}, {Richmond}, {Rivetta}, {Rockosi}, {Ruthmansdorfer}, {Sandford},
  {Schlegel}, {Schneider}, {Sekiguchi}, {Sergey}, {Shimasaku}, {Siegmund},
  {Smee}, {Smith}, {Snedden}, {Stone}, {Stoughton}, {Strauss}, {Stubbs},
  {SubbaRao}, {Szalay}, {Szapudi}, {Szokoly}, {Thakar}, {Tremonti}, {Tucker},
  {Uomoto}, {Vanden Berk}, {Vogeley}, {Waddell}, {Wang}, {Watanabe},
  {Weinberg}, {Yanny}, \& {Yasuda}}]{york2000}
{York}, D.~G., {Adelman}, J., {Anderson}, Jr., J.~E., {et~al.} 2000, \aj, 120,
  1579, \dodoi{10.1086/301513}

\bibitem[{{Zabludoff} {et~al.}(1996){Zabludoff}, {Zaritsky}, {Lin}, {Tucker},
  {Hashimoto}, {Shectman}, {Oemler}, \& {Kirshner}}]{zabludoff1996}
{Zabludoff}, A.~I., {Zaritsky}, D., {Lin}, H., {et~al.} 1996, \apj, 466, 104,
  \dodoi{10.1086/177495}

\end{thebibliography}



\appendix

\section{The HDSG maps and example spectra}

\renewcommand{\thefigure}{\ref{cluster_HDS_galaxies} (Cont.)}
\addtocounter{figure}{-1}
\begin{figure*}
\includegraphics[width=.43\textwidth]{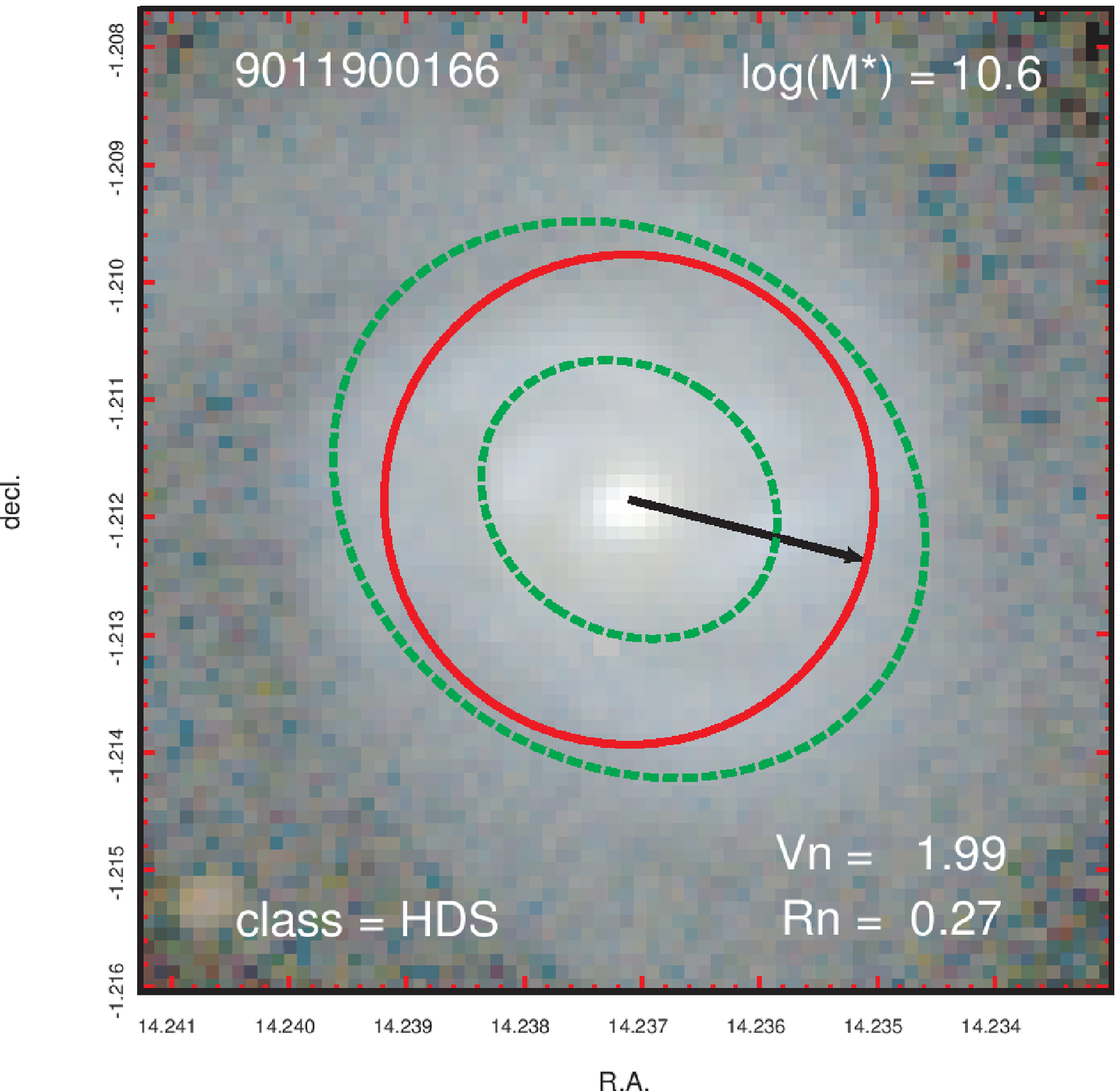}
\includegraphics[width=.55\textwidth]{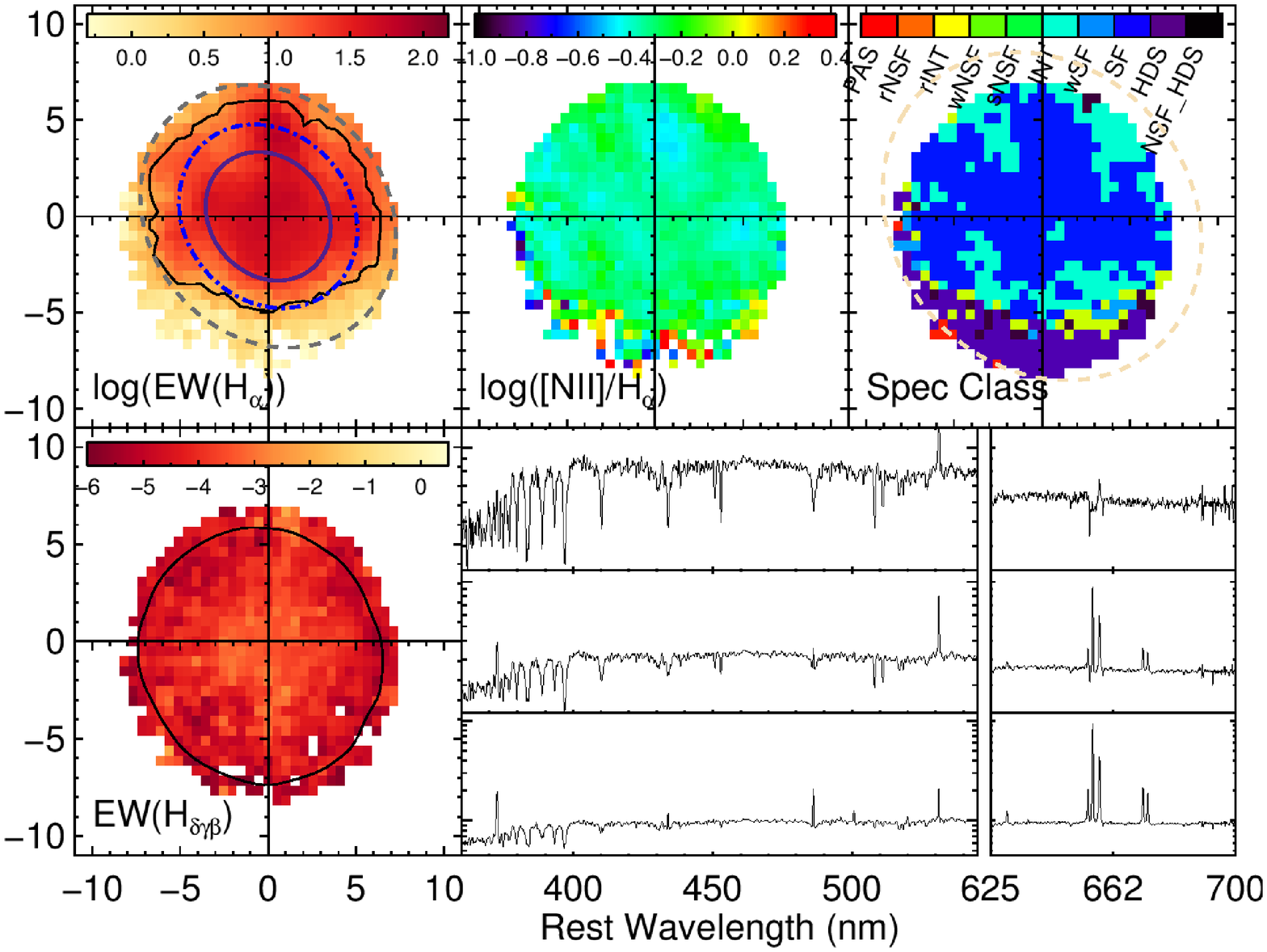}
\\
\includegraphics[width=.43\textwidth]{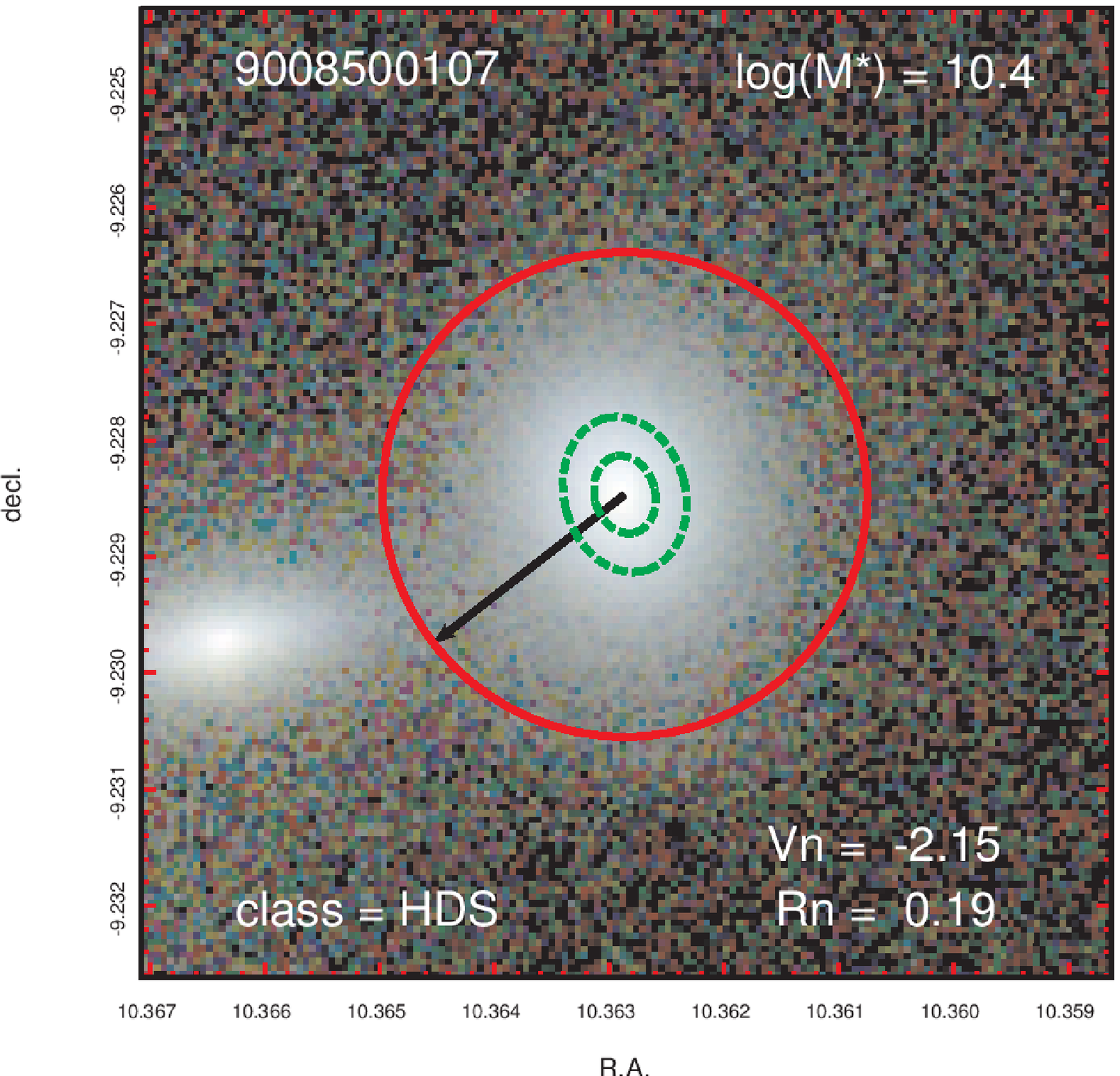}
\includegraphics[width=.55\textwidth]{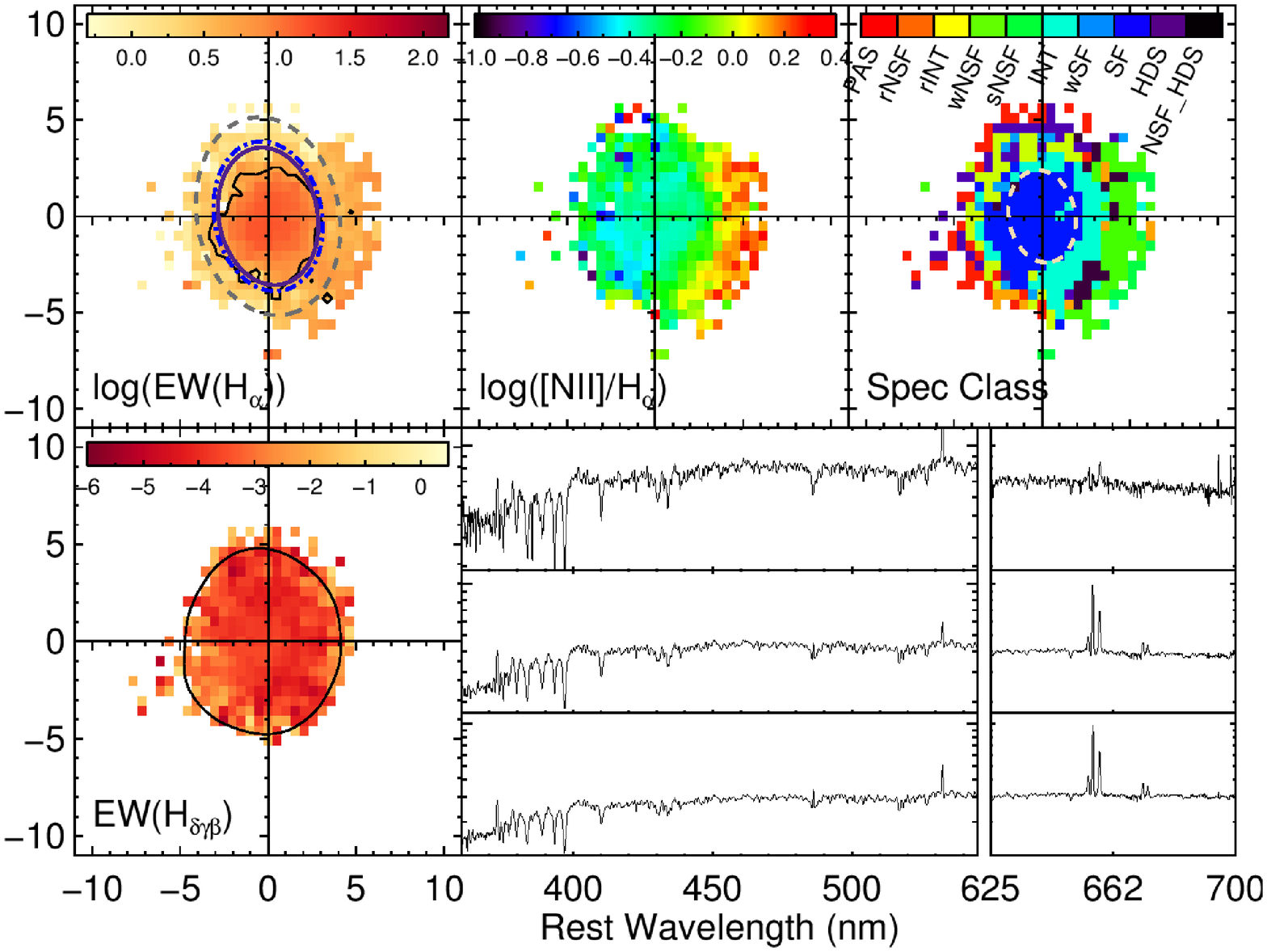}\\
\\
\includegraphics[width=.43\textwidth]{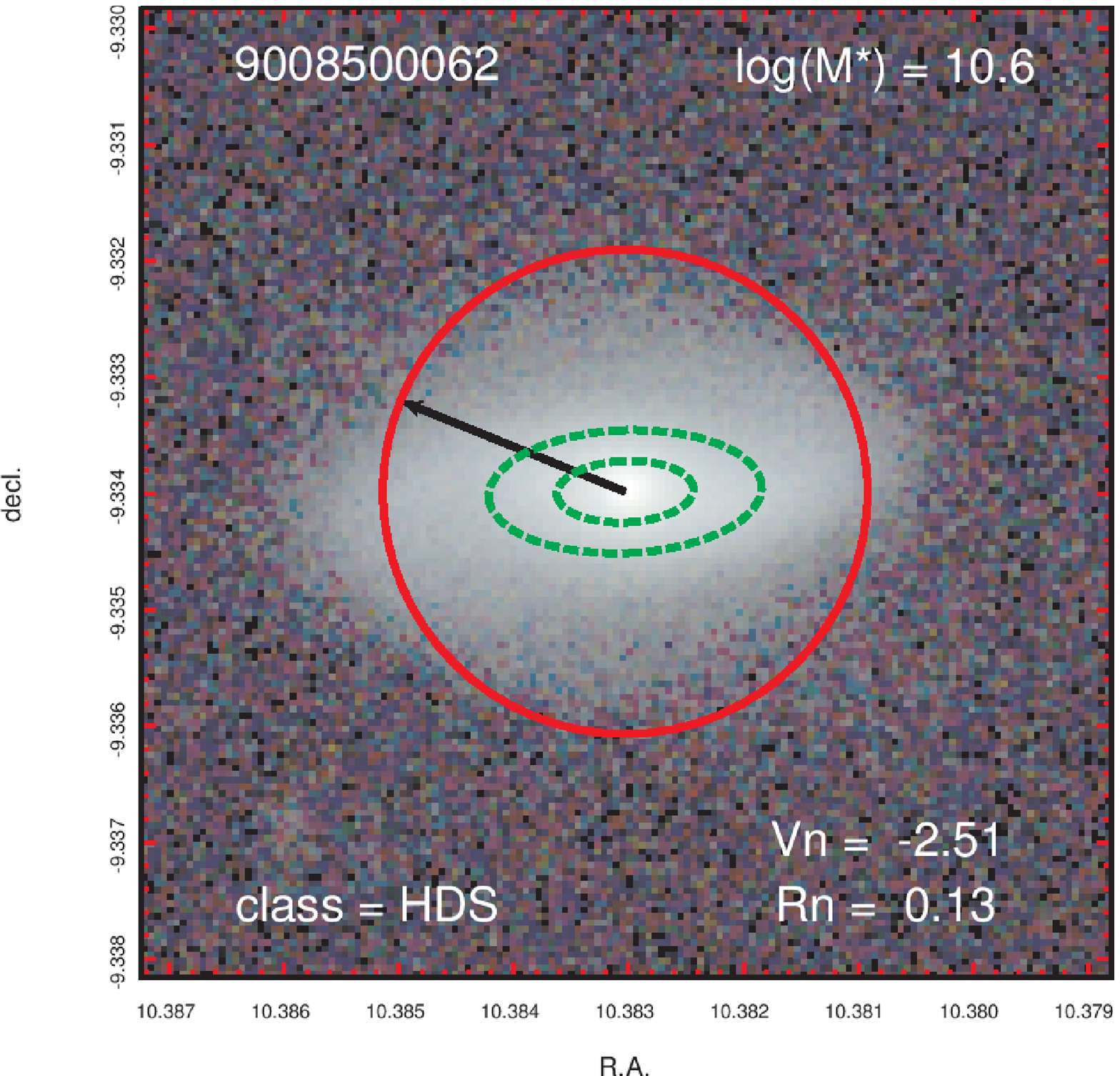}
\includegraphics[width=.55\textwidth]{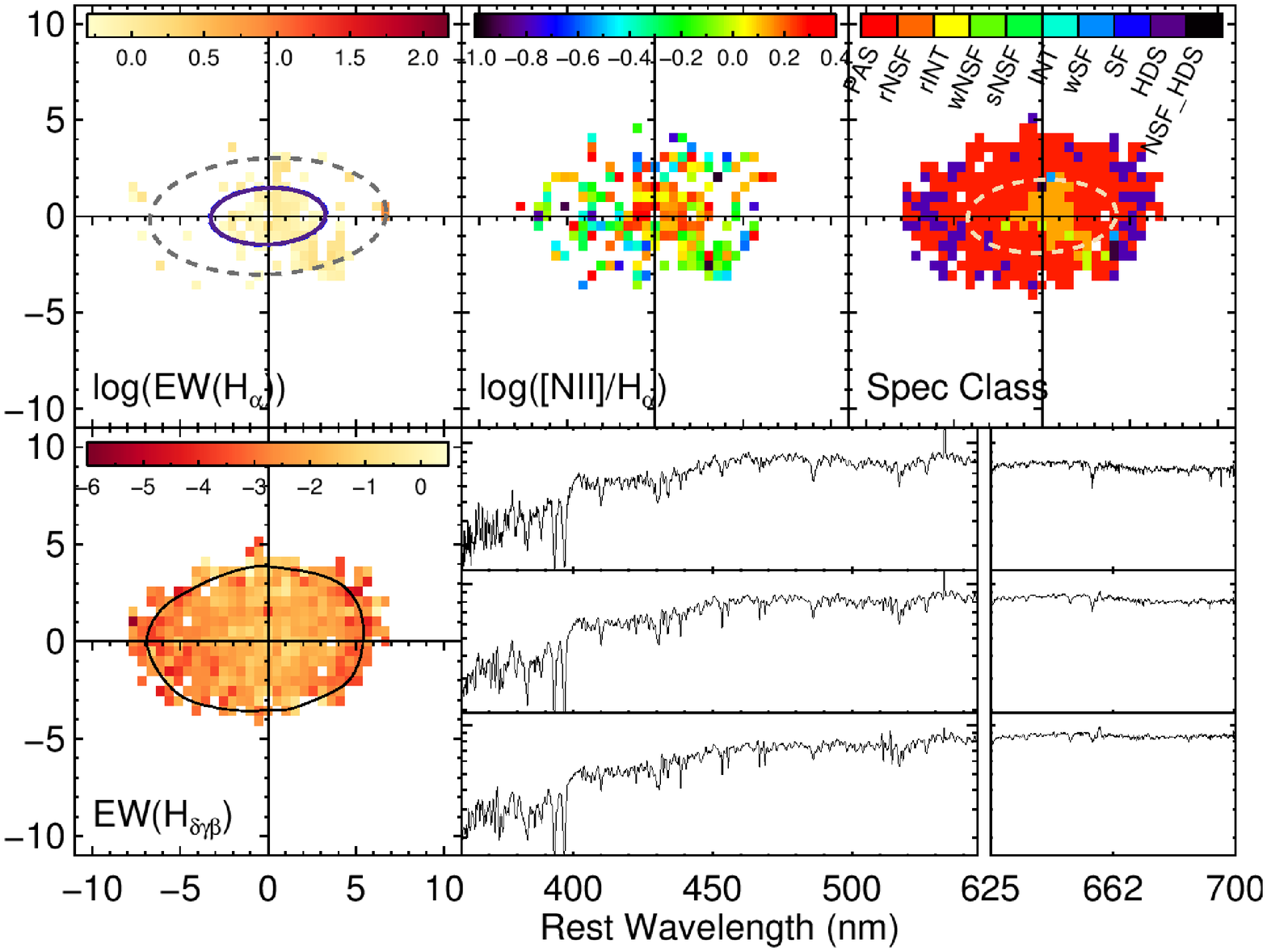}
\caption{}
\end{figure*}

\renewcommand{\thefigure}{\ref{cluster_HDS_galaxies} (Cont.)}
\addtocounter{figure}{-1}
\begin{figure*}
\includegraphics[width=.43\textwidth]{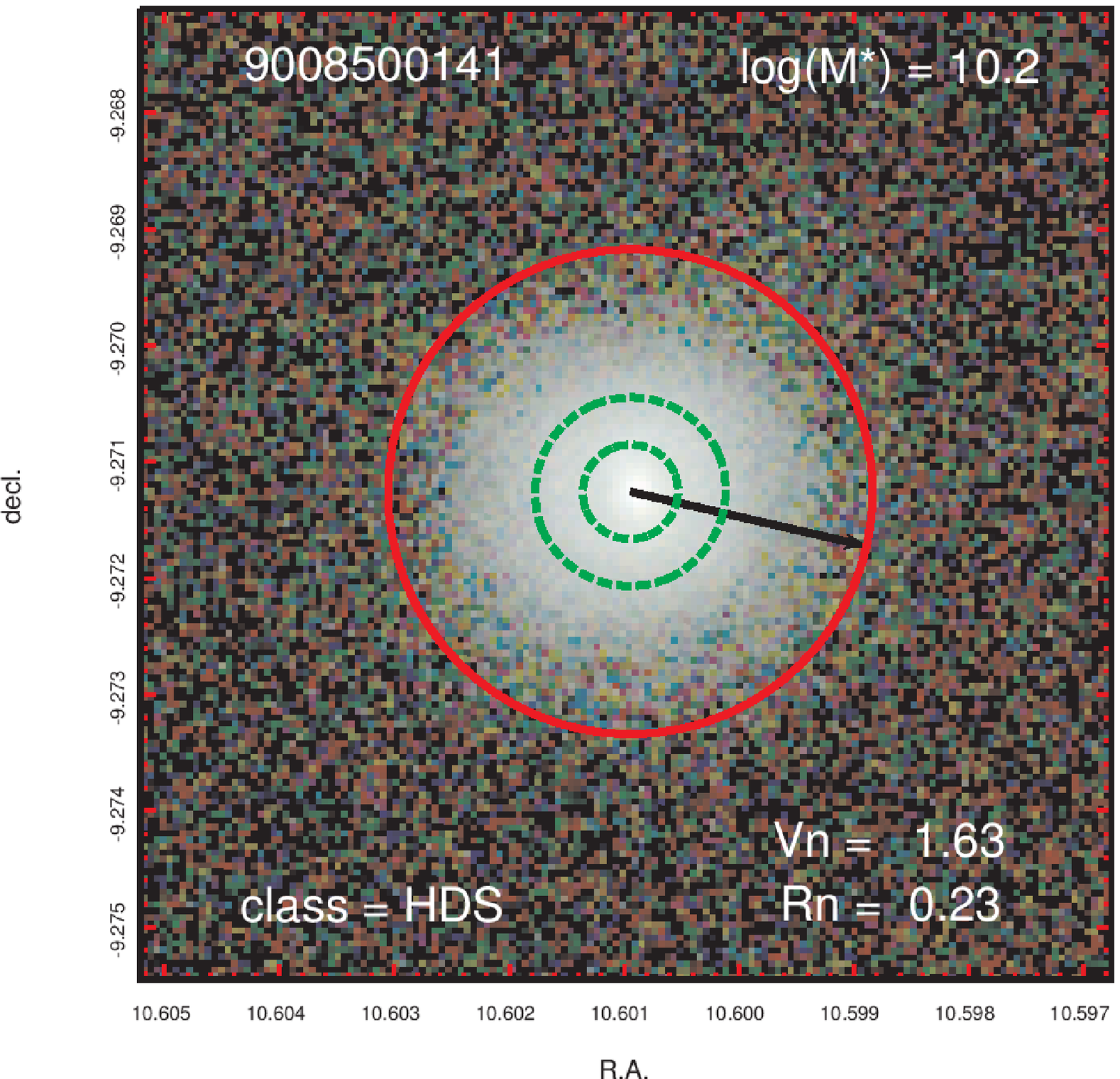}
\includegraphics[width=.55\textwidth]{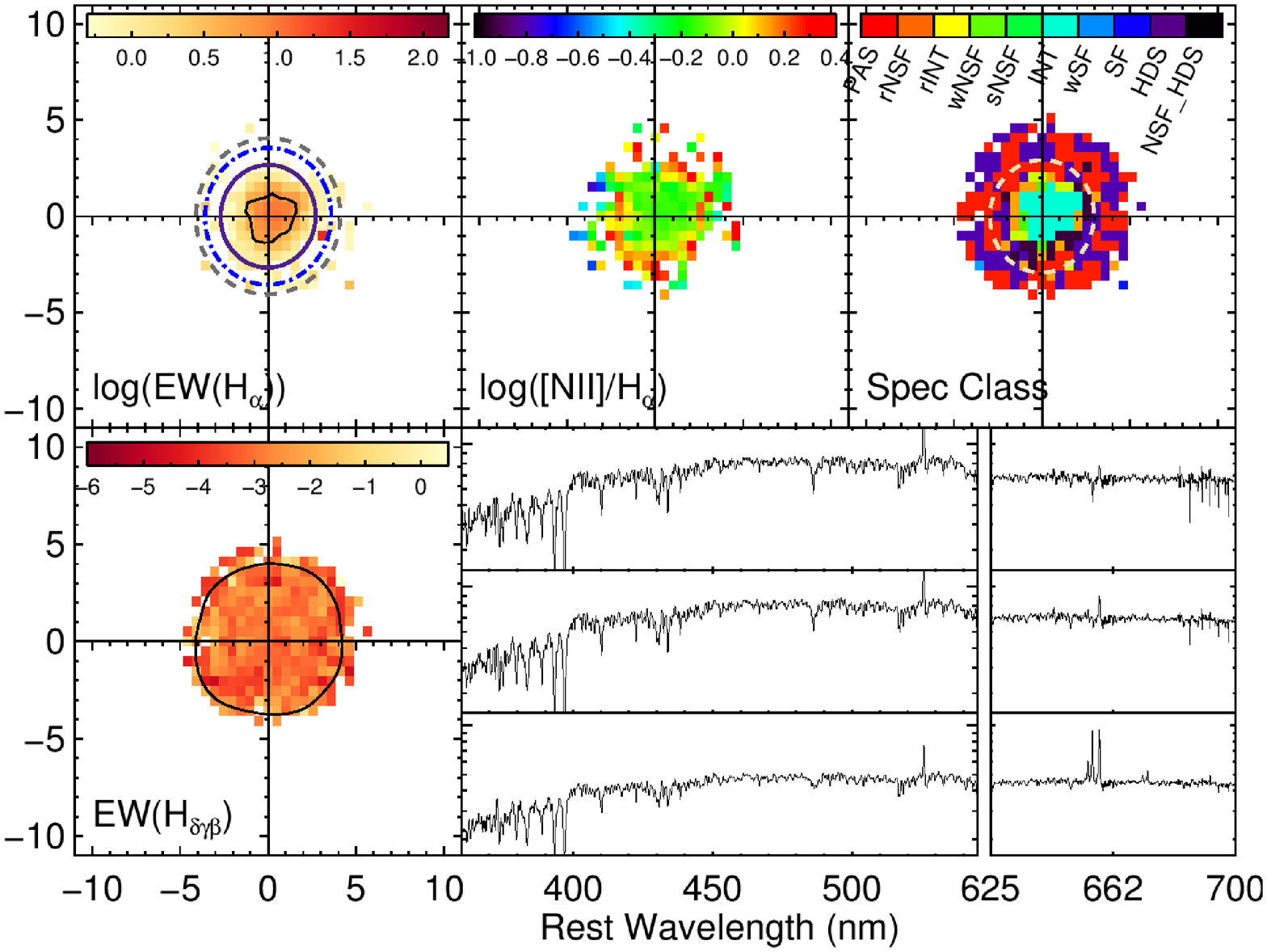}
\\
\includegraphics[width=.43\textwidth]{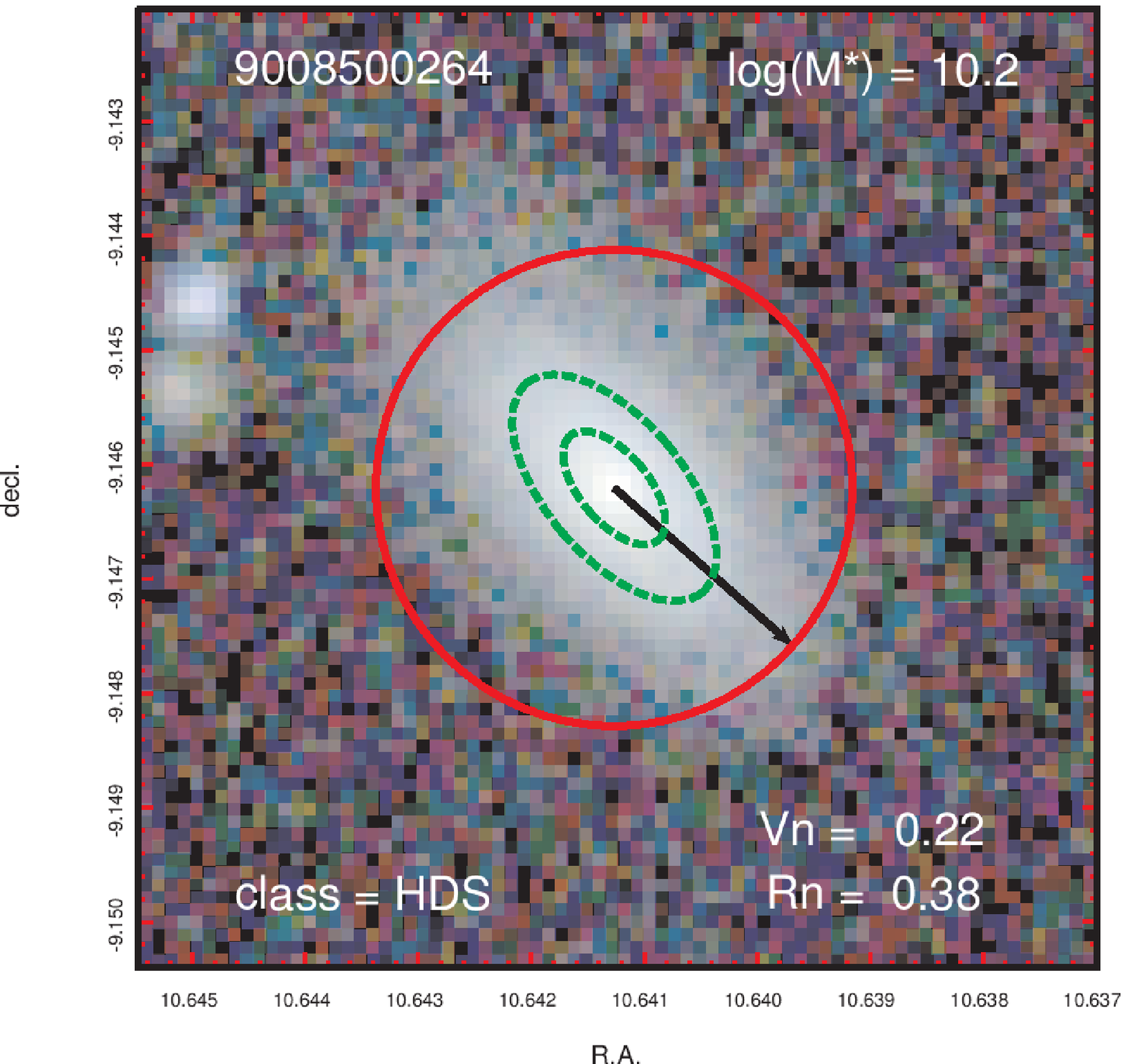}
\includegraphics[width=.55\textwidth]{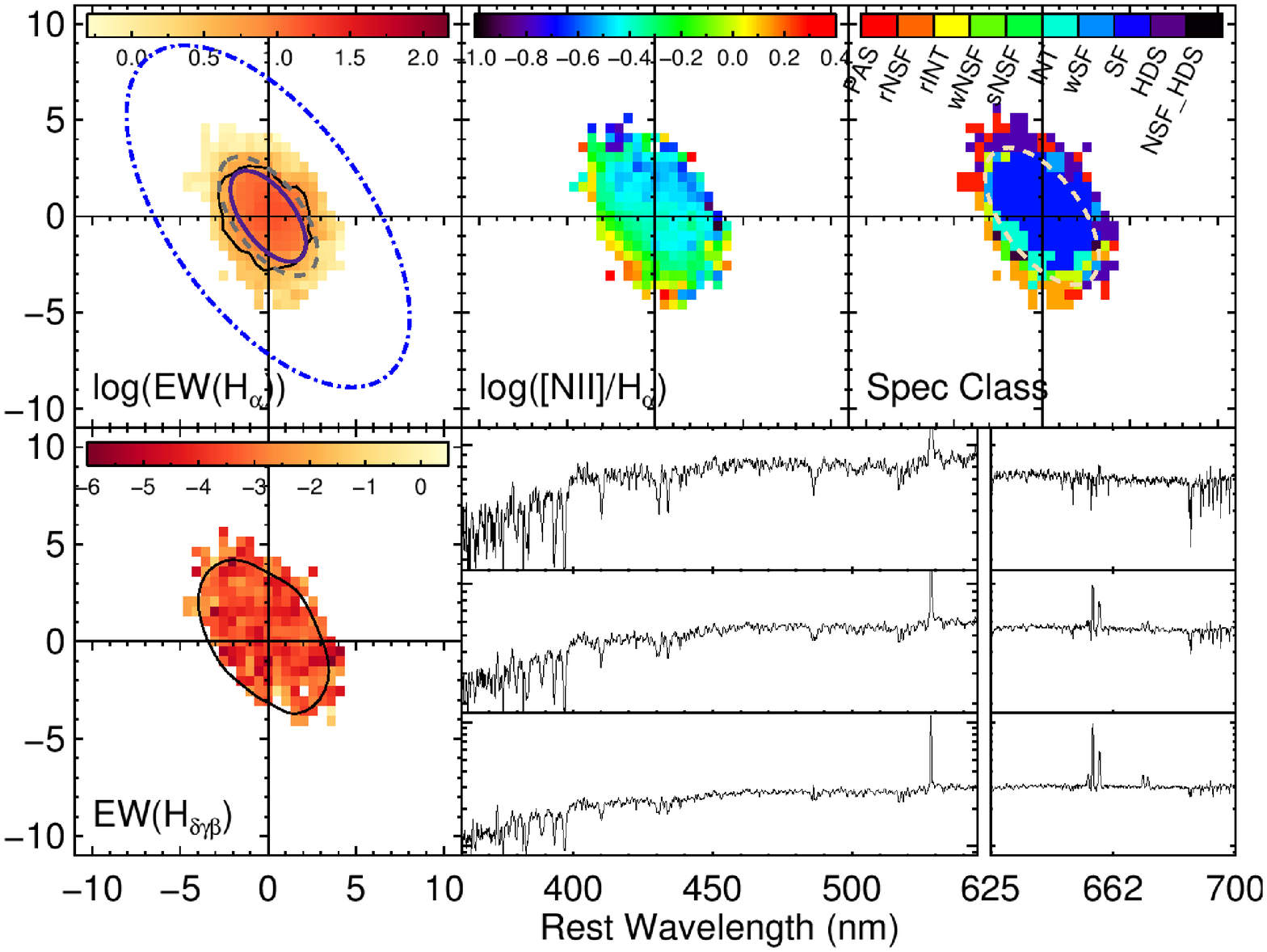}
\\
\includegraphics[width=.43\textwidth]{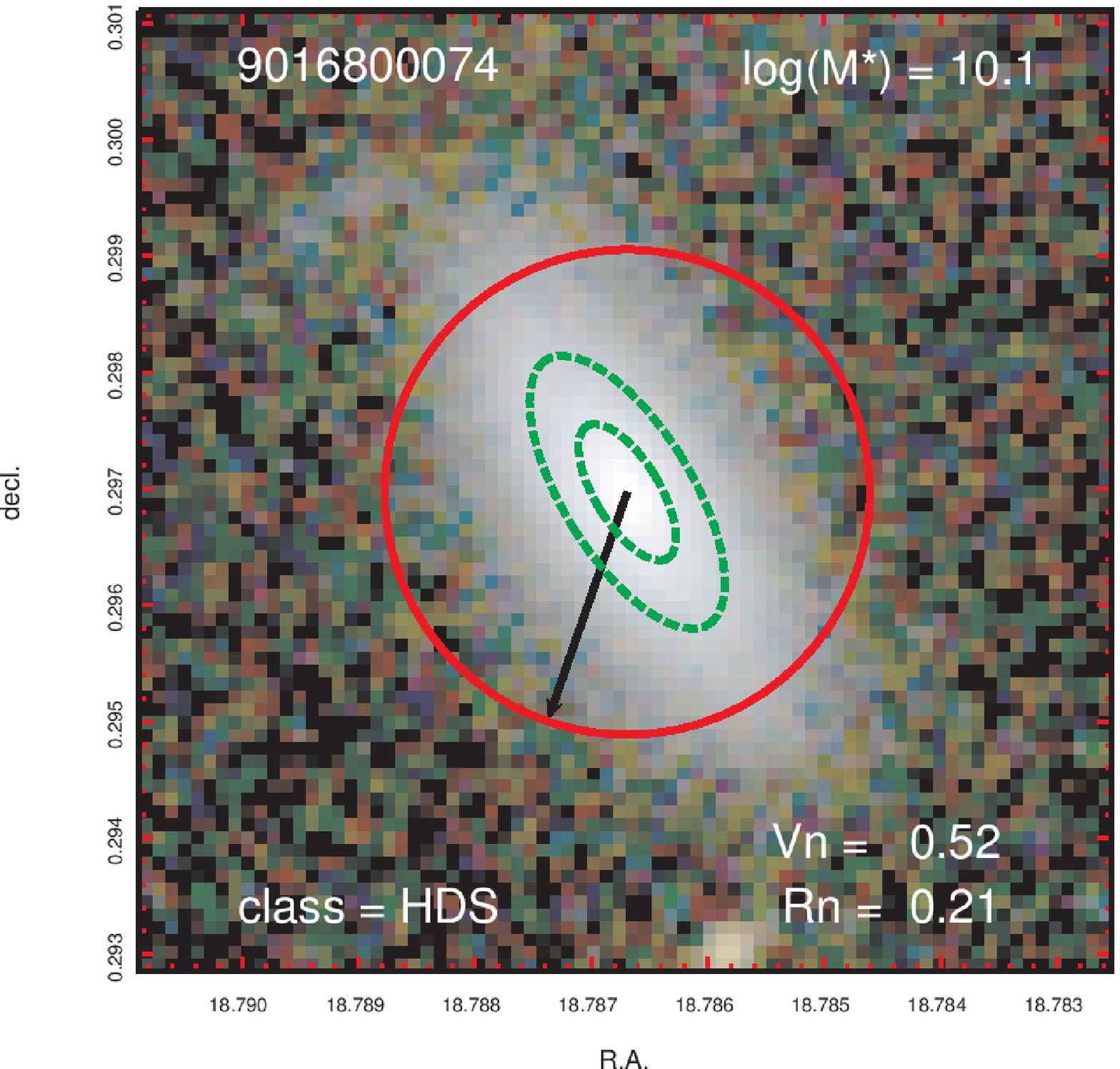}
\includegraphics[width=.55\textwidth]{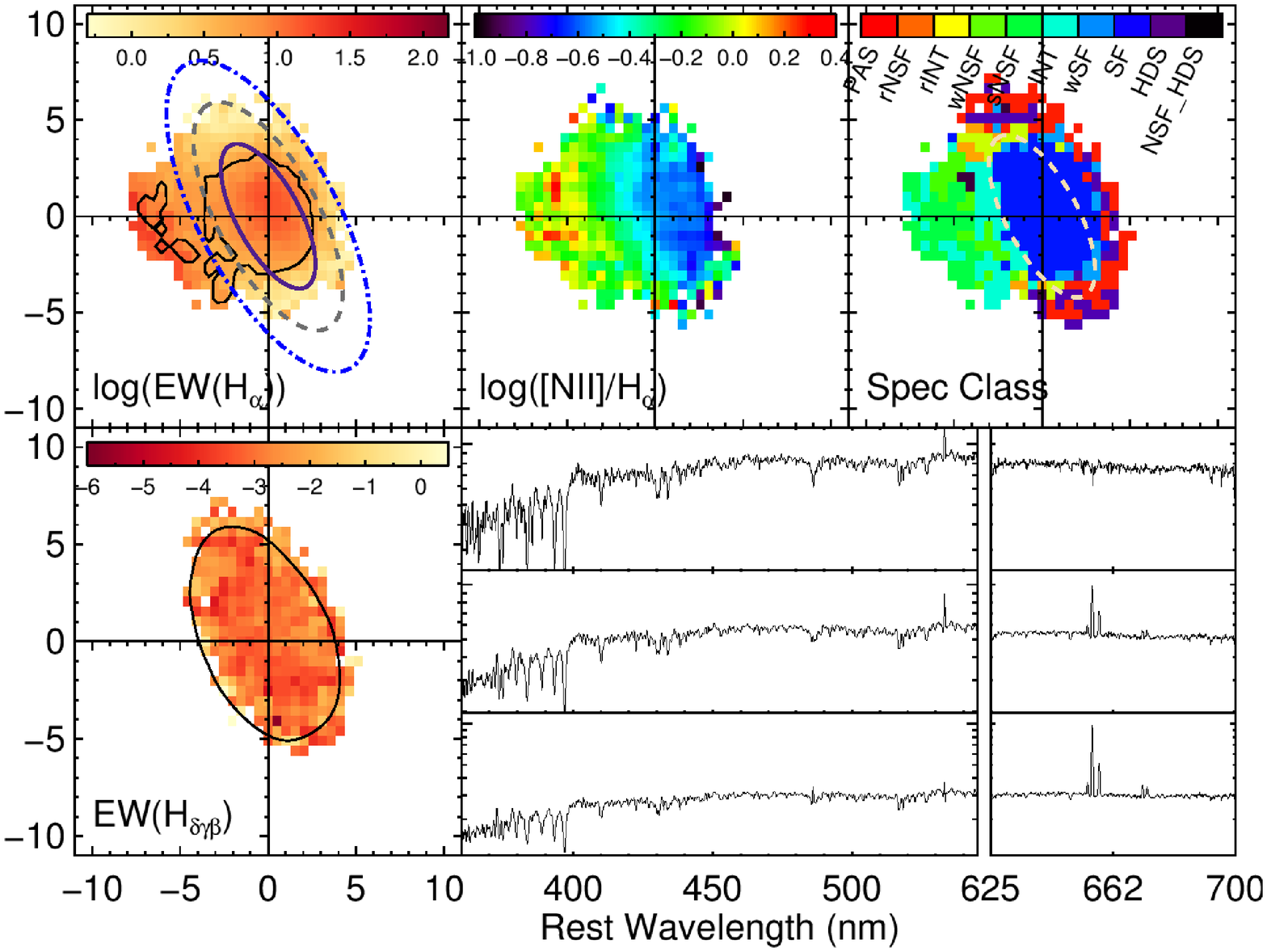}
\caption{}
\end{figure*}

\renewcommand{\thefigure}{\ref{cluster_HDS_galaxies} (Cont.)}
\addtocounter{figure}{-1}
\begin{figure*}
\includegraphics[width=.43\textwidth]{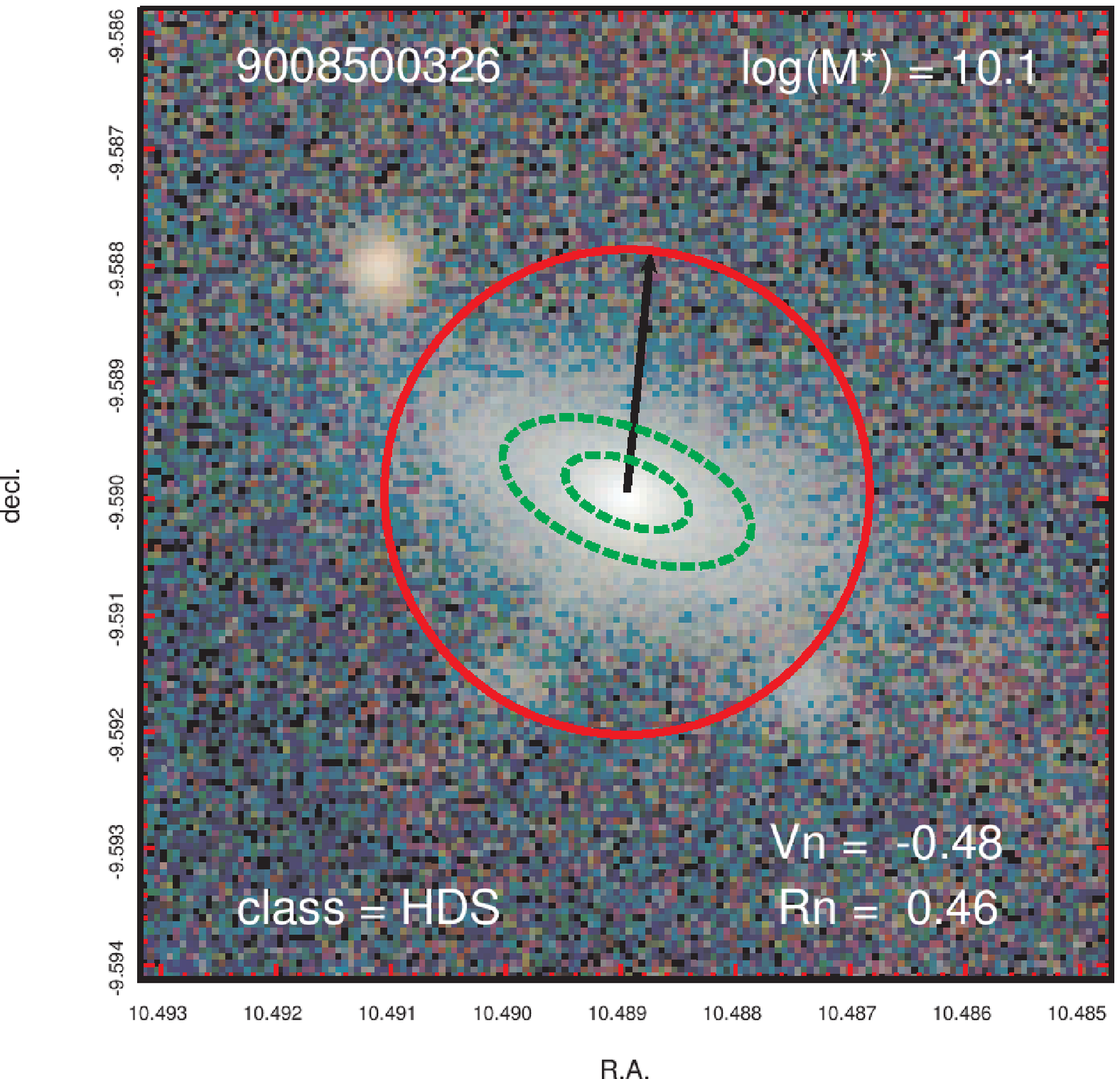}
\includegraphics[width=.55\textwidth]{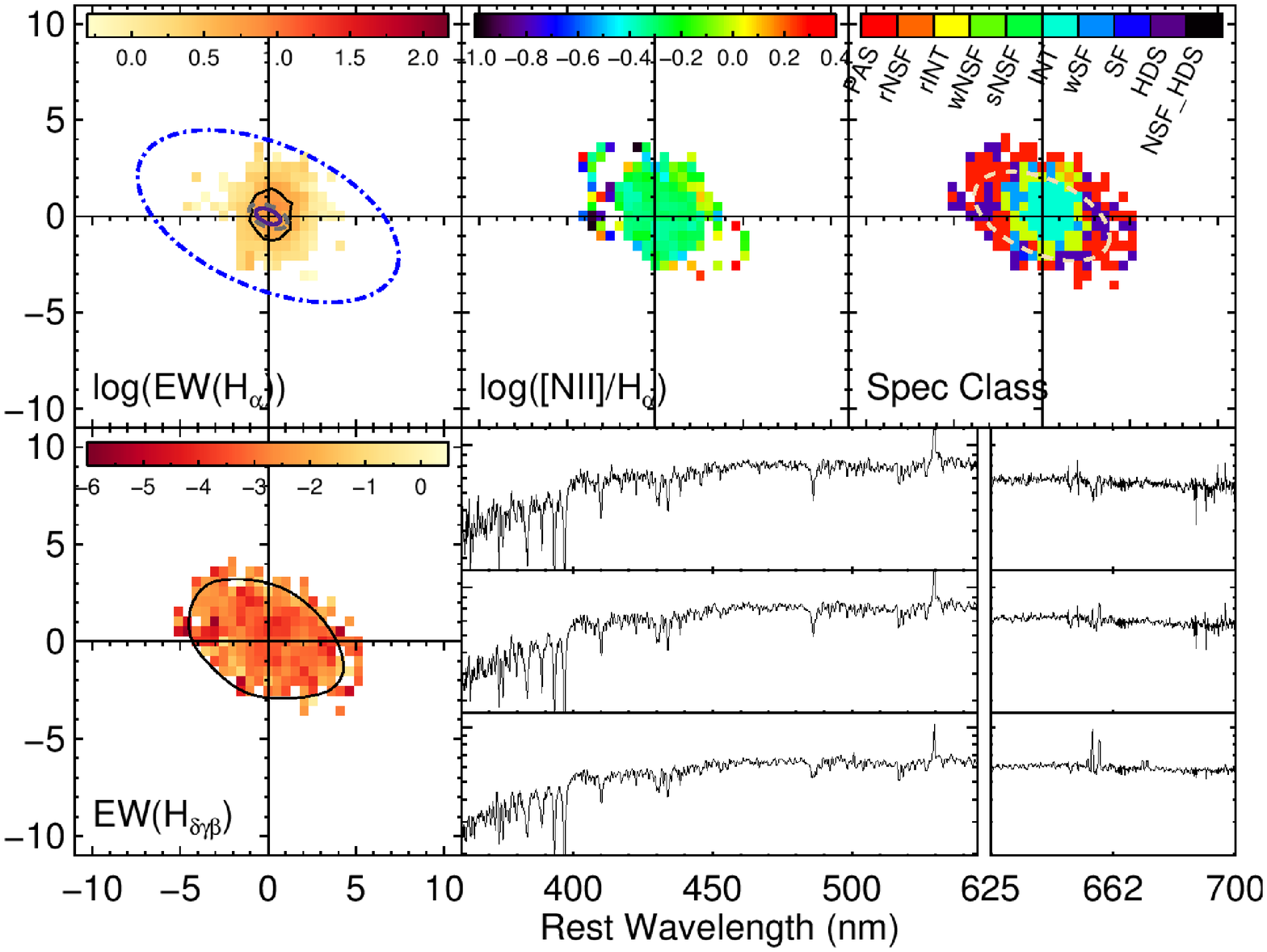}\\
\\
\includegraphics[width=.43\textwidth]{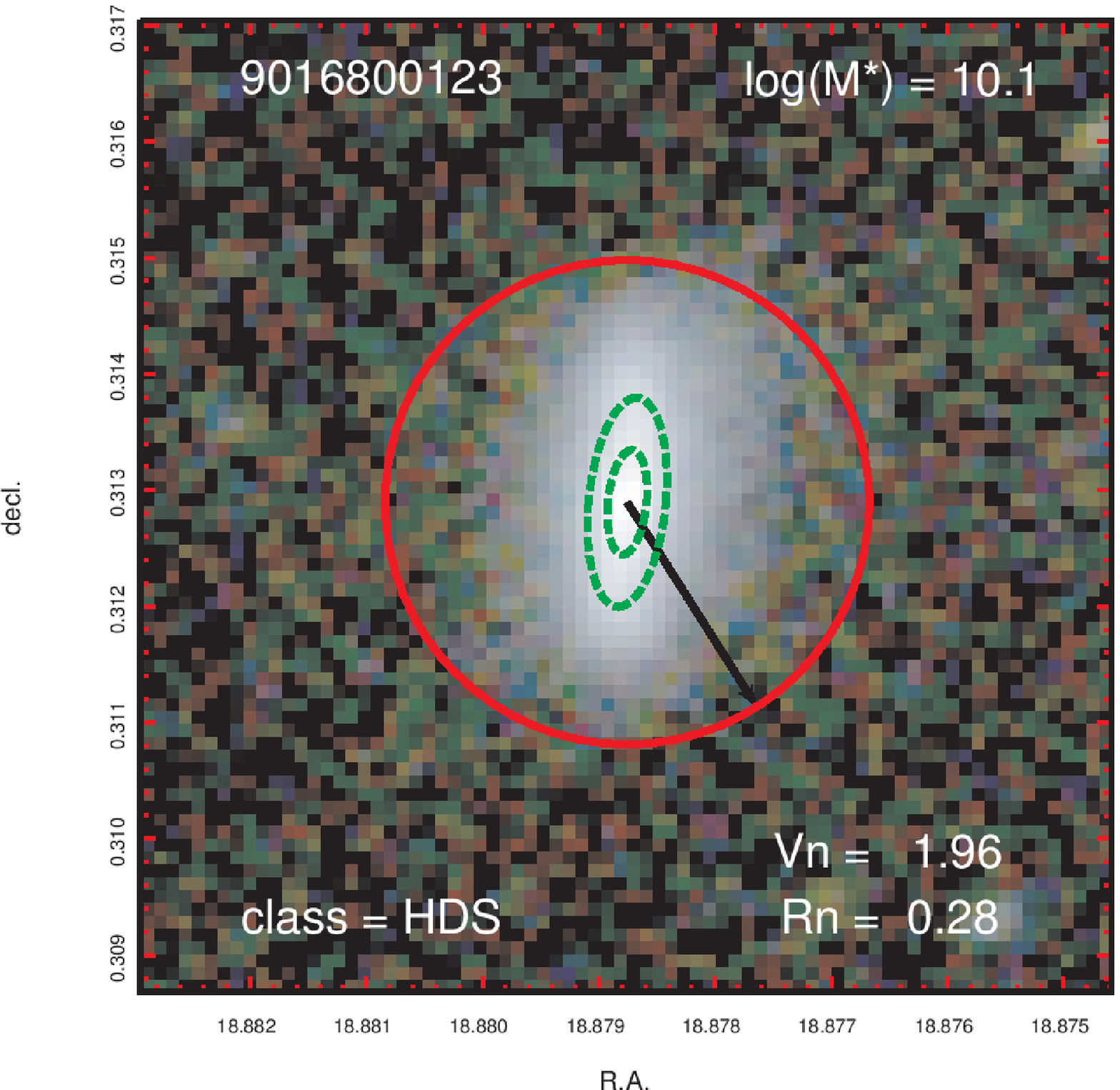}
\includegraphics[width=.55\textwidth]{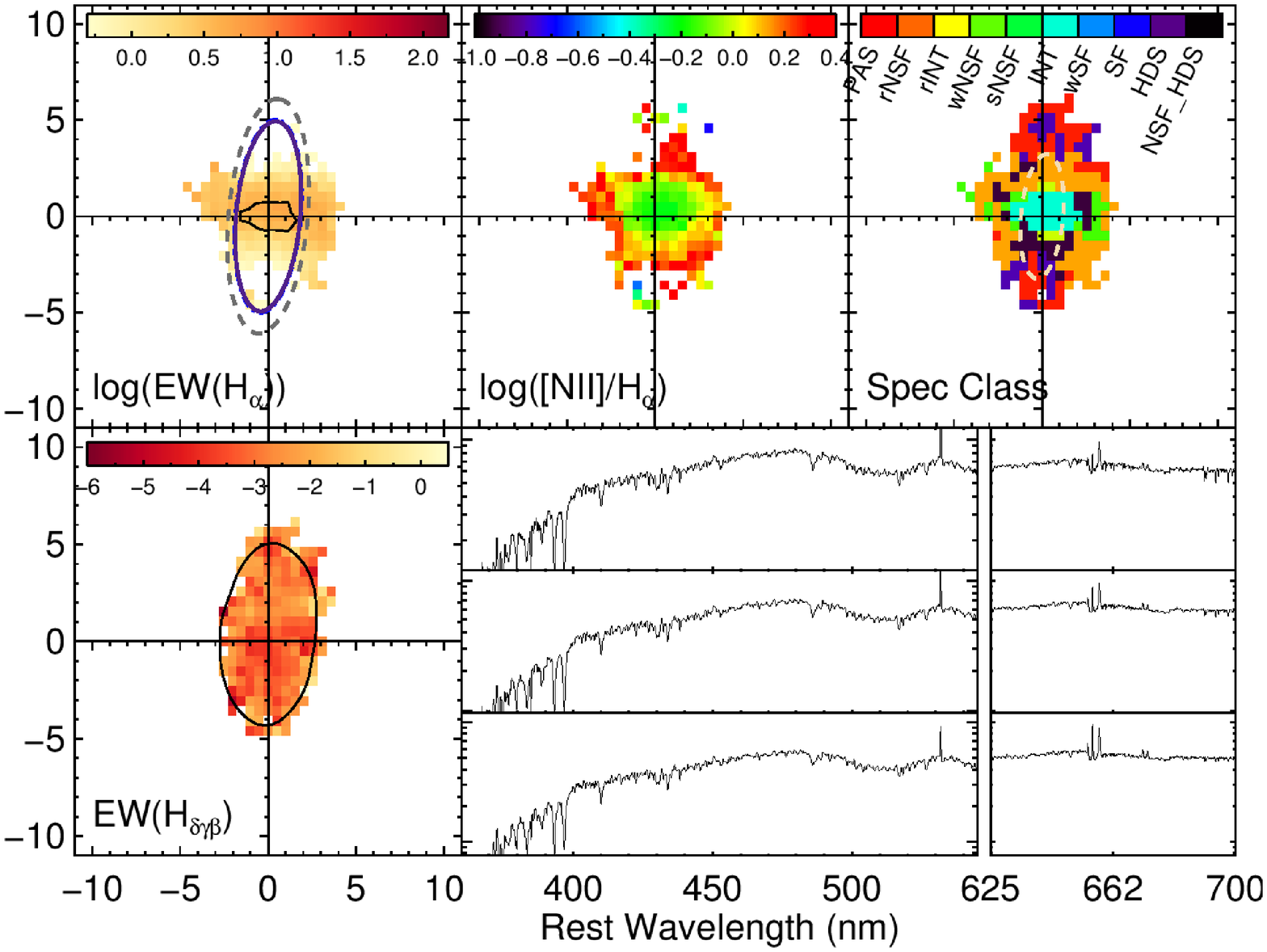}\\
\\
\includegraphics[width=.43\textwidth]{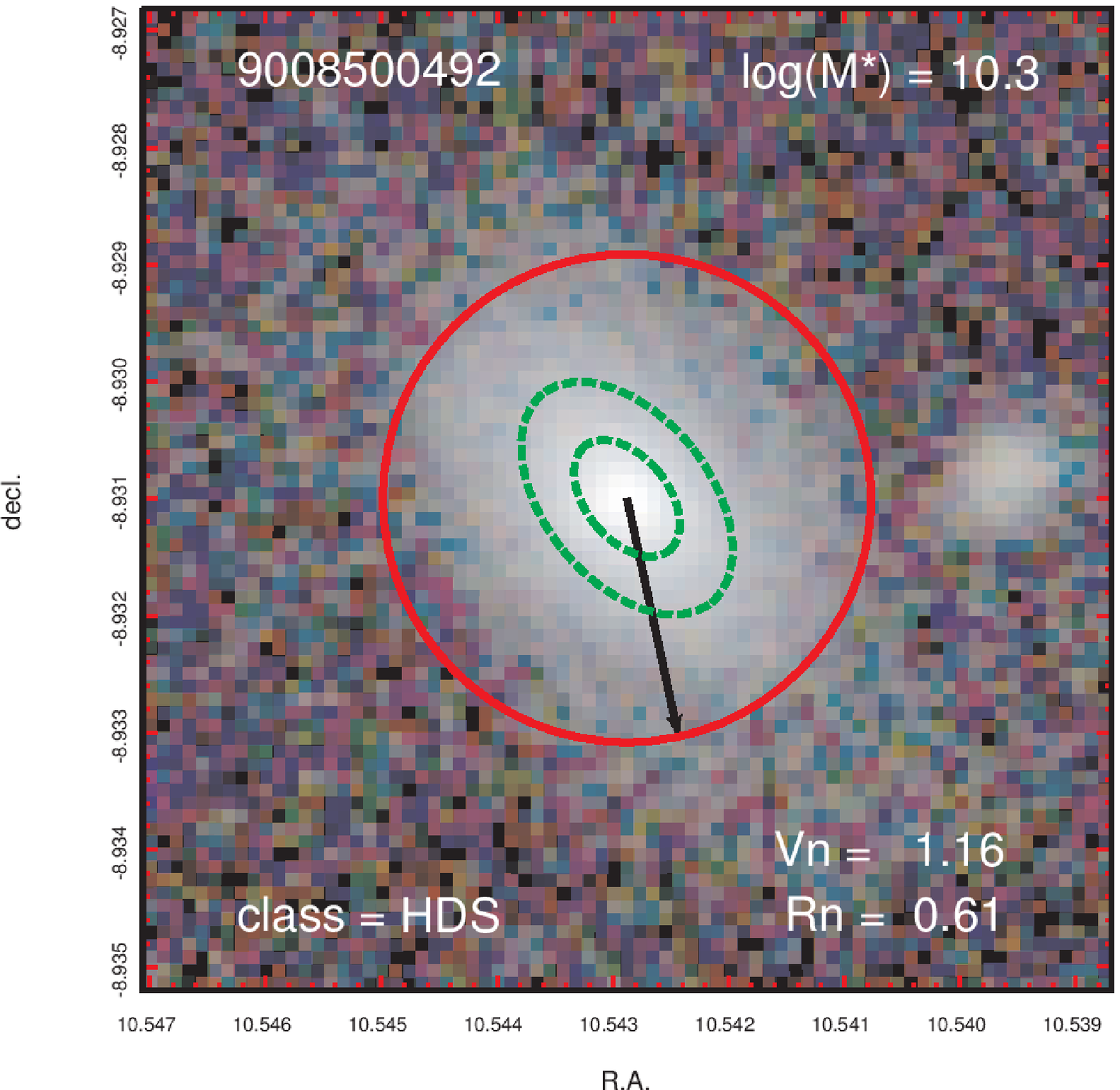}
\includegraphics[width=.55\textwidth]{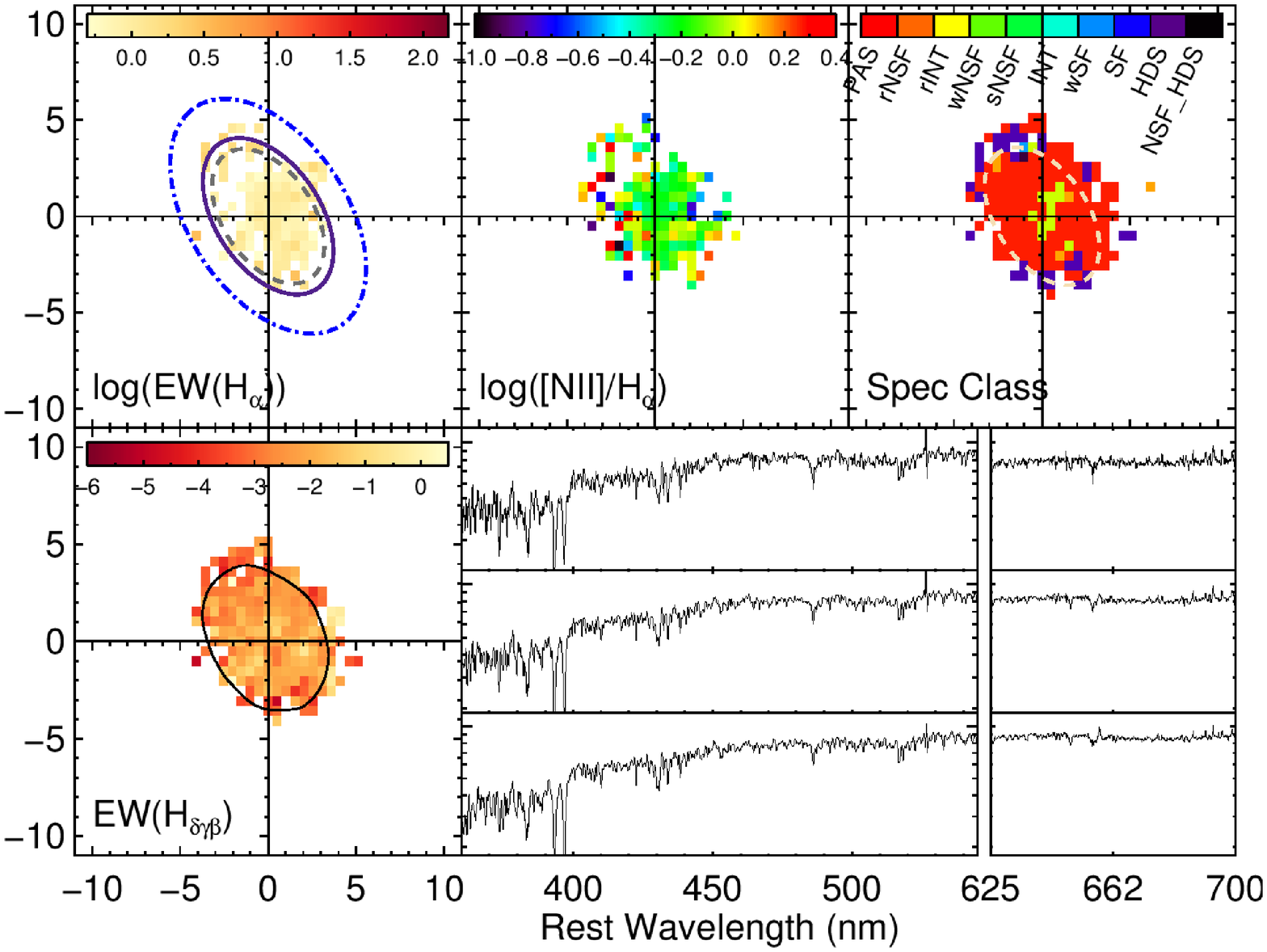}
\\
\caption{}
\end{figure*}

\renewcommand{\thefigure}{\ref{cluster_HDS_galaxies} (Cont.)}
\addtocounter{figure}{-1}
\begin{figure*}
\includegraphics[width=.43\textwidth]{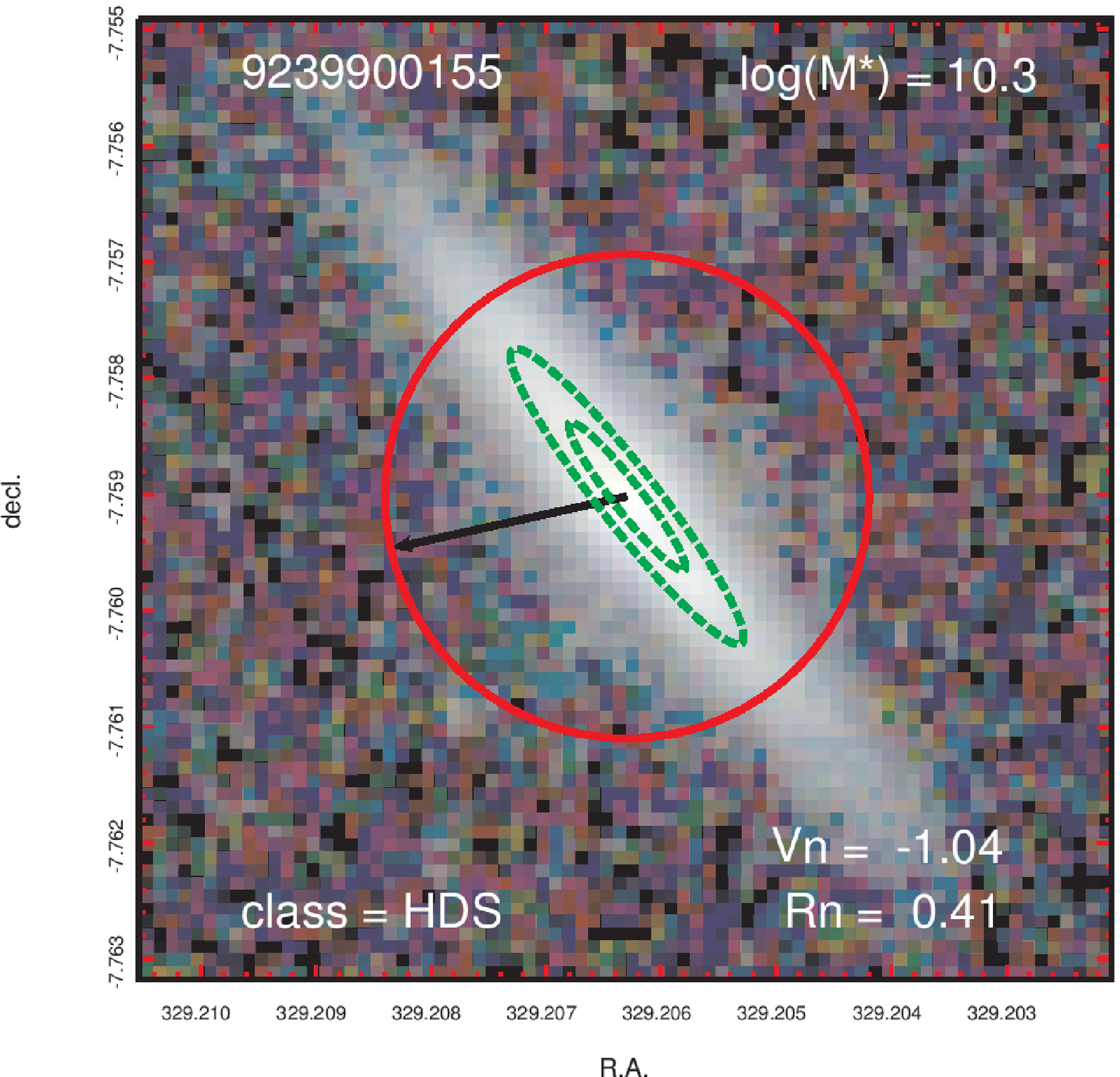}
\includegraphics[width=.55\textwidth]{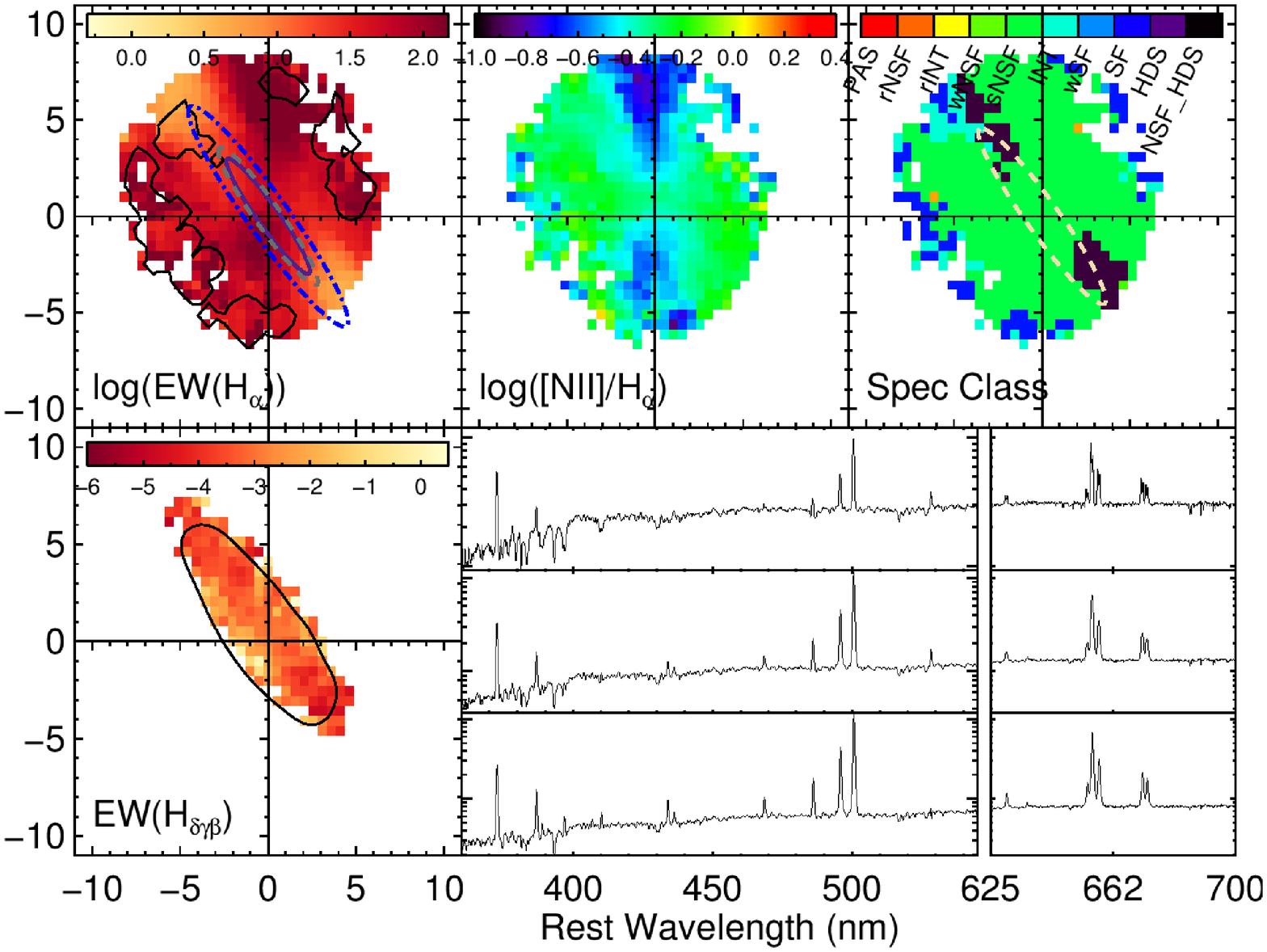}
\\
\includegraphics[width=.43\textwidth]{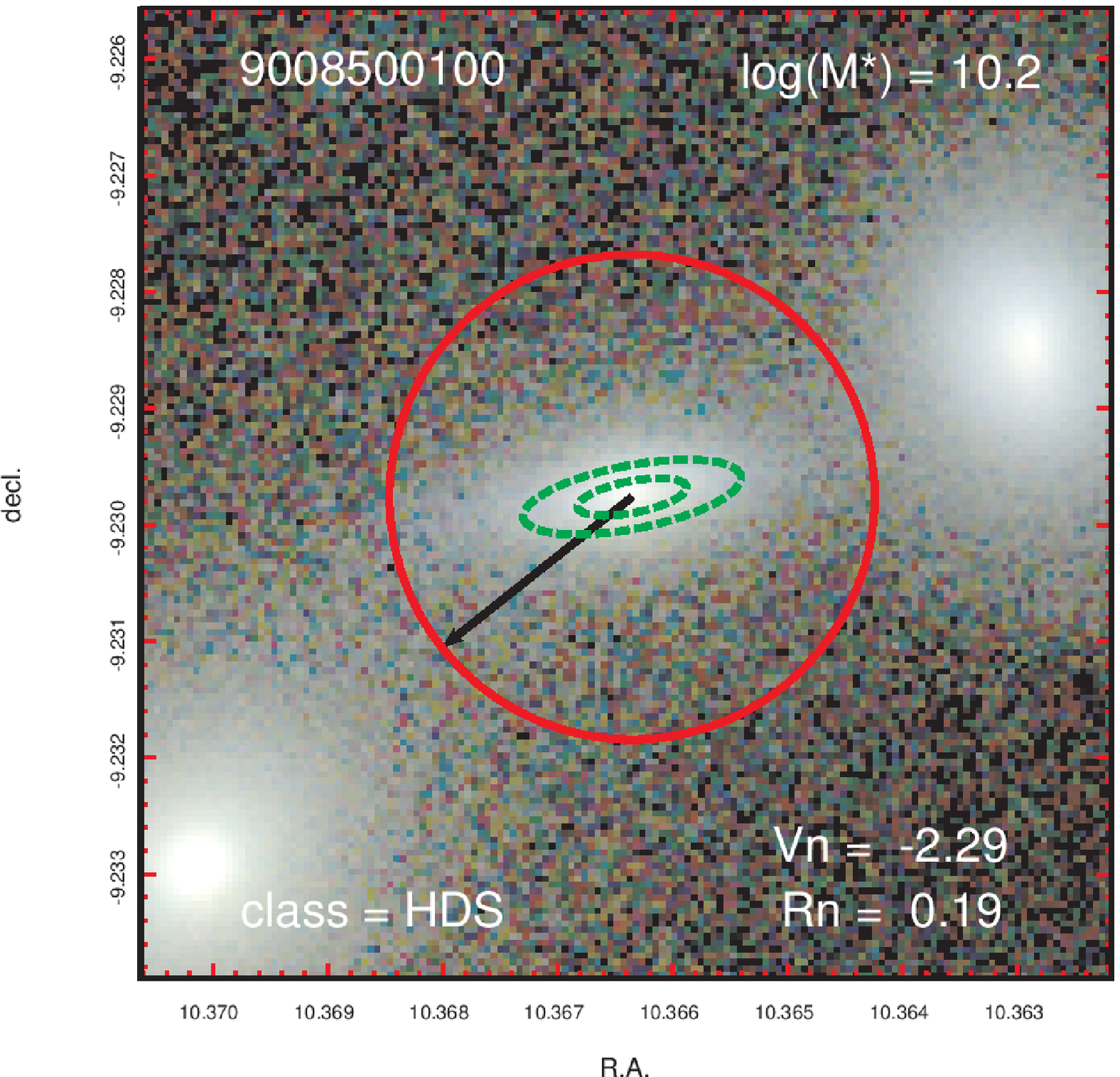}
\includegraphics[width=.55\textwidth]{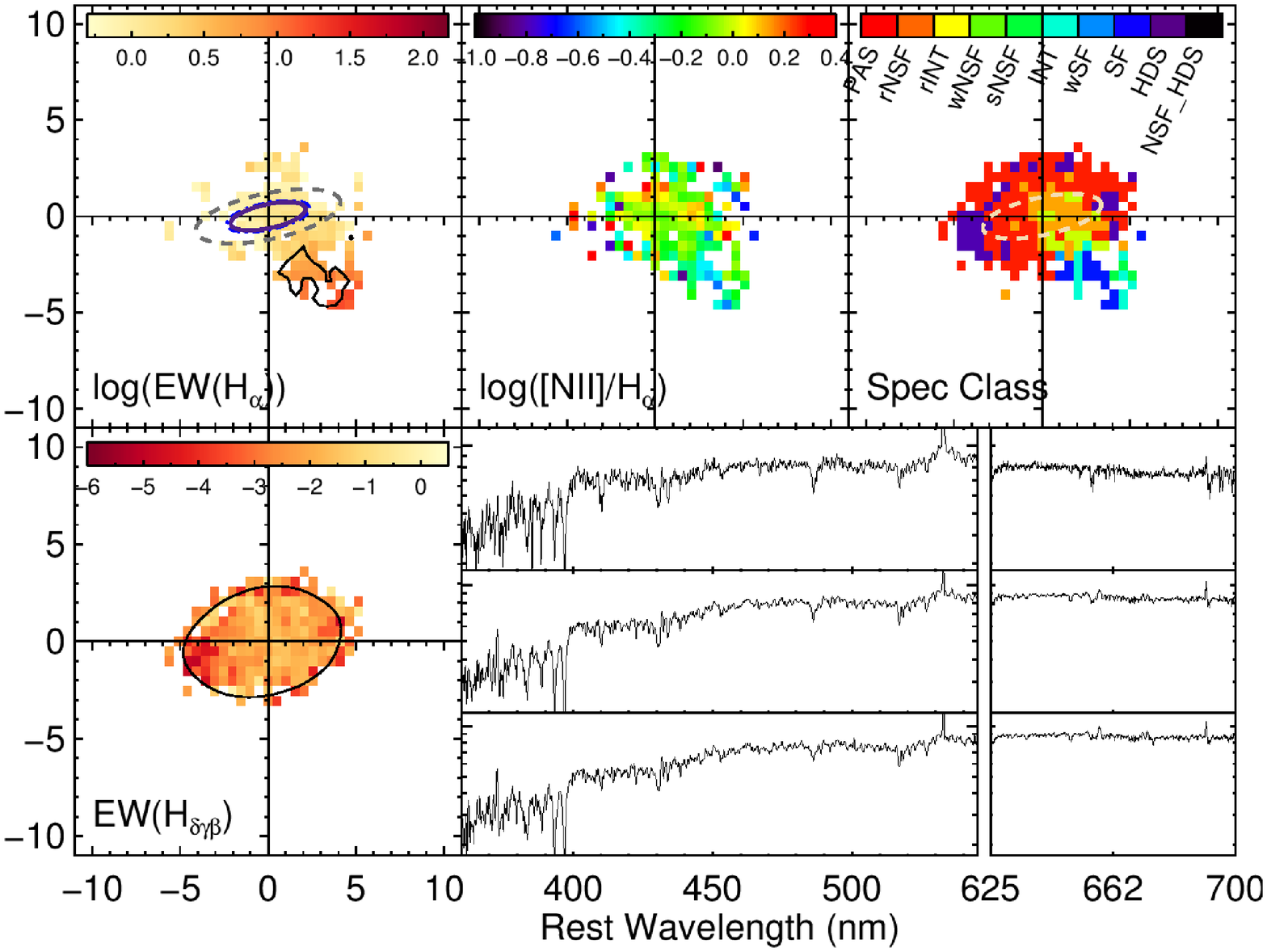}\\
\\
\includegraphics[width=.43\textwidth]{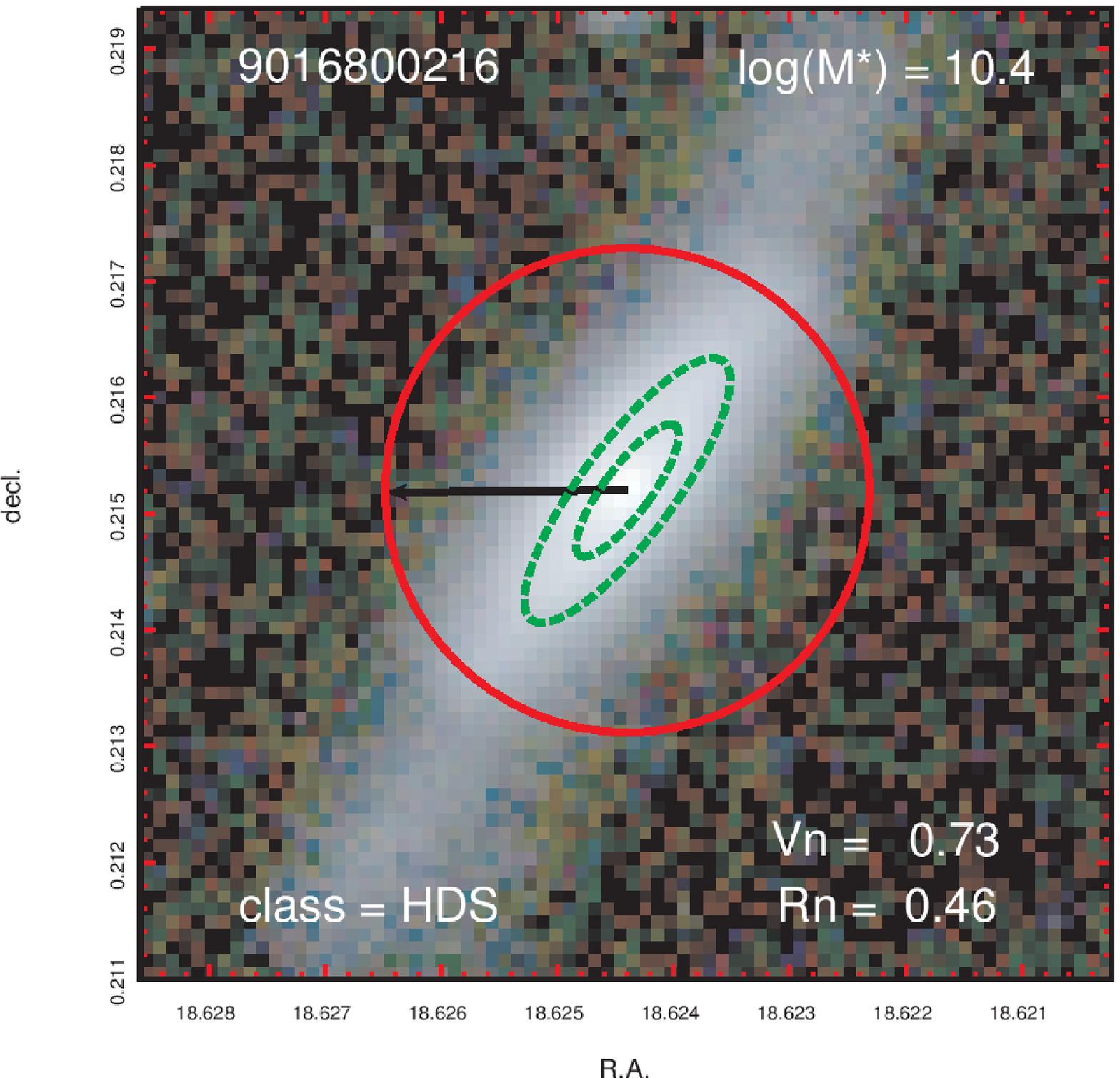}
\includegraphics[width=.55\textwidth]{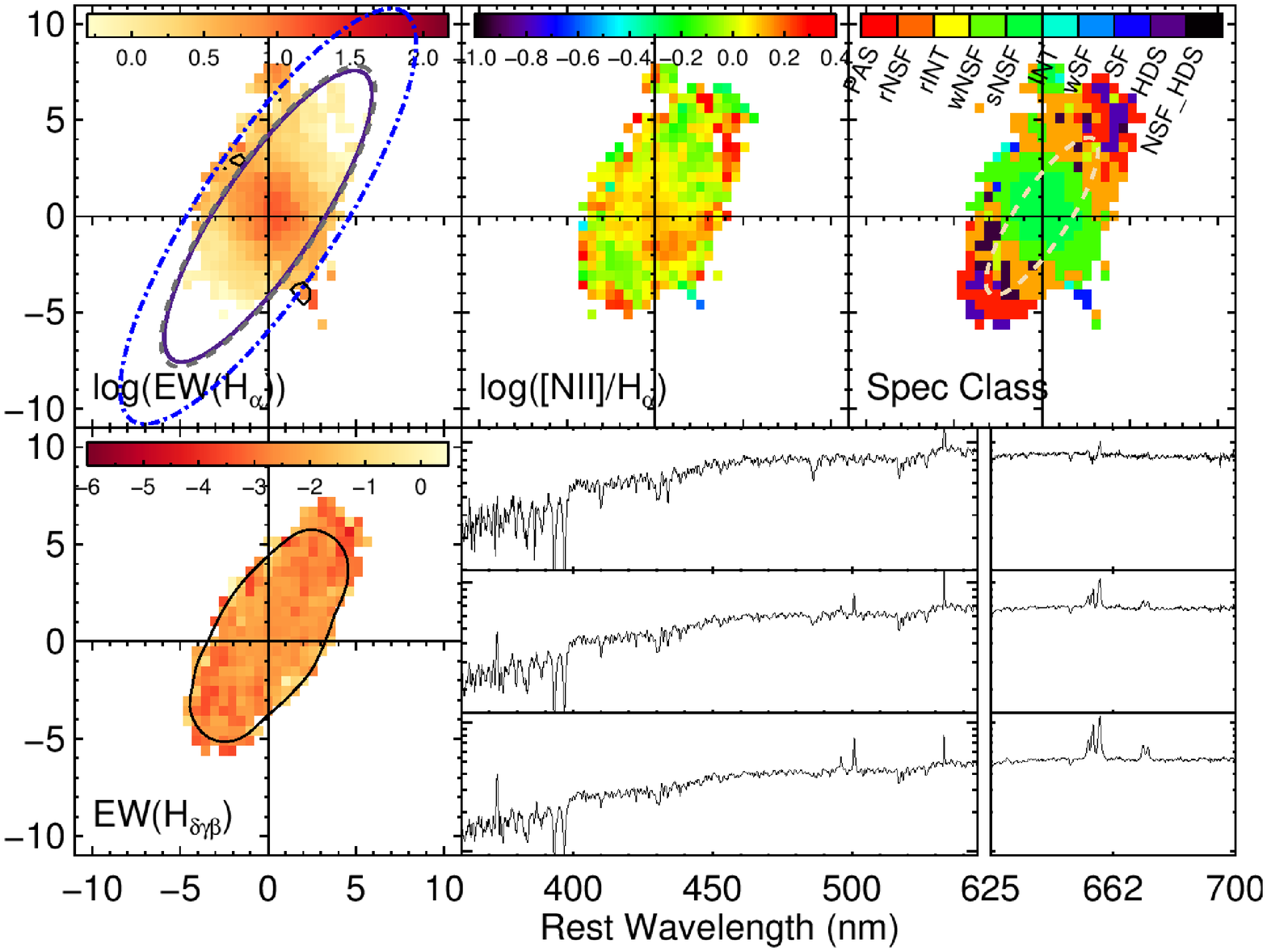}
\\
\caption{}
\end{figure*}

\renewcommand{\thefigure}{Figure~\ref{cluster_HDS_galaxies} (Cont.)}
\addtocounter{figure}{-1}
\begin{figure*}
\includegraphics[width=.43\textwidth]{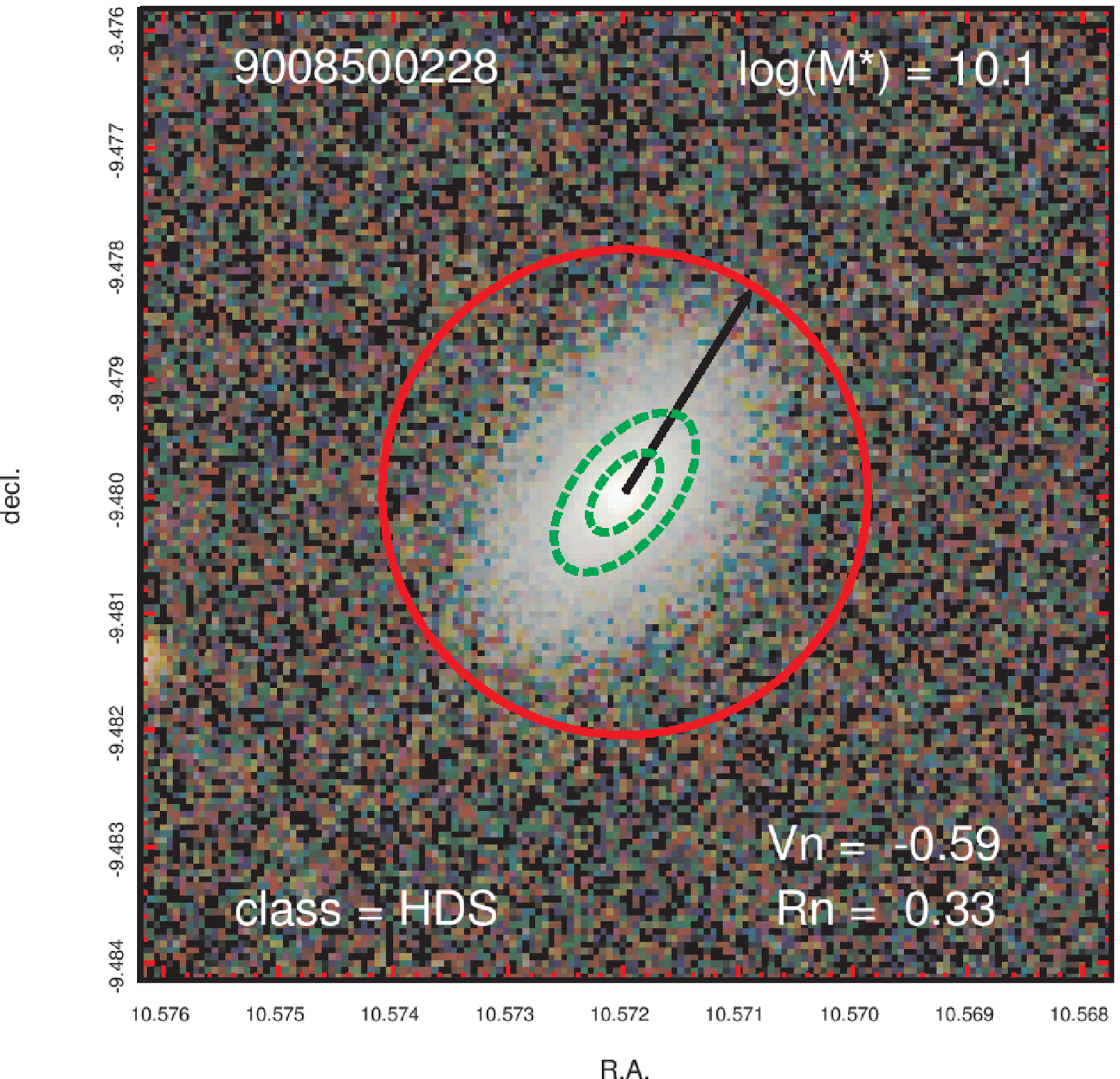}
\includegraphics[width=.55\textwidth]{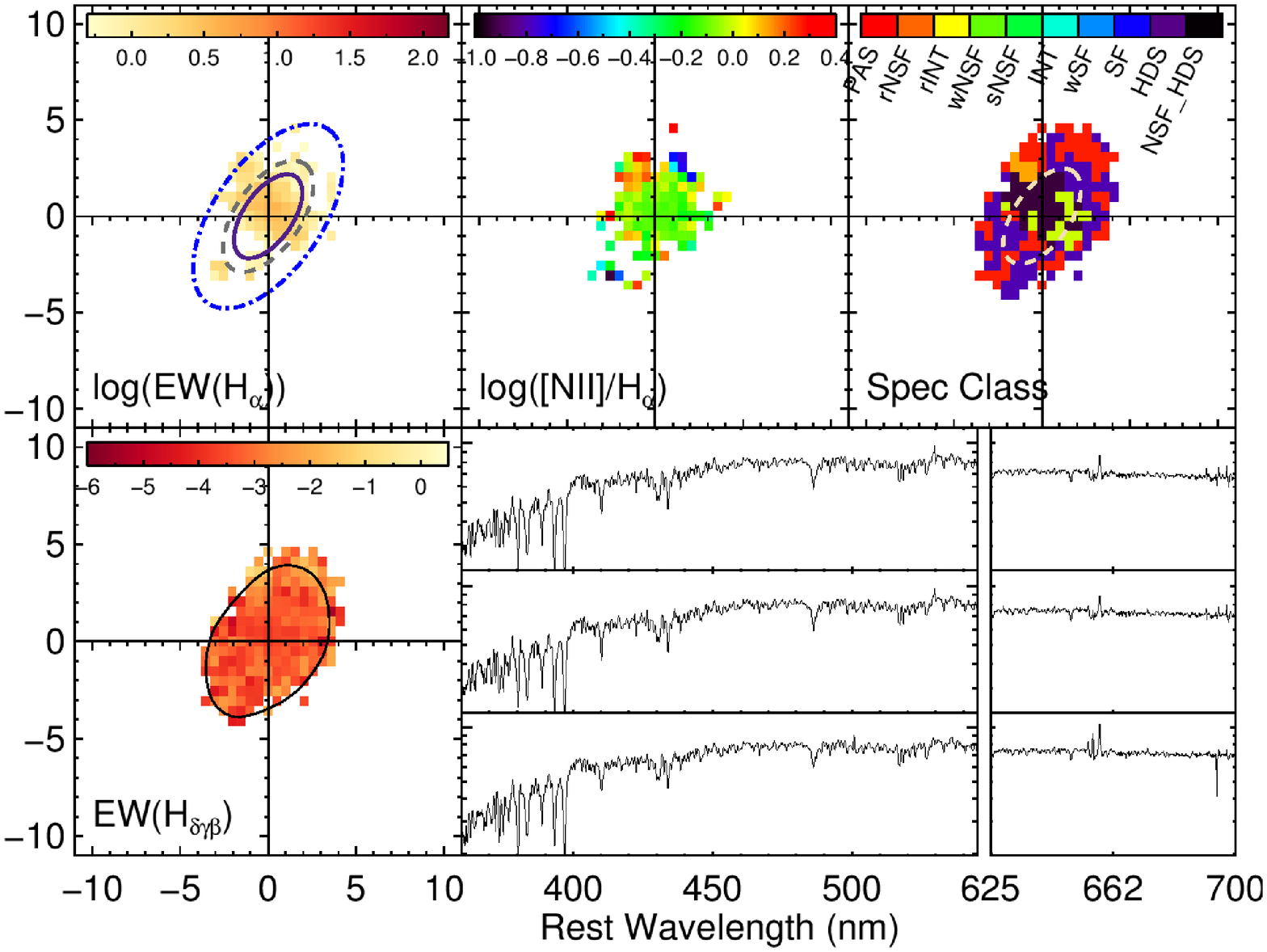}\\
\\
\includegraphics[width=.43\textwidth]{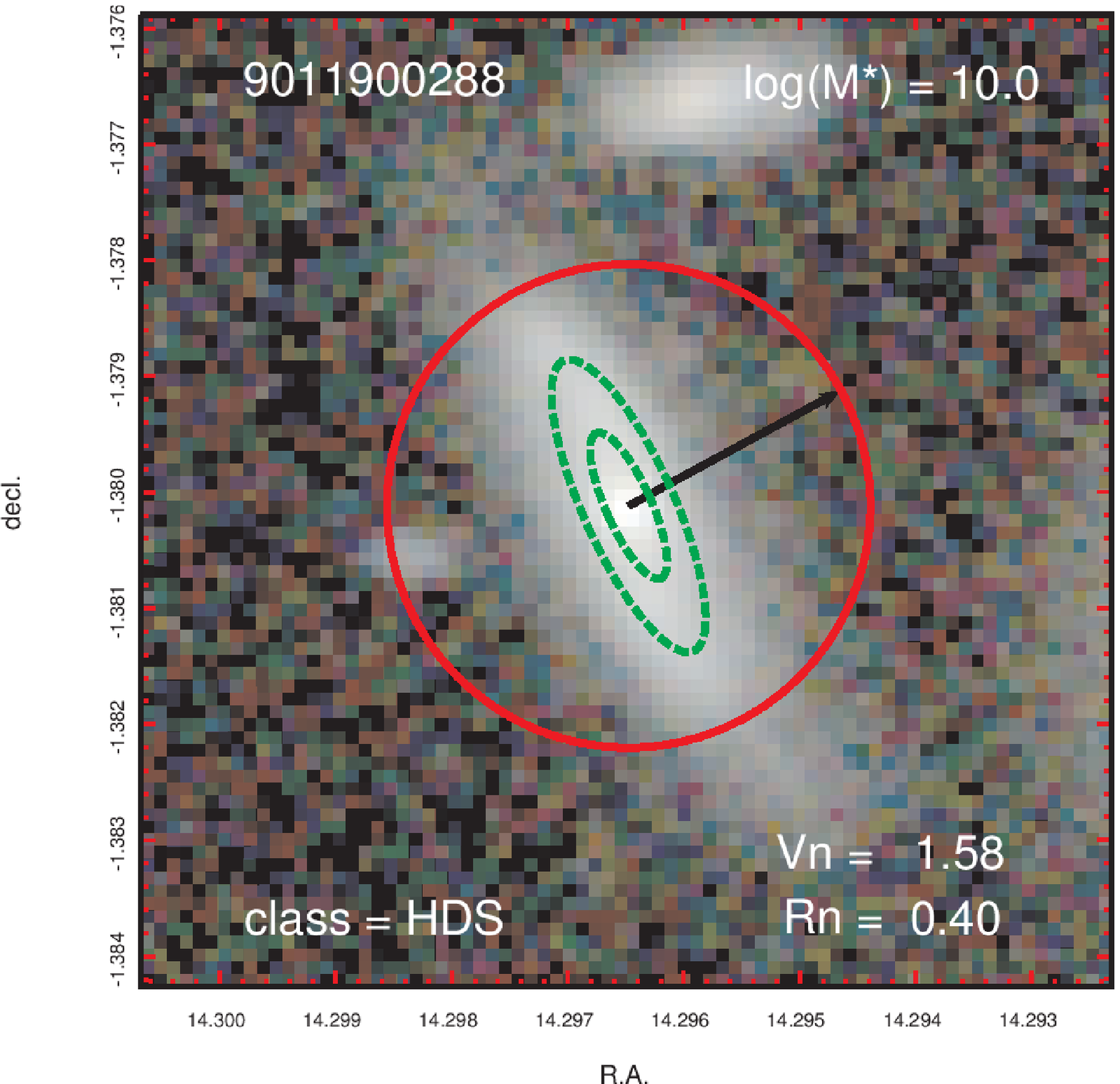}
\includegraphics[width=.55\textwidth]{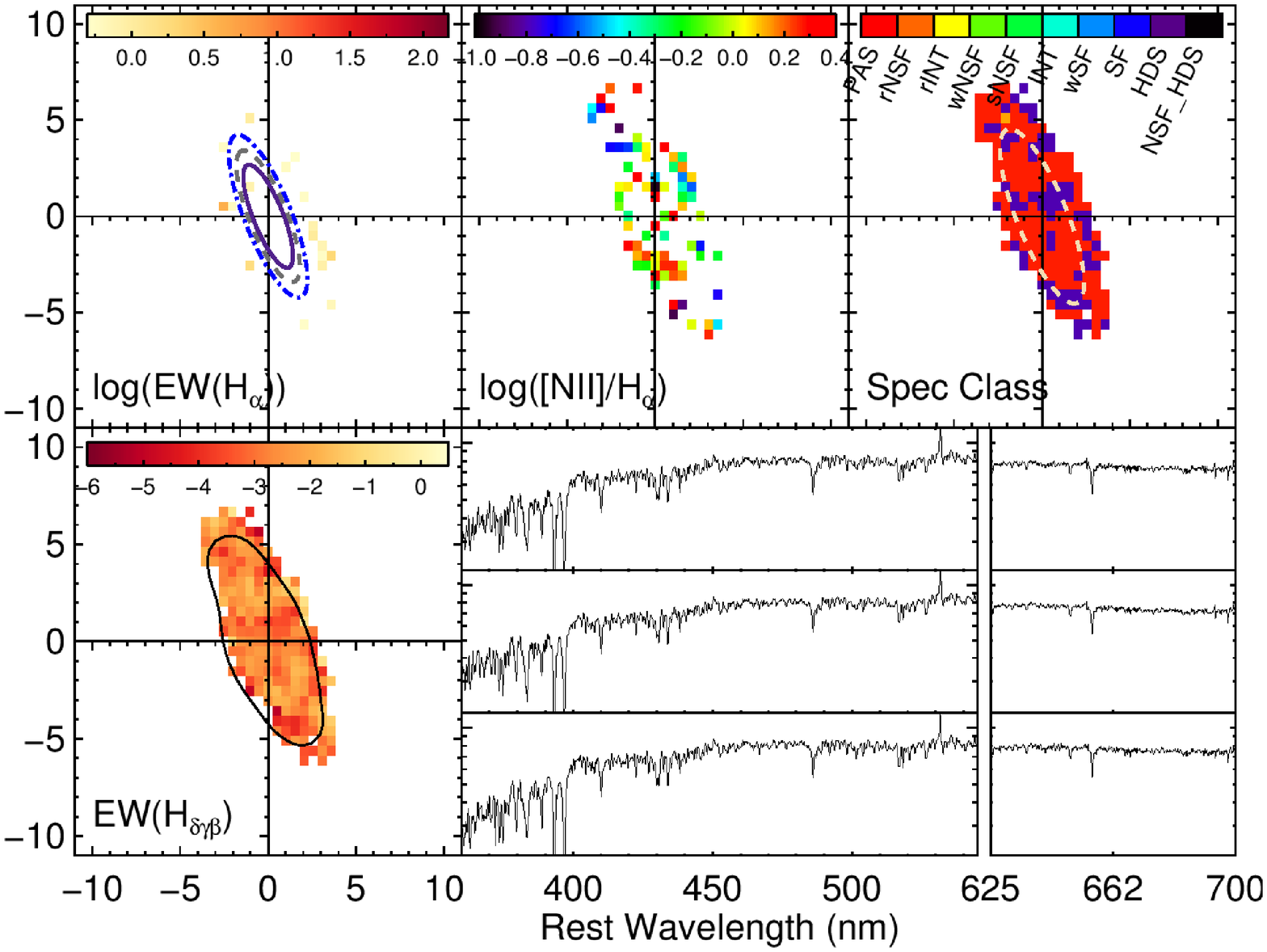}
\\
\includegraphics[width=.43\textwidth]{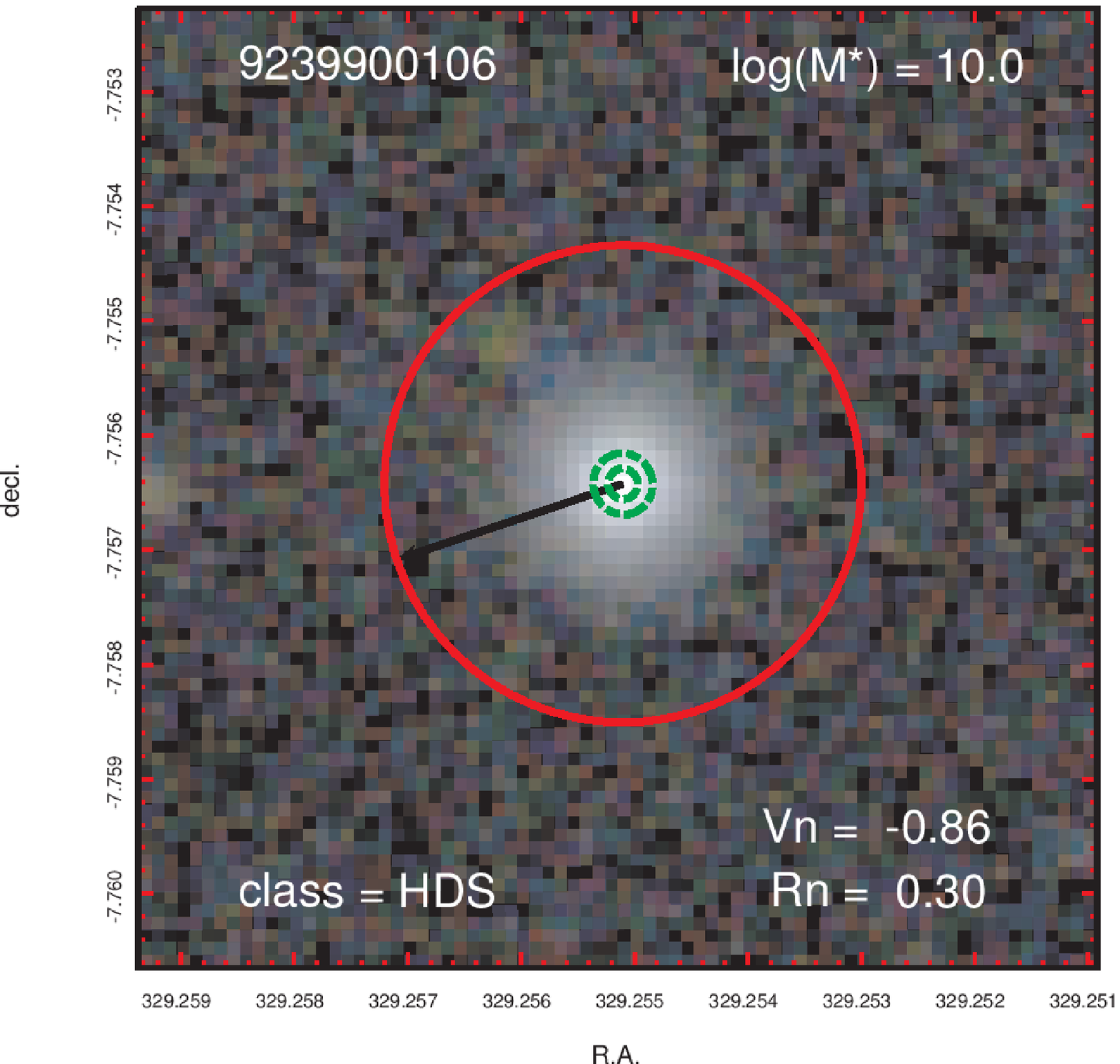}
\includegraphics[width=.55\textwidth]{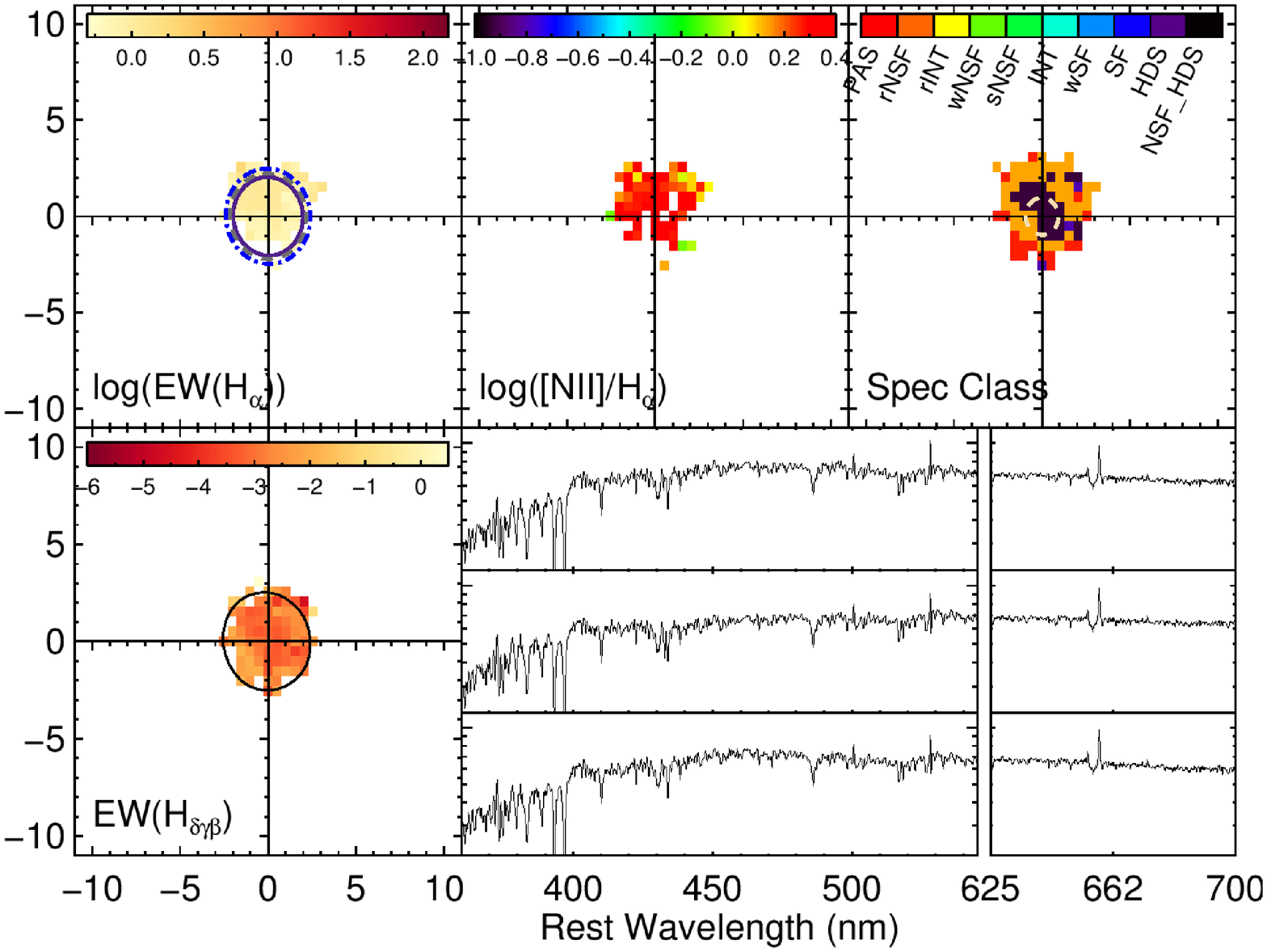}
\caption{}
\end{figure*}

\renewcommand{\thefigure}{\ref{GAMA_HDS_galaxies} (Cont.)}
\addtocounter{figure}{-1}
\begin{figure*}
\includegraphics[width=.43\textwidth]{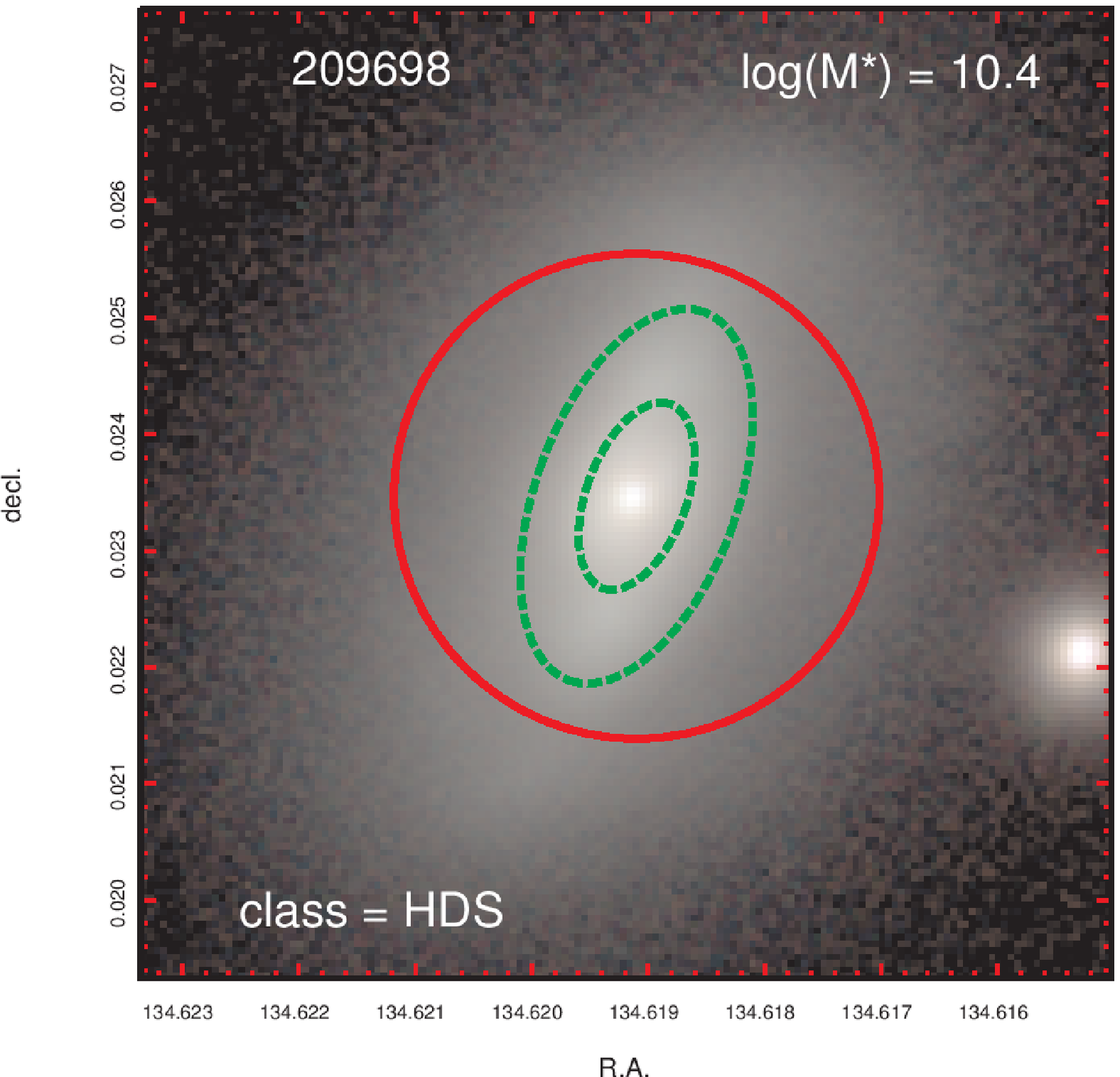}
\includegraphics[width=.55\textwidth]{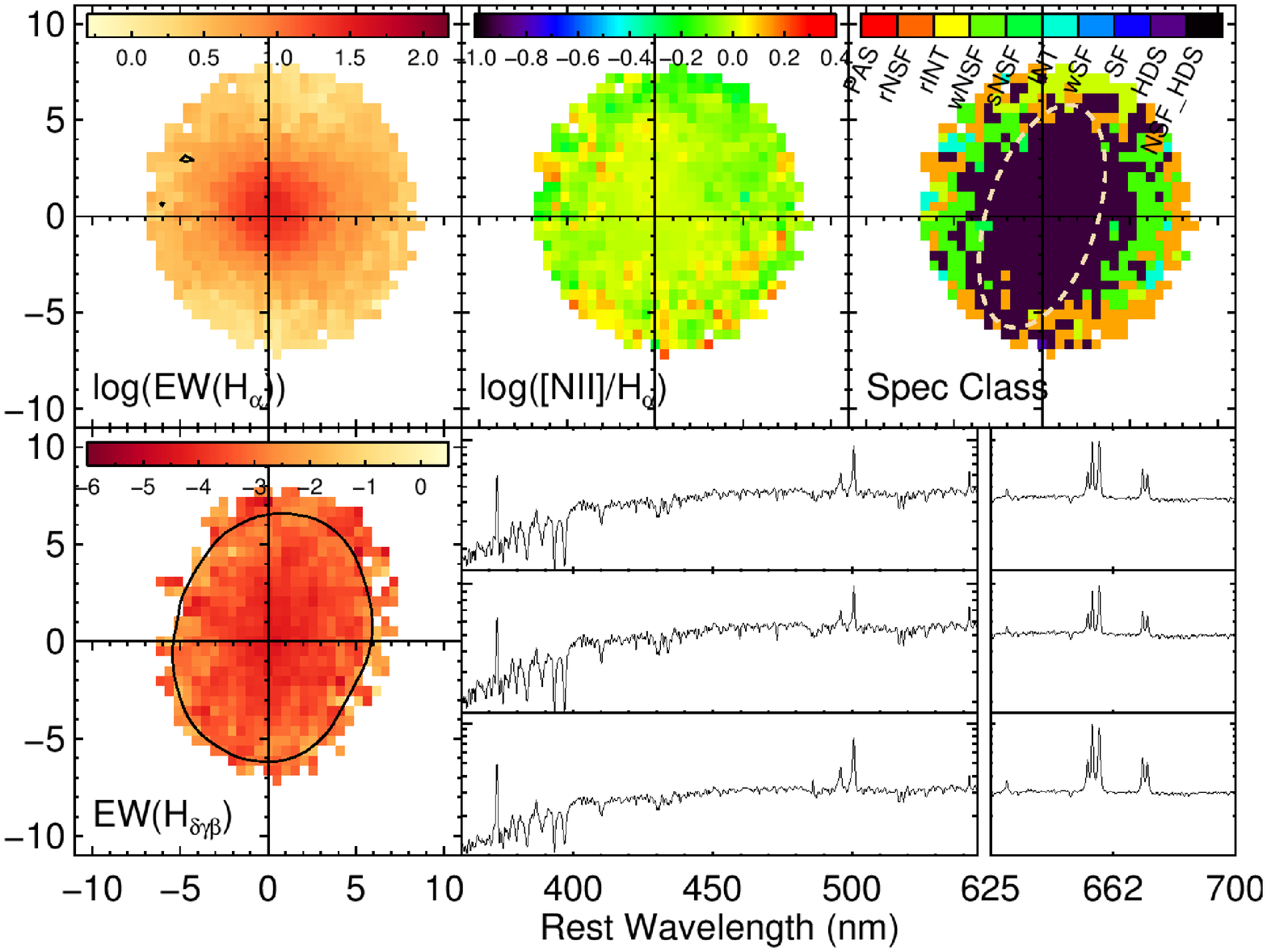}\\
\\
\includegraphics[width=.43\textwidth]{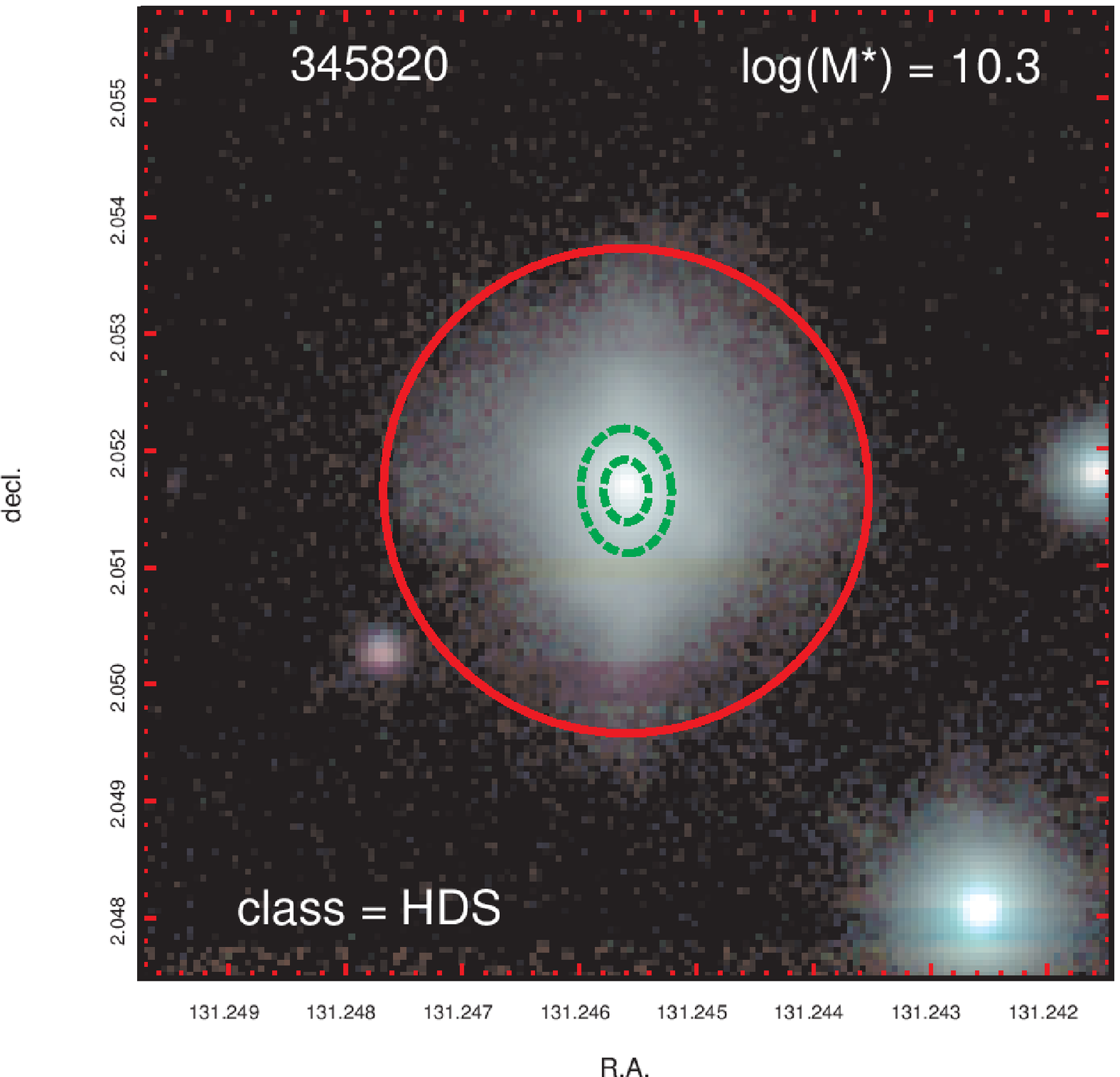}
\includegraphics[width=.55\textwidth]{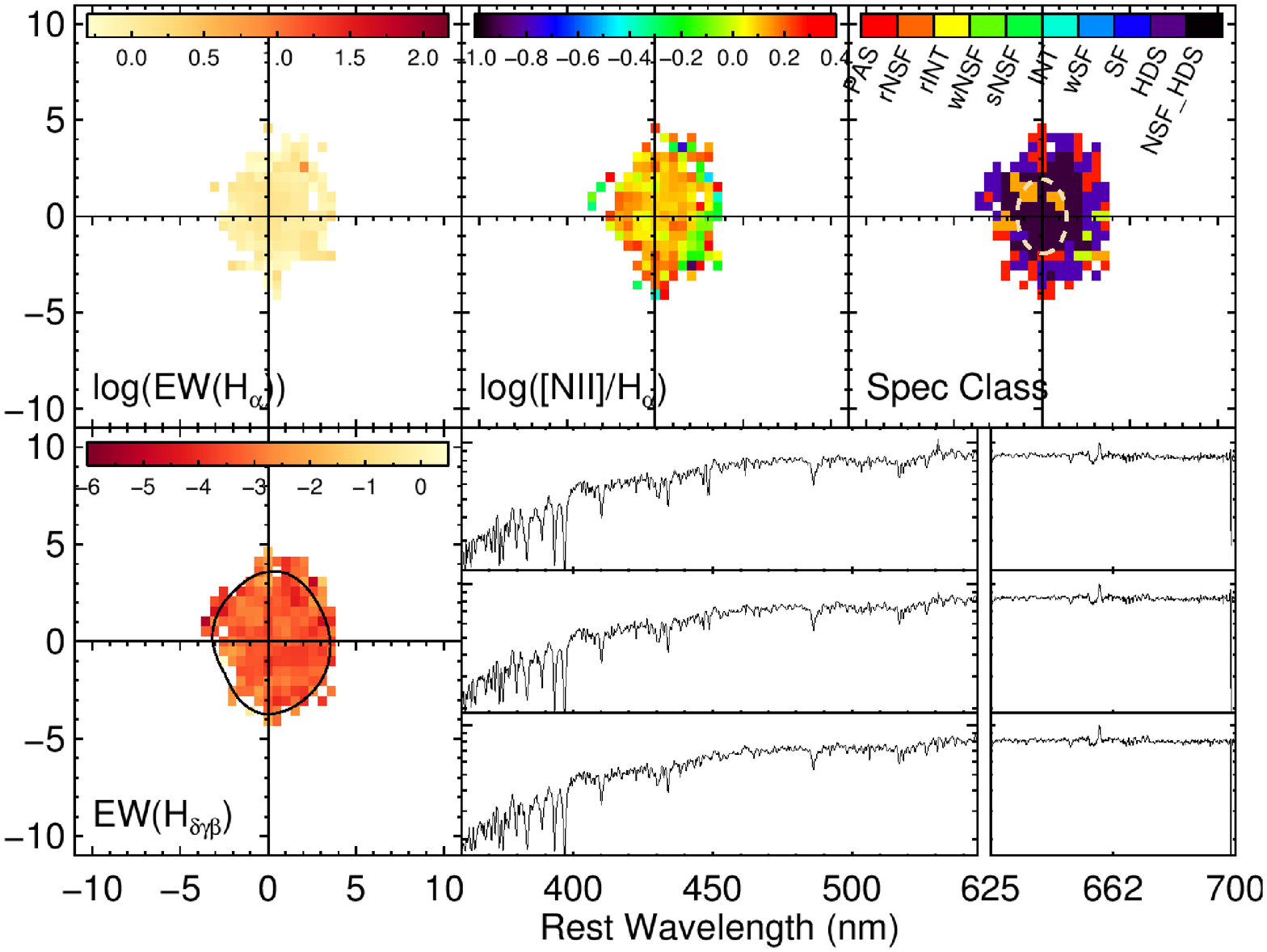}\\
\\
\includegraphics[width=.43\textwidth]{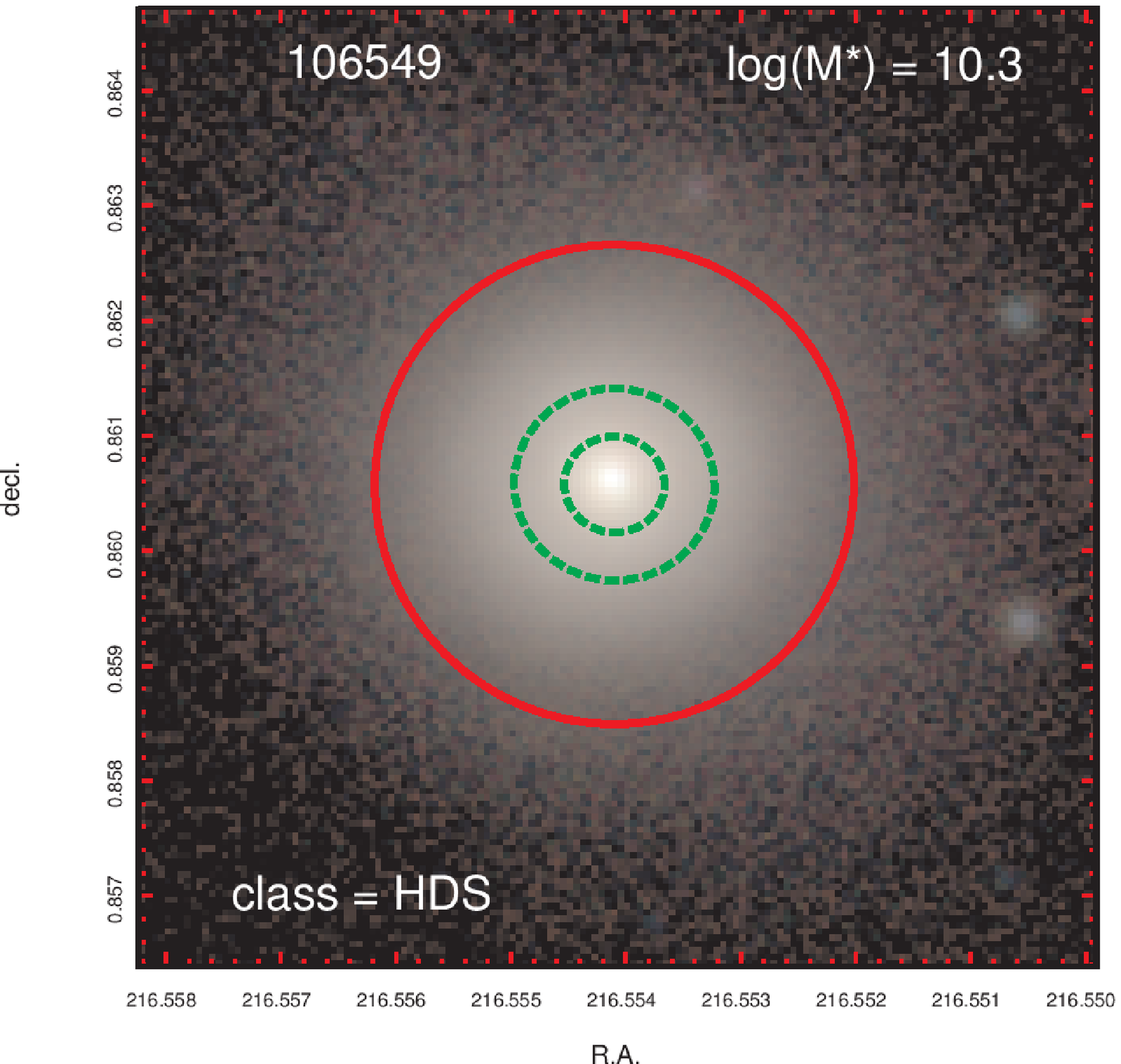}
\includegraphics[width=.55\textwidth]{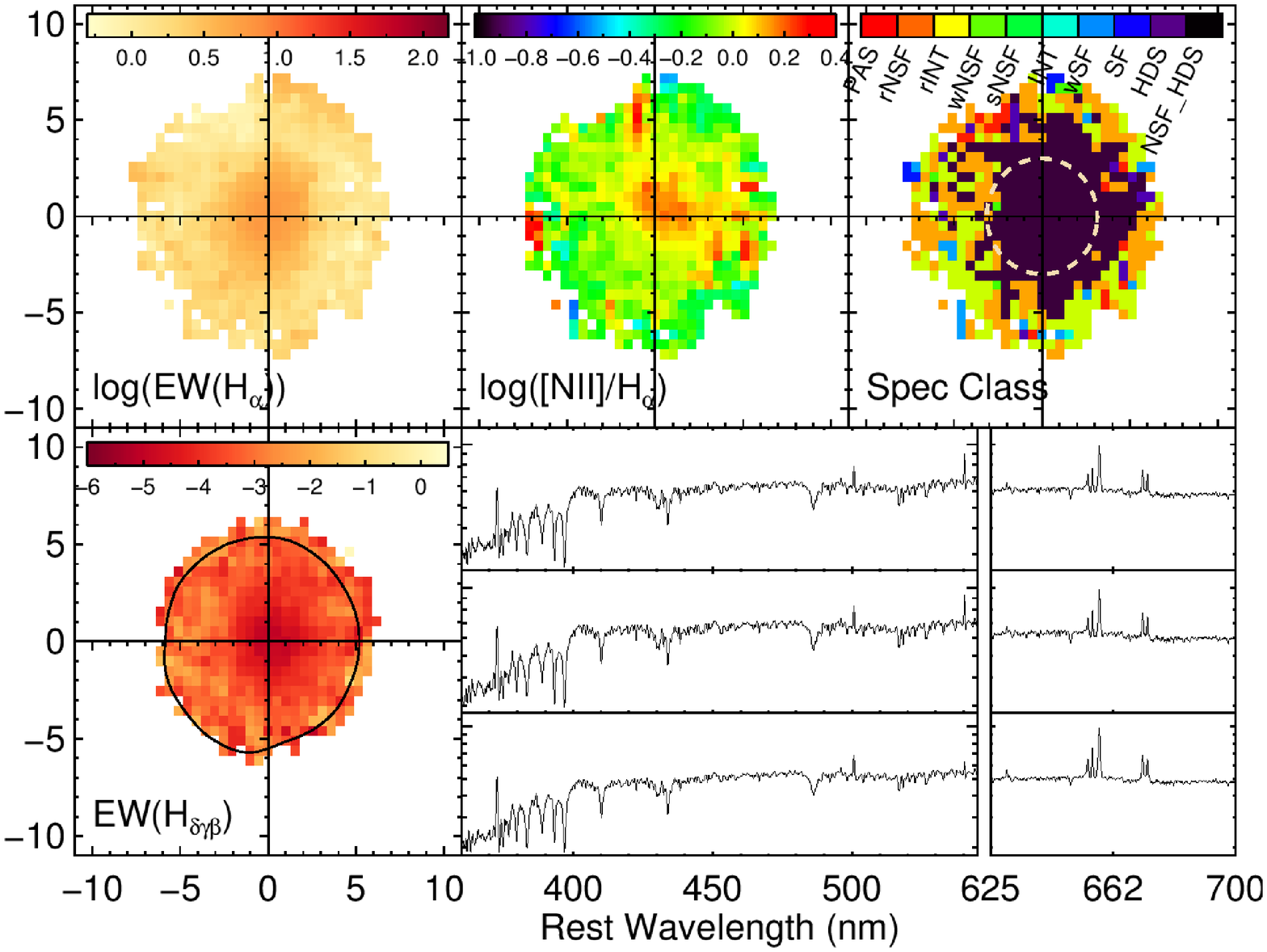}
\caption{}
\end{figure*}

\renewcommand{\thefigure}{\ref{GAMA_HDS_galaxies} (Cont.)}
\addtocounter{figure}{-1}
\begin{figure*}
\includegraphics[width=.43\textwidth]{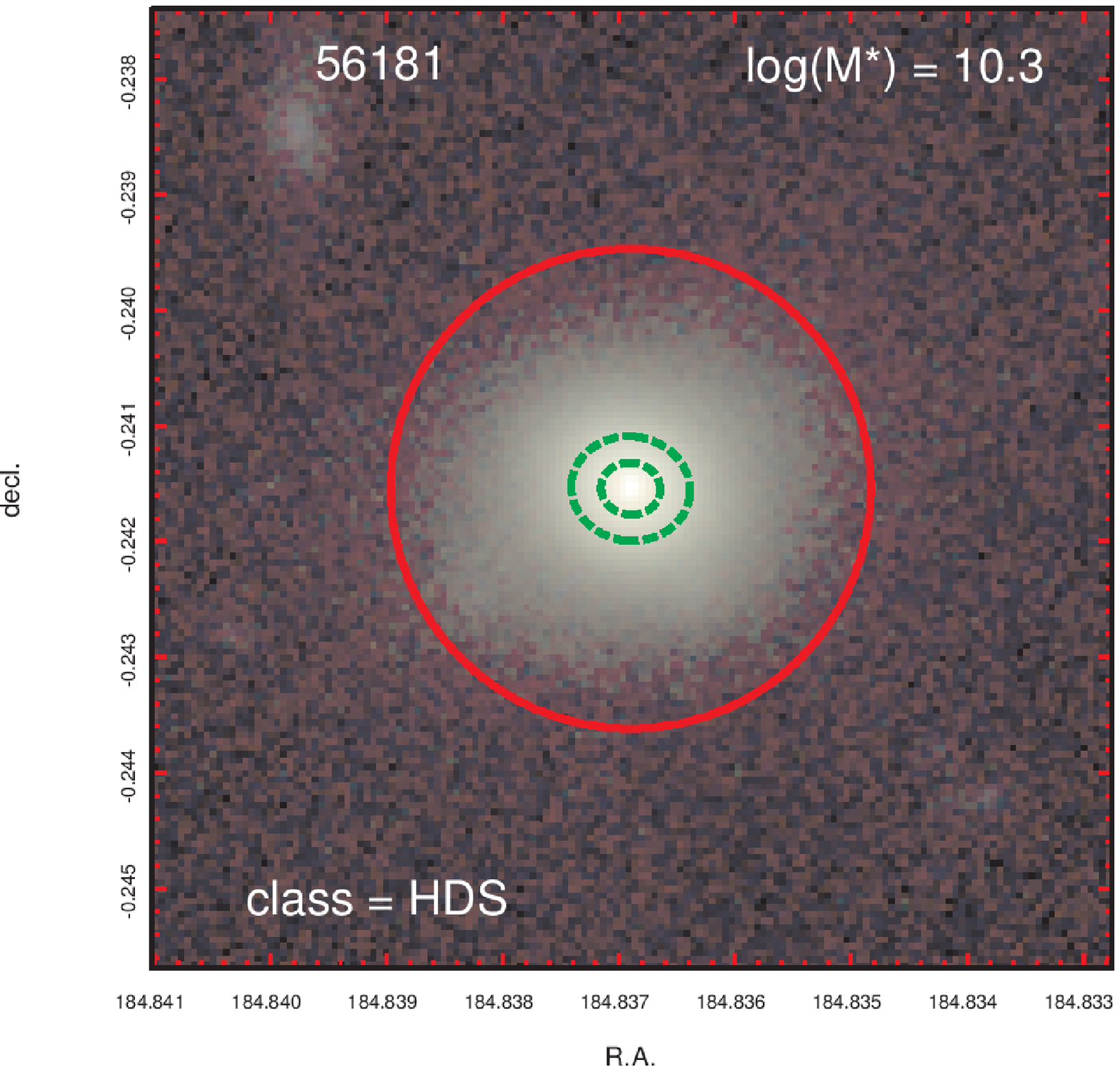}
\includegraphics[width=.55\textwidth]{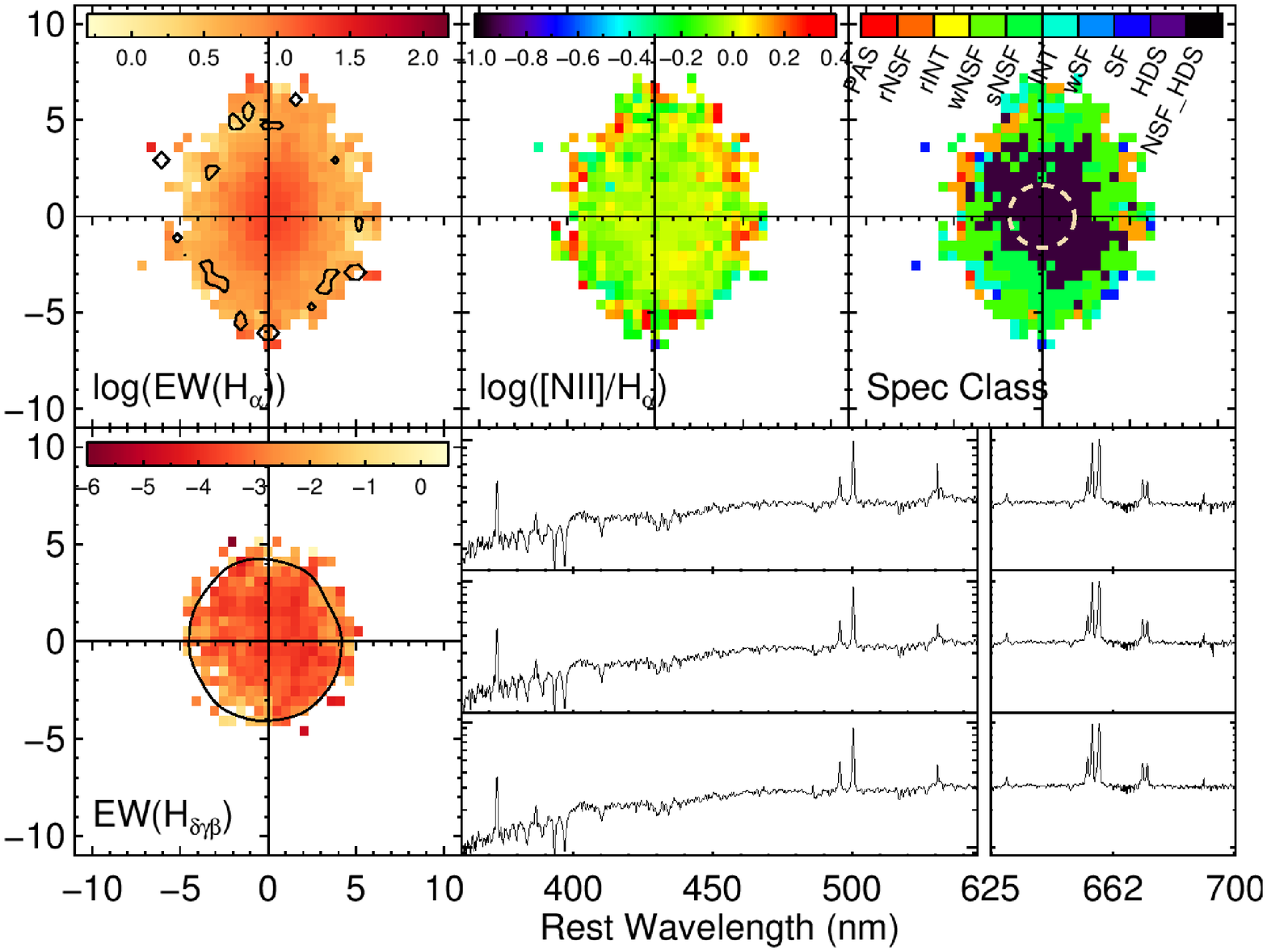}
\\
\includegraphics[width=.43\textwidth]{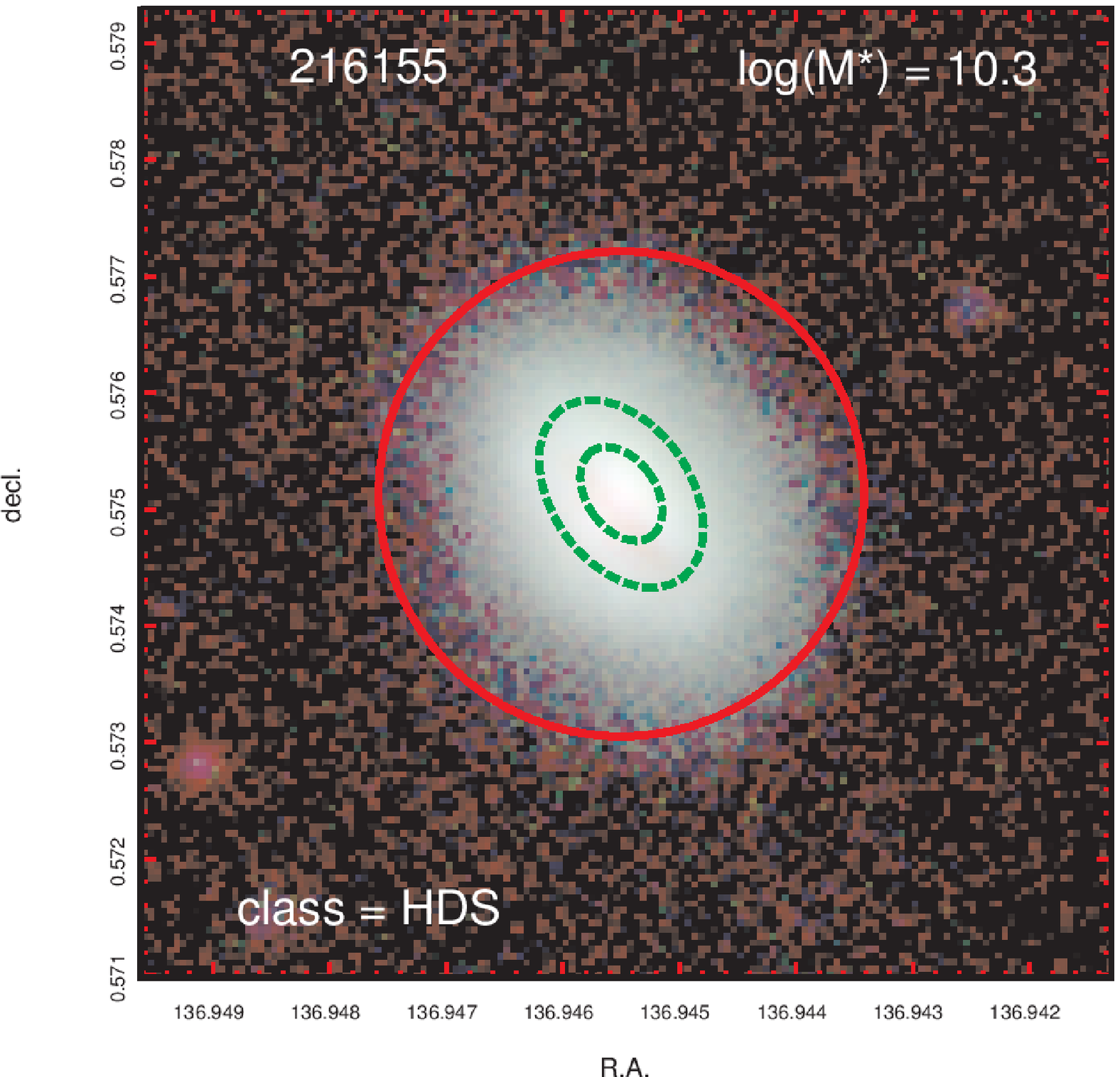}
\includegraphics[width=.55\textwidth]{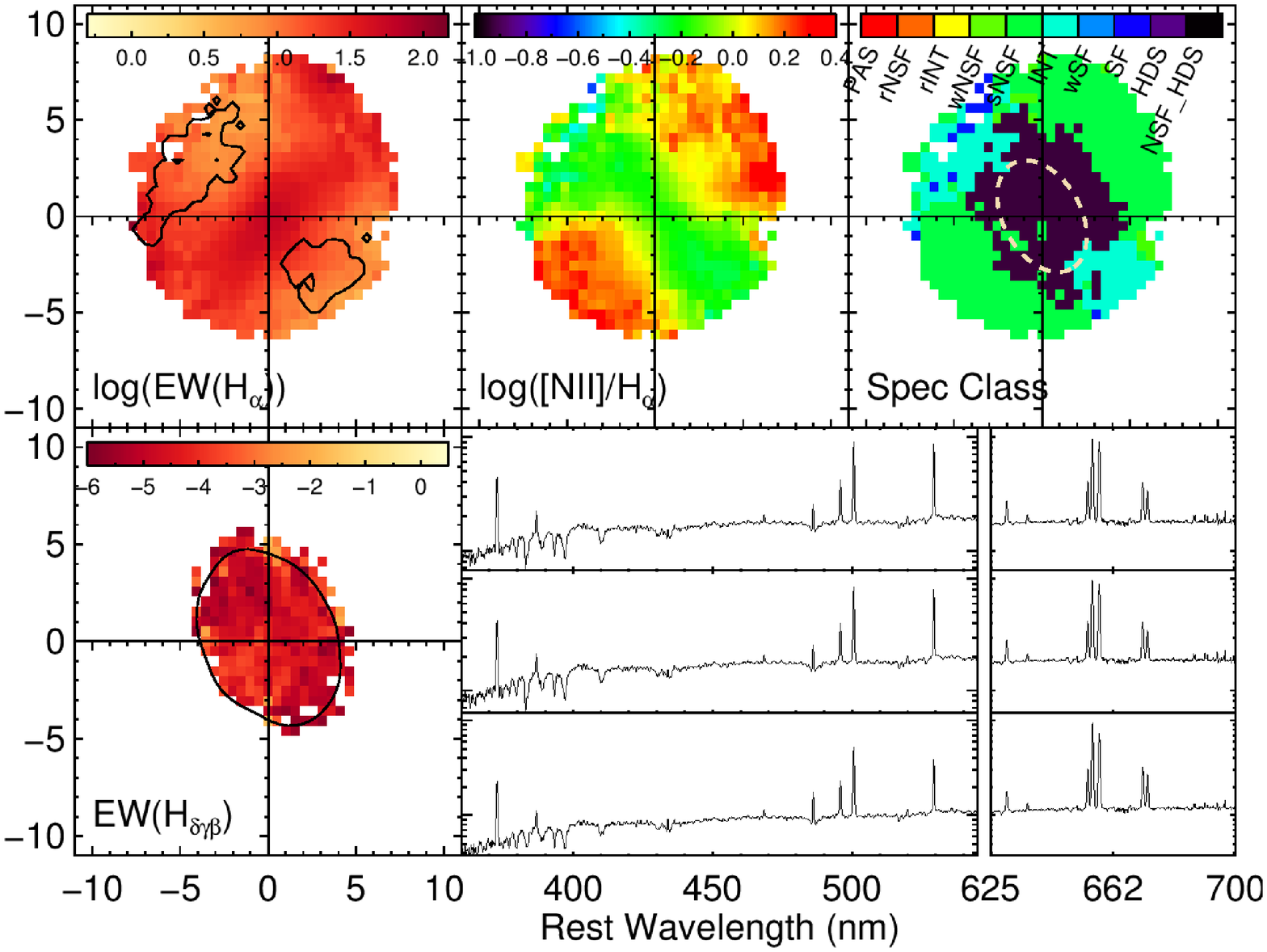}
\\
\includegraphics[width=.43\textwidth]{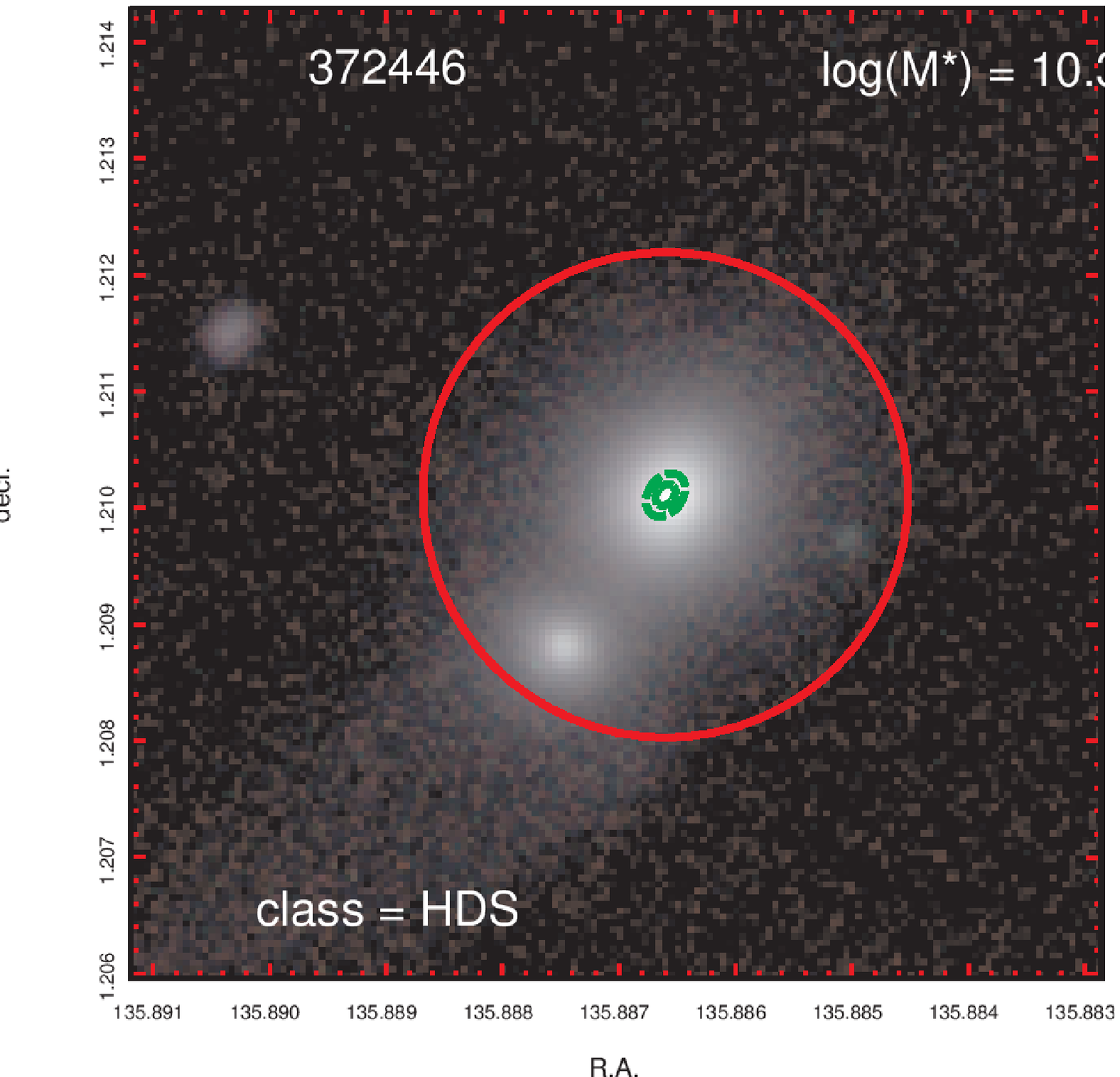}
\includegraphics[width=.55\textwidth]{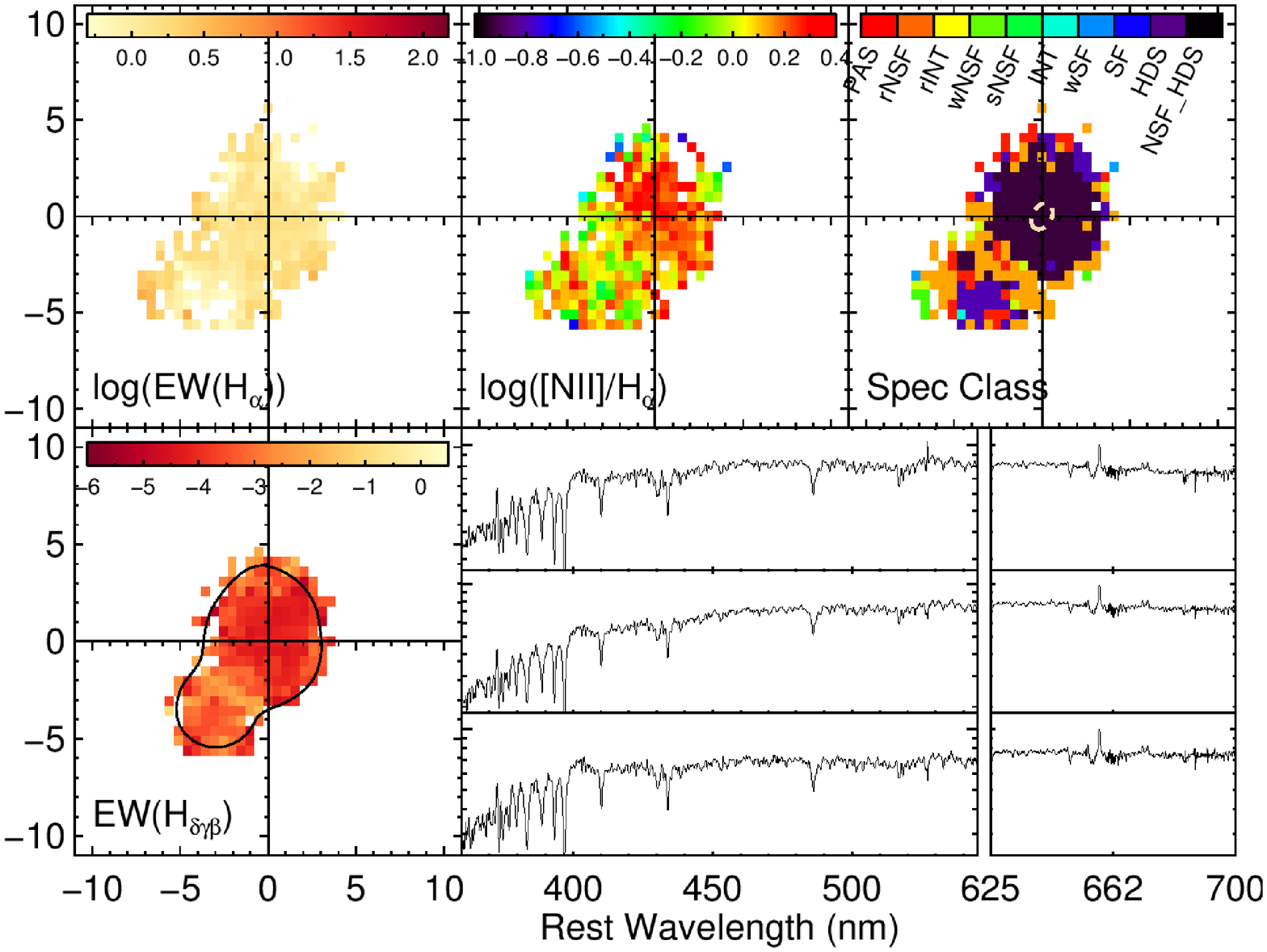}

\caption{}
\end{figure*}

{\it Facilities:} \facility{AAT (SAMI)}

\label{lastpage}

\end{document}